\documentclass[10pt,twocolumn,twoside,cleanfoot]{asme2ej}
\usepackage{amsmath}
\usepackage{amssymb}
\usepackage{color}

\usepackage{graphics,graphicx,amssymb}
\usepackage{subfigure}

\newcommand{\xvec}[1]{\mbox{\boldmath$#1$}}

\begin{document}

%\selectlanguage{english}

\newcommand{\blue}[1]{\textcolor[rgb]{0,0,1}{#1}}
\newcommand{\red}[1]{\textcolor[rgb]{1,0,0}{#1}}
\renewcommand{\blue}[1]{\textcolor[rgb]{0,0,1}{#1}}
\renewcommand{\red}[1]{\textcolor[rgb]{1,0,0}{#1}}
\newcommand{\green}[1]{\textcolor[rgb]{0.5,0.8,0.5}{#1}}
\newcommand{\purple}[1]{\textcolor[rgb]{0,0,1}{#1}}

\newcommand{\FIG}[1]{\purple{Fig.~\ref{#1}}}

\newcommand{\SEC}[1]{\blue{Sec.~\ref{#1}}}
\newcommand{\THEORY}{\red{Sec.~\ref{Theory}}}

%\title{On the nature and uses of three-dimensional chaotic advection}
%\title{On the nature and uses of three-dimensional Lagrangian transport and chaotic advection}
%\title{Three-dimensional chaotic advection and its application in industry and beyond}
%\title{Three-dimensional Lagrangian transport in industry and beyond}
%\title{Faces and applications of three-dimensional Lagrangian transport and chaotic advection}
%\title{Three-dimensional Lagrangian transport and chaotic advection in technology and beyond}
%\title{Three-dimensional Lagrangian transport and chaotic advection}
%\title{Lagrangian transport and chaotic advection in three-dimensional flows}
\title{Lagrangian transport and chaotic advection in three-dimensional laminar flows}

\author{Michel Speetjens
\affiliation{
%Eindhoven University of Technology\\
%Mechanical Engineering\\
Energy Technology \& Fluid Dynamics\\
Eindhoven University of Technology\\
PO Box 513, 5600 MB Eindhoven\\
The Netherlands\\
m.f.m.speetjens@tue.nl
}
}
\author{Guy Metcalfe
\affiliation{
School of Engineering\\
Swinburne University of Technology\\
Hawthorn VIC 3122\\
Australia\\
gmetcalfe@swin.edu.au
}	
}
\author{Murray Rudman
\affiliation{
%MAXIMA---Monash Academy for Cross \& Interdisciplinary Mathematical Applications\\
Mechanical \& Aerospace Engineering\\
Monash University\\
Clayton VIC 3800\\
Australia\\
murray.rudman@monash.edu
    }	
}

\maketitle

\begin{abstract}

{\it
Transport and mixing of scalar quantities in fluid flows is ubiquitous in industry and Nature. Turbulent flows promote efficient transport and mixing by their inherent randomness. Laminar flows lack such a natural mixing mechanism and efficient transport is far more challenging. However, laminar flow is essential to many problems and insight into its transport characteristics of great importance. Laminar transport, arguably, is best described by the Lagrangian fluid motion (``advection'') and the geometry, topology and coherence of fluid trajectories. Efficient laminar transport being equivalent to ``chaotic advection'' is a key finding of this approach.

The Lagrangian framework enables systematic analysis and design of laminar flows. However, the gap between scientific insights into Lagrangian transport and technological applications is formidable primarily for two reasons. First, many studies concern two-dimensional (2D) flows yet the real world is three dimensional (3D). Second, Lagrangian transport is typically investigated for idealised flows yet practical relevance requires studies on realistic 3D flows.

The present review aims to stimulate further development and utilisation of know-how on 3D Lagrangian transport and its dissemination to practice. To this end 3D practical flows are categorised into canonical problems. First, to expose the diversity of Lagrangian transport and create awareness of its broad relevance. Second, to enable knowledge transfer both within and between scientific disciplines. Third, to reconcile practical flows with fundamentals on Lagrangian transport and chaotic advection. This may be a first incentive to structurally integrate
the ``Lagrangian mindset'' into the analysis and design of 3D practical flows.
}

\end{abstract}

%\tableofcontents

%\clearpage
%\input{Introduction_arXiv.tex}

\section{Introduction}
\label{Intro}

The scope of this review is transport and mixing of scalar quantities such as additives, chemical species, heat and nutrients in realistic three-dimensional (3D) fluid flows under {\it laminar
flow conditions}. This flow regime sets in for a Reynolds number $Re = UL/\nu$ below the threshold of turbulence and is common to many systems and processes in
industry and Nature due to high fluid viscosities $\nu$, small length scales $L$ and/or low velocities $U$. Industrial examples are found abundantly in fluids processing and span a wide range of scales from conventional food or polymer processing \cite{Harnby1997,Todd2004,Thakur2003,Wein09,Erwin2011} to emerging technologies as e.g process intensification and micro-fluidics \cite{Stone2004,Squires2005,Laporte07,Kockmann2007,Becht07,Mills08,Kjeang2009,Wu2009,Harmsen2010,Lutze2010,Suh2010,Oh2012b,Ward2015,Karimi2016}.
Further technological applications include (at first glance) less obvious systems as Darcy representations of flow and transport
in porous media, relevant e.g. for {\it in situ} mining, enhanced oil recovery, geothermal energy extraction or groundwater remediation \cite{Brown2012,Chen2015,Seredkin2016,Guo2016,Cunningham2002,Goltz2012}, as well
as continuum descriptions of granular flows \cite{Ottino2000,Meier2007,Schlick2016}.
% \red{and magneto-hydrodynamic (MHD) approximations of plasmas.}

Scalar transport in laminar flows is also key to many systems beyond industry and technology. Consider, for example, transport of oxygen or pathogens
in physiological flows \cite{Schelin2009,Tsuda_inevitable_2011,Henry2012,Pittman2013,Ferrua2014,Koslover2017} and geophysical flow problems
as plate tectonics driven by mantle convection \cite{COLTICE2017120} and dispersion of nutrients or spreading of pollutants in
oceans \cite{Wiggins2005,Dijkstra2005,Samelson2013,Prants2017}.\footnote{Oceanographic flows typically are turbulent yet often admit approximation by Euler flows or spatio-temporal averaged flows obtained via e.g. RANS or LES. Such approximations are {\it deterministic} in the sense of being robustly reproducible and thus effectively behave as laminar flows. The underlying turbulent flows, on the other hand, are {\it stochastic} in being realisations from an ensemble of states and thus are inherently unpredictable.
}

%\blue{
%The primary focus of this review are industrial and technological applications. Scalar transport specifically under {\it laminar} flow conditions %namely has become (and continuous to be) increasingly relevant due to the emergence and ongoing expansion of technologies as process
% intensification and micro-fluidics. \blue{Moreover, this review will demonstrate that laminar transport is at the heart also of unexpected %applications and processes. Beyond industry: in particular physiological and cardiovascular flows also ... laminar ... } These fields are ... %outside the reach of traditional (thermal) fluids engineering, which concentrates primarily on macroscopic systems and turbulent flows. Moreover, %fluids engineering and system/process ... However, traditional (thermal) fluids engineering concentrates primarily on turbulent flows. Moreover, %fluids engineering and system/process design hitherto still rely largely upon empirical studies and practical experience and to a lesser extent %upon case-specific numerical and experimental investigations. Profound insight into the underlying fundamental transport mechanisms and its %integration in fluids engineering is imperative for further technological advancement.
%}

Scalar transport in flows generically involves an interplay of the physical motion of the fluid (``advection'') with
molecular mechanisms as diffusion and chemical reactivity. Advection is, given its pivotal role in scalar transport, the
focus of this review and for laminar flow conditions, arguably, best viewed from the Lagrangian perspective of the motion
of individual fluid parcels. The stochastic nature of {\it turbulent} flows gives rise to Brownian motion of fluid parcels
that typically admits expression as an effective diffusion (``turbulent diffusion'') \cite{Dimotakis2005}. This is absent
in {\it laminar} flows and, in consequence, the structure of Lagrangian fluid trajectories is essential to the transport
characteristics. Moreover, the connection between Eulerian velocity field and Lagrangian motion may be non-trivial and
counter-intuitive in that extremely tangled fluid trajectories can coexist with very simple velocity fields.
Hence transport studies in laminar flows should on account of these fundamental characteristics concentrate on the {\it Lagrangian} transport of fluid parcels.

The most striking disconnect between flow and Lagrangian transport is the intriguing phenomenon
of ``chaotic advection'' that, in a ``rough--and--ready'' definition, concerns the rapid deformation of
material fluid elements into highly ramified and filamented structures. This promotes efficient transport reminiscent
of the rapid dispersion by stochastic fluctuations in turbulent flows and thus renders chaotic advection the laminar counterpart
to turbulent mixing. Chaotic advection has first been demonstrated in the seminal study of Hassan Aref in the
early 1980s \cite{Aref1984} and can occur for very general (and often deceptively simple) flows \cite{Aref_development_2002,Aref2017}.
%\footnote{\red{Nonlinear (chaotic) dynamics of vortices in \cite{Aref1983} precursor to notion of chaotic advection of fluid parcels advanced in \cite{Aref1984}.}}

\begin{figure*}
\centering

\subfigure[Industrial batch mixer]{\includegraphics[width=0.6\columnwidth,height=12em]{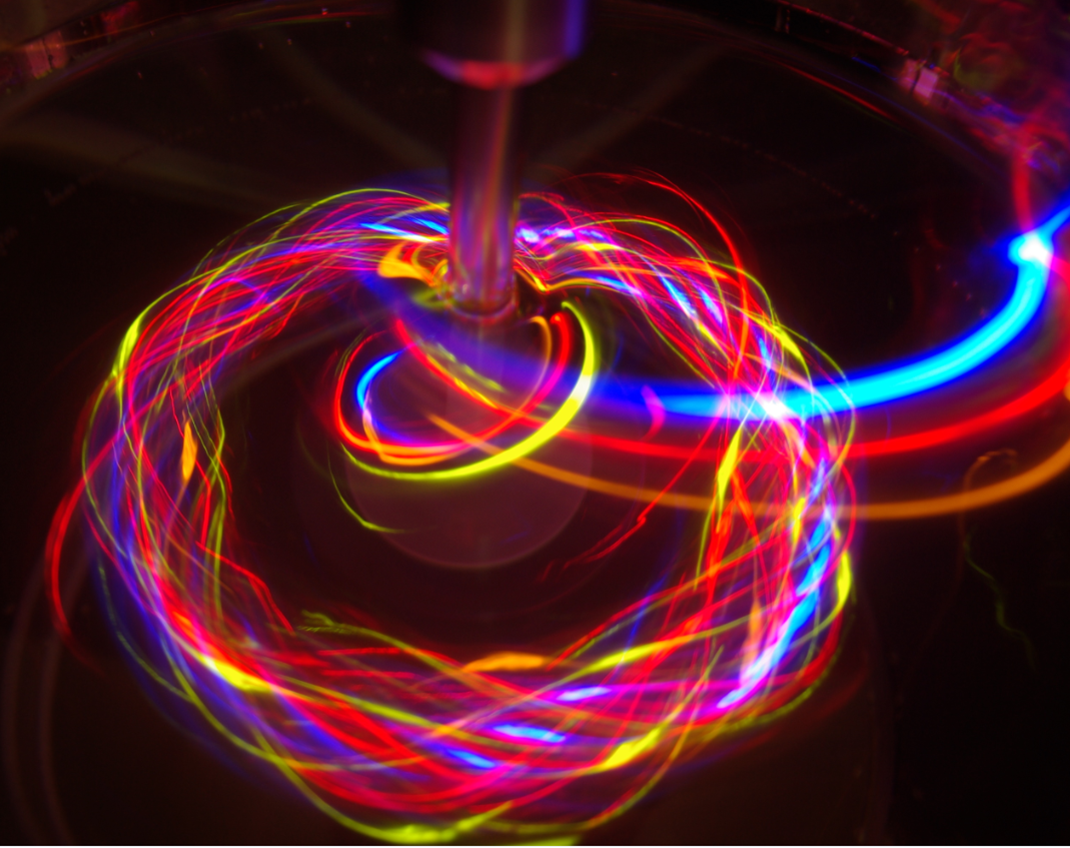}\label{fig:intro_examplesa}
}
\hspace*{6pt}
\subfigure[Aortic aneurysm]{\includegraphics[width=0.6\columnwidth,height=12em]{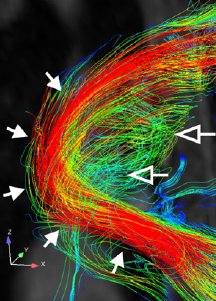}\label{fig:intro_examplesc}
}
\hspace*{6pt}
\subfigure[Coral polyps]{\includegraphics[width=0.6\columnwidth,height=12em]{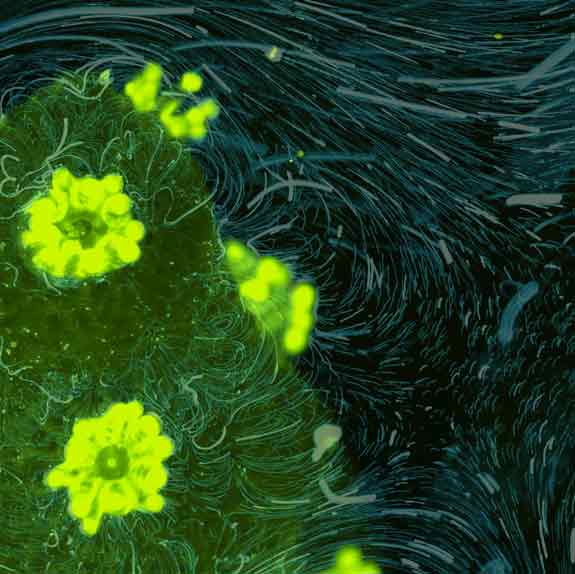}\label{fig:intro_examplesb}
}

\caption{Instances of Lagrangian (chaotic) transport in 3D flows: (a) impeller-driven (chaotic) trajectories in laboratory model for industrial batch mixers visualised by fluorescent tracers (reproduced from \cite{Wang_clustering_2014}); (b) vortical (chaotic)
trajectories in human aortic aneurysm (open arrows) driven by main aortic flow (closed arrows) visualised by {\it in vivo} phase-contrast magnetic-resonance imaging (reproduced from \cite{Markl_heart_2011}); (c) vortical (chaotic) trajectories in cilia-driven flow around (explants of) coral polyps visualised by fluorescent tracers (reproduced from \cite{Stocker2012}).}
\label{fig:intro_examples}
\end{figure*}

%
Consider as an illustration of the versatility of Lagrangian transport phenomena a few examples in completely different settings. \FIG{fig:intro_examplesa} gives 3D trajectories (visualised by fluorescent tracer particles) in the steady flow inside a cylindrical container stirred by an impeller \cite{Wang_clustering_2014}. The system serves as laboratory model for industrial batch mixers and the Lagrangian transport -- in particular the accomplishment of global 3D chaotic advection -- evidently is crucial to the functionality and performance of such devices. The disordered nature of the trajectories suggests chaos yet dedicated Lagrangian analysis is necessary to conclusively establish this. The web of trajectories, by virtue of continuity and mass conservation,
%
%\red{, though invisible to the untrained eye,}
namely contains hidden ``pathways'' and coherent
structures that geometrically guide the transport. Relevant here are in particular toroidal (i.e. donut-shaped) material surfaces that may emerge
around the impeller axis, which, if indeed occurring, act as barriers to chaotic trajectories and thus compromise the mixing performance.
Such structures cannot be inferred from the velocity field nor are directly visible, meaning they elude conventional analyses and engineering. Hence many mixer designs and operating conditions likely are sub-optimal.

%... may result in poor or sub-optimal performance if not accounted for in design ... %engineers unaware of this ...

A non-technological example is found in physiology. \FIG{fig:intro_examplesc} gives the 3D Lagrangian trajectories inside a human aortic aneurysm (balloon-like malformations of blood vessels) visualised by {\it in vivo} measurements using phase-contrast
cardio-vascular magnetic-resonance imaging \cite{Markl_heart_2011}. This reveals vortical structures (indicated by open arrows) akin to the batch mixer in \FIG{fig:intro_examplesa} driven by the main bloodstream (closed arrows) and thus suggests comparable (chaotic) dynamics due to a similar ``hidden organisation'' of the fluid trajectories into chaotic tangles and toroidal barriers. However, unlike mixers, chaotic advection has a potentially adverse impact by tending to entrap blood-borne platelets and substances implicated in thrombosis and atherosclerosis \cite{Schelin2009}.
%Hence, chaotic advection is in fact undesired here.
Hence Lagrangian transport is a key factor in these vascular diseases and investigation of the pathways and coherent structures embedded in the
fluid trajectories may shed more light on their origin and causes.

A biological counterpart exists in the cilia-driven flow at polyps in coral reefs visualised (for polyp explants) by fluorescent particles in \FIG{fig:intro_examplesb} \cite{Stocker2012,Shapiro2014,Shapiro2016}. The ``stirring'' by the cilia aims, similar to the batch mixers, at creating favourable transport conditions
near the coral surface for the benefit of photosynthesis and suppression of adverse effects as carbon fixation and pathogen invasion.
However, {\it what} conditions are in fact ``favourable'' (this must not necessarily be chaos; transport barriers as mentioned before may very well benefit
these processes) and {\it how} Nature accomplishes this remains unclear. Lagrangian transport analyses may again contribute to the understanding of these phenomena.

The examples in \FIG{fig:intro_examples} give a first flavour of the richness and universality of Lagrangian transport and demonstrate that seemingly different systems have much more in common than meets the eye. The fluid trajectories are namely organised into elementary ``building blocks'' that form the network of transport pathways and barriers according to general ``rules'' depending only secondarily on the particulars of the flow (forcing). This implies unifying structures that transcend specific flow configurations (as well as scientific and engineering disciplines) and thus enable
a universal strategy for analysis of Lagrangian transport by the geometry, topology and coherence of fluid trajectories.
%
%Moreover, this enables generalisation ... translation of insights between systems.

%\red{Above: Lagrangian transport necessary for insight; moreover, enables unified approach. Hence transport in laminar flows must explicitly %concentrate on the Lagrangian transport of fluid parcels. ... }

The relevance and usefulness of a (cross-disciplinary) unified Lagrangian approach towards laminar transport is evident from the above. Moreover, efficient laminar transport being synonymous to chaotic advection is in itself recognised in the fluid-dynamics community \cite{Aref2017}.
%
% and the Lagrangian framework in principle offers a way to systematically analyse, optimise and design (industrial) fluid systems involving %chaotic as well as non-chaotic transport.
%
However, the gap between scientific research and insights into (chaotic) Lagrangian transport and technological applications is still formidable primarily for two reasons.
{\it First}, scientific studies on Lagrangian transport to date mainly concern two-dimensional (2D) flows. The real world is three dimensional (3D), on the other hand, yet 3D Lagrangian transport remains elusive. The additional spatial degree of freedom greatly increases the dynamical richness and geometric complexity and a comprehensive theoretical and conceptual framework, in contrast with the basically complete picture of 2D Lagrangian transport, is therefore still non-existent in 3D \cite{Wiggins2010,Aref2017}.
{\it Second}, Lagrangian transport is typically investigated via theoretical and computational studies and often concerns idealised flow situations that are difficult (or even impossible) to create in laboratory experiments. Dedicated experimental studies on Lagrangian transport, crucial for (conclusively) establishing physical meaningfulness and practical relevance of results, remain scarce, though.
The only systematic technological application to date exists in a subclass of industrial and micro-fluidic mixers that admit reconciliation with 2D flows and the associated theory \cite{Ottino2004}. However, dissemination even of the proven principle of chaotic advection for such devices to the practicing engineering community remains a challenge.

The growing importance and urgency to close the gap between fundamentals and applications motivates this review. Its principal aim is to stimulate further utilisation and development of know-how on 3D Lagrangian transport for
technological applications and its dissemination to practitioners in industry and beyond. To this end 3D practical flows to which (non-)chaotic
Lagrangian transport is essential are categorised into canonical problems so as to (i) identify analogies and
similarities, (ii) establish connections with configurations in fundamental 3D Lagrangian studies, (iii) outline a unified Lagrangian framework for 3D transport studies and (iv) isolate challenges specifically regarding applications.
The focus will be primarily on flows and applications in industry and technology with excursions into life
sciences and on occasion beyond. Furthermore, the canonical flow problems will be exemplified and represented by
experimental(ly-realisable) cases to ensure practical relevance and robustness of phenomena.

The Lagrangian framework and the concept of chaotic advection have, expanding on Aref's pioneering work \cite{Aref1984}, been developed in the 1980s and 1990s mainly for 2D time-periodic flows (as approximations for inline industrial mixers) using dynamical-systems methods as Hamiltonian mechanics and vector-field topology \cite{Salmon1988,Ottino_Kinematics_1989,Ottino1990,Meiss1992,Wiggins1992,Arnold1992,Morrison1998}.
Two important extensions of the Lagrangian framework in the early 2000s include granular-media flows that admit a continuum description \cite{Ottino2000,Meier2007}
and deterministic descriptions of large-scale oceanographic flows \cite{Wiggins2005,Dijkstra2005}.
Recent and ongoing efforts focus on generalisations of the Lagrangian approach to aperiodic and finite-time flows \cite{Mezic2013,Haller2015} and the interaction
between (chaotic) advection and other transport physics (e.g. diffusion, reaction, propulsion, particle inertia) \cite{Metcalfe_beyond_2012,Aref2017,Villermaux2018}.
%
%\red{... mostly 2D and fundamental with weak link to applications ... However, many studies on Lagrangian (chaotic) transport continue to ... 2D flows ... state-of-the-art on chaotic
%advection and identification of 3D as ``new frontier '' ... \cite{Aref2017}. ... Fundamental/scientific research on fundamentals of chaotic advection including 3D \cite{Aref2017}.}
%
%The present review expands on this body of work by linking developments and progress ... to 3D practical flows and is organised as follows.\\\\

The present review expands on the above body of work by explicitly positioning this in the context of 3D practical flows and is organised as follows. \SEC{LagrangianView}
introduces the general concepts and methods of Lagrangian transport studies. \SEC{Categorisation} categorises 3D practical flows into canonical problems in order
to, first, identify analogies and similarities and, second, establish connections with fundamental 3D Lagrangian studies. \SEC{Theory} outlines a unified Lagrangian
framework for 3D transport studies by reconciling the flow categories and canonical problems with theory and fundamentals on Lagrangian transport.
\SEC{Commercialisation} investigates the degree of dissemination and technological utilisation of the concept of chaotic advection via a survey on relevant patents and commercial applications. Concluding
remarks including an overview of challenges exposed by this review are in \SEC{Conclusions}.

%\clearpage
%\section{\red{Intro -- Guy}}
%\input{Introduction_OriginalVersionGuy.tex}

%\input{LagrangianApproach_arXiv.tex}

%\section{The Lagrangian perspective on transport and mixing}
\section{Lagrangian approach to transport and mixing}
\label{LagrangianView}

%\subsection{General principles}
%\label{LagrangianView1}

\begin{figure}
\centering
\includegraphics[width=\columnwidth,height=0.4\columnwidth]{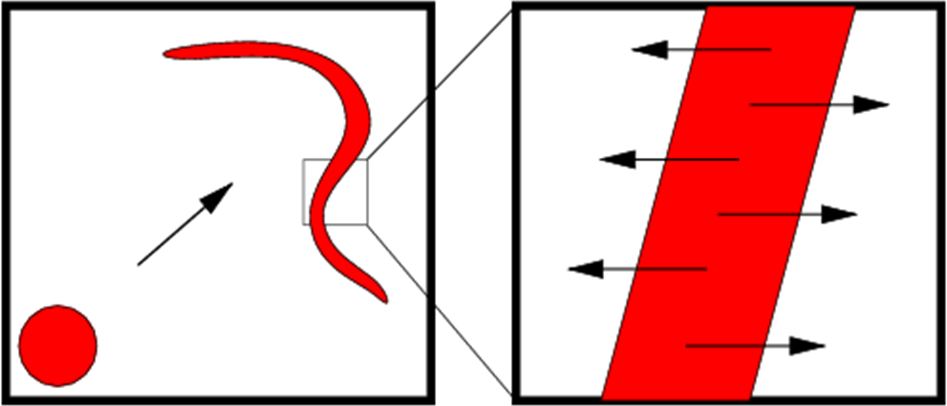}

\setlength{\unitlength}{\columnwidth}
\begin{picture}(1,0.05)
\put(0.,0.0){\footnotesize (a) Advection}
\put(0.5,0.0){\footnotesize (b) Diffusion}
\end{picture}

\caption{Transport of scalar quantity $C$ (red) in fluid flows by interplay of advection of material region (left) and diffusion across its interface (right). Arrows indicate transport by respective mechanisms.}
\label{SketchAdvDif}
\end{figure}
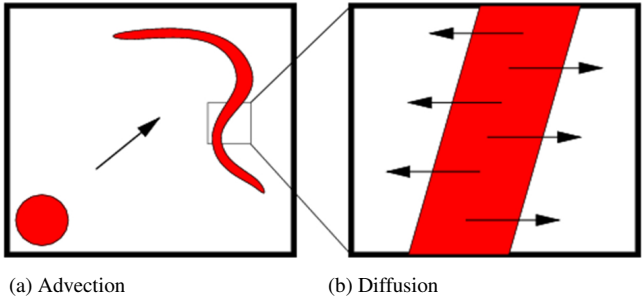

In the absence of chemical reactions, transport of a scalar quantity $C$ in fluid flows occurs by an interplay of two physical
mechanisms: advection by the fluid motion and molecular diffusion along concentration gradients (\FIG{SketchAdvDif}).\footnote{Advection and diffusion of heat is usually denoted ``convection'' and ``conduction'', respectively.}
%
%\blue{Advection plays a twofold role in this process by (i) global distribution and (ii) creation of steep gradients ... ; diffusion ... }
%
The corresponding Eulerian evolution is governed by the advection-diffusion equation
\begin{eqnarray}
\frac{dC}{dt} = \frac{\partial C}{\partial t} + \vec{u}\cdot\vec{\nabla} = \frac{1}{Pe}\vec{\nabla}^2 C,\quad
C(\vec{x},0) = C_0(\vec{x}),
\label{ADE}
\end{eqnarray}
with $\vec{u}$ the velocity field and $Pe = U L/\alpha$ the well-known P\'{e}clet number. Here $U$ and $L$ are characteristic velocity and
length scales, respectively, and $\alpha$ is the scalar diffusivity. The present review concerns systems with $Pe\gg 1$, which holds
for many practical systems, meaning that scalar transport is dominated by advection and \eqref{ADE} effectively reduces
to $dC/dt = 0$. Thus scalar $C$ is passively advected by fluid parcels, i.e. $C(\vec{x}(t)) = C(\vec{x}_0)$,
with $\vec{x}(t)$ the current parcel position governed by the
%Lagrangian equations of motion (``kinematic equation'')
kinematic equation
%and corresponding formal solution
%
\begin{eqnarray}
\frac{d\vec{x}}{dt} = \vec{u}\left(\vec{x}(t),t\right)
%,\quad
\quad\Rightarrow\quad
\vec{x}(t) = \vec{\Phi}_t(\vec{x}_0),
\label{KinEq}
\end{eqnarray}
and flow $\vec{\Phi}_t$ its formal solution describing the Lagrangian motion of a parcel released at $\vec{x}(0) = \vec{x}_0$.
Hence, $dC/dt = 0$ and $d\vec{x}/dt = \vec{u}$ are the equivalent Eulerian and Lagrangian representations of scalar advection. The fluid is
assumed incompressible, implying solenoidal flow ($\vec{\nabla}\cdot\vec{u}=0$), which has fundamental ramifications for Lagrangian transport.

\begin{figure}
\centering

\subfigure[First half of each period $p$.]{\includegraphics[width=0.48\columnwidth,height=0.42\columnwidth]{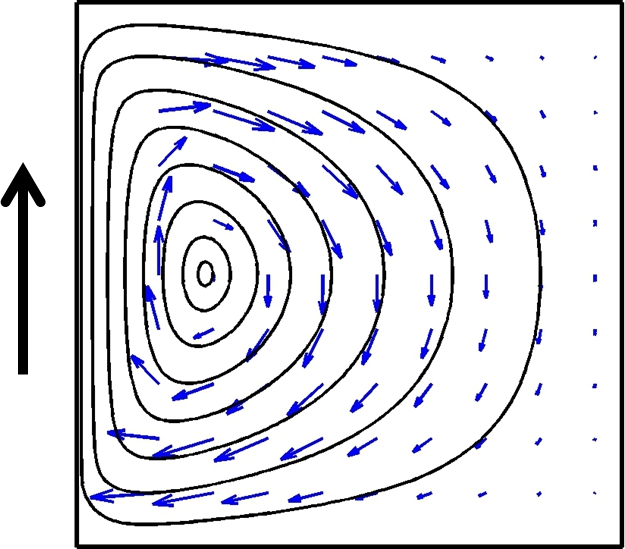}\label{LidDrivenCavity_a}}
\hspace{3pt}
\subfigure[Second half of each period $p$.]{\includegraphics[width=0.48\columnwidth,height=0.42\columnwidth]{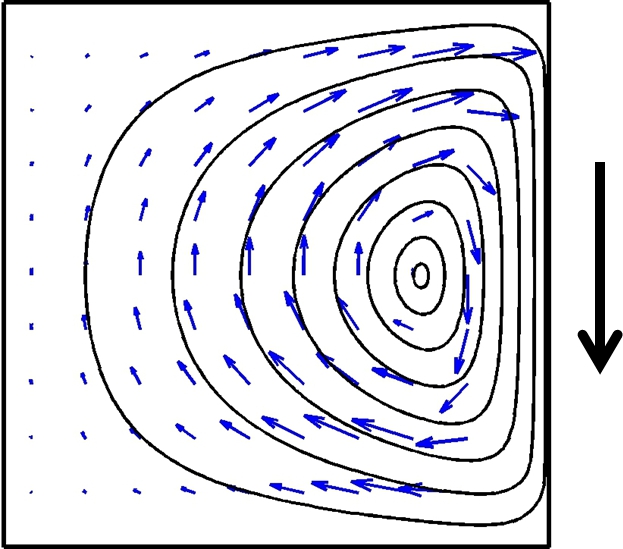}\label{LidDrivenCavity_b}}

\caption{Time-periodic 2D lid-driven cavity flow $\vec{u}(\vec{x},t+pT) = \vec{u}(\vec{x},t)$ due to alternate steady translation of left and right
wall during first ($0\leq t'\leq T/2$) and second ($T/2 \leq t'\leq T$) half, respectively, of each period $p$ with time interval $0\leq t'\leq T$.
Heavy black arrows indicate translation direction; blue arrows and closed curves indicate velocity vectors and streamlines, respectively.
}
\label{LidDrivenCavity}
\end{figure}

The exposition hereafter adopts the Lagrangian perspective on scalar transport with kinematic equation \eqref{KinEq} as its cornerstone.
The latter defines a generically nonlinear dynamical system and enables investigation of advective transport by the geometry and topology of the Lagrangian fluid trajectories, i.e. the ``Lagrangian flow topology'', which is composed of elementary structures denoted ``Lagrangian coherent structures'' (LCSs) \cite{Ottino_Kinematics_1989,Haller2015,Aref2017}.\footnote{Term ``LCS'' originally denotes specific classes of coherent structures
in aperiodic flows \cite{Haller2015}. Here it denotes coherent structures in general.
}
The nature of LCSs is described and illustrated by the time-periodic flow
%$\vec{u}(\vec{x},t) = \vec{u}(\vec{x},t+T)$
\begin{eqnarray}
%\vec{u}(\vec{x},t) = \vec{u}(\vec{x},t+pT),
\vec{u}(\vec{x},t+pT) = \vec{u}(\vec{x},t),
\label{TimePeriodic2D}
\end{eqnarray}
with $T$ the period time and $p$ the period, in a 2D cavity driven by the alternate steady motion of the left and right walls. \FIG{LidDrivenCavity} gives the translation direction of each wall and the corresponding
velocity vectors and streamline patterns obtained by numerical solution of the non-dimensional Navier-Stokes equations. Note that the steady flow during the second half of each period is a reorientation of that of the first half. This composition of time-periodic flows from reorientations of a steady base flow is common practice both in scientific studies on mixing and industrial applications.

\begin{figure}
\centering

\subfigure[Initial state]{\includegraphics[width=0.4\columnwidth,height=0.36\columnwidth]{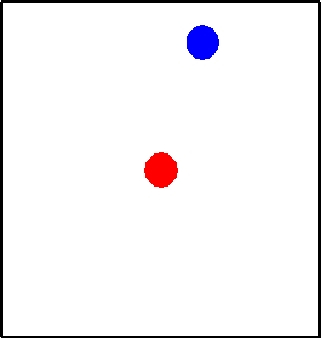}}
\hspace{12pt}
\subfigure[After 1 period]{\includegraphics[width=0.4\columnwidth,height=0.36\columnwidth]{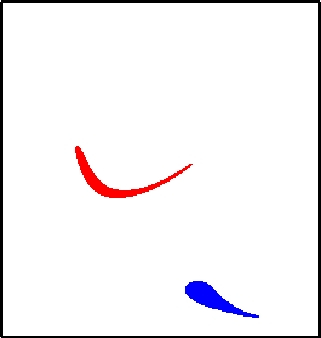}}

\subfigure[After 3 periods]{\includegraphics[width=0.4\columnwidth,height=0.36\columnwidth]{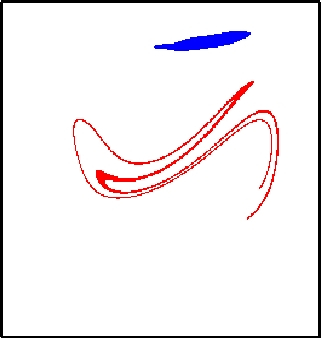}}
\hspace{12pt}
\subfigure[After 6 periods]{\includegraphics[width=0.4\columnwidth,height=0.36\columnwidth]{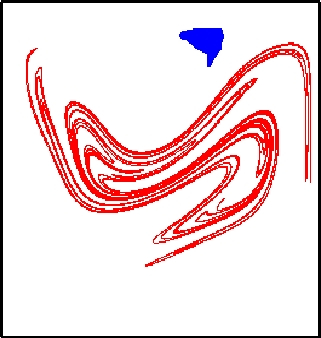}}

\caption{Non-chaotic versus chaotic advection illustrated by evolution of blue and red material elements, respectively, in 2D time-periodic lid-driven cavity.}
\label{IllustrateChaos}
\end{figure}

The scalar advection illustrated in \FIG{SketchAdvDif}(a) may take place in two fundamentally different ways: non-chaotic advection versus chaotic
advection. This is demonstrated in \FIG{IllustrateChaos} for the advection of blue and red material elements in the 2D lid-driven cavity, revealing a dramatic difference in behaviour. The blue element undergoes a moderate shear-like deformation that yields (at most) linear stretching of the interface. This signifies {\it non-chaotic} advection. The red element, on the other hand, exhibits strong deformation due to repeated stretching and folding, resulting in exponential elongation of the interface. This is the hallmark of {\it chaotic} advection \cite{Ottino_Kinematics_1989}.

Chaotic advection promotes efficient mixing in laminar flows by (i) rapid global scalar distribution (\FIG{IllustrateChaos}) and (ii)
creation of large ``working areas'' and steep gradients for diffusion (\FIG{SketchAdvDif}(b))
%
%\blue{due to steep concentration gradients induced}
by exponential interface stretching.
Hence, it is the laminar counterpart to efficient mixing
by turbulent diffusion in turbulent flows. However, unlike turbulence, chaotic advection is not a natural mixing mechanism of the
flow and in fact can be challenging to accomplish. This is a direct consequence of the fundamental disconnect between the Eulerian velocity
field and the Lagrangian transport in laminar flows, allowing complex fluid trajectories to coexist with simple flow fields.
Compare to this end the simple step-wise flows in \FIG{LidDrivenCavity} with the intricate Lagrangian advection in \FIG{IllustrateChaos}. Thus the Lagrangian approach is essential to describe and understand scalar transport in laminar flows.

The Lagrangian flow topology underlying scalar advection in time-periodic flows (as the lid-driven cavity) admits visualisation by so-called ``stroboscopic maps'' of tracer particles. Such maps visualise the motion of a tracer released at $\vec{x}_0$ via the sequence of positions at the end of each period, i.e.
\begin{eqnarray}
\mathcal{S}(\vec{x}_0)=\{\vec{x}_0,\vec{\Phi}_T(\vec{x}_0),\vec{\Phi}_T^2(\vec{x}_0),\dots\},
\label{StroboMap}
\end{eqnarray}
with
\begin{eqnarray}
\vec{x}_{p} = \vec{\Phi}_T(\vec{x}_{p-1}) = \vec{\Phi}_T^p(\vec{x}_0),
\label{StroboMap2}
\end{eqnarray}
the period-wise mapping, as if illuminated by a stroboscope. This is a common technique also often referred to as ``Poincar\'{e} sectioning'' \cite{Aref2017}.

\begin{figure}
\centering

%\begin{tabular}{cc}
\subfigure[Stroboscopic map]{\includegraphics[width=0.47\columnwidth,height=0.46\columnwidth]{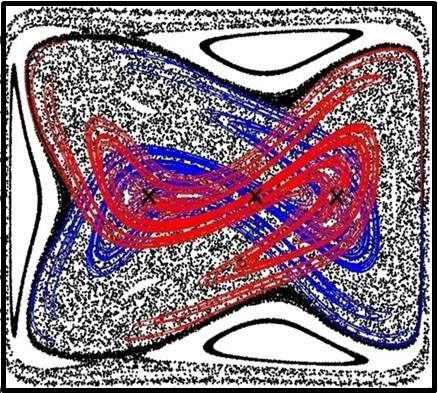}\label{OrderWithinChaosa}}
\hspace*{3pt}
\subfigure[Initial material element]{\includegraphics[width=0.47\columnwidth,height=0.46\columnwidth]{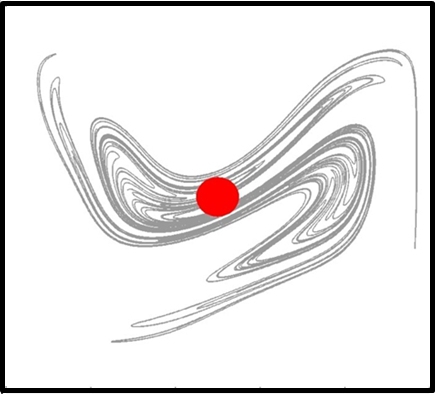}\label{OrderWithinChaosb}}
\subfigure[After 2 periods]{\includegraphics[width=0.47\columnwidth,height=0.46\columnwidth]{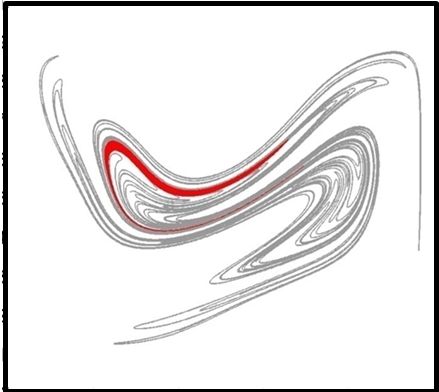}\label{OrderWithinChaosc}}
\hspace*{3pt}
\subfigure[After 6 periods]{\includegraphics[width=0.47\columnwidth,height=0.46\columnwidth]{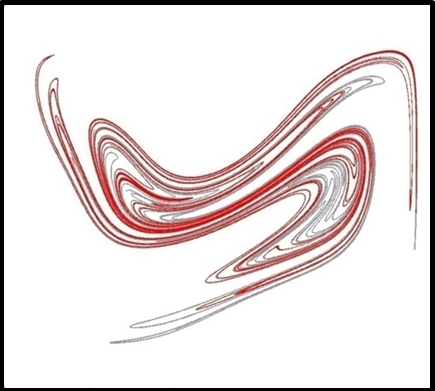}\label{OrderWithinChaosd}}

\caption{Typical Lagrangian flow topology and transport in 2D time-periodic flows demonstrated by lid-driven cavity: (a) stroboscopic map
including stable (blue) and unstable (red) manifolds of chaotic sea; (b-c) convergence of material element (red) on asymptotic mixing pattern
demarcated by unstable manifold (grey).}
\label{OrderWithinChaos}
\end{figure}

\FIG{OrderWithinChaosa} gives the Lagrangian flow topology visualised by the combined stroboscopic maps (black dots) of a number of tracer particles released on the vertical line $x=0.5$. This exposes a first kind of LCSs, viz. islands (white circular patches), embedded in ``chaotic seas'' (densely filled regions). Such islands systematically return to their initial position after a given number of periods and act as transport barriers by entrapping material regions during their excursion through the flow domain. The blue element in \FIG{IllustrateChaos} is trapped in the island near the top wall that cyclically returns to its initial position after 3 periods through intermediate positions at the island near the bottom and left walls after the first and second periods, respectively.
%
%\blue{Important to note is that islands are a direct consequence of the
%solenoidality of the velocity field ($\vec{\nabla}\cdot\vec{u}=0$); such entities are generically absent in compressible flows.
%}
The centres of the islands are periodic points, i.e. material points cyclically returning to the initial position after $p\geq 1$
periods following
%
%$\vec{x}_0 = \vec{\Phi}^p_T(\vec{x}_0)$,
%
\begin{eqnarray}
\vec{x}_0 = \vec{\Phi}^p_T(\vec{x}_0),
\label{PeriodicPoint}
\end{eqnarray}
commonly denoted {\it elliptic points} (\SEC{CriticalPoints}) \cite{Ottino_Kinematics_1989}. Periodic points for $p>1$ always emerge as sets of $p$ material points
\begin{eqnarray}
\mathcal{X}_p(\vec{x}_0) =
%\chi_p(\vec{x}_0) =
\{\vec{x}_0,\vec{\Phi}_T(\vec{x}_0),\dots,\vec{\Phi}^{p-1}_T(\vec{x}_0)\},
\label{PeriodicPoint2}
\end{eqnarray}
through which each element of $\mathcal{X}_p$ cyclically progresses in the course of time.

A second kind of LCSs in \FIG{OrderWithinChaosa} concerns structures termed ``manifolds'', which are associated with another type of periodic points known as {\it hyperbolic points} (\SEC{CriticalPoints}) \cite{Ottino_Kinematics_1989}. At each hyperbolic point (marked with a cross in \FIG{OrderWithinChaosa}), a stable (blue) and unstable (red) manifold exists that wind their way through the chaotic sea.
%
%concerns the {\it stable} (blue) and {\it unstable} (red) manifolds in the chaotic sea associated with periodic points \eqref{PeriodicPoint}
%of the {\it hyperbolic} type located at the crosses.
%
(The type of a periodic point is determined by the local deformation characteristics; \SEC{CriticalPoints}.)
These manifolds are ``special'' material curves that delineate the principal transport directions in this sea and thus dictate the chaotic advection \cite{Aref2017}. The red element in \FIG{IllustrateChaos} coincides with
the chaotic sea and its chaotic transport -- characterised by the repeated stretching and folding -- is determined by the unstable
manifold. \purple{Figs.~\ref{OrderWithinChaos}(b-d)} demonstrate this by the rapid convergence of the red element on the unstable
manifold (coloured in grey); essentially the same convergence occurs for {\it any} material element released in the chaotic sea, exposing the unstable manifold as its fundamental mixing pattern (or ``mixing template'' \cite{Aref2017}).\footnote{Refer to \blue{www.youtube.com/watch?v=B3dwryNgPXY} for
an experimental demonstration of chaotic advection and the emergence of a mixing pattern due to the unstable manifold.}

Solenoidal 2D velocity fields ($\vec{\nabla}\cdot\vec{u}=0$) exclusively admit elliptic/hyperbolic points and associated
islands/manifolds \cite{Ottino_Kinematics_1989}. Hence these LCS are the fundamental ``building blocks'' of Lagrangian flow
topologies of 2D incompressible flows.
%
%Moreover, solenoidality imposes a particular structure on kinematic equation \eqref{KinEq} that makes unsteadiness
% a necessary (yet not sufficient) condition for chaotic advection in 2D flows (\SEC{Hamiltonian2D}).
%
%Moreover, solenoidality renders kinematic equation \eqref{KinEq} a {\it Hamiltonian system} and thus makes unsteadiness
%a necessary (yet not sufficient) condition for chaotic advection in 2D flows.
%
Moreover, solenoidality in 2D imposes a {\it Hamiltonian structure} on kinematic equation \eqref{KinEq} that makes
unsteadiness a necessary (yet not sufficient) condition for chaotic advection in 2D flows. This Hamiltonian
structure and its ramifications for the dynamics are detailed in \SEC{Hamiltonian2D}.
Time-periodic 2D flows -- here represented by the lid-driven cavity -- meet this requirement and their Lagrangian flow
topologies therefore typically consist of arrangements of islands and chaotic seas as in \FIG{OrderWithinChaosa} \cite{Aref2017}.

The Lagrangian flow topology of 3D flows (both steady and unsteady) admits an essentially similar decomposition into distinct LCSs yet
encompasses a larger set of such ``building blocks'' and corresponding interaction scenarios.
%
% The principal difference with 2D flows exists in a greater set of LCSs and, in consequence, a far richer topology and
%dynamics.
%
Furthermore, the beforementioned Hamiltonian structure of kinematic equation \eqref{KinEq} is retained only for certain conditions (refer to \SEC{Hamiltonian3Dsteady}, \SEC{Hamiltonian3Dunsteady} and \SEC{Towards3D}).
Thus 3D flows in general exhibit much greater topological complexity and, in consequence, far richer dynamics than their
2D counterparts.
%
%3D flow topologies nonetheless also typically comprise of both LCSs that obstruct and LCSs that promote global transport, meaning that (3D) laminar flows generically are inefficient mixing flows.
%
LCSs enable systematic and rigorous exposure of this topology and associated dynamics and therefore are indispensable for
in-depth transport studies and characterisations. Furthermore, LCSs, given their robustness, can (in principle) be instrumentalised for specific purposes. Efficient mixing requires systematic ``destruction'' of transport barriers as the islands in 2D; containment of polluted groundwater, on the other hand, requires establishment of such barriers in the subsurface flow. This can (basically) be realised by appropriate flow
conditions.
%under the premise of sufficient insight into the LCSs and their response to the flow forcing.
\SEC{Categorisation} below reconciles the Lagrangian framework and the utilisation of LCSs as ``tools'' for
analysis, characterisation and engineering 3D practical flows to which Lagrangian transport is essential.

%LCSs in 3D nonetheless also either obstruct or promote global transport and, given their robustness, can (in principle) be
%instrumentalised for specific purposes. Efficient mixing requires systematic ``destruction'' of transport barriers (as e.g.
%islands in 2D); containment of polluted groundwater, on the other hand, requires establishment of such barriers in the
%subsurface flow. This can (basically) be realised by appropriate flow conditions.
%

%\clearpage

%\section{A tour of experiments on 3D Lagrangian transport}
%\section{Lagrangian transport in 3D practical flows}
\section{Lagrangian transport in 3D practical flows}
\label{Categorisation}

\subsection{Introduction}
\label{Intro2}

The examples in \FIG{fig:intro_examples} clearly demonstrate the diversity and relevance of 3D Lagrangian transport yet applications of concepts and
insights remain limited due to the still formidable gap between theory and practice for reasons mentioned in \SEC{Intro}. The only systematic technological application is found in the subclass of industrial and micro-fluidic mixers
that, for ``proper'' conditions, admit reconciliation with 2D flows and the associated theory (\SEC{Ducts}).
%
% (which often admit representation by a 2D unsteady flow and thus also connect well with existing theory; \SEC{Ducts}.)
%
Julio Ottino and co-workers laid the groundwork for mixing technology based on Lagrangian chaos in such 2D analogons in the late 1980s (resulting in his classical
textbook \cite{Ottino_Kinematics_1989}) and the framework thus developed has since found frequent application in engineering sciences \cite{Hobbs1997,Szalai2003,Ottino2004,Stremler2004,Stone2004,Paul2004,Squires2005,Metcalfe_ram_2006,Zumbrunnen2006,Kockmann2007,Wu2009,Tiwari2009,Suh2010,Zumbrunnen2011,Meijer2012,Hessel2012,Ghanem2014,Cullen2015}.
The Ottino group extended this framework to granular flows that behave as a continuum \cite{Ottino2000,Meier2007,Schlick2016}.
%
%However, such studies mainly concern 2D (simplifications of) mixing flows yet further (technological) development increasingly calls for more
%faithful and essentially 3D approaches. Moreover, the dissemination to the practicing engineering community remains a major challenge.
%

Employment of the Lagrangian framework for transport studies beyond industry and technology occurs primarily in the geophysical
sciences. Large-scale oceanographic and atmospheric flows, due to their quasi-2D nature, namely also connect well with existing
2D theory \cite{Wiggins2005,Dijkstra2005,Samelson2013,Prants2017}. Development of Lagrangian methods specifically for
aperiodic flows is in fact strongly motivated by such systems \cite{Mezic2010,Haller2015,Budisic2016}.

Dissemination of the above chaos-based mixing technology to the practicing engineering community poses a major challenge in its own right.
%
%\blue{in part due to the above theory--practice gap yet also due to the generally conservative nature of industry}.
%
Many recent handbooks on fluids processing, despite overwhelming evidence of its crucial role in (at least) inline mixers, mention chaotic advection only perfunctorily and continue to rely on properties of the Eulerian velocity field\footnote{This may lead to erroneous results. A common misconception is e.g. that vortices imply efficient
mixing. Reconsider to this end the lid-driven cavity. The flow consists of two alternating
vortices (\FIG{LidDrivenCavity}) yet mixing is inefficient (\FIG{IllustrateChaos}) due to islands in the Lagrangian flow topology (\FIG{OrderWithinChaosa}).} or standard metrics as the ``coefficient of variation'' (CoV) \cite{Kraume2010,Dickey2010,Erwin2011,Freund2011}.\footnote{The CoV is in essence the ``intensity of segregation'' introduced by Danckwerts in the early 1950s \cite{Danckwerts1952}. An important shortcoming is arbitrariness of the cell size to determine the scalar concentration, which may cause non-physical effects
as artificial diffusion and thus yield misleading results \cite{Alberini2014}. More sophisticated measures seek to overcome this yet their practical utilisation is scarce to non-existent \cite{Aref2017}.}
%
%Standard practice in industry is design and optimisation via parametric studies using the CoV.
%
Two leading manufacturers of static mixers, viz. Sulzer AG (Winterthur, Switzerland) and Chemineer (Dayton, USA), still heavily lean on the CoV to
design and optimise their Sulzer SMX and Kenics KMX-V mixers, respectively.\footnote{Refer to brochures ``Mixing and Reaction Technology'' ({\tt sulzer.com}) and ``Kenics KMX-V Static Mixer'' ({\tt chemineer.com}).}
Similarly, the dedicated Sulzer CFD tool for laminar mixing adopts the CoV as mixing measure and functionalities for mixing diagnostics (other than visual inspection of particle paths) are absent altogether in the popular commercial CFD package {\tt ANSYS-FLUENT} \cite{FLUENTmanual,Moser2006}.

%Furthermore, commercial CFD packages as {\tt ANSYS-FLUENT} and {\tt COMSOL} offer no dedicated mixing modules incorporating Lagrangian %capabilities as e.g. Poincar\'{e} sectioning and also the Sulzer CFD tool for laminar mixing flows still utilises the CoV as principal metric %\cite{Moser2006}. Hence, many industrial mixers very likely are sub-optimal designs.

Development and design by Lagrangian concepts enables technological leaps, though.
Consider as a case in point the Rotated Arc Mixer (RAM) to be introduced in \SEC{Ducts}. The RAM is to date the only commercially-available industrial mixer that from
its inception is designed entirely on the basis of scientific insights into Lagrangian transport and the notion of chaotic advection (\SEC{RAM}). This sound scientific basis
allows it to substantially outperform conventional inline mixers both in energy consumption and mixing quality (\SEC{RAM}). A survey of such ``commercialisation'' of chaos-based mixing technology is given in \SEC{Commercialisation}. This includes the above Kenics mixer, which has
been developed well before the advent of the concept of chaotic advection and thus is an indirect ``commercialisation'' of chaos (\SEC{kenics}).

Application of Lagrangian concepts has similar potential for technological development in other (industrial) fields as well as in non-technological
disciplines as life sciences. The physiological and biological examples in \FIG{fig:intro_examples} clearly highlight the central role of Lagrangian
transport in such processes and the inherent need for LCS-based analyses to expose the underlying mechanisms. Said fields and disciplines may thus
greatly benefit from a Lagrangian ansatz.
%
% and the present review aims to stimulate ...\\\\
%

One of the key aims of this review is to lay the groundwork for describing and analysing transport and mixing in 3D
practical flows by the Lagrangian ansatz. To this end, the following (heuristic) categorisation of such flows into canonical problems is adopted:
\begin{itemize}
%\item periodic duct flows (\SEC{Ducts});
\item[$\bullet$] flows in ducts (\SEC{Ducts});
%\item flows in closed containers (\SEC{Containers});
%\item flows in closed domains (\SEC{Containers});
\item[$\bullet$] flows in containers (\SEC{Containers});
\item[$\bullet$] flows in drops (\SEC{Drops});
\item[$\bullet$] flows in webs (\SEC{Webs}).
\end{itemize}
%
%This hinges on fundamental similarities in configurations and transport characteristics that unite (seemingly different) practical flows and
%thus establishes an important bridge between theoreticians/physicists and engineers/practitioners. This facilitates
%application and dissemination of scientific insights to practice (see e.g. the RAM). Conversely, this enables systematic identification of theoretical/fundamental
%challenges specifically regarding applications. Moreover, reduction to canonical problems may promote cross-fertilisation and cross-disciplinarity.
%The batch mixer and aneurysm in \FIG{fig:intro_examples} e.g. both concern cases of Lagrangian transport in closed domains (\SEC{Containers}) and transport
%within blood vessels is similar to that in inline (micro-)mixers (\SEC{Ducts}). Such analogies may enable application of mixing technology to cardiovascular
%research and treatment.
%
This hinges on the fact that analogies and similarities exist between seemingly different problems and thus provides a framework
that enables (i) connection with configurations in fundamental studies on 3D Lagrangian transport, (ii) development of generic
strategies and methods utilising Lagrangian concepts and (iii) isolation of challenges specifically regarding applications. Moreover,
reduction to canonical problems promotes cross-fertilisation and multi-disciplinarity. The batch mixer and aneurysm in \FIG{fig:intro_examples}
e.g. both concern transport in containers (\SEC{Containers}) and transport within blood vessels is similar to that in inline micro-mixers (\SEC{Ducts}). Such analogies may enable application of mixing technology to cardio-vascular research and treatment.

%similarities in configurations and transport characteristics that unite (seemingly different) practical flows and
%thus establishes an important bridge between theoreticians/physicists and engineers/practitioners. This facilitates
%application and dissemination of scientific insights to practice (see e.g. the RAM). Conversely, this enables systematic identification of theoretical/fundamental
%challenges specifically regarding applications.

%\subsection{Periodic duct flow}
\subsection{Flow in ducts}
\label{Ducts}

%\subsubsection{Introduction}
\subsubsection{General configuration}
\label{DuctsConfig}

Flows in ducts are assumed {\it periodic} in space unless stated otherwise. This assumption makes these problems tractable
by allowing representation of the complete system by a periodic subsection and is a valid approximation in important
industrial devices as the beforementioned inline mixers. Using this approximation, such periodic duct flows consist of a duct of infinite length and (in principle) arbitrary cross-section composed of repeated duct segments of length $L$.
% with period inlet ($z=0$) and outlet ($z=L$).
Each duct segment $p$ accommodates a 3D steady flow that is a periodic repetition of that in the first segment $p=1$, i.e.
%$\vec{u}(x,y,z+pL)=\vec{u}(x,y,z)$,
%
\begin{eqnarray}
\vec{u}(x,y,z+pL)=\vec{u}(x,y,z),
\label{SpatiallyPeriodic3D}
\end{eqnarray}
rendering the duct flow the spatially-periodic counterpart to the time-periodic flow \eqref{TimePeriodic2D}. The duct segments are, in
turn, partitioned into $N$ axial ``cells'' of length $\Delta = L/N$. Each cell $k$ accommodates a 3D steady flow that is a reorientation of flow $\vec{v}$ in the first cell $k=1$, i.e.
%
%\begin{eqnarray}
%\vec{u}(r,\theta,z+k\Delta)=\vec{v}\left(r,\theta-(k-1)\Theta,z\right),
%\label{SpatiallyPeriodic3D2}
%\end{eqnarray}
%
%with $1\leq k\leq N$ and $\Theta$ the reorientation angle such that $N\Theta$ is commensurate with $2\pi$.
%
\begin{eqnarray}
\vec{u}(r,\theta,z+k\Delta)=\vec{v}\left(r,\theta-\Theta_R(z),z\right),
\label{SpatiallyPeriodic3D2}
\end{eqnarray}
with axial reorientation following
\begin{eqnarray}
\Theta_R(z) = \Theta\sum_{k=1}^{N-1}\mathcal{H}(z - kL),
\label{SpatiallyPeriodic3D2x}
\end{eqnarray}
$\Theta$ the cell-wise reorientation angle such that $N\Theta$ is commensurate with $2\pi$ and $\mathcal{H}$ the Heaviside function \cite{Speetjens_symmetry_2006}.

\subsubsection{Industrial and technological relevance}
\label{DuctFlowIndustrial}

\begin{figure}
\centering
%\subfigure[...]{\includegraphics[width=0.9\columnwidth,height=0.6\columnwidth]{./RAM1.jpg}}

%\subfigure[Schematic of RAM]
\subfigure[RAM as representative open-duct flow]
{\includegraphics[width=\columnwidth,height=0.4\columnwidth]{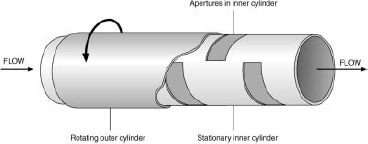}\label{RAM1a}}

\subfigure[Reorientation in RAM]{\includegraphics[width=0.37\columnwidth,height=0.37\columnwidth]{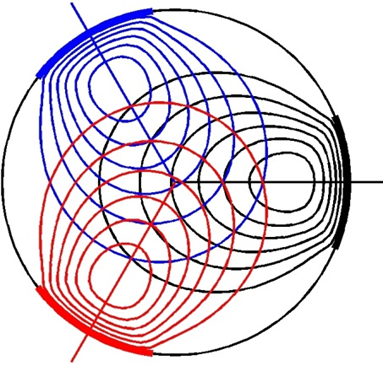}\label{RAM1b}}
%\subfigure[Schematic of PPM]
\subfigure[PPM as representative partitioned-duct flow]
{\includegraphics[width=0.62\columnwidth,height=0.4\columnwidth]{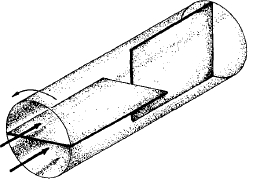}\label{RAM1c}}

\caption{Rotated Arc Mixer (RAM) and Partitioned-Pipe Mixer (PPM) as representative {\it open-duct} and {\it partitioned-duct} inline mixing
flows, respectively:
%Transverse flow driven by rotating (outer) cylinder ...
(a) reorientation
%of transverse flow
%driven by rotating outer cylinder
in RAM by angular offset of apertures in inner cylinder (reproduced from \cite{Metcalfe_ram_2006}); (b) reoriented transverse streamline portraits in RAM (colour distinguishes aperture); (c) reorientation
in PPM by splitting and recombination of axial flow via alternating horizontal/vertical internal plates (reproduced from \cite{Kusch_Experiments_1992}).}
\label{RAM1}
\end{figure}

Periodic duct flows constitute the generic configuration for industrial devices involving
{\it continuous} processing or treatment of laminar fluid streams.\footnote{Flow in such
devices normally settles on a repeated pattern and thus effectively becomes
periodic after a short transient following the inlet \cite{Baskan2017}.}
Applications are numerous and include \cite{Harnby1997,Todd2004,Thakur2003,Stone2004,Squires2005,Becht07,Kockmann2007,Wein09,Kjeang2009,Erwin2011,Laporte07,Mills08,Wu2009,Suh2010,Harmsen2010,Lutze2010,Oh2012b,Ward2015,Karimi2016}:
\begin{itemize}

\item[$\bullet$] inline mixers for food processing;

\item[$\bullet$] extruders for polymer processing;

\item[$\bullet$] compact inline mixers for process intensification;

\item[$\bullet$] inline micro-mixers and throughflow compartments in micro-fluidic (lab-on-a-chip) devices.

\end{itemize}
The purpose is mixing of a continuous axial throughflow at velocity $u_z$ (typically pressure-driven) by systematic reorientation of
the transverse flow $(u_x,u_y)$ via e.g. static internal mixing elements \cite{Harnby1997,Thakur2003,Wein09}.
\FIG{RAM1} demonstrates this principle for the RAM introduced above \cite{Metcalfe_ram_2006}. Here the transverse flow is driven by viscous drag
exerted by a rotating outer cylinder via apertures in the stationary inner cylinder (\FIG{RAM1a}); the cell-wise reorientation angle $\Theta$
following \eqref{SpatiallyPeriodic3D2x} is the angular offset of consecutive apertures accomplishes systematic downstream reorientation
of the transverse flow following \FIG{RAM1b}. Each tube section containing an aperture defining one cell and shown
black/blue/red transverse streamline patterns correspond with flow \eqref{SpatiallyPeriodic3D2} in cells $k=1,2,3$,
respectively, for $\Theta = 2\pi/3$ of duct segments of $N=3$ cells.

The RAM is inspired by the Partitioned-Pipe Mixer (PPM) shown in \FIG{RAM1c} \cite{Kusch_Experiments_1992}. The PPM
achieves reorientation by the combined effect of a rotating cylinder and alternating horizontal/vertical static internal
elements that repeatedly split and recombine the axial throughflow. The PPM served as model problem for the pioneering work on industrial mixers by the Ottino group (\SEC{Intro2}).
%
%The RAM and PPM capture the essentials of two kinds of periodic duct flows. The RAM represents configurations with {\it open} ducts, i.e. devoid of internal walls, which rely on transverse mixing by flow reorientation only; the PPM represents configurations with {\it partitioned} ducts, i.e. including internal walls as mixing baffles, which, besides
%reorientation, further employ splitting and recombination of the axial throughflow for transverse mixing.
%
%The RAM and PPM respectively capture the essentials of two kinds of periodic duct flows:
%The RAM and PPM respectively capture the essentials of the following two kinds of periodic duct flows:
The RAM and PPM capture the essentials of two kinds of periodic duct flows relevant in the present context:
%The RAM and PPM capture the essentials of the two relevant kinds of periodic duct flows here:
%
\begin{itemize}
\item[$\bullet$] {\it Open duct flows}: consist of a duct devoid of internal walls and rely on reorientation by the transverse boundary forcing as primary mixing mechanism.
 \item[$\bullet$] {\it Partitioned duct flows}: consist of a duct partitioned by internal walls and rely on reorientation
 by splitting and recombination of the axial flow as primary mechanism.
\end{itemize}
%
%The RAM and PPM belong to the former and latter class, respectively.
Moreover, the RAM and PPM have both been the subject of in-depth
(experimental) Lagrangian transport analyses and thus are well-suited to bridge the theory--practice gap. Hence, the RAM and PPM are adopted hereafter as the two representative periodic duct flows.

%\subsubsection{Industrial and technological relevance}
%\label{DuctFlowIndustrial}

\paragraph{Generic flow topology}

\begin{figure}
\centering
\subfigure[3D flow topology]{\includegraphics[width=0.95\columnwidth]{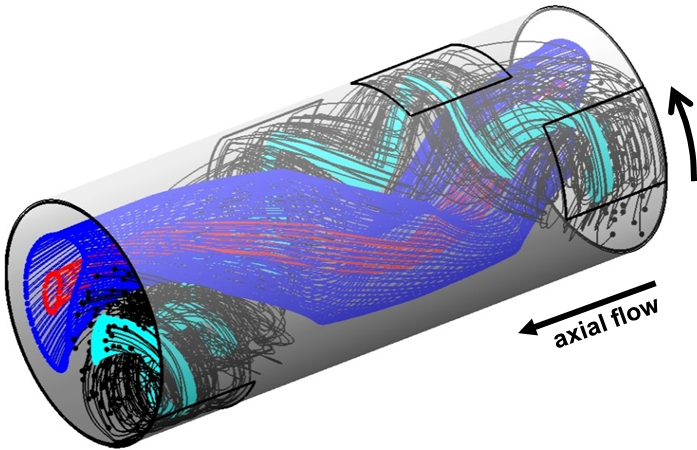}\label{RAM2a}}
\subfigure[Cross-section at inlet]{\includegraphics[width=0.42\columnwidth,height=0.4\columnwidth]{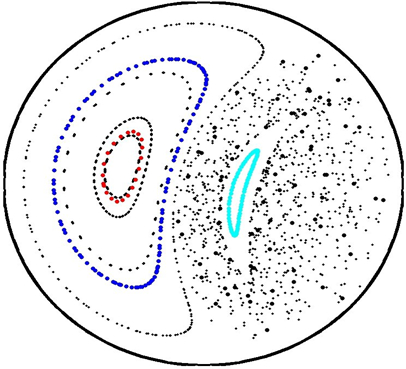}\label{RAM2b}}\hspace*{12pt}
\subfigure[Globally-chaotic topology]{\includegraphics[width=0.42\columnwidth,height=0.4\columnwidth]{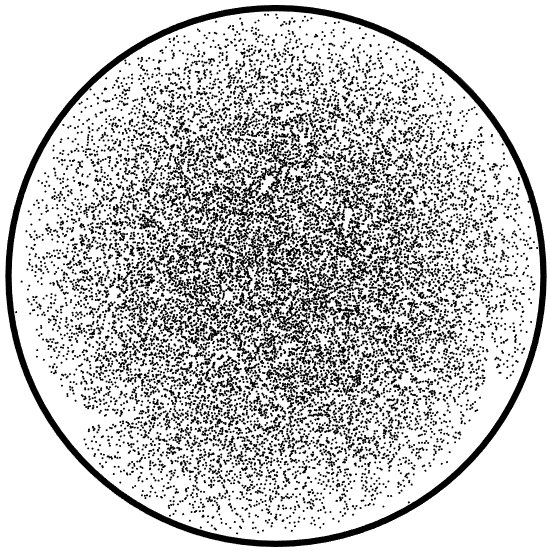}\label{RAM2c}}

\caption{Generic composition of 3D flow topology of periodic duct flows demonstrated for RAM: (a) stream tubes (colour) embedded in
chaotic environment (black) for sub-optimal RAM; (b) corresponding cross-section at inlet; (c) globally-chaotic topology for optimised RAM. Created using analytical 2.5D RAM flow from \cite{Speetjens_symmetry_2006}.}
\label{RAM2}
\end{figure}

The generic 3D flow topology of periodic duct flows is demonstrated in \FIG{RAM2a} for the RAM. It typically consists of families of concentric stream tubes running from inlet to
outlet of the duct segment (3 specimens highlighted in colour) embedded in chaotic streamlines (black). \FIG{RAM2b} gives the corresponding axial cross-section at the inlet
and reveals a transverse topology reminiscent of the stroboscopic map of the lid-driven cavity in \FIG{OrderWithinChaosa}: islands in a chaotic sea. This reflects the
fundamental property that 3D steady duct flows (in principle) are dynamically equivalent to 2D time-periodic flows and kinematic equation \eqref{KinEq} admits a (formal) transformation $\mathcal{F}$ following
\begin{eqnarray}
\frac{d}{dt}
\left[\begin{array}{c}x\\y\\z\end{array}
\right] =
\left[\begin{array}{c}u_x\\u_y\\u_z\end{array}\right]
%\quad\quad\stackrel{\mathcal{F}}{\Rightarrow}\quad\quad
\quad\stackrel{\mathcal{F}}{\Rightarrow}\quad
\frac{d}{d\tau}
\left[\begin{array}{c}\zeta_1\\\zeta_2\end{array}
\right] =
\left[\begin{array}{c}\nu_1\\\nu_2\end{array}\right],
%
%\left[\frac{dx}{dt},\frac{dy}{dt},\frac{dz}{dt}\right] = (u_x,u_y,u_z)
%\Rightarrow
%\left[\frac{d \zeta_1}{d \tau},\frac{d\zeta_2}{d \tau}\right] = (\nu_1,\nu_2)
%
\label{Steady3DvsUnsteady2D}
\end{eqnarray}
such that the transverse and axial Lagrangian motion corresponds with transport in the 2D domain $(\zeta_1,\zeta_2)$ and periodic evolution
in ``time'' $\tau$, respectively.
The LCSs relate as follows: periodic points, islands and 1D manifolds (i.e. curves) in 2D materialise as axially reconnecting streamlines, tubes and 2D manifolds
(i.e. surfaces), respectively, in 3D (\SEC{ClosedTrajectories}).

%
%Transformation $\mathcal{F}$ in essence translates the axial flow into an evolution in time; former and latter are
%in one-to-one correspondence in case of uni-directional axial flow (i.e. $u_z>0$ everywhere).\\\\

The Lagrangian framework developed by the Ottino group (\SEC{Intro2}) implicitly relies on transformation \eqref{Steady3DvsUnsteady2D} following
\SEC{HamiltonianFormDuctFlows} and studies on the RAM explicitly
demonstrate its existence (at least) for generic open-duct flows \cite{Speetjens_symmetry_2006,Speetjens2014}. Localised phenomena such as reversed axial flow or
internal recirculations may restrict this transformation -- and the equivalence with 2D dynamics -- to a ``net throughflow region'' yet
this is inconsequential for the functionality of open-duct mixers \cite{Speetjens2014}.

The equivalence (of at least said net throughflow region) with 2D unsteady flows means that 3D steady duct flows can basically always be designed to achieve global chaotic advection. Parametric studies enable systematic isolation of the appropriate operating conditions and thus are a common approach for optimisation and design of actual devices \cite{Hobbs1997,Szalai2003,Ottino2004,Stremler2004,Stone2004,Paul2004,Squires2005,Metcalfe_ram_2006,Zumbrunnen2006,Kockmann2007,Wu2009,Tiwari2009,Suh2010,Zumbrunnen2011,Hessel2012,Ghanem2014,Cullen2015}.
\FIG{RAM2c} gives the stroboscopic map for the RAM optimised by this approach, yielding a chaotic state devoid of stream tubes and, inherently, efficient mixing.
\begin{figure}
\centering

\subfigure[RAM: poor (top) versus good (bottom) mixing.]{\includegraphics[width=\columnwidth,height=0.4\columnwidth]{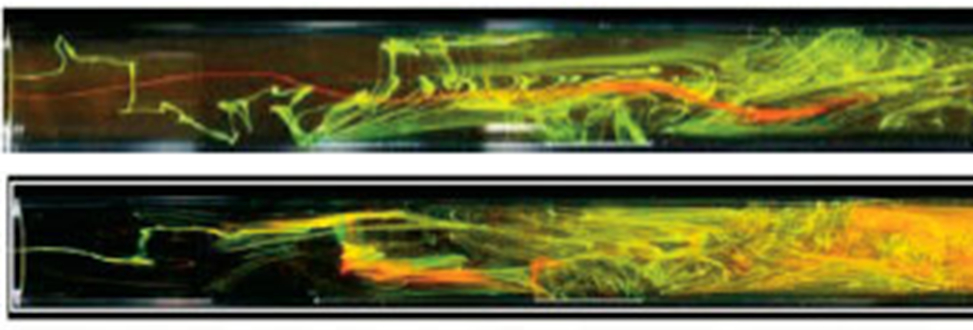}\label{PPM_RAMa}}

\subfigure[PPM: poor (top) versus good (bottom) mixing.]{\includegraphics[width=\columnwidth,height=0.4\columnwidth]{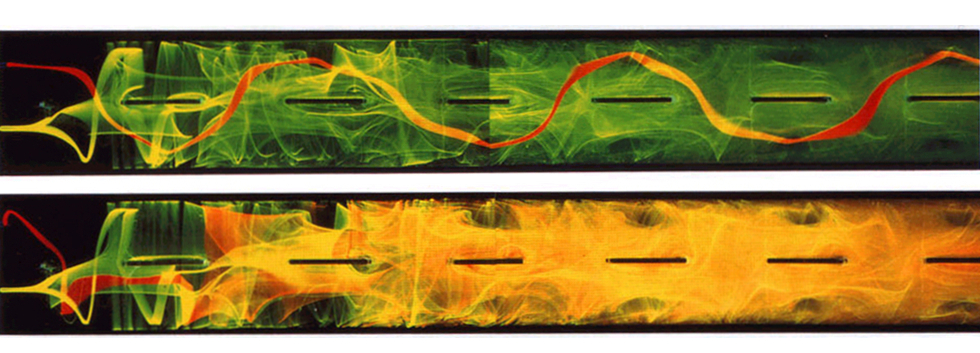}\label{PPM_RAMb}}

\caption{Mixing performance versus flow topology demonstrated experimentally by dye visualisation: (a) RAM (adapted from \cite{Metcalfe_ram_2006}); (b)
PPM (adapted from \cite{Kusch_Experiments_1992}). Poor (good) mixing concerns topologies with(out) streamtubes; side view with axial flow from left to right.
}
%
%of the impact of the flow topology on the mixing performance of the RAM (top) and the partitioned-pipe mixer (bottom). Axial flow is from left to
%right. Adapted from \cite{Kusch_Experiments_1992,Metcalfe_ram_2006}.}
\label{PPM_RAM}
\end{figure}

The generic composition of 3D Lagrangian flow topologies according to \FIG{RAM2a} and the impact upon the transport is demonstrated experimentally via dye visualisation for the RAM in \FIG{PPM_RAMa} and the PPM  in \FIG{PPM_RAMb} \cite{Kusch_Experiments_1992,Metcalfe_ram_2006}.
Two situations are considered for each system, viz. a poor-mixing case (top) with cross-sectional stroboscopic map containing
islands similar to \FIG{RAM2b} and a good-mixing case (bottom) exhibiting global chaos as in \FIG{RAM2c}. (Stroboscopic maps of
the PPM are essentially similar as in \FIG{RAM2} \cite{Kusch_Experiments_1992,Meleshko1999}.) The dye visualisations in both RAM and PPM indirectly expose the coexisting tubes and chaotic regions of the poor-mixing case via the entrapment of the red dye and transverse spreading of the green dye, respectively. The global chaotic advection of the good-mixing case is clearly demonstrated by the transverse spreading of both the red and green dye and the resulting formation of a homogeneous (yellowish) mixture.

\begin{figure}
\centering

\subfigure[Simulated downstream evolution]{\includegraphics[width=0.49\columnwidth,height=0.47\columnwidth]{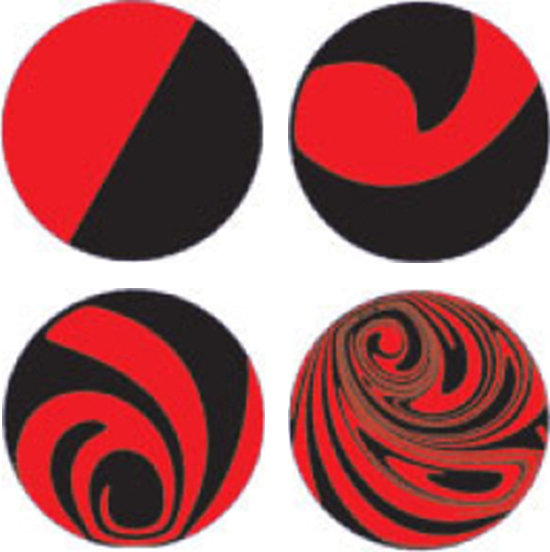}\label{RAM3a}}
\subfigure[Experimental visualisation]{\includegraphics[width=0.49\columnwidth,height=0.47\columnwidth]{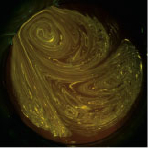}\label{RAM3b}}

\caption{Typical cross-sectional mixing patterns in RAM: (a) simulated downstream evolution of binary (red/black) concentration distribution (reproduced from \cite{Singh2008}); (b) corresponding experimental visualisation by fluorescent dye (reproduced from \cite{Metcalfe_ram_2006}).}
\label{RAM3}
\end{figure}

Chaotic advection implies cross-sectional mixing patterns with downstream progression similar to \FIG{IllustrateChaos}. This is demonstrated in \FIG{RAM3a} by the simulated evolution of a binary concentration distribution in the RAM, revealing the characteristic stretching and folding imparted by the unstable manifold (not shown) following \FIG{OrderWithinChaos} \cite{Singh2008}. Experimental visualisation using fluorescent dye illuminated by a transverse light sheet near the outlet of the RAM set-up of \FIG{PPM_RAMa} exposes (for comparable operating
conditions) an essentially similar cross-sectional mixing pattern (\FIG{RAM3b}) \cite{Metcalfe_ram_2006}. This experimentally validates
the key mechanism underlying chaotic advection in inline mixers and indirectly visualises the governing LCS(s), viz. the unstable manifold(s).

\paragraph{Open-duct flows: micro-mixers using patterned walls}
%\paragraph{Open-duct mixers: micro-mixer with cross-baffles}
%\paragraph{RAM-type mixers: micro-mixer with internal cross-baffles}
%\paragraph{Micro-mixer based on wall texture/...}

\begin{figure}
\centering

\subfigure[Configuration]{\includegraphics[width=\columnwidth,height=0.5\columnwidth]{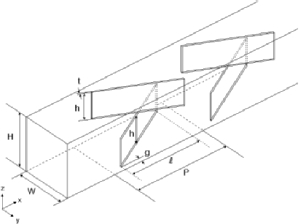}\label{MicroMixerSuh2007a}}
\subfigure[Simulated (left) vs. experimental (right) cross-sectional mixing pattern.]{
\includegraphics[width=0.45\columnwidth,height=1.2\columnwidth]{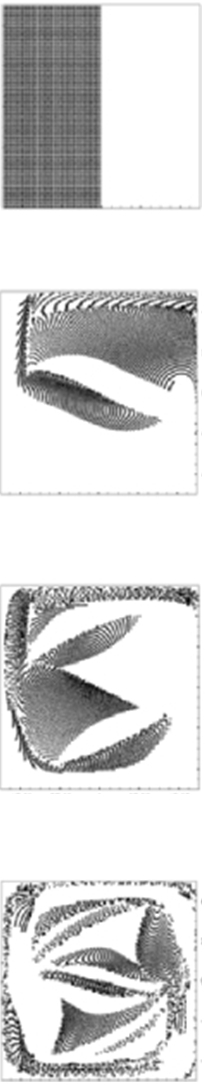}
\hspace*{12pt}
\includegraphics[width=0.45\columnwidth,height=1.2\columnwidth]{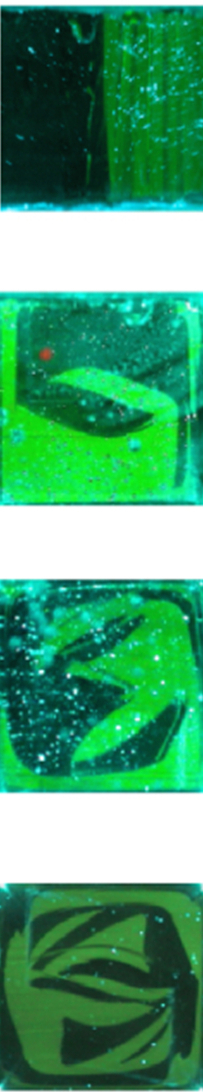}
\label{MicroMixerSuh2007b}}

%\caption{Mixing pattern in micro-mixer using cross-baffles (panel a): simulated stroboscopic maps (left) versus fluorescent dye
%visualisation (right). Reproduced from \cite{Suh2007}.}

\caption{
Open-duct micro-mixer using patterned wall for reorientation: (a) rectangular channel with wall-mounted cross-baffles; (b)
cross-sectional mixing pattern according to simulated downstream cross-sections of streamlines of black tracers
(left) versus experimental visualisation by fluorescent dye (right). Adapted from \cite{Suh2007}.
}

\label{MicroMixerSuh2007}
\end{figure}

Inline mixers of the open-duct type are, apart from the RAM, found mainly in micro-fluidics and usually accomplish
transverse flow and reorientation by patterned duct walls.\footnote{Reorientation by static wall structures instead of wall motion (as e.g. in the
RAM) is inconsequential for classification as an open-duct flow.} Consider as an example the square micro-channel with cross-baffles
extending from top and bottom channel walls shown in \FIG{MicroMixerSuh2007a} \cite{Suh2007}. The cross-baffles merely ``deform'' the duct
boundary without partitioning the domain and thus the system indeed constitutes an open-duct flow. This holds, irrespective of the geometric complexity, for any wall pattern and duct shape.

\FIG{MicroMixerSuh2007b} gives a typical cross-sectional mixing pattern in the above micro-mixer according to the simulated evolution of a patch of
black tracer particles (left) versus experimental visualisation by fluorescent dye via transverse light sheets (right). This reveals the
same progressive stretching and folding as in the RAM (\FIG{RAM3}) and thus demonstrates (i) the universality of Lagrangian (chaotic) transport phenomena
of open-duct flows and (ii) the intrinsic similarity in mixing characteristics of (practical) flows belonging to this class. Moreover, the close
agreement between computational and experimental results further substantiates the physical validity and meaningfulness of the Lagrangian concept.

Other inline micro-mixers adopting patterned walls for reorientation -- thus defining open-duct
flows with basically the same transport characteristics as above -- include the ``staggered-herringbone mixer'' (using grooved walls) and
the ``C-shaped serpentine micromixer'' (using C-shaped duct segments) and numerous variations on these designs \cite{Stroock2004,Wu2009}. State-of-the-art
micro-manufacturing technologies make commercial fabrication of such devices increasingly feasible and may thus become a key enabler for the practical
application of chaos-based micro-mixing technology \cite{Farshchian2017}.

%\paragraph{An industrial static mixer ... baffles ...}
%\paragraph{Partitioned-duct mixers: Quatro static mixer}
%\paragraph{Partitioned-duct mixers: mixers using static mixing elements}
\paragraph{Partitioned-duct flows: static mixers using baffles}
%\paragraph{PPM-type mixers: the Quatro static mixer}

Inline mixers of the partitioned-duct type have a long history as static mixers in the traditional processing
industry. These devices rely on partitioning of the flow domain by internal static mixing elements (``baffles'') for reorientation
and the Kenics mixer is perhaps the best known incarnation of this principle. Its original design dates back to the
1960s \cite{Kenicspatent1965} and to this day finds widespread utilisation in fluids engineering (\SEC{kenics}) \cite{Harnby1997,Todd2004}. Modified Kenics designs have even been implemented in micro-mixers \cite{Kim2004b}. Hence, the Kenics mixer, arguably, is the archetypical
static mixer and on that grounds served as model for the PPM \cite{Ottino_Kinematics_1989}.

\begin{figure}
\centering

\includegraphics[width=0.9\columnwidth,height=0.6\columnwidth]{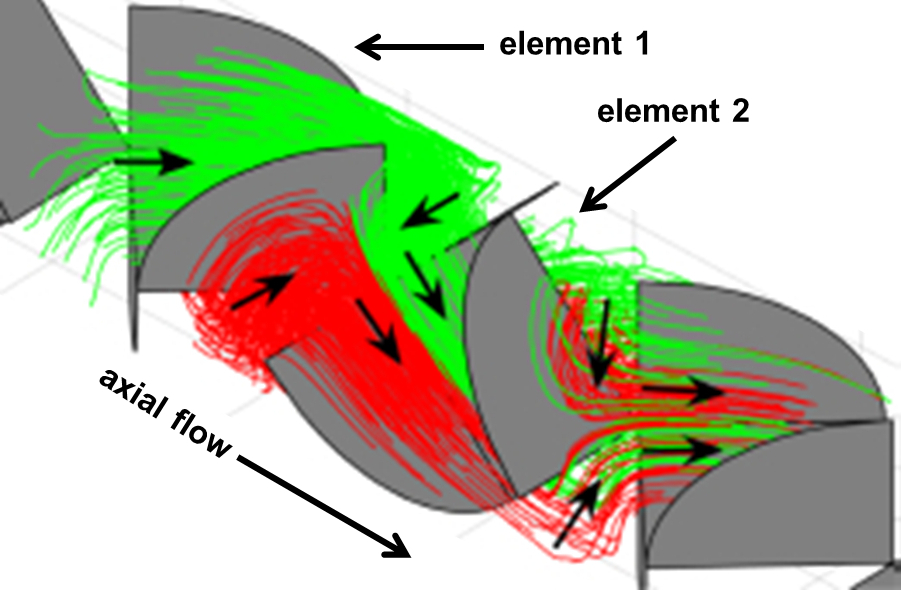}

%\caption{3D streamline pattern in the Quatro mixer measured by 3DPTV (inlet: $z=0$; downstream flow in $-z$ direction): experiments (blue) vs. simulations (red). Reproduced from \cite{Jilisen2013}.}

%\caption{Splitting and recombination of 3D internal streamline pattern by internal mixing elements in Quatro mixer measured by 3DPTV. Red and green streamlines represent axial flow separated by %indicated mixing element; arrows within streamlines indicate local flow direction. Adapted from \cite{Baskan2017}.}

\caption{Partitioned-duct mixer using internal mixing elements for reorientation demonstrated for Quatro mixer by 3D internal streamlines measured
by 3DPTV. Red/green streamlines represent axial flow separated by element 1; arrows within streamlines indicate local flow direction. Adapted from \cite{Baskan2017}.}

\label{Quatro1}
\end{figure}

The Quatro mixer (Primix BV, Mijdrecht, The Netherlands) is a static mixer reminiscent of the Kenics mixer \cite{Jilisen2013}.
Its mixing elements consist of chevron-shaped central plates with perpendicular elliptical segments extending to the cylinder wall and are alternately reversed and rotated about the cylinder axis (grey elements 1 and 2 in \FIG{Quatro1}). This yields a periodic repetition of pairs of mixing elements analogous to the consecutive pairs of reordered plates in the PPM.

\begin{figure}
\centering

%\subfigure[Evolution (left to right) from inlet via mixing elements 1, 3 and 5.]
%\subfigure[Transverse cuts (left to right) from inlet via mixing elements 1,3,5.]
\subfigure[Transverse cross-sections from inlet (left) via mixing elements 1,3,5.]
{\includegraphics[width=\columnwidth,height=0.25\columnwidth]{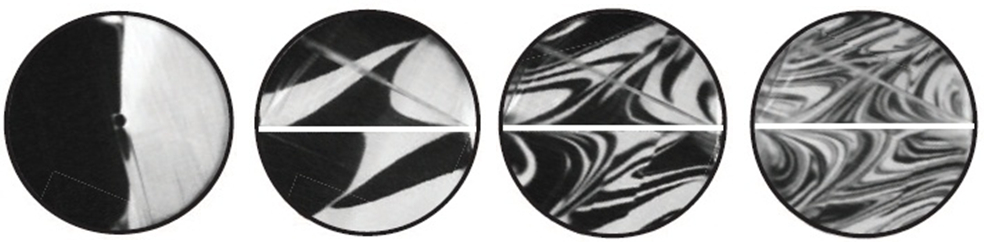}\label{Quatro2a}}
\subfigure[Longitudinal cross-sections in mixing elements 1,3,5.]{\includegraphics[width=\columnwidth,height=0.43\columnwidth]{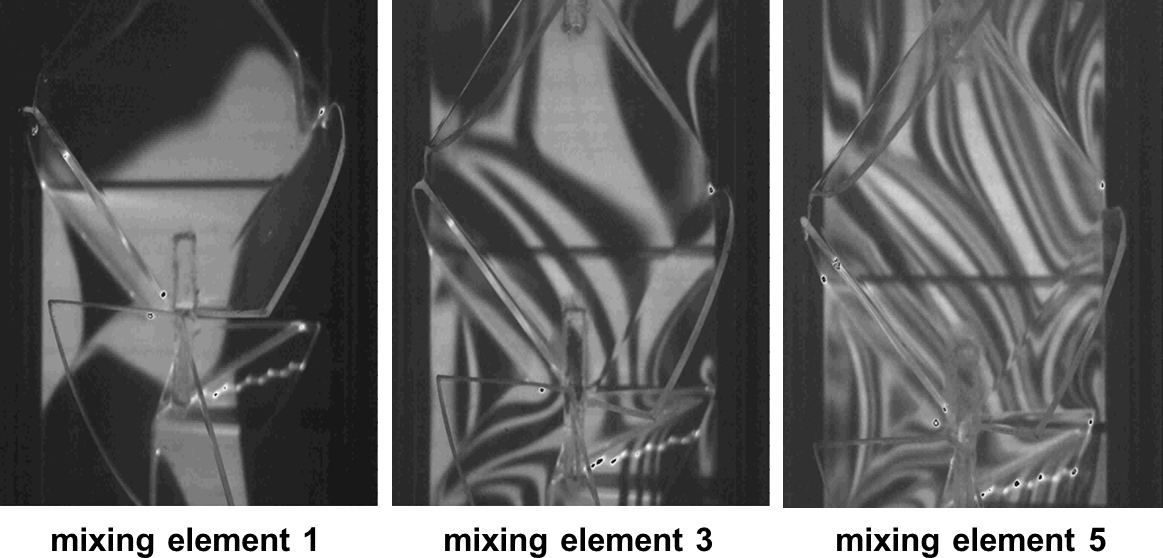}\label{Quatro2b}}

\caption{Evolution of 3D mixing pattern in Quatro mixer visualised by 3DLIF: (a) transverse cross-sections at indicated elements (adapted from \cite{Baskan2017}); (b) corresponding longitudinal cross-sections.}
\label{Quatro2}
\end{figure}

\FIG{Quatro1} demonstrates the transverse mixing within each duct segment by horizontal and vertical splitting (and subsequent recombination) of the axial throughflow
by the central plates of elements 1 and 2, respectively, via the 3D streamline pattern measured with 3D Particle Tracking Velocimetry (3DPTV) \cite{Jilisen2013,Baskan2017}.
%
%Red and green distinguishes the fluid streams separated by mixing element 1; the black arrows
%indicate the mean local flow direction. The red and green streams, in turn, ... splitting by element 2 ...
%
Mixing element 1 splits the incoming fluid into two separate streams (indicated in red and green) that, in turn, undergo further splitting by element 2, resulting in a ``mixture'' of red and green streamlines at the segment exit. Repetition of this process by consecutive duct segments yields ever finer striations as visualised in \FIG{Quatro2a} by transverse cuts of the 3D distribution of a fluorescent dye (bright) injected at the inlet and measured by 3D Laser-Induced Fluorescence (3DLIF) \cite{Baskan2017}. Here mixing elements 1,3,5 are the upstream elements (i.e. element 1 in \FIG{Quatro1}) in the consecutive duct segments 1,2,3, respectively, and shown cuts are halfway the central horizontal plate (white bar).

The cross-sectional mixing pattern in \FIG{Quatro2a} exhibits stretching and folding (signifying chaotic advection) akin to the RAM (\FIG{RAM3}) and the micro-mixer (\FIG{MicroMixerSuh2007}) and thus the Quatro mixer, notwithstanding the entirely different designs of the devices,  accomplishes efficient mixing by essentially the same Lagrangian mechanism. However, the underlying 3D fluid motion is markedly different.
%
% for these open-duct mixers and the (partitioned-duct) Quatro mixer.
%
% owing to the working principles.
%
Said open-duct mixers effectively coil up streamlines reminiscent of twirling spaghetti around a fork, illustrated in particular by the black chaotic streamlines in \FIG{RAM2a}, giving rise to cross-sectional swirling (\FIG{RAM3}). The (partitioned-duct) Quatro mixer, on the other hand, progressively ``slices'' the downstream fluid stream not unlike a meat grinder, as visualised in \FIG{Quatro2b} by longitudinal
cuts of the 3D dye distribution in the indicated mixing elements (top view of central horizontal plate). This manifests itself in the cross-sectional filamentation shown in \FIG{Quatro2a}.

Inline static mixers, regardless of baffle shape and fluid rheology, exhibit basically the
same Lagrangian transport characteristics as the Quatro mixer. Consider to this end e.g. the qualitative resemblance of the cross-sectional mixing patterns in \FIG{Quatro2a} with the dye visualisations of those for shear-thinning fluids in Kenics static mixers \cite{Alberini2014}. The wide variety of baffles employed in industry is nonetheless designed primarily by empirical studies and case-specific numerical simulations using standard mixing measures such as the CoV (\SEC{Intro2}). Design and optimisation of static mixers using Lagrangian concepts, on the other hand, is rare and mainly restricted to studies in engineering sciences on the beforementioned Kenics and Sulzer mixers \cite{Hobbs1997,Szalai2003,Kumar2008,Meijer2012}. This strongly suggests that many mixer designs are in fact sub-optimal and industry may greatly benefit from dedicated engineering tools as e.g. post-processing modules in a (commercial) CFD package for determination of cross-sectional stroboscopic maps. The latter namely enables systematic and conclusive isolation of the operating conditions that yield global chaotic advection as in \FIG{RAM2c}.

An open fundamental issue concerns the impact of the internal walls on the Lagrangian framework. The existing approach implicitly relies on transformation \eqref{Steady3DvsUnsteady2D} and
studies to date basically confirm its validity. However, internal walls may, besides the beforementioned reversed axial flow and recirculation, introduce singularities that (locally)
break the link with 2D unsteady systems and thus pave the way to essentially 3D phenomena (\SEC{HamiltonianFormDuctFlows}). In particular LCSs emerging from critical points on the internal
boundaries (instead of periodic points \eqref{PeriodicPoint} and corresponding reconnecting streamlines in the flow interior) are likely to play a (decisive) role in the transport characteristics on grounds of the analogy with transport in porous media. The latter namely involves similar splitting and recombination of throughflows by internal walls and is in fact dominated by
such LCSs (\SEC{WebsConfig}). Hence investigation of transformation \eqref{Steady3DvsUnsteady2D} for partitioned flow domains and wall-induced LCSs according to \SEC{WebsConfig} may strengthen the theoretical foundation and yield important insights with possible new applications.

%\paragraph{Chaotic serpentine micro-mixer (CSM)}

%\paragraph{Branching ducts: chaotic serpentine micro-mixer}
%\paragraph{PPM-type mixers: chaotic serpentine micro-mixer}
%\paragraph{Partitioned-duct mixers: micro-mixers using branching ducts}
\paragraph{Partitioned-duct flows: branching micro-channels}

\begin{figure}
\centering

\subfigure[Single duct segment]{\includegraphics[width=0.9\columnwidth,height=0.6\columnwidth]{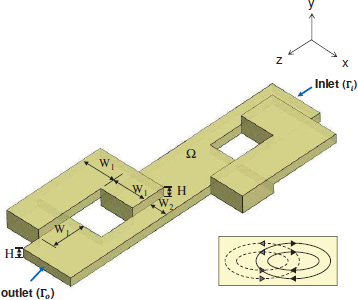}\label{ChaoticSerpentineMicroMixer_a}}

\subfigure[3D interface between species]{\includegraphics[width=0.8\columnwidth,height=0.8\columnwidth]{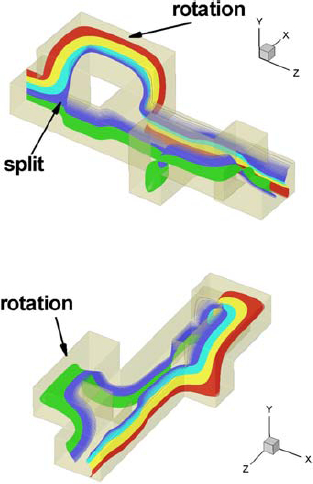}\label{ChaoticSerpentineMicroMixer_b}}

\subfigure[Typical cross-sectional mixing pattern (inlet to outlet)]{\includegraphics[width=\columnwidth,height=0.3\columnwidth]{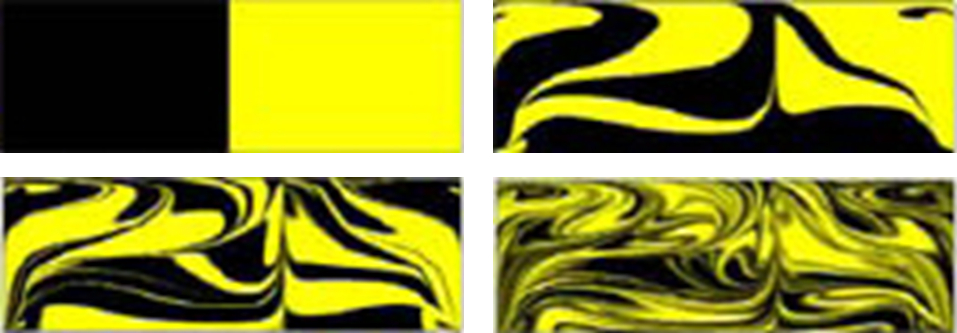}\label{ChaoticSerpentineMicroMixer_c}}

%\caption{Chaotic serpentine micro-mixer ... panel b: colour indicates vertical position at the inlet (red: top; blue:bottom). Panel a: reproduced/adapted from \cite{Kang2014}; panels b,c: %reproduced/adapted from \cite{Kang2009}.}

\caption{Partitioned-duct micro-mixer using branching channels for reorientation demonstrated by chaotic serpentine micro-mixer: (a) single duct segment (reproduced from \cite{Kang2014}); (b) simulated splitting and recombination of 3D material interface (colour distinguishes vertical position at inlet); (c) simulated evolution of segment-wise cross-sectional mixing pattern.
Panels (b,c) adapted from \cite{Kang2009}.
}
\label{ChaoticSerpentineMicroMixer}
\end{figure}

The partitioned-duct principle finds application also in micro-fluidic inline mixers yet, owing to the difficult fabrication
of microscopic baffles, typically in the form of branching and reconnecting channels. Consider for illustration the
chaotic serpentine micro-mixer (CSM) \cite{Kang2009,Kang2014}. \FIG{ChaoticSerpentineMicroMixer_a} shows one duct segment
comprising of two branching-reconnecting cells that each are composed of a pair of reversed L-shaped channels. Thus each cell in
the CSM, despite distributing the fluid streams over two physically separate channels, gives rise to basically the same splitting and recombination of the axial throughflow as imparted by each mixing element in the Quatro mixer. \FIG{ChaoticSerpentineMicroMixer_b}
demonstrates this by the simulated foliation of the internal 3D material interface separating the incoming fluid streams, revealing
a ``slicing'' as in \FIG{Quatro1} and \FIG{Quatro2b}. The corresponding cross-sectional mixing pattern is shown in \FIG{ChaoticSerpentineMicroMixer_c} and exhibits similar filamentation as in \FIG{Quatro2a}. Hence, transport in the CSM, despite completely different designs and length scales, indeed emanates from essentially the same Lagrangian mechanisms as the Quatro mixer and, inherently, any partitioned-duct mixer.

%\paragraph{Aperiodic duct flow: T-shaped micro-mixer}
%\paragraph{Hybrid inline mixers}
\paragraph{Aperiodic duct flows}

%\blue{
%A class of inline mixers intimately related to the above devices exists in {\it aperiodic} duct flows. Its Lagrangian transport characteristics, apart from strictly not defining periodic duct flows, %namely close resembles that of periodic duct flows: (i) also transverse mixing of axial throughflow and thus in principle 2D unsteady flow according to transformation \eqref{Steady3DvsUnsteady2D}; %(ii) also cross-sectional mechanism of stretching and folding.}\\\\
%

\begin{figure}
\centering

\subfigure[Configuration]{\includegraphics[width=0.76\columnwidth,height=0.4\columnwidth]{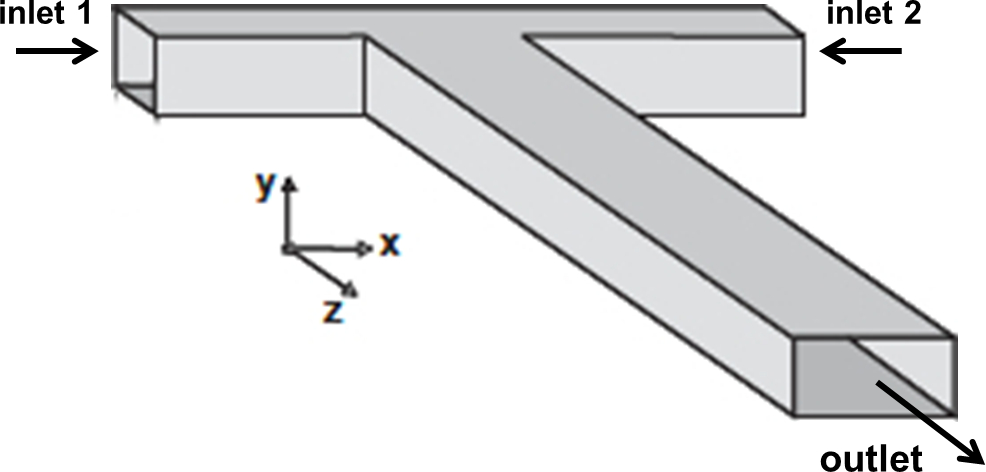}\label{TshapedMicroMixer1a}}
\subfigure[3D vortical flow induced at junction]{\includegraphics[width=\columnwidth,height=0.6\columnwidth]{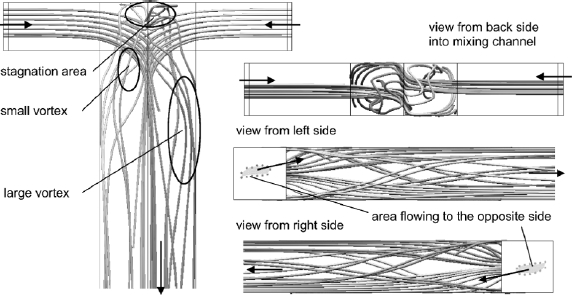}\label{TshapedMicroMixer1b}}

%\caption{T-shaped micro-mixer ... ... Panel a reproduced from \cite{Hoffmann2006}; panel b reproduced from \cite{Kockmann2007}.}

\caption{T-shaped micro-mixer as representative aperiod duct flow: (a) configuration of double inlet and single outlet (reproduced from \cite{Hoffmann2006}); (b) transverse mixing by 3D vortical flow (illustrated by typical 3D streamline pattern) induced at junction (reproduced from \cite{Kockmann2007}).
}

\label{TshapedMicroMixer1}
\end{figure}

A class of inline mixers intimately related to the above devices exists in {\it aperiodic} duct flows, that is, axial throughflows without the periodic structure
following \eqref{SpatiallyPeriodic3D} and \eqref{SpatiallyPeriodic3D2}. Consider for illustration the T-shaped micro-mixer in \FIG{TshapedMicroMixer1a}: two fluid streams
are brought together via separate inlets and undergo transverse mixing by a vortical flow in one common outlet duct (reminiscent of open-duct flows) induced at the junction
in case of sufficiently strong fluid inertia (\FIG{TshapedMicroMixer1b}) \cite{Hoffmann2006,Kockmann2007}.

\begin{figure}
\centering

\subfigure[Experimental 3D mixing pattern near junction]{\includegraphics[width=\columnwidth,height=0.7\columnwidth]{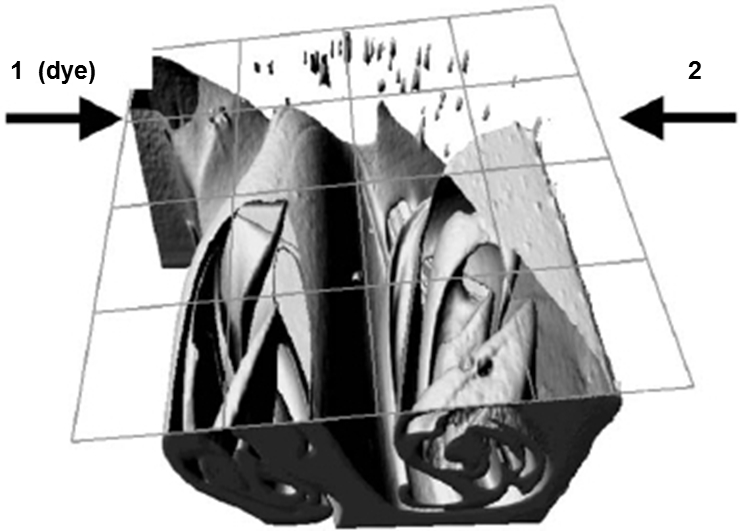}\label{TshapedMicroMixer2a}}
%\subfigure[Typical cross-sectional mixing pattern]{\includegraphics[width=0.7\columnwidth,height=0.7\columnwidth]{./TshapedMicroMixer3.jpg}\label{TshapedMicroMixer2b}}
%\subfigure[Typical cross-sectional mixing pattern]{\includegraphics[width=0.7\columnwidth,height=0.7\columnwidth]{./TshapedMicroMixer3Reflected.jpg}\label{TshapedMicroMixer2b}}

\subfigure[Dye pattern directly (left) \& further (right) downstream of junction]{
\includegraphics[width=0.48\columnwidth,height=0.2\columnwidth]{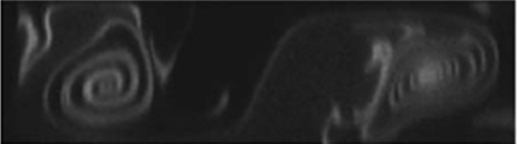}
\hspace*{3pt}
\includegraphics[width=0.48\columnwidth,height=0.2\columnwidth]{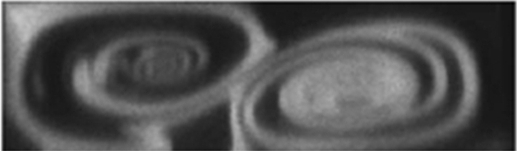}
\label{TshapedMicroMixer2b}
}

\subfigure[Simulated pattern directly (left) \& further (right) downstream of junction]{
\includegraphics[width=0.48\columnwidth,height=0.2\columnwidth]{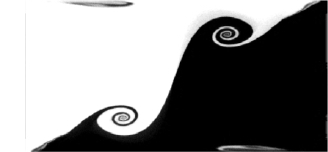}
\hspace*{3pt}
\includegraphics[width=0.48\columnwidth,height=0.2\columnwidth]{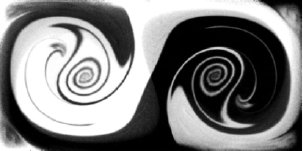}
\label{TshapedMicroMixer2c}
}

\caption{Typical 3D mixing pattern in T-shaped micro-mixer: (a) experimental visualisation
by fluorescent dye near junction using $\mu$-LIF; (b) corresponding cross-sectional patterns downstream of junction (adapted from \cite{Hoffmann2006}); (c) simulated cross-sectional
mixing patterns downstream of junction (adapted from \cite{Kockmann2007}).
}
\label{TshapedMicroMixer2}
\end{figure}

Said vortical flow (termed ``engulfment flow'' in \cite{Hoffmann2006,Kockmann2007} and occurring if the Reynolds number $Re$ exceeds a certain threshold) is characterised by two adjacent counter-rotating longitudinal swirling flows that intensify in downstream direction and each cross the vertical axial midplane of the duct (\FIG{TshapedMicroMixer1b}). This yields a 3D mixing pattern consisting of two longitudinal ``mixing rolls'' visualised in \FIG{TshapedMicroMixer2a} by iso-contours of a 3D dye distribution -- and corresponding cross-sectional evolution (\FIG{TshapedMicroMixer2b}) -- using $\mu$-LIF \cite{Hoffmann2006}. Numerical simulations for similar flow conditions (\FIG{TshapedMicroMixer2c}) reveals a roll-wise ``stretching and folding'' pattern reminiscent of the RAM (\FIG{RAM3}). This
implies similar transport characteristics as the periodic counterparts and thus strongly suggests that the nature of the axial evolution, i.e. periodic versus aperiodic, is of
secondary importance to this process.

The (spatial) aperiodicity has in fact primarily theoretical/conceptual ramifications by considerably limiting the usefulness of the existing Lagrangian machinery for transport
analysis due to its strong reliance on stroboscopic maps and LCSs associated with periodic points \eqref{PeriodicPoint} (\SEC{LagrangianView}). This limitation is of even greater
significance for geophysical flows and in particular the latter spurred scientific efforts, with George Haller and co-workers at the vanguard, to generalise the notion of coherence imparted by ``special'' Lagrangian entities (i.e. LCSs) to aperiodic flows \cite{Haller2015}.\footnote{Term ``Lagrangian coherent structures'' (LCSs) in fact stems from these efforts and here denotes such entities in both periodic and aperiodic flows.}
Relevant here are particularly so-called {\it attracting LCSs}, i.e. the aperiodic counterparts to the {\it unstable} manifolds in \FIG{OrderWithinChaosa}, which in a similar manner as the latter govern stretching and folding -- and thus chaotic advection -- in aperiodic flows \cite{Haller2015} (\SEC{CriticalPoints}). Isolation of attracting LCSs in the T-shaped micro-mixer is (to the best of our knowledge) outstanding yet their identification in an analogous flow in life sciences, i.e. an artery bifurcation (\FIG{HumanArteryBifurcation} in \SEC{DuctFlowNonIndustrial}), provides compelling evidence of their existence. This, in turn, underpins the above conjecture that the transport characteristics of (a)periodic duct flows are in essence the same.
However, generalisation of the Lagrangian framework to a universal and unambiguous approach for {\it aperiodic} and, intimately related, {\it finite-time} (transient) flows is
still in progress (\SEC{Conclusions}).

%The resemblance between the cross-sectional mixing patterns of the T-shaped micro-mixer and the RAM suggests the presence
%of hyperbolic LCSs in the former. \blue{The identification of such hyperbolic LCSs in an analogous flow in life sciences, i.e. an
%artery bifurcation (\SEC{DuctFlowNonIndustrial}), underpins this conjecture.}

%\begin{figure}
%\centering
%\subfigure[Typical simulated mixing pattern]{\includegraphics[width=\columnwidth,height=0.7\columnwidth]{Figures/SerpentineMicroMixer1.jpg}}
%\subfigure[Visualised mixing pattern]{\includegraphics[width=\columnwidth,height=0.4\columnwidth]{Figures/SerpentineMicroMixer2.jpg}}
%\caption{... serpentine laminating micro-mixer. Reproduced from (graphical abstract of) \cite{Kim2004}.}
%\label{SerpentineMicroMixer1}
%\end{figure}

\subsubsection{Relevance beyond industry}
\label{DuctFlowNonIndustrial}

Lagrangian transport and 3D chaotic advection in (a)periodic duct flows is also at the heart of several non-industrial configurations
%
%Thus the canonical ... as illustrated below by way of examples.
%
%This is illustrated below by way of examples.
%
and thus implies fundamental similarities with the above industrial systems. This is illustrated below by way of examples.
%
%This implies similarities and analogies with the above systems

%3D Lagrangian transport and chaotic advection in (a)periodic duct flows may also occur in non-industrial flow configurations.
%Instances of 3D Lagrangian transport and chaotic advection in (a)periodic duct flows may also occur in non-industrial flow configurations. This is illustrated below by way of examples.

\paragraph{Cardio-vascular flows}
%
%\paragraph{Physiological flows}

\begin{figure}
\centering

\subfigure[Configuration]{\includegraphics[width=0.8\columnwidth,height=0.8\columnwidth]{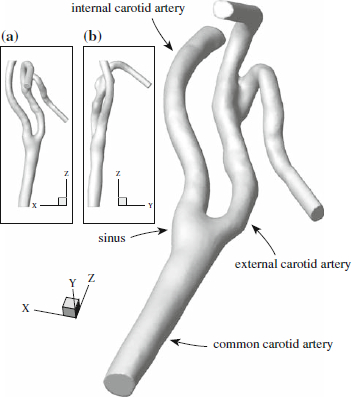}\label{HumanArteryBifurcation_a}}
\subfigure[Cross-sections of attracting ($T<0$) and repelling ($T>0$) LCSs.]{\includegraphics[width=\columnwidth,height=0.6\columnwidth]{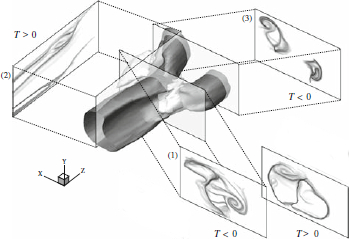}\label{HumanArteryBifurcation_b}}

%\caption{2D cross-sections of 3D attracting ($T<0$) and repelling ($T>0$) LCSs in the human carotid artery bifurcation. Adapted from \cite{Vetel2009}.}

\caption{Human carotid artery bifurcations as physiological analogy to aperiodic T-shaped micro-mixers:
%
%Aperiodic duct flows in physiology analogous to T-shaped micro-mixer:
%
(a) configuration; (b) flow topology consisting of attracting and repelling LCSs. Adapted from \cite{Vetel2009}.
}
\label{HumanArteryBifurcation}
\end{figure}

An important physiological occurrence of {\it aperiodic} duct flows is found in human carotid artery bifurcations (\FIG{HumanArteryBifurcation_a}). Carotid arteries supply oxygenated blood to the head and neck and their bifurcations are known as a site of chronic atherosclerotic plaque formation by the deposition of fat dissolved in the blood stream on the artery walls.
%
%This process closely resembles a ``reversed'' T-shaped micro-mixer by concerning Lagrangian transport of dissolved fat (prior to deposition on the artery walls) in the blood flow entering a single
%duct and exiting via two ducts (implying 3D vortical flow akin to \FIG{TshapedMicroMixer1} yet in opposite direction).
%
The corresponding Lagrangian transport closely resembles that in the T-shaped micro-mixer yet reversed in time in that the blood flow {\it approaches} a bifurcation via
a single duct and {\it exits} via two ducts (implying 3D vortical flow akin to \FIG{TshapedMicroMixer1} yet in opposite direction).
Experimental 3D steady velocity fields constructed in \cite{Vetel2009}
from 2D cross-sectional measurements with stereo Particle Image Velocimetry (PIV) in a laboratory model indeed reveal vortical motion at the bifurcation similar to the
T-shaped micro-mixer (\FIG{TshapedMicroMixer1b}).
The {\it attracting} LCSs, computed using the experimental flow, are given in \FIG{HumanArteryBifurcation_b} in the transverse
cross-sections 1 and 3 (indicated by $T<0$) and exhibit similar stretching and folding as their periodic counterparts, i.e. the {\it unstable} manifolds, in \FIG{OrderWithinChaosa}.
Shown also are the corresponding {\it repelling} LCSs (indicated by $T<0$) in cross-section 1 and stream-wise cross-section 2, which are the aperiodic
counterparts to the {\it stable} manifolds in \FIG{OrderWithinChaosa}.) Existence of these entities -- and their similarity to manifolds in chaotic seas -- strongly suggests chaotic advection in the artery bifurcation and, by analogy, also in the T-shaped micro-mixer.
Moreover, the implication of chaotic advection with atherosclerosis in carotid arteries is consistent with its purported role in blood-clot and plaque
formation in aneurysms (\SEC{Intro2} and \SEC{ContainersNonIndustrial}). Hence, chaotic advection could be, by tending to augment the residence time of blood-borne
substances \cite{Schelin2009}, a central player in the development of vascular diseases. Further exploitation of the resemblance with industrial inline mixers and application of the corresponding (scientific) knowledge on (chaotic) Lagrangian
transport may contribute to the understanding of these processes.

\begin{figure}
\centering

\subfigure[Tokamak reactor with magnetic field lines and flux surfaces.]{\includegraphics[width=0.9\columnwidth,height=0.5\columnwidth]{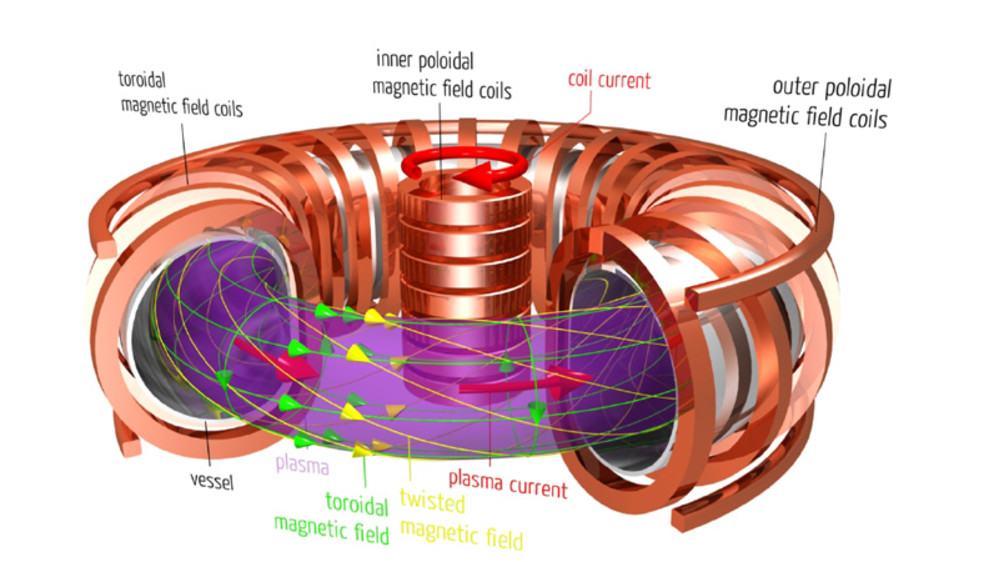}\label{Tokamaka}}

\subfigure[Simplified NTM topology]{\includegraphics[width=0.5\columnwidth,height=0.35\columnwidth]{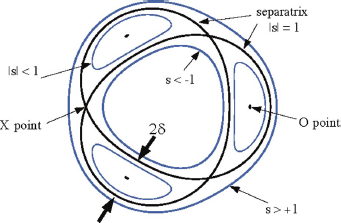}\label{Tokamakb}}
\hspace*{6pt}
\subfigure[Realistic NTM topology]{\includegraphics[width=0.45\columnwidth,height=0.42\columnwidth]{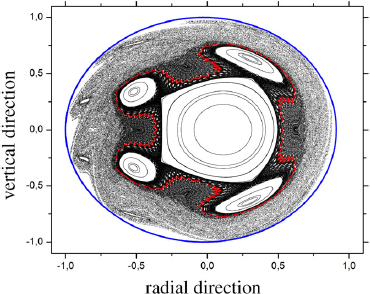}\label{Tokamakc}}

%\caption{... Tokamak fusion plasmas ... Panel a: image courtesy of Max-Planck Intitut f\"{u}r Plasmaphysik; panel b reproduced from \cite{Boozer2005}; panel c reproduced from \cite{Caldas2012}.}

\caption{Tokamak fusion reactors as MHD analogy to periodic open-duct flows: (a) configuration including magnetic field lines (green/yellow) and flux surfaces (purple) (reproduced from Max-Planck Intitut f\"{u}r Plasmaphysik); (b)
simplified cross-sectional NTM topology of magnetic islands only (reproduced from \cite{Boozer2005}) ; (c) realistic topology with islands
and chaotic regions (reproduced from \cite{Caldas2012}).
}

\label{Tokamak}
\end{figure}

\paragraph{Tokamak fusion reactors} A remarkable analogy exists between {\it periodic} duct flows $\vec{u}$ and the magnetic fields $\vec{B}$ of magneto-hydrodynamic (MHD) descriptions of plasmas in tokamak fusion reactors (\FIG{Tokamaka}). First, a
periodic duct is topologically a torus and thus equivalent to the interior of said reactors. Second, flow and magnetic fields are both solenoidal, i.e. $\vec{\nabla}\cdot\vec{u}=0$ and $\vec{\nabla}\cdot\vec{B}=0$. Third, the toroidal-poloidal magnetic forcing imposes similar net flux and reorientation as the axial-transverse forcing in periodic ducts.
Thus the tokamak admits a transformation $\mathcal{F}$ following \eqref{Steady3DvsUnsteady2D} of magnetic field $\vec{B}$ and toroidal
reference frame $(\sigma,\eta,\varphi)$ such that poloidal $(\sigma,\eta)$ and toroidal ($\varphi$) dynamics correspond with canonical
space $(\zeta_1,\zeta_2)$ and time $\tau$, respectively.\footnote{This holds equally for the so-called ``flux coordinates'' often employed
in tokamaks; these are namely intimately related to the toroidal frame \cite{Boozer2005}.}
These commonalities render periodic duct flows and magnetic fields in MHD tokamak models topologically and kinematically equivalent \cite{Meiss1992,Hazeltine1992,Boozer2005}.
%
%and thus enable transference of insights on the properties of said fields \blue{(\SEC{PeriodicDomain})} \cite{Meiss1992,Hazeltine1992,Boozer2005}.

The objective in tokamak fusion plasmas is accomplishment of so-called ``magnetic confinement'', that is, a magnetic topology consisting entirely
of one global family of concentric tubes (denoted ``magnetic flux surfaces'' and indicated schematically by the purple torus in \FIG{Tokamaka}) similar to the stream tubes in the RAM (\FIG{RAM2a}). However, this remains a major
challenge and is in fact the principal hindrance to the envisaged application of fusion plasmas as heat source for power plants (and safe alternative to nuclear {\it fission}) \cite{Boozer2005}.

Magnetic confinement may be broken by the disintegration of certain flux surfaces through the nonlinear process of reconnection,
yielding metastable states denoted ``neo-classical tearing modes'' (NTMs) that precede global instability of the
magnetic field \cite{Hazeltine1992,Boozer2005}.
Such NTMs are often investigated in an oversimplified manner, though.
First, their (poloidal) magnetic topology is typically assumed to entirely consist of one central island with multiple
satellite islands arranged by ``o-points'' and ``x-points'' following \FIG{Tokamakb} \cite{Boozer2005}. (Former and latter are the magnetic equivalents to elliptic and hyperbolic
points, respectively, in flow topologies.)
The analogy with periodic duct flows strongly suggests far more complex NTM topologies similar to \FIG{RAM2}. Both
numerical studies and experimental evidence support this \cite{Pires2005,Marcus2008,Onofri2011,Caldas2012}. \FIG{Tokamakc} gives a typical simulated poloidal NTM topology that
indeed exhibits the expected characteristics; the red curve is a shearless transport barrier and is (considered) key to the survival of the remaining flux surfaces \cite{Caldas2012}.
Second, transport by chaotic magnetic field lines is commonly regarded a stochastic turbulent-diffusion-like mechanism (``magnetic turbulence'') devoid of spatial organisation \cite{Boozer2005,Lopes2006}. However, chaotic advection in mixing flows implies a (though highly complex) well-defined spatial structure imparted by manifolds (\FIG{OrderWithinChaos}) and thus fundamentally contradicts this key assumption. Dedicated studies on chaotic magnetic field lines are indicative of such a spatial structure in NTM topologies \cite{Rover1999,Lopes2006}.

The above strongly suggests that incorporation of the true (3D) topological structure of the magnetic field and insights into (chaotic) Lagrangian transport is imperative for in-depth MHD-based analyses of tokamak fusion plasmas and magnetic confinement. The established link with periodic duct flows offers a way
to achieve this.

\subsection{Flow in containers}
%\subsection{Flow in closed containers}
%\subsection{Flow inside closed domains}
\label{Containers}

%\subsubsection{Introduction}
\subsubsection{General configuration}
\label{ContainersConfig}

%\red{
%- Start with part below on spheroidal domains as second canonical domain shape next to tori and, in consequence, flows in containers as second %case next to periodic duct flows.\\
%- container with solid wall (rigid/elastic) so as to differentiate from drops?\\
%}

Containers are defined as 3D flow domains confined by a closed surface (i.e. flux across the boundaries is absent) that are topologically equivalent to spheres. This includes, besides the latter, common geometries as cylinders and cubes. The key difference with periodic ducts
exists in the nature of the domain. Periodic ducts and containers namely are topologically a {\it torus} and a {\it sphere}, respectively,
which
%(depending on circumstances)
can have fundamental consequences for the Lagrangian flow topology and associated transport properties

Flow and transport in periodic ducts and containers constitute the two principal canonical problems in the present categorisation. The
surface classification theorem namely states that orientable\footnote{I.e. with a uniquely defined normal. Examples of non-orientable surfaces are the M$\ddot{\mbox{o}}$bius strip and the Klein Bottle.} closed surfaces are topologically equivalent to the surface of either a connected sum of tori or a single sphere \cite{Alexandroff1961}. Hence, tori and spheres are the fundamental building blocks of 3D (realistic) flow domains and Lagrangian transport in any such configuration thus intimately relates to that in periodic ducts and/or containers.

%{\it spheroidal} ... {\it toroidal} ...\\\\

%driven by surface forces (e.g. moving wall segments or impellers) and/or body forces (e.g. buoyancy or magnetic forcing). Furthermore, both %steady and unsteady confined flows are considered due to their relevance to practical applications.

\subsubsection{Industrial and technological relevance}
\label{ContainersIndustrial}

Flows in containers are the generic configuration for industrial devices involving
{\it batch} processing or treatment of fluid bodies under laminar conditions. Practical applications include \cite{Harnby1997,Todd2004,Grumann2005,Erwin2011,Mills08,Gopalakrishnan20105309,Strohmeier2015,Petkovic2017}:
\begin{itemize}

\item[$\bullet$] bio-reactors;

\item[$\bullet$] food processing;

\item[$\bullet$] electro-magnetic materials processing;

\item[$\bullet$] batch micro-mixers and processing chambers in micro-fluidic (lab-on-a-chip) devices.

\end{itemize}
The confinement by the closed boundary imparts a flow (topology) that generically involves a global circulation and consists of (multiple) vortical structures. Consider as a first
illustration the impeller-driven flow in \FIG{fig:intro_examplesa}; here the toroidal tracer paths reveal a primary circulation about the impeller and a secondary circulation
transverse to this. However, unlike periodic duct flows, the flow (topology) in containers may vary significantly with driving mechanism (i.e. surface versus body forcing) and
time signature (i.e. steady versus unsteady). Hence, this flow class is discussed below in terms of the following basic practical configurations:
\begin{itemize}
\item[$\bullet$] steady flow driven by {\it body} forcing;

\item[$\bullet$] steady flow driven by {\it surface} forcing;

\item[$\bullet$] unsteady flow.

\end{itemize}
This classification is somewhat pragmatic by leaning on relevant {\it practical} characteristics of container flows and is adopted here for reasons of accessibility.
A more rigorous (yet at this point inscrutable) differentiation can be made on the basis of {\it dynamic} characteristics and emerges naturally in the course of the following discussion.

%\paragraph{Steady impeller-driven flow}
%\paragraph{Steady flow driven by impeller}
\paragraph{Steady flow driven by body forcing}
%\paragraph{Steady flow driven by \textit{local} body forcing}

The abovementioned impeller-driven flow is archetypal of a steady container flow driven by body forcing; the immersed stirring device namely
effectively acts as a (local) body force. This is the most common configuration in industrial batch mixers and thus of great practical relevance.

\begin{figure}
\centering

\subfigure[Configuration]{\includegraphics[width=0.36\columnwidth,height=0.5\columnwidth]{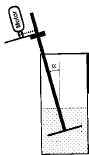}\label{StirredTankFountain_a}}
\hspace*{6pt}
\subfigure[KAM torus in 3D flow topology]{\includegraphics[width=0.58\columnwidth,height=0.5\columnwidth]{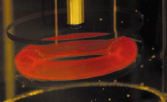}\label{StirredTankFountain_b}}
\subfigure[Experimental cross-section]{\includegraphics[width=0.49\columnwidth,height=0.32\columnwidth]{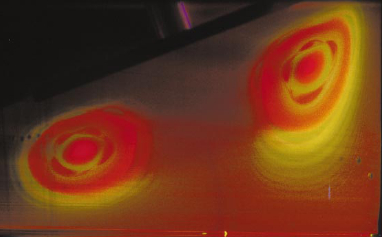}\label{StirredTankFountain_c}}
\subfigure[Computational cross-section]{\includegraphics[width=0.49\columnwidth,height=0.32\columnwidth]{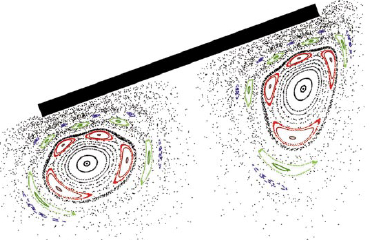}\label{StirredTankFountain_d}}

%\caption{... stirred tank ... tilted disk ... Adapted from \cite{Fountain_3d_2000}.}

\caption{Impeller-driven steady flow in a cylindrical container as physical model for industrial batch mixers: (a) configuration with tilted disk as
impeller; (b) typical KAM torus in 3D flow topology visualised by red fluorescent dye; (c) vertical cross-section of 3D flow topology visualised
by red/green dye; (d) corresponding simulated topology visualised by cross-sections of 3D streamlines. Adapted from \cite{Fountain_3d_2000}.}

\label{StirredTankFountain}
\end{figure}

The study by \cite{Fountain_3d_1998,Fountain_3d_2000} adopts the flow inside a cylindrical container driven by a tilted rotating disk following \FIG{StirredTankFountain_a} as
physical model for industrial batch mixers. (The specific domain and stirrer geometry are of secondary importance.) \FIG{StirredTankFountain_b} visualises the
Lagrangian transport using fluorescent dye and reveals, consistent with the
toroidal paths in \FIG{fig:intro_examplesa}, a 3D flow topology comprising of nested closed stream tubes (one specimen highlighted by red dye) encircling the
rotation axis of the impeller. The corresponding vertical cross-section, visualised in \FIG{StirredTankFountain_c} by red and green dye, and intersections of
3D computational streamlines (\FIG{StirredTankFountain_d}) imply an internal structure resembling the transverse topology of duct flows (\FIG{RAM2b}): islands in
a chaotic sea. This has the fundamental implication that, despite topologically different flow domains, the 3D Lagrangian motion and flow topology of the impeller-driven flow are dynamically similar to periodic duct flows. The similarity is particularly apparent upon comparison with the tokamak fusion reactor in \FIG{Tokamak}. The cross-sectional and
azimuthal dynamics in the impeller-driven flow are equivalent to the poloidal/toroidal dynamics of the tokamak and, in turn, to the
transverse/axial dynamics of periodic duct flows. This implies for the impeller-driven flow a transformation \eqref{Steady3DvsUnsteady2D} of the cylindrical reference frame $(r,\theta,z)$ such that canonical space $(\zeta_1,\zeta_2)$ and time $\tau$ correspond with $rz$-plane and azimuthal component $\theta$, respectively. (Refer to \SEC{HamiltonianFormCirculatoryFlows} and \SEC{Conditions3Ddynamics} for the conditions permitting this transformation.) The (closed) stream tubes as shown in \FIG{RAM2a} and \FIG{StirredTankFountain_b} and the
magnetic flux surfaces in \FIG{Tokamaka} thus constitute equivalent LCSs and are denoted ``Kolmogorov-Arnold-Moser (KAM) tori'' in
dynamical-systems terminology \cite{Ottino_Kinematics_1989,Boozer2005}. This nomenclature is adopted hereafter to indicate such entities.

\begin{figure}
\centering

\subfigure[Single Rushton impeller]{
\includegraphics[width=0.31\columnwidth,height=0.4\columnwidth]{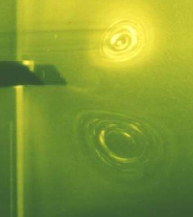}
\includegraphics[width=0.31\columnwidth,height=0.4\columnwidth]{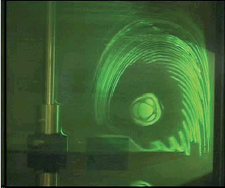}
\label{StirredTank2a}}
%
%\subfigure[Single Rushton %impeller]{\includegraphics[width=0.32\columnwidth,height=0.4\columnwidth]{Figures/StirredTankParticleClustering1.jpg}\label{StirredTank2b}}
%
\subfigure[3 Rushton impellers]{\includegraphics[width=0.32\columnwidth,height=0.4\columnwidth]{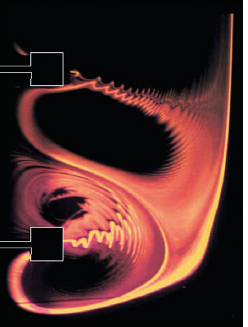}\label{StirredTank2c}}

%\caption{Impact of stirrer geometry on mixing properties in bioreactor ... Reproduced/adapted from \cite{Wang_clustering_2014,Alvarez-Hernandez2002,Alvarez2005}.}

\caption{Impact of impeller geometry on transport properties of impeller-driven flow in cylindrical container: (a) cross-sectional mixing pattern (left) and
close-up (right) for single-impeller case visualised experimentally by fluorescent dye (adapted from \cite{Alvarez2005}); (b) dye visualisation for 3-impeller case (reproduced from \cite{Alvarez-Hernandez2002}).}

\label{StirredTank2}
\end{figure}

Lagrangian flow topologies consisting of KAM tori embedded in chaos is generic for steady flows driven by body forcing. The particular forcing and/or domain geometry may (significantly) affect the emergence and arrangement of KAM tori and other LCSs, though. Consider to this end the impact of the impeller shape and positioning on the flow topology of the stirred container in \FIG{StirredTankFountain_a}. A single impeller consisting of blades and rotating about the cylinder axis (i.e. a so-called ``Rushton impeller'' widely used in industry) results in two ``stacked'' families of KAM tori as visualised in \FIG{StirredTank2a} (left) by the transverse cross-section of a 3D distribution of tracer particles \cite{Alvarez2005}. Detailed dye visualisation under comparable operating conditions reveals the upper tori family and its chaotic environment (\FIG{StirredTank2a}; right) \cite{Wang_clustering_2014}. The latter corresponds with the ``ripples'' extending from the impeller, which exhibit the characteristic stretching-and-folding pattern created by manifolds akin to \FIG{OrderWithinChaos}.
A configuration of 3 Rushton impellers yields a sizeable chaotic region, visualised by the dye concentration at the upper/middle impellers in \FIG{StirredTank2c}, enveloping multiple tori families of variable extent (black islands)
\cite{Alvarez-Hernandez2002}. The impellers induce similar ``rippling'' (signifying chaotic advection) as the single-impeller case.

%The ``ripples'' extending from the impellers exhibit the characteristic stretching-and-folding pattern created by manifolds akin to %\FIG{OrderWithinChaos} and thus further substantiates the dynamic analogy with periodic duct flows.
%
%\FIG{StirredTank2c} visualises the 3D extent of typical chaotic trajectories near the upper impeller under comparable operating conditions %\cite{Alvarez2005}.

The above experimental analyses reveals a dramatic impact of the impeller geometry on the mixing properties. Stirring by disks rotating about
the cylinder axis yields only KAM tori; tilting of the rotation axis is essential to induce chaos yet this remains highly localised (\FIG{StirredTankFountain}) \cite{Fountain_3d_2000}. The Rushton impellers, on the other hand, yield significant chaotic regions (\FIG{StirredTank2}) and, potentially, global chaos. The generic composition of the flow topology (i.e. KAM tori and chaos) nonetheless is essentially the same in all cases.

\begin{figure}
\centering
\subfigure[Joule heating only]{\includegraphics[width=0.47\columnwidth,height=0.5\columnwidth]{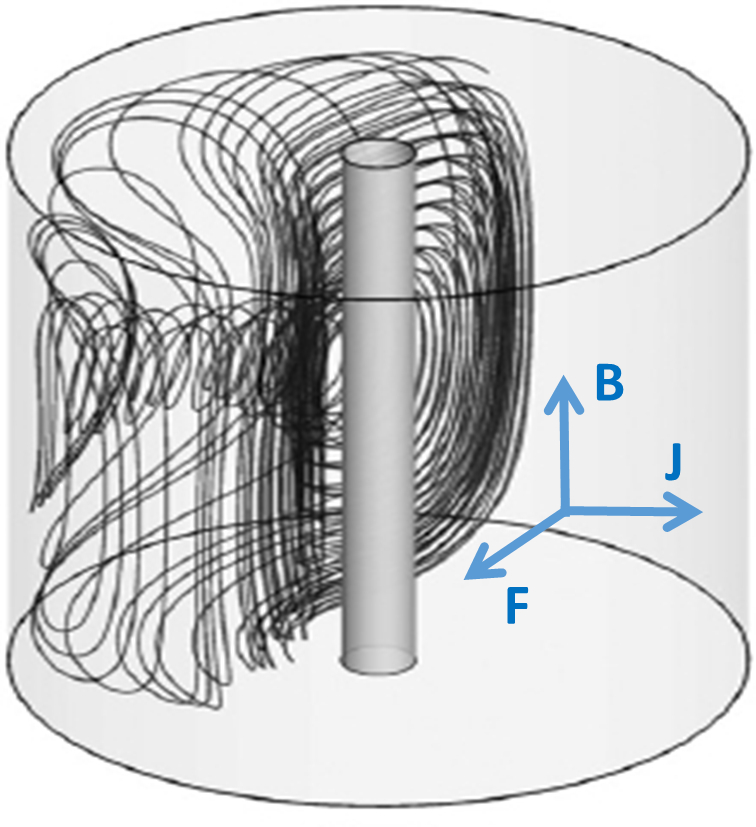}\label{GlassMelt1a}}
\hspace*{6pt}
\subfigure[Steady magnetic field]{\includegraphics[width=0.47\columnwidth,height=0.5\columnwidth]{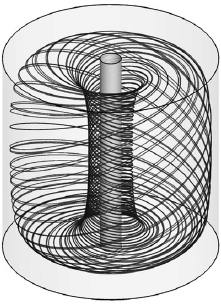}\label{GlassMelt1b}}

\caption{Electro-magnetic steady stirring of glass melt in cylindrical container by Lorentz force $\vec{F}$ due to radial current $\vec{J}$ between
central rod and cylinder wall and vertical magnetic field $\vec{B}$ : (a) one torus of two tori families symmetric about vertical midplane for Joule heating only ($\vec{B} = \vec{0}$); (b) one
torus of single tori family centered on cylinder axis for electro-magnetic case ($\vec{B} \neq \vec{0}$). Adapted from \cite{Gopalakrishnan20105309}.}
\label{GlassMelt1}
\end{figure}

The generality of KAM tori as key LCSs in 3D steady container flows -- and the dynamic analogy with periodic duct flows -- admits further substantiation by an entirely different (industrial) example involving (global) body forcing: electro-magnetic stirring of glass melts \cite{Gopalakrishnan20105309}.\footnote{Electro-magnetic stirring is in fact a common technique for processing of liquid metals. However, this occurs (unlike glass melts) under {\it turbulent} flow conditions and thus is beyond the present scope \cite{Davidson1999}.}
This relies on buoyancy (induced by Joule heating via an electric current) in conjunction with
%an actively-controlled
a Lorentz force (induced via interaction of said current with an external magnetic field). \FIG{GlassMelt1} gives the
particular configuration for a cylindrical container: a central (gray rod) and outer (cylinder wall) electrode set
up a radial electric current $\vec{J}$; external magnets (not shown) create a vertical magnetic field $\vec{B}$ that
(together with $\vec{J}$) yield
%an azimuthal (volumetric) Lorentz force $\vec{F} = \vec{J}\times\vec{B}$.
a (volumetric) Lorentz force
\begin{eqnarray}
\vec{F} = \vec{J}\times\vec{B},
\end{eqnarray}
with a dominant azimuthal component.
Joule heating only ($\vec{B} = \vec{0}$) results for sufficiently strong $\vec{J}$ in a steady buoyancy-driven circulatory flow with a topology that consists entirely of two families of tori symmetric about
the vertical midplane following \FIG{GlassMelt1a} (shown is one torus of one family described by a single streamline). Activation of the magnetic field ($\vec{B} \neq \vec{0}$) changes this completely into a steady swirling flow and corresponding topology of one global family of tori centered on the central electrode (\FIG{GlassMelt1b}) \cite{Gopalakrishnan20105309}. KAM tori nonetheless persist as the topological building blocks and their arrangement is similar to that of the individual families in the above impeller-driven flows (\FIG{StirredTankFountain} and \FIG{StirredTank2}). However, an essential difference with the latter is (at least in the considered parameter ranges) the absence of (local) chaotic advection; this requires unsteady electro-magnetic stirring and is addressed below.\footnote{This is largely inherent in the particular configuration, though. Buoyancy alone may (in 3D steady container flows) namely already yield intricate flow topologies with multiple tori families and chaotic regions \cite{Contreras2017}.}
%
%\begin{figure}
%\centering
%\subfigure[Configuration]{\includegraphics[width=0.47\columnwidth,height=0.4\columnwidth]{Figures/GlassMelt1.jpg}\label{GlassMelt_a}}
%
%\hspace*{6pt}
%
%\subfigure[Joule heating only]{\includegraphics[width=0.47\columnwidth,height=0.47\columnwidth]{Figures/GlassMelt2.jpg}\label{GlassMelt_b}}
%\subfigure[Steady magnetic field]{\includegraphics[width=0.47\columnwidth,height=0.47\columnwidth]{Figures/GlassMelt3.jpg}\label{GlassMelt_c}}
%
%\hspace*{6pt}
%
%\subfigure[Oscillating magnetic field]{\includegraphics[width=0.47\columnwidth,height=0.47\columnwidth]{Figures/GlassMelt4.jpg}\label{GlassMelt_d}}

%\caption{Electro-magnetic stirring of glass melt ... Reproduced/adapted from \cite{Gopalakrishnan20105309}.}
%\label{GlassMelt}
%\end{figure}

\begin{figure}
\centering

\subfigure[Initial particle distribution]{\includegraphics[width=0.45\columnwidth,height=0.5\columnwidth]{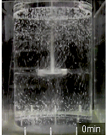}\label{StirredTankParticleClustering_b}}
\hspace*{6pt}
\subfigure[Final particle distribution]{\includegraphics[width=0.45\columnwidth,height=0.5\columnwidth]{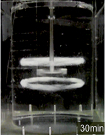}\label{StirredTankParticleClustering_c}}

%\caption{Impeller-driven ``unmixing'' of inertial particles suspended in flow in cylindrical container by clustering in KAM tori of flow topology. Reproduced from \cite{Wang_clustering_2014}.}
\caption{Impeller-driven ``unmixing'' of inertial particles in fluid flow in cylindrical container by clustering in KAM tori of flow topology. Experimental visualisation using polystyrene particles suspended
in glycerine. Reproduced from \cite{Wang_clustering_2014}.}

\label{StirredTankParticleClustering}
\end{figure}

The above concerns practical utilisations of Lagrangian transport seeking to accomplish chaotic advection for the purpose of efficient mixing and dispersion.
KAM tori, in their capacity as barriers to fluid transport, are detrimental to this process. However, industrial batch processing often aims at solid-liquid
separation in suspensions of finite-size particles for e.g. waste-water treatment or biorefinery. KAM tori offer a way to realise such ``unmixing'' for (at least) small particles slightly lighter than the bulk fluid by exploiting the fact that they closely (yet not exactly) follow the fluid parcels. The weak deviation in Lagrangian motion namely has a remarkable effect by causing such particles to migrate towards KAM tori and thus cluster
there. This ``unmixing'' is demonstrated experimentally in \FIG{StirredTankParticleClustering} for the above steady flow driven by a single Rushton impeller via the clustering of polystyrene particles suspended in glycerin in the KAM tori at the impeller (\FIG{StirredTank2a}) \cite{Wang_clustering_2014}. \FIG{fig:intro_examplesa} shows individual particle trajectories in this process.

The particle behaviour in \FIG{StirredTankParticleClustering} stems from the fact that the Lagrangian dynamics, though also described by a kinematic equation as \eqref{KinEq}, is
determined by a particle velocity $\vec{v}$ instead of the fluid velocity $\vec{u}$. The fundamental difference between both fields is that the former is non-solenoidal, i.e. $\vec{\nabla}\cdot\vec{v}\neq 0$, and its (asymptotic) dynamics, in consequence, is governed by so-called ``attractors'', that is, LCSs associated with $\vec{v}$ onto which the particles converge in the course of time. Such entities are absent in solenoidal flows and thus non-existent for the Lagrangian flow topology underlying the fluid motion. However, an intriguing connection occurs in that the attractors for the particular kind of particles considered here are in fact the KAM tori \cite{Cartwright2002}. This gives rise to the ''unmixing'' of particles shown in \FIG{StirredTankParticleClustering}. General criteria for particle clustering in LCSs are developed in \cite{Haller_clustering_2010}.
%

%\paragraph{Steady wall-driven flow}
\paragraph{Steady flow driven by surface forcing}

Flows in a cylindrical cavity due to rotating or translating lids are representative examples of surface-driven flows and are 3D counterparts of
the 2D lid-driven cavity flow introduced in \SEC{LagrangianView}. The flow and transport characteristics depend essentially on the specific lid
motion (i.e. rotating versus translating) and time signature (i.e. steady versus unsteady).

\begin{figure}
\centering
\subfigure[$Re=1350$]{\includegraphics[width=0.48\columnwidth,height=0.4\columnwidth]{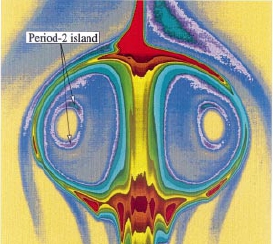}}
\subfigure[$Re=2100$]{\includegraphics[width=0.48\columnwidth,height=0.4\columnwidth]{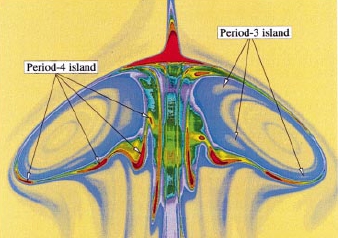}}
\caption{Lagrangian flow topology in steady swirling flow in cylinder driven by {\it rotation} of bottom wall. Shown are vertical cross-sections of KAM tori and higher-order tori (indicated by period-2/3/4 islands) for given $Re$ as visualised experimentally by level sets of fluorescence intensity. Reproduced from \cite{Sotiropoulos_breakdown_2002}.}
\label{VortexBreakdownBubble}
\end{figure}

Steady {\it rotation} of the bottom lid gives rise to a swirling flow and associated flow topology comparable to the steady impeller-driven and electro-magnetically-driven flows in \FIG{StirredTankFountain} and \FIG{GlassMelt1b}, respectively: a family of KAM tori centered on the cylinder axis and surrounded by chaos. This is demonstrated experimentally in \FIG{VortexBreakdownBubble} by the time-averaged
light intensity emitted by a fluorescent dye in a vertical cross-section; its level sets (differentiated by colour) expose the characteristic island
structure including localised LCSs as the indicated period-2/3/4 island chains \cite{Sotiropoulos_breakdown_2002}. Moreover, this reveals a progressive breakdown
of tori into chaos with increasing Reynolds number $Re$.

\begin{figure}
\centering

\subfigure[$Re=0$]{\includegraphics[width=0.48\columnwidth,height=0.48\columnwidth]{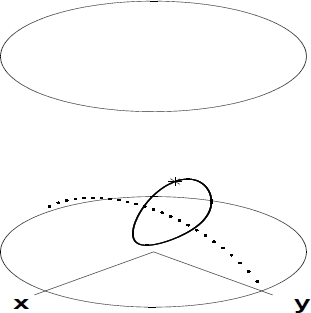}\label{SteadyCylinderFlow_0a}}
\subfigure[$Re=10$]{\includegraphics[width=0.48\columnwidth,height=0.48\columnwidth]{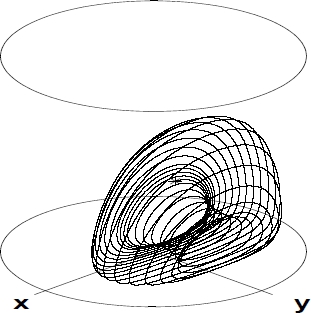}\label{SteadyCylinderFlow_0b}}
\subfigure[$Re=100$]{\includegraphics[width=0.48\columnwidth,height=0.48\columnwidth]{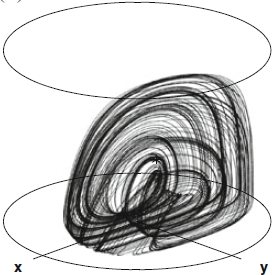}\label{SteadyCylinderFlow_a}}
\subfigure[Cross-section with plane $x=0$]{\includegraphics[width=0.48\columnwidth,height=0.48\columnwidth]{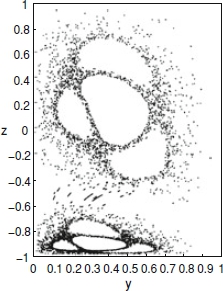}\label{SteadyCylinderFlow_b}}
%
%\caption{Lagrangian flow topology in the cylinder flow driven by steady translation of the bottom wall in $+x$-direction ... typical single streamline versus $Re$ ... Reproduced from \cite{Speetjens2001}.}

\caption{Lagrangian flow topology in steady cylinder flow driven by {\it translation} of bottom wall in $+x$-direction: (a-c) emergence and progressive disintegration of two families of tori symmetric
about midplane $y=0$ with increasing $Re$ as visualised by simulated single streamline; (d) vertical cross-section with plane $x=0$ exposing KAM tori embedded in chaos. Reproduced from \cite{Speetjens2001}.}

\label{SteadyCylinderFlow}
\end{figure}

Steady {\it translation} of the bottom lid yields a circulatory flow consisting of two symmetric vortices similar to the buoyancy-driven flow
in \FIG{GlassMelt1a} \cite{Shankar1997,Shankar_cavity_2000}). \FIG{SteadyCylinderFlow} demonstrates the emergence of the corresponding flow topology by a typical
streamline in case of translation in $+x$-direction. The non-inertial limit $Re=0$ yields a closed streamline (\FIG{SteadyCylinderFlow_0a}) that, upon
introduction of fluid inertia, becomes non-closed and describes a torus as shown for $Re=10$ in \FIG{SteadyCylinderFlow_0b}. Thus two global families of
tori separated by the symmetry plane $y=0$ are created for ``sufficiently small'' $Re$. The streamline becomes chaotic upon further
increasing $Re$, as demonstrated for $Re=100$ in \FIG{SteadyCylinderFlow_a}, signifying breakdown of certain tori and resulting in KAM tori embedded in chaos
(outlined by the corresponding cross-section with the plane $x=0$ in \FIG{SteadyCylinderFlow_b}) similar as the above rotating-lid case (\FIG{VortexBreakdownBubble}).
The flow field and streamline pattern are validated experimentally by 3D PTV in \cite{Speetjens_stokes_2004,Znaien2012}.
Single-wall forcing of {\it cubical} cavities yields essentially the same flow topology as in \FIG{SteadyCylinderFlow}, implying, similar as the previous impeller-driven flows, an only secondary importance of the specific geometry \cite{Shankar_cavity_2000}.
Moreover, forcing by multiple walls or wall segments typically induces bifurcations into more complex topologies yet nonetheless with a generic composition of
(multiple) families of tori embedded in chaos \cite{Shankar_cylinder_1998,Lackey_cavity_2006}.

The transition of the flow topology from a global family of closed streamlines in the non-inertial limit $Re=0$ to two global families of tori, as
demonstrated for a generic streamline in \FIG{SteadyCylinderFlow}, happens for any $Re=0^+$. Moreover, other minute perturbations of the non-inertial limit
as e.g. asymmetric forcing have the same effect \cite{Shankar_cylinder_1998,Speetjens2011}. Thus the response of the steady cylinder flow to fluid inertia in fact reflects
universal behaviour. Flow topologies in {\it any} system consisting entirely of closed streamlines namely constitute a singular limit in that the slightest perturbation
induces transition into (multiple families of) tori that robustly survive a finite perturbation strength (here $Re\lesssim 10$) \cite{Cartwright_chaotic_1996,Mezic_invariant_2001}.
The impeller-driven flow in \FIG{StirredTankFountain} exhibits this behaviour e.g. as a function of the tilt angle $\alpha$: aligned impellers ($\alpha=0$) yield closed circular streamlines about the cylinder axis; tilting ($\alpha\neq 0$) results in tori formation as in \FIG{SteadyCylinderFlow}. Hence, given realistic systems are inherently imperfect, tori are the LCSs that make-up the ``true'' non-chaotic baseline topology of both the cylinder flow and the other cases considered so far as well as most of the configurations discussed hereafter. Flows in micro-droplets are, for instance, a further manifestation of this scenario (\SEC{DropsIndustrial}).

\begin{figure}
\centering

\subfigure[Working principle]{\includegraphics[width=0.9\columnwidth,height=0.25\columnwidth]{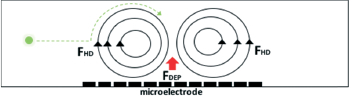}\label{ACEO1a}}
\subfigure[Simulated 3D flow topology]{\includegraphics[width=0.9\columnwidth,height=0.6\columnwidth]{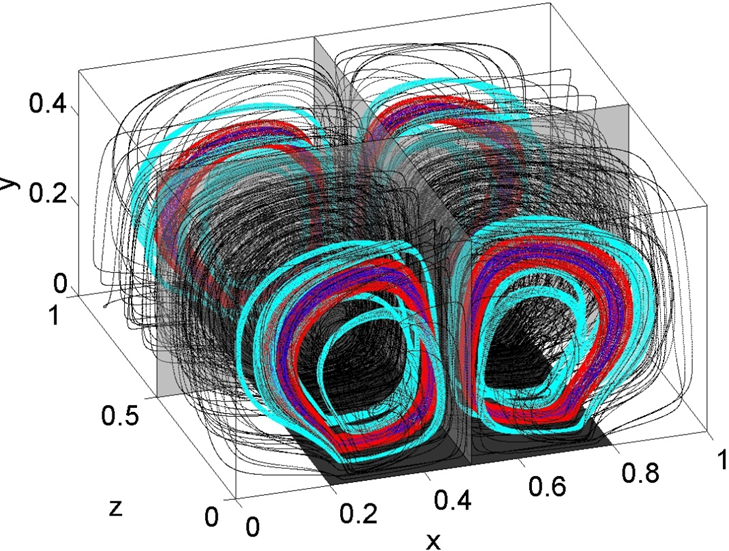}\label{ACEO1b}}
%\subfigure[Cross-section plane $x=0.25$]{\includegraphics[width=0.7\columnwidth,height=0.4\columnwidth]{Figures/ACEO2.jpg}\label{ACEO1c}}
\subfigure[Experimental 3D streamline portrait]{\includegraphics[width=0.9\columnwidth,height=0.6\columnwidth]{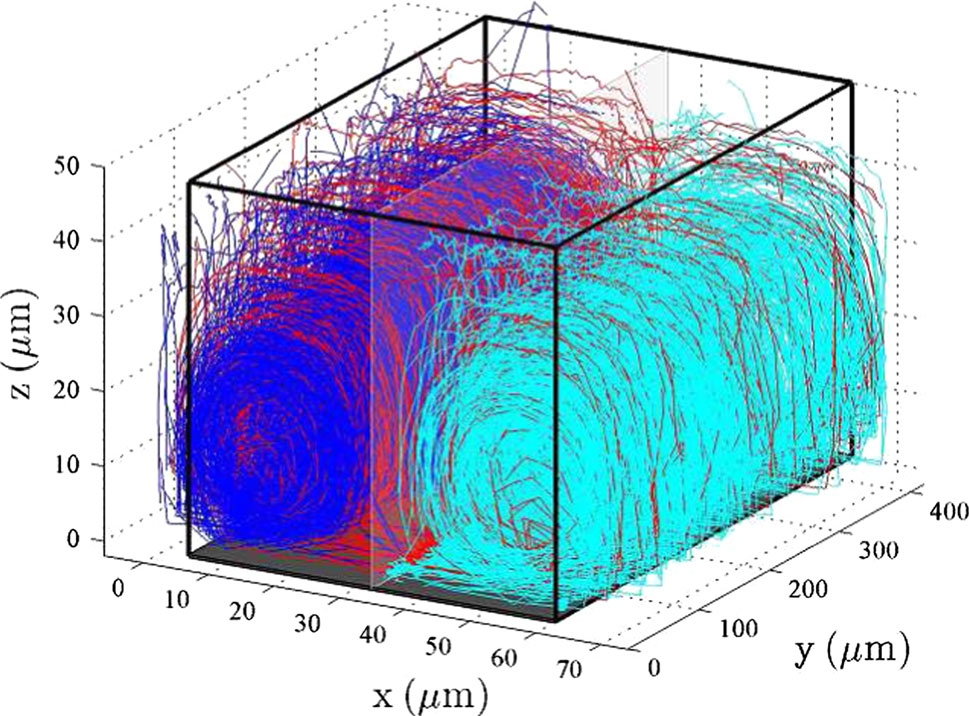}\label{ACEO2}}

%\caption{Steady ``container'' flow driven by surface forcing: trapping of nano-particles in vortical structures in ACEO-driven micro-fluidic flow. Panel a: adapted from \cite{Salafi2017}; panel b: adapted from %\cite{Speetjens2011}.}

\caption{Trapping of nano-particles in ACEO-driven vortical micro-fluidic flow as practical instance of steady lid-driven container flow: (a) particle trapping by interplay of centrifugal ($\vec{F}_{HD}$) and
dielectrophoretic ($\vec{F}_{DEP}$) forces (adapted from \cite{Salafi2017}); (b) simulated 3D flow topology comprising of KAM tori (coloured) embedded in chaos (black) (adapted from \cite{Speetjens2011}); (c) 3D streamline pattern measured by astigmatism $\mu$-PTV (reproduced from \cite{Liu2014}).
}
\label{ACEO1}
\end{figure}

A practical instance of 3D steady container flows driven by translational surface forcing, though not immediately apparent, exists in micro/nano-particle separation in micro-fluidic channel flows driven by AC electro-osmosis (ACEO) \cite{Salafi2017}. \FIG{ACEO1a} illustrates the working principle: micro-electrodes at the channel bottom induce a slip velocity in the electric double layer (EDL), setting up vortices that trap and concentrate particles by the resulting centrifugal forces ($\vec{F}_{HD}$). The local dielectrophoretic force ($\vec{F}_{DEP}$) at the electrode centre supports this process by pushing particles into the bulk flow. The ACEO-driven flow is dynamically equivalent to said container flows: (i) separatrices between adjacent vortices effectively compartmentalise the channel into confined domains; (ii) the EDL has a negligible thickness and the slip velocity thus effectively acts as an elastic translating wall. \FIG{ACEO1b} gives the 3D flow topology associated with each electrode (dark patch) \cite{Speetjens2011}. This exposes a composition of KAM tori (3 specimens highlighted in colour) and chaotic streamlines (black) within each of the compartments bounded by the separatrices (transparent planes) that is indeed similar to the lid-driven container flow in \FIG{SteadyCylinderFlow}.\footnote{The ACEO-driven flow is non-inertial ($Re=0$); here asymmetric slip velocity induces the universal tori formation from closed streamlines and their (local) breakdown into chaos demonstrated in \FIG{SteadyCylinderFlow} \cite{Speetjens2011}.}
Measurements of the 3D streamline pattern by astigmatism $\mu$-PTV (\FIG{ACEO2}) provides the first experimental validation of this structure \cite{Liu2014}. Here dark/light blue streamlines remain within the subdomains divided by the separatrix at $x=35$ $\mu$m (transparant plane). Red streamlines cross this separatrix due to experimental imperfections that, akin to tori formation from closed streamlines, act as natural perturbations of an idealised state. Essentially the same perturbation-induced separatrix crossing may occur in droplet flows (\SEC{DropsIndustrial}) and convection cells (\SEC{WebsIndustrial}); its underlying mechanisms are elaborated in \SEC{DropsIndustrial}.

The ACEO-driven flow is situated in a duct and thus admits transport characteristics of both duct and container flows. Imposition of an
axial ($x$-wise) pressure gradient of ``appropriate'' strength sets up a throughflow that passes over the vortical structures shown
in \FIG{ACEO1b} and thus yields a flow topology comprising of (diminished) tori at the electrode and stream-wise tubes as in \FIG{RAM2a} surrounded by a chaotic region \cite{Speetjens2011}. Tracers in this chaotic region exhibit both container-flow and duct-flow dynamics by prolonged entrapment in the vortical structure during their excursion from inlet to outlet. Similar hybrid behaviour occurs also in the flow inside an aneurysm shown in \FIG{fig:intro_examplesc} and is examined further in \SEC{ContainersNonIndustrial}.

%\red{Demonstrated in \FIG{CerebralAneurysm1b} for aneurysms.}

%\paragraph{Unsteady wall-driven flow}
%\paragraph{Unsteady container flows}
\paragraph{Unsteady flows}

\begin{figure}
\centering
%
%\subfigure[Impeller-driven non-Newtonian fluid]{\includegraphics[width=0.53\columnwidth,height=0.6\columnwidth]{Figures/StirredTankNonNewtonian.jpg}\label{Unsteady1a}}
\subfigure[Stirred non-Newtonian fluid]{\includegraphics[width=0.48\columnwidth,height=0.55\columnwidth]{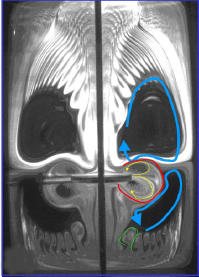}\label{Unsteady1a}}
\hspace*{3pt}
%
%\subfigure[EM stirring of glass melt]{\includegraphics[width=0.47\columnwidth,height=0.58\columnwidth]{./GlassMelt4.jpg}\label{Unsteady1b}}
\subfigure[EM stirring of glass melt]{\includegraphics[width=0.48\columnwidth,height=0.48\columnwidth]{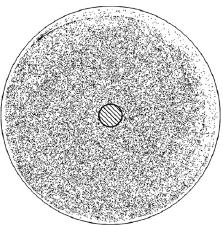}\label{Unsteady1b}}

\caption{Impact of unsteady flow on Lagrangian dynamics: (a) emergence of ``rippled'' structures (green) around tori (blue/yellow) signifying chaos in
rheology-induced time-periodic cylinder flow driven by rotating disk (red indicates high-shear region) visualised experimentally
by dye (adapted from \cite{Arratia2005}); (b) onset of global chaos due to time-periodic electro-magnetic (EM) stirring of glass melt visualised by
computational stroboscopic map (reproduced from \cite{Gopalakrishnan20105309}).}
%
%experimental visualisation of mixing pattern in rheology-induced {\it time-periodic} cylinder
%flow for {\it steady} stirring by rotating disk (red indicates high-shear region) exposing ``rippled'' structures (green) signifying chaos
%signifying chaos (green indicates lobe demarcated by corresponding manifolds)
%and enveloping tori (blue/yellow) (adapted from \cite{Arratia2005});
%
\label{Unsteady1}
\end{figure}

Dependence of the flow $\vec{u}$ on time $t$ expands the phase space of kinematic equation \eqref{KinEq}
from the 3D physical domain to the 4D space-time domain. This affords greater dynamic freedom and generically tends to promote chaotic advection.
%(Unsteadiness is in fact a necessary condition for chaos in 2D flows; \SEC{LagrangianView}.)
Reconsider for illustration the flow inside a cylindrical container driven by a rotating disk (\FIG{StirredTankFountain}). Stirring of a Newtonian fluid yields a
steady flow and corresponding flow topology consisting of KAM tori embedded in chaos (\FIG{StirredTankFountain}). Tilting of the disk is
imperative to attain this state; alignment with the cylinder axis yields a topology devoid of chaos and comprising entirely of
tori \cite{Fountain_3d_2000}. (Recall that this constitutes the ``true'' baseline topology emanating from a singular state consisting of closed streamlines.)
Non-Newtonian fluid rheology may fundamentally change the latter by (for {\it steady} stirring) inducing {\it time-periodic} oscillations and, in consequence, richer dynamics. \FIG{Unsteady1a}
demonstrates this for a fluid exhibiting both visco-elastic and shear-thinning behaviour stirred by a rotating disk aligned with the cylinder axis \cite{Arratia2005}.
Shown is a typical instantaneous dye distribution in the vertical cross-section that visualises the mixing pattern
and thus outlines the LCSs in the corresponding stroboscopic map as in \FIG{OrderWithinChaos}. The emergence of ``rippled'' structures (green) enveloping islands (blue/yellow)
implies a 3D flow topology of KAM tori and chaos basically similar to that emanating from the tilted disk (\FIG{StirredTankFountain_c}) and
the single Rushton impeller (\FIG{StirredTank2a}) for Newtonian fluids.
%(Blue/yellow indicates two tori families; green indicates lobe demarcated by manifolds of chaotic sea.)
Thus unsteadiness promotes chaos in a comparable way as a non-trivial geometry in said steady flows. Here
in particular the high-shear region around the impeller (indicated by red in \FIG{Unsteady1a}) is the physical cause of the rheology-induced time-periodicity.

Introduction of time-periodicity in the electro-magnetic stirring of glass melts by an oscillating magnetic field in \FIG{GlassMelt1} has essentially the
same effect: breakdown of the global family of tori in \FIG{GlassMelt1b} into chaos. \FIG{Unsteady1b} demonstrates this by the resulting global dispersion of 25,600 tracers released on a small line segment of length $l/R = 0.02$ in the axial midplane (top view) \cite{Gopalakrishnan20105309}.
First experimental evidence of 3D chaos in unsteady flows such as those in \FIG{GlassMelt1b} has been obtained by direct measurement of 3D stroboscopic maps of a single tracer via combined 2DPTV in two projections in a time-periodic swirling flow driven by alternate eccentric rotation of upper and lower endwalls \cite{Miles1995three}.

%\FIG{Unsteady1b} gives a typical 3D trajectory in the chaotic sea together with the instantaneous vertical velocity in the axial midplane \cite{Gopalakrishnan20105309}.

%\begin{figure}
%\centering
%
%\subfigure[...]{\includegraphics[width=0.47\columnwidth,height=0.47\columnwidth]{Figures/GlassMelt4.jpg}\label{GlassMelt2a}}
%
%\hspace*{6pt}
%
%\subfigure[...]{\includegraphics[width=0.47\columnwidth,height=0.47\columnwidth]{Figures/GlassMelt5.jpg}\label{GlassMelt2b}}
%
%\caption{Electro-magnetic stirring of glass melt ... time-periodic ... Reproduced/adapted from \cite{Gopalakrishnan20105309}.}
%\label{GlassMelt2}
%\end{figure}

The above container flows, notwithstanding substantial differences in configuration and forcing, in essence all concern cases
in which the Lagrangian dynamics are inextricably linked to the emergence and breakdown of KAM tori. This renders these flows dynamically analogous both to one another and to periodic duct flows with behaviour that, in general, is similar to that of 2D unsteady systems.
The link with 2D systems is broken only near possible singularities in transformation \eqref{Steady3DvsUnsteady2D} and/or upon development of 3D defects in KAM tori (\SEC{Conditions3Ddynamics}).

\begin{figure*}
\centering
\subfigure[Basic levels of (chaotic) Lagrangian dynamics]{\includegraphics[width=0.65\linewidth,height=0.45\columnwidth]{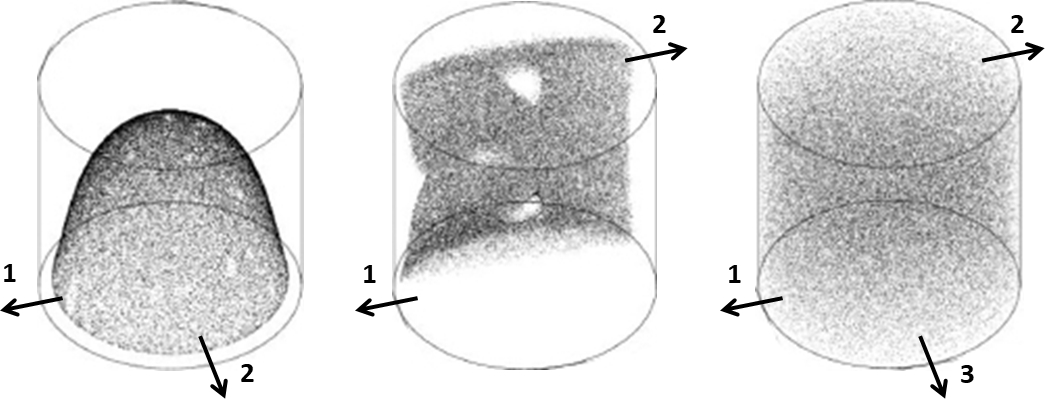}\label{UnsteadyChaos1_a}}
\hspace*{6pt}
\subfigure[Manifolds of isolated periodic points.]{\includegraphics[width=0.3\linewidth,height=0.45\columnwidth]{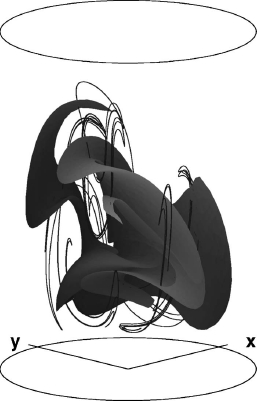}\label{UnsteadyChaos1_b}}
%
%\caption{Examples of essentially 3D unsteady phenomena in realistic flows demonstrated with time-periodic lid-driven (translating/rotating) cavity flows. Panel a adapted from
%\cite{Malyuga_cylinder_2002}; panel b reproduced from \cite{Speetjens_inertia_2006}.}

%\caption{Essentially 3D Lagrangian dynamics
\caption{Richness of Lagrangian dynamics in realistic 3D unsteady (container) flows demonstrated for time-periodic cylinder flow driven by step-wise steady
translations of endwalls: (a) basic dynamic levels exposed by time-periodic Stokes flow ($Re=0$) due to given forcing protocols (numbered arrows indicate
period-wise sequence of wall translations) visualised by computational stroboscopic map of single tracer (adapted from \cite{Malyuga_cylinder_2002}); (b) 1D (curve) and 2D
(surface) manifold of isolated periodic point in global chaotic sea ensuing from breakdown of invariant spheroids in panel (a) (left) by fluid inertia ($Re=100$)
(reproduced from \cite{Speetjens_inertia_2006}).}

\label{UnsteadyChaos1}
\end{figure*}

However, 3D unsteady flows admit far greater richness beyond that shown above and its exploration is still incomplete \cite{Wiggins2010,Aref2017}. \FIG{UnsteadyChaos1_a} demonstrates this by 3D time-periodic flows constructed from reorientations
of the steady lid-driven cylinder flow in \FIG{SteadyCylinderFlow} (for Stokes limit $Re=0$) by piece-wise steady translation of endwalls according to certain protocols. (Numbered arrows
indicate the sequence of periodic wall translations.) Shown stroboscopic maps correspond with a {\it single} tracer released in a given flow and expose intensifying ``levels'' of (chaotic)
dynamics: 2D chaos restricted to spheroidal surfaces (left); quasi-2D chaos in layers (centre); global 3D chaos (right) \cite{Malyuga_cylinder_2002}. (Dye visualisations experimentally
validate these basic ``dynamic levels'' of 3D Lagrangian transport \cite{Speetjens_stokes_2004,Anderson_cube_2006}.
%Moreover, first experimental evidence of 3D chaos as
%in \FIG{UnsteadyChaos1_a} (right) is obtained by direct measurement of 3D Poincar\'{e} sections of a single tracer by combined
%2DPTV in two projections in a time-periodic swirling flow driven by alternate eccentric rotation of upper and lower endwalls %\cite{Miles1995three}.
)
The second case concerns a topology of KAM tori (outlined by the ``holes'' in the stroboscopic map) embedded in chaos and thus belongs to the beforementioned class. The first and third cases, on the other hand, reflect essentially different situations.

%Lagrangian motion is in \FIG{UnsteadyChaos1_a} (left) restricted to {\it spheroidal} (instead of {\it toroidal}) invariant surfaces (shown is one specimen of a continuous family) and thus fundamentally distinct from that on KAM %tori.
Lagrangian motion is in \FIG{UnsteadyChaos1_a} (left) restricted to spheroids (shown is one specimen of a continuous family) and in consequence qualitatively deviates from that on KAM tori. (Refer to \SEC{Hamiltonian3Dunsteady} for the mechanisms behind the spheroids and the intrinsically 2D dynamics.)
%
%Lagrangian motion is in \FIG{UnsteadyChaos1_a} (left) restricted to {\it spheroidal} (instead of {\it toroidal}) invariant surfaces and thus
%fundamentally distinct.
%
Tori generically accommodate spiralling trajectories (as e.g. the magnetic field lines in \FIG{Tokamaka}) due to an interplay of poloidal and toroidal circulation.
%Spheroids preclude such double circulation on topological grounds and dictates dynamics equivalent to that of 2D container flows.
%Topological properties preclude such double circulation on spheroids and yield dynamics equivalent to that of 2D container flows.
The fundamentally different topology of spheroids precludes such double circulation and dictates dynamics equivalent to that of 2D confined
unsteady flows (\SEC{InvariantSurfaces}). Hence spheroids generically contain islands and chaos as e.g. the lid-driven cavity in \FIG{OrderWithinChaos}.\footnote{This is a consequence of the domain topology. Brouwer's well-known fixed-point theorem states that time-periodic flows in {\it convex} domains have (at least one) periodic point(s) following \eqref{PeriodicPoint} \cite{Brouwer1911}. Spheroids and 2D confined domains (as the lid-driven cavity) are convex domains and, by virtue of said theorem, both have flow topologies consisting of periodic points and associated LCSs. Moreover, periodic points preclude double circulation.}

Global chaos as in \FIG{UnsteadyChaos1_a} (right) sets in after total breakdown of transport-obstructing LCSs
as e.g. KAM tori or spheroids in response to some perturbation. The topology of such states generically
consists of (multiple) isolated periodic points \eqref{PeriodicPoint} and associated manifolds that, due to the 3D nature of the flow, materialise as surface--curve pairs. This is exemplified in \FIG{UnsteadyChaos1_b} by the 2D stable (convoluted surface) and 1D unstable (convoluted curve) manifolds of an isolated periodic point (not shown) in the globally-chaotic state that ensues from disintegration of said spheroids due to significant fluid inertia ($Re=100$) \cite{Speetjens_inertia_2006}.
%
%\paragraph{Interlude: tori versus spheres}
%
%

\paragraph{Intermezzo: toroidal versus spheroidal ``route to chaos''}
%\paragraph{\red{Intermezzo: toroidal versus spheroidal baseline topology}}

The particular ``route to chaos'' depends on the kind of LCSs constituting the baseline topology. The above findings imply
that -- from a {\it dynamical} perspective -- two canonical cases can be distinguished:
%chaos emanating from disintegration of (i) tori and (ii) spheroids
baseline topologies consisting of (i) tori and (ii) spheroids.\footnote{Shown examples suggest that baseline topologies in realistic flows generically consist of either tori or spheres yet in principle more complex situations are possible. Studies on kinematic models emulating 3D time-periodic flows demonstrate that a range of closed invariant surfaces of intricate shape accommodate essentially the same (chaotic)
dynamics as encountered on spheroids \cite{Gomez_invariant_2002}. Similar studies demonstrate that tori in 3D systems with non-trivial coexistence and nesting of multiple
tori families behave the same as in simpler (canonical) configurations \cite{Dombre1986,Rypina2015}.
%Hence, tori and spheroids ... Moreover, they are the only invariant surfaces found in realistic flows and thus the two generic/canonical cases.
}
Disintegration of KAM tori in both steady and unsteady 3D flows often (including the previous examples) follows the quasi-2D route dictated by their equivalence with 2D unsteady systems: progressive breakdown of tori into chaos with increasing perturbation according to well-defined scenarios (\SEC{Hamiltonian2D}).
%This results in the characteristic cross-sectional flow topologies comprising of a central family of islands (surviving tori) surrounded by %island chains and chaos (disintegrated tori) as shown in e.g. \FIG{Tokamakb}, \FIG{VortexBreakdownBubble} and \FIG{SteadyCylinderFlow_b}.
%
However, 3D systems may furthermore follow two essentially 3D ``routes to chaos'' associated with tori and spheroids. This is elaborated below.
%
%Two essentially 3D ``routes to chaos'' associated with KAM tori and spheres are elaborated below.

\begin{figure}
\centering
\subfigure[Resonance-induced dispersion (RID) for toroidal baseline]{
\includegraphics[width=0.54\columnwidth,height=0.5\columnwidth]{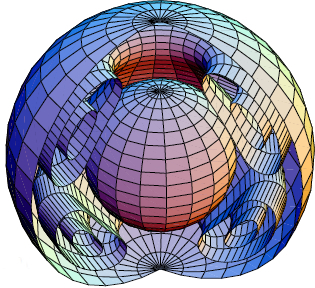}
\includegraphics[width=0.42\columnwidth,height=0.5\columnwidth]{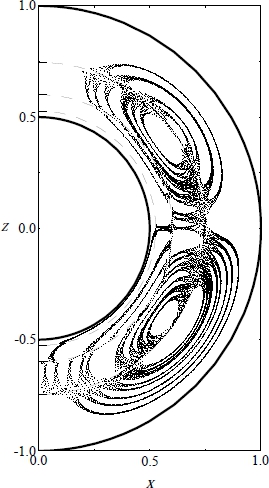}
\label{UnsteadyChaos2_a}}
\subfigure[Resonance-induced merger (RIM) for spheroidal baseline]{
\includegraphics[width=0.48\columnwidth,height=0.48\columnwidth]{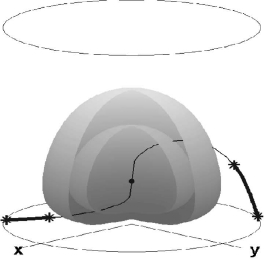}
\includegraphics[width=0.48\columnwidth,height=0.48\columnwidth]{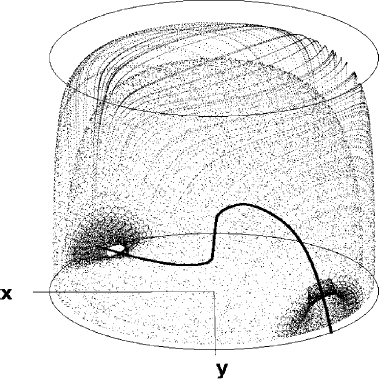}
\label{UnsteadyChaos2_b}}
\caption{3D ``routes to chaos'' for baseline topologies of tori versus spheroids: (a) global dispersion in time-periodic flow between two concentric spheres due to defects in tori (left) visualised by cross-sectional stroboscopic map of single tracer (right) (reproduced from \cite{Cartwright_chaotic_1996}); (b) formation of intricate LCSs in time-periodic lid-driven cylinder flow by merger of truncated tubes
forming on elliptic (heavy) segments of periodic lines and shells forming on chaotic spheroid sectors (left) due to weak fluid inertia ($Re=0.1$) visualised by 3D stroboscopic map of single tracer (right) (reproduced from \cite{Speetjens_inertia_2006}).
}
\label{UnsteadyChaos2}
\end{figure}

{\it Tori} may under certain conditions develop local defects that enable ``jumping'' of tracers between tori and, in consequence global
dispersion.
Such defects (locally) break the link with 2D unsteady systems and thus cause essentially 3D behaviour (\SEC{Conditions3Ddynamics}).
The study in \cite{Cartwright_chaotic_1996} investigates this for the time-periodic flow inside the 3D annular region between two concentric spheres driven by their co-axial rotation about axis $\vec{r} = \cos(\alpha)\vec{e}_z + \sin(\alpha)\vec{e}_x$, with $\alpha$ alternating between $\alpha = 0$ and $\alpha = 1^\circ$. The base flow ($\alpha = 0$) consists (for sufficiently small $Re>0$) of a primary ($\theta$-wise) circulation about the $z$-axis ($\vec{r} = \vec{e}_z$) and a weak secondary circulation in the $rz$-plane \cite{Bajer1992}. The corresponding  streamlines describe two families of tori (with a dense poloidal winding) centred on the $z$-axis and symmetric about plane $z=0$ following \FIG{UnsteadyChaos2_a} (left), implying a flow topology qualitatively the same as the steady cylinder flow in \FIG{SteadyCylinderFlow_0b}. The time-periodic flow, due to the only minute reorientation of the base flow during rotation about the tilted axis ($\alpha = 1^\circ$), yields a stroboscopic map consisting of basically the same tori as the base flow (\FIG{UnsteadyChaos2_a}; left). The principal difference is the emergence of defects in the tori that enable said ``jumping'' as demonstrated in \FIG{UnsteadyChaos2_a} (left) by the cross-sectional stroboscopic map of a {\it single} tracer. This results in a gradual global dispersion due to intermittent dynamics comprising of circulation {\it along} and jumping {\it between} tori at local defects (indicated by dashed semi-circles).

The mechanism causing the defects is the local synchronisation of the dynamics with the time-periodic forcing: the poloidal ($\phi$) and toroidal ($\varphi$) angles of certain tracers (in the local toroidal reference frame) evolve with rotations $\omega = \omega(\vec{x}_0)$ and $\gamma = \gamma(\vec{x}_0)$ relating via rational fractions, i.e.
%
%The mechanism underlying the defects is the local synchronisation of the dynamics with the time-periodic forcing: certain tracers (in the local %toroidal reference frame) have poloidal and toroidal angles $\phi$ and $\varphi$, respectively, that evolve with rotations $\omega = %\omega(\vec{x}_0)$ and $\gamma = \gamma(\vec{x}_0)$ relating via rational fractions, i.e.
%
\begin{eqnarray}
\phi_p = \phi_0 + p\omega,\quad
\varphi_p = \varphi_0 + p\gamma,\quad
\omega/\gamma = m/n,
\label{Resonance}
\end{eqnarray}
where $m$ and $n$ are integers. They describe trajectories that reconnect after $n$ toroidal windings. Tracers exhibiting such ``resonance'' prevent neighbouring trajectories from densely filling the entire torus and thus cause local defects (refer to \SEC{Conditions3Ddynamics} for the basic mechanisms).
Here this occurs in surfaces termed ``resonance sheets''; these sheets are unstable and under arbitrarily small perturbation disintegrate into localised chaotic zones that interrupt the poloidal circulation within tori and thus enable random jumps between them. This results in global transport following \FIG{UnsteadyChaos2_a} (right) and
is aptly termed ``Resonance-Induced Dispersion'' (RID) \cite{Cartwright_chaotic_1996}. The actual jumps in RID are facilitated by localised versions of global manifolds similar to
\FIG{UnsteadyChaos1_b} corresponding with isolated periodic points \eqref{PeriodicPoint} that remain from the disintegrated resonance sheets \cite{Mezic_invariant_2001}.

%\red{Resonances preclude global transformations following \eqref{Steady3DvsUnsteady2D} with the toroidal coordinate acting as canonical time $\tau$ and
%thus break the link with 2D unsteady systems. This rigorously establishes the essentially 3D nature of RID.
%}

Circumstantial experimental evidence of RID is obtained by the dispersion properties of fluorescent tracer particles in convection cells (\SEC{WebsIndustrial}) \cite{Solomon_resonant_2003}.
Moreover, RID has a direct counterpart in 3D steady flows in global dispersion following \FIG{UnsteadyChaos2_a} (right) enabled by local defects in tori induced by poloidal/toroidal evolutions
\begin{eqnarray}
\phi(t) = \phi_0 + \omega\, t,\quad
\varphi(t) = \varphi_0 + \gamma\, t,
\label{Resonance2}
\end{eqnarray}
with $\omega$ and $\gamma$ forming rational fractions following \eqref{Resonance} \cite{Vainchtein_resonant_2006}.
%and $\tau$ the streamline-specific period time of a toroidal cycle \cite{Vainchtein_resonant_2006}.
%
The notion of resonance furthermore generalises to defects in tori caused by significant slow-down of Lagrangian motion.\footnote{
%Local slow-down of Lagrangian motion is in fact the true cause for such defects.
%
The survival of tori subject to weak perturbation relies on the so-called ``averaging principle'', which assumes motion
along trajectories that (i) causes tracer positions to densely fill them over time and (ii) is significantly larger than the drift induced by the perturbation \cite{Arnold1978}. These assumptions break down near resonances as well as in slow-motion regions and yield local motion in all coordinate directions in the order of the perturbation strength. This results in essentially 3D dynamics that is no longer restricted to tori.
Thus slow motion is a generalisation of ``true'' resonances by (though not strictly precluding) strongly impeding the filling of entire trajectories and, inherently, tori formation. This is further
explained in \SEC{Conditions3Ddynamics}.
}
%
%that is significantly slower
%
%significantly slower motion along trajectories.
%
%near (isolated) periodic/stagnation points and/or
%
%separating material surfaces.\footnote{\red{In fact ``averaging principle'' ... }}
%
This e.g. underlies the separatrix crossing in the ACEO-driven flows, droplets flows (\SEC{DropsIndustrial}) and convection cells (\SEC{WebsIndustrial}).

{\it Spheroids} may also develop local defects due to resonances yet fundamental disparities exist with tori on grounds of the different topology and dynamics. Key is the formation of periodic lines through coalescence of isolated periodic points \eqref{PeriodicPoint} in the case
of spheroids. These lines intersect the spheroids as illustrated in \FIG{UnsteadyChaos2_b}
(left) for the two-step forcing by the bottom wall in \FIG{UnsteadyChaos1_a} (left). (Such points -- and thus the resulting periodic lines -- always exist on account of Brouwer's fixed-point theorem. Tori, on the other hand, are generically devoid of periodic points, which sets them fundamentally apart from spheroids \cite{Speetjens_inertia_2006}.) The impact of periodic lines on the response to weak perturbation depends fundamentally on the nature of the associated LCSs: chaotic seas (demarcated by manifolds
of the individual points on hyperbolic line segments) survive as thin shells; islands (centered on individual points of elliptic line segments) coalesce into tubes. Resonance here
ensues from the isolated degenerate periodic points (termed ``parabolic points'') that partition the periodic line into elliptic and hyperbolic segments and (instead of causing local
defects in otherwise complete tori as in RID) results in {\it truncation} of the {\it entire} tube. Shells and truncated tubes thus formed merge into intricate LCSs and this process is
therefore termed ``Resonance-Induced Merger'' (RIM). RIM is also an essentially 3D phenomenon by, similar to RID, breaking the link
with 2D unsteady systems (again refer to \SEC{Conditions3Ddynamics} for the basic mechanisms).

\FIG{UnsteadyChaos2_b} (right) demonstrates RIM due to perturbation of the flow topology in \FIG{UnsteadyChaos2_b} (left) by weak fluid inertia ($Re=0.1$), with heavy and normal curves indicating elliptic and hyperbolic line segments, respectively, separated by the isolated parabolic points (stars) that underlie resonance here \cite{Speetjens_inertia_2006,Speetjens_merger_2006}. Shown is the 3D stroboscopic map of a {\it single} tracer outlining an inner and outer shell connected by two tubes centred on the elliptic line segments. The LCS thus visualised (and formed by merger of shells and truncated tubes) is one member of a family of such LCSs that are arranged concentrically according to the truncated tubes.
Hence RIM ``creates'' new LCSs from existing ones; RID, on the other hand, (partially) destroys LCSs. Moreover, RIM, unlike the random jumps in RID, is a predictable and systematic phenomenon
in that families of truncated tubes form on elliptic line segments and merge with shells at tube ends at the well-defined positions of parabolic points. (Segmentation of periodic lines therefore
is key to RIM; it cannot occur for single-type lines.) The observation of RIM in other flow configurations as well as direct isolation of periodic lines and tube segments by experimental
stroboscopic maps using 3DPTV strongly suggests that RIM is a universal and robust phenomenon in flows with spheroidal LCSs and segmented periodic
lines \cite{Pouransari2010,Moharana2013,Znaien2012,Wu2014}.

Important to note is that RIM, though creating rather than destroying LCSs, nonetheless is on a ``route to chaos''. Said entities namely, instead of being strictly invariant, effectively constitute regions of prolonged (yet not indefinite) entrapment for (tens of) thousands of periods that in the long run admit global dispersion by accumulated transverse drifting. Moreover, incomplete shells may form that exhibit ``leakage'' of tracers into 3D chaotic zones \cite{Pouransari2010,Speetjens_inertia_2006,Speetjens_merger_2006}.

The route to 3D chaos from spheroids, in stark contrast with tori, is (to the best of our knowledge) hardly explored in the literature. However, the above findings as well as the considerations below suggest that this scenario may
%-- in a technological context -- be relevant specifically for container flows consisting of reoriented base flows.
be relevant specifically for unsteady container flows consisting of reoriented base flows.

%direct 3D ... from 2D unsteady systems.

%\begin{figure}
%\centering
%\subfigure[Resonance-induced dispersion]{\includegraphics[width=0.49\columnwidth,height=0.65\columnwidth]{Figures/RID1.jpg}\label{UnsteadyChaos2_a}}
%\subfigure[Resonance-induced merger]{\includegraphics[width=0.49\columnwidth,height=0.65\columnwidth]{Figures/RIM1.jpg}\label{UnsteadyChaos2_b}}
%
%\caption{Essentially 3D ``routes to chaos'' associated with \red{KAM tori and spheres} ... Panel (a,b) reproduced respectively from \cite{Cartwright_chaotic_1996}, \cite{Speetjens_inertia_2006}. %\red{Panel b: make 1 + 2 panels using Fig. 1a and 5? Or make full pagewidth, with panel b including Fig. 1a and 5a,b?}}
%\label{UnsteadyChaos2}
%\end{figure}
%

\paragraph{Unsteady flows revisited}

\begin{figure}
\centering
\subfigure[Configuration]{\includegraphics[width=0.8\columnwidth,height=0.32\columnwidth]{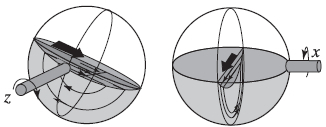}\label{SphericalTumbler_a}}
%\subfigure[...]{\includegraphics[width=0.9\columnwidth,height=0.6\columnwidth]{Figures/SphericalTumbler2.jpg}\label{SphericalTumbler_b}}
\subfigure[Equal rotation rates]{\includegraphics[width=0.48\columnwidth,height=0.28\columnwidth]{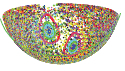}\label{SphericalTumbler_b}}
\hspace*{3pt}
\subfigure[Different rotation rates]{\includegraphics[width=0.48\columnwidth,height=0.28\columnwidth]{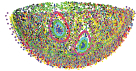}\label{SphericalTumbler_c}}
\caption{Signatures of RIM in mixing of granular media by time-periodic flow in half-full spherical
tumbler driven by alternating rotation axes: (a) configuration; (b) 2D (chaotic) dynamics within spheroids for equal rotation rates; (c) tube and
shell formation due to slight deviation in rotation rates. Adapted from \cite{Christov2014}.}

\label{SphericalTumbler}
\end{figure}

Mixing of granular media in spherical tumblers is an industrial application that (under the premise of continuum-like flow)
exhibits spheroidal LCSs and (signatures of) RIM \cite{Christov2014}. The configuration is given schematically in \FIG{SphericalTumbler_a}
and consists of a time-periodic flow in a half-full sphere driven by alternate rotation about $x$ and $z$-axes. Equal rotation rates
yield a flow topology comprising of spheroids and periodic lines and, in consequence, 2D dynamics similar to that in the
lid-driven cylinder flow in \FIG{UnsteadyChaos1_a} (left). \FIG{SphericalTumbler_b} demonstrates this by the computational stroboscopic map of multiple tracers (distinguished by colour) released within a given spheroid; green/blue curves are
periodic lines consisting entirely of hyperbolic/elliptic points and intersect the spheroid at the chaotic sea and
centre of the two opposite islands (outlined by circular orbits). Slight deviation in rotation rates triggers a similar
response as weak fluid inertia in the cylinder flow: the chaotic sea forms a thin shell; islands of adjacent spheroids
coalesce into tubes. \FIG{SphericalTumbler_c} demonstrates former and latter by weak drifting transverse to the spheroid
and incipient spiralling about the elliptic line (blue), respectively, resulting in a ``fuzzier'' stroboscopic map.
These dynamics are clear signs of RIM and, though actual tube-shell merger as in \FIG{UnsteadyChaos2_b} is not investigated in
\cite{Christov2014}, thus strongly suggest that the spherical tumbler follows this route to 3D chaos upon perturbation by varying rotation rates. An important outstanding question is whether the periodic lines (in appropriate parameter regimes) admit the segmentation into
elliptic and hyperbolic segments that is essential to RIM.

%Moreover, RIM may be more/less pronounced based on e.g. segmentation, extent of islands ... \cite{Pouransari2010}.\\\\

%\begin{itemize}
%\item[$\bullet$] Signatures of RIM yet less pronounced as in above cylinder flow (\FIG{UnsteadyChaos2_b}).
%RIM very delicate process; well-defined structures form only for sufficiently weak perturbation; become ``fuzzy''
%if too strong \cite{Speetjens_inertia_2006}. Moreover, segmentation is essential. Now only intersection
%of both lines instead of sizable segment.
%\end{itemize}
%

Unsteady transport phenomena as exemplified above are yet to find their way into analysis, design and technological development of other
industrial applications. However, accomplishment of 3D global chaos as in \FIG{UnsteadyChaos1_a} (right) by time-periodic agitation of
industrial batch mixers may significantly boost performance compared to conventional steady stirring that (implicitly) relies on
effectively 2D chaos due to breakdown of tori (\FIG{StirredTankFountain} and \FIG{StirredTank2}). Novel concepts such as glass processing by
time-periodic electro-magnetic stirring (\FIG{Unsteady1b}) take important first steps in this direction.

Micro-fluidic devices and emerging technologies that involve batch processing may particularly benefit from (insights into) unsteady
Lagrangian transport and chaos due to the high degree of flexibility and control offered by micro-fluidic forcing techniques and
domain shapes. Proper electrode layouts and (time-periodic) activation in ACEO-driven flows e.g. enable systematic reorientations of the steady ``base flow''
in \FIG{ACEO1b} so as to create various flow and transport conditions in micro-channels similar to the cylinder flow (\FIG{UnsteadyChaos1})
including (in principle) delicate processes such as RIM (\FIG{UnsteadyChaos2_b}). Suspended magnetic nano-particles facilitate
well-controlled body forcing reminiscent of the above electro-magnetic glass processing for basically any liquid and find
first applications for batch mixing in microfluidic platforms for biomedical and diagnostic purposes \cite{Grumann2005,Petkovic2017}.
The latter employ time-periodic magnetic forcing to accomplish chaotic advection and incorporation of essentially 3D insights
may enable further (technological) development of such devices. Moreover, this may enable new (diagnostic) functionalities
by systematically creating and utilising LCSs for e.g. particle trapping and/or separation similar to \FIG{StirredTankParticleClustering}.

\subsubsection{Relevance beyond industry}
\label{ContainersNonIndustrial}

%\blue{
%Discussed below are a number of non-industrial flow situations in which 3D Lagrangian transport and chaotic advection \red{in container flows} plays a central
%role. The overview again does not attempt to be exhaustive. ... concerns primarily physiological flows ... \red{Key difference: container with elastic instead of rigid wall.}
%}

Lagrangian (chaotic) transport in 3D container flows also play a central role in many (seemingly disconnected) non-industrial flow situations and
thus have fundamental similarities with the systems considered above. This is, as for duct flows (\SEC{DuctFlowNonIndustrial}), again illustrated via examples.

\paragraph{Aneurysms}

\begin{figure}
\centering

\subfigure[Streamline portrait of 3D mean flow]{
\includegraphics[width=0.49\columnwidth,height=0.45\columnwidth]{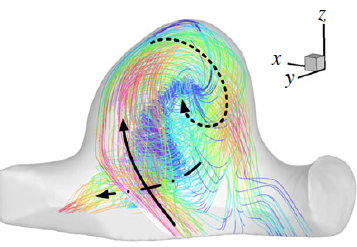}
\includegraphics[width=0.49\columnwidth,height=0.45\columnwidth]{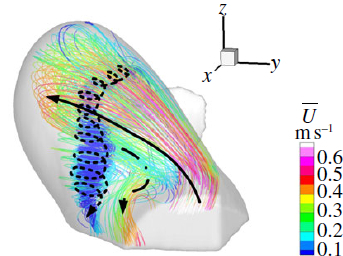}
\label{CerebralAneurysm1a}}
%\subfigure[...]{\includegraphics[width=0.49\columnwidth,height=0.45\columnwidth]{./CerebralAneurysmExperimental2.jpg}\label{CerebralAneurysmc}}
\subfigure[Hybrid duct/container-flow character of Lagrangian dynamics]{\includegraphics[width=0.9\columnwidth,height=0.65\columnwidth]{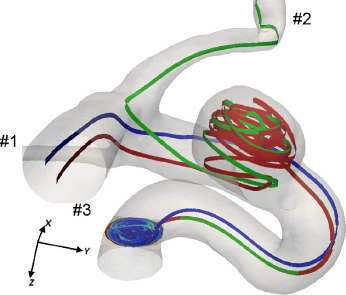}\label{CerebralAneurysm1b}}

%\caption{Transport of blood-borne substances in cerebral aneurysm as physiological instance of transport in time-periodic container flow: (a)
\caption{Flow in cerebral aneurysm as physiological instance of time-periodic lid-driven container flow: (a) streamline portrait of 3D mean flow
over cardiac cycle exposing inflow jet (solid arrow), vortical internal flow (dotted arrow) and outflow (dashed arrow) (reproduced from \cite{Yagi2013}); (b) hybrid duct/container-flow character of Lagrangian dynamics demonstrated by 3 tracer paths (distinguished by colour) released at inlet cross-section (reproduced from \cite{Zavodszky2015}).
}
\label{CerebralAneurysm1}
\end{figure}

Aneurysms are balloon-like malformations of blood vessels that, upon rupturing, lead to internal bleeding with potentially
lethal consequences. Transport of blood-borne substances is a key player in this process and in particular chaotic
advection (induced by the anomalous shape of aneurysms) has been implicated in the promotion of cardiac diseases (\SEC{Intro2} and
\ref{DuctFlowNonIndustrial}).
The typical flow in aortic aneurysms is visualised experimentally in \FIG{fig:intro_examplesc} by 3D Lagrangian trajectories
and consists of a vortical structure within the aneurysm (open arrows) driven by the passing aortic flow (closed arrows) \cite{Markl_heart_2011}.
\FIG{CerebralAneurysm1a} substantiates this generic structure by the 3D streamline portrait of the cyclic mean velocity field in a
laboratory replica of a cerebral aneurysm constructed from 2D stereo-PIV measurements \cite{Yagi2013}. This exposes a similar
vortical interior flow (dotted arrow) as well as the local in/outflow (solid/dashed arrow) at the attachment to
the cerebral artery that drives the former.
Thus the flow in the aneurysm resembles the above time-periodic lid-driven cylinder flow by consisting of a reorientation
of a circulatory base flow akin to \FIG{SteadyCylinderFlow} forced at its boundary; a non-essential difference is that the flood flow varies continuously
(instead of step-wise) during each cardiac cycle.
This resemblance implies unsteady 3D (chaotic) transport and its exploitation offers a way to investigate and unravel this process
in aneurysms so as to deepen insights into their origins and implicated vascular diseases as thrombosis
and atherosclerosis (\SEC{Intro2}).

A further analogy exists in the larger configuration of aneurysm and blood vessel combined. This situation is comparable to the previous ACEO-driven flow in conjunction with a main throughflow, implying similar interactions between subregions of the vortical flow in the aneurysm and the main
blood stream and, in consequence, similar hybrid behaviour including both duct-flow and container-flow characteristics.
The computational study by \cite{Zavodszky2015} investigates such 3D Lagrangian transport in the 3D time-periodic flow in a cerebral aneurysm induced
by the cardiac cycle using a realistic geometry based on CT angiography. \FIG{CerebralAneurysm1b} shows typical Lagrangian trajectories (distinguished
by colour) of three initially close fluid parcels released at the inlet of the single ingoing blood vessel. (The cross-sectional distribution at the
inlet gives the residence time of a dense set of parcels released here.) The parcel dynamics nicely demonstrates (i) said hybrid behaviour of prolonged
entrapment in the vortical structure inside the aneurysm during excursion from inlet to outlet and (ii) its essentially chaotic nature by the rapid
separation of the trajectories. Chaotic trajectories in the above ACEO-driven flow exhibit exactly the same behaviour \cite{Speetjens2011}.

\FIG{CerebralAneurysm1b} exposes a further ramification of chaotic advection specific for the aneurysm flow: deflection of trajectories into different outgoing blood vessels (labelled 1--3).
Such random distribution over multiple branches is in fact an important mechanism for both local and global dispersion in (random) porous media (\SEC{WebsIndustrial}) including
cardio-vascular networks (\SEC{WebsNonIndustrial}).

%Further example of ... translation of insights between (at first glance) different systems.

\paragraph{Lung alveoli}

\begin{figure}
\centering

\subfigure[Schematic structure of airways in lungs]{\includegraphics[width=\columnwidth,height=0.22\columnwidth]{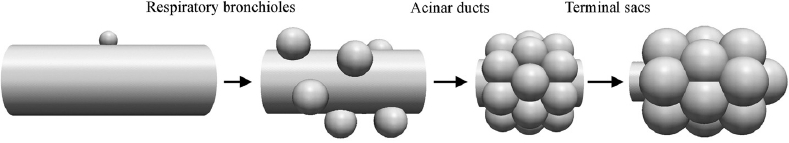}\label{LungAlveolia}}
\subfigure[Instantaneous 3D streamlines]{\includegraphics[width=0.49\columnwidth,height=0.45\columnwidth]{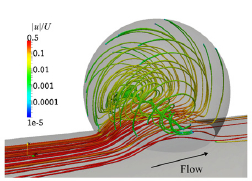}\label{LungAlveolib}}
\subfigure[Cross-sectional mixing pattern]{\includegraphics[width=0.49\columnwidth,height=0.45\columnwidth]{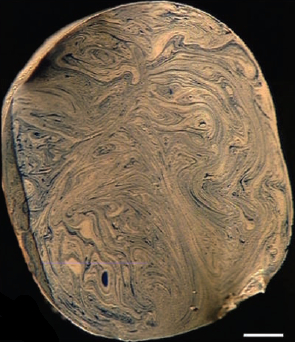}\label{LungAlveolic}}

%\caption{Aerosol mixing and deposition in lungs  ...  ... Panels a,b reproduced/adapted from \cite{Sznitman2013}; Panel c reproduced from %\cite{Tsuda2002} (bar = 500$\mu$m).}

\caption{Aerosol mixing and deposition in lungs as physiological instance of transport in time-periodic
container flow: (a) structure of airways; (b) simulated instantaneous 3D streamline portrait in lung alveolus (adapted from \cite{Sznitman2013}); (c) cross-sectional mixing pattern in alveolus visualised experimentally by polymerised liquids of different colour (bar = 500$\mu$m) (reproduced from \cite{Tsuda2002}).
}

\label{LungAlveoli}
\end{figure}

Flow and transport within the lungs, though physiologically entirely different, intimately relates to the above aneurysms. Air enters the
lungs via the respiratory bronchioles, passes through the acinar ducts and ends in the terminal sacs (\FIG{LungAlveolia}). These airways
become increasingly populated with hemispheroidal protrusions denoted ``alveoli'' that release air into the bloodstream. The internal
flow is primarily driven by the passing ductal flow and the configuration thus resembles that in aneurysms and, in turn, the time-periodic lid-driven cylinder flow. Compare to this end the simulated instantaneous 3D streamline pattern inside an alveolus in \FIG{LungAlveolib} with \FIG{SteadyCylinderFlow} and \FIG{CerebralAneurysm1a}. The similarity in flow configuration furthermore implies interaction between alveolic
and throughflow akin to that in aneurysms and, inherently, hybrid dynamics in the chaotic regions as demonstrated in \FIG{CerebralAneurysm1b}.

Aerosol mixing and deposition in the alveoli is essential to the overall functioning of the lungs and motivated a range of studies on Lagrangian transport in the pulmonary
acinus \cite{Tsuda1995,Butler1997,Tsuda2002,Tsuda_inevitable_2011,Sznitman2013}. This gave the key insight that chaotic advection, in stark contrast with its adverse effect in aneurysms, is vital for the kinematic irreversibility necessary to achieve net aerosol transport during each breathing cycle.
\FIG{LungAlveolic} experimentally validates stretching and folding -- signifying chaotic advection -- in
the airways by visualisation the cross-sectional mixing pattern using two polymerised liquids of different colour driven through the
bronchial tree of rat lungs by mechanical ventilation \cite{Tsuda2002}.
The majority of these studies considers 3D steady flow in an axi-symmetric alveolus wrapped around a
bronchiolus, implying a toroidal geometry and, inherently, essentially 2D unsteady dynamics equivalent to that of periodic duct
flows (\SEC{Ducts}). Moreover, the explicit links with Lagrangian concepts established in \cite{Tsuda_inevitable_2011} also adopt this configuration.

The true system, notwithstanding the evident importance of said studies, concerns 3D unsteady dynamics in a container flow driven by translational
boundary forcing according to the analogy with aneurysms made before, however.
Thus flow and Lagrangian dynamics in alveoli must (on topological grounds) in fact also resemble that of the time-periodic lid-driven
cylinder flow.
%
%, suggesting that, besides chaos, ... .\\\\
%
%The asymmetry of the latter about plane $x=0$ depends on $Re$, suggesting that, besides chaos, the irreversibility may (at least in part) ... %fluid inertia.\\\\
%
This is supported by computational studies on (instantaneous) 3D streamlines and Lagrangian transport of (finite-size) aerosols in spheroidal
configurations \cite{Haber2003,Kumar2009,SemmlerBehnke2012,Henry2012,Sznitman2013}.
%
%Hence flow and Lagrangian dynamics in alveoli must (on topological grounds) in fact be akin to that of the cylinder flow in \FIG{SteadyCylinderFlow}.
%
The coherent structures in the instantaneous streamline patterns denoted ``Russian nesting dolls'' in \cite{Henry2012} in particular are directly
equivalent to the toroidal streamlines in \FIG{SteadyCylinderFlow_a}, implying a cross-sectional stroboscopic map of the corresponding time-periodic
flow akin to \FIG{SteadyCylinderFlow_b}.
Hence, as for aneurysms, in-depth investigations of the 3D alveolic transport based on the analogy with
3D lid-driven container flows may significantly deepen insights.

\paragraph{Gastric digestion}

%A further physiological example of 3D time-periodic flow in an essentially spheroidal domain driven by boundary movement exists in the
%digestive process inside the stomach. Goal is efficient mixing of the gastric contents ...

\begin{figure}

\subfigure[Configuration]{\includegraphics[width=0.49\columnwidth,height=0.5\columnwidth]{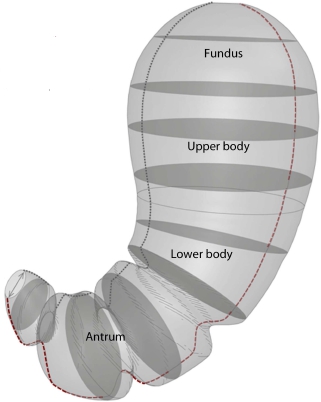}\label{Stomach_a}}
\subfigure[Typical 3D instantaneous streamlines]{\includegraphics[width=0.5\columnwidth,height=0.45\columnwidth]{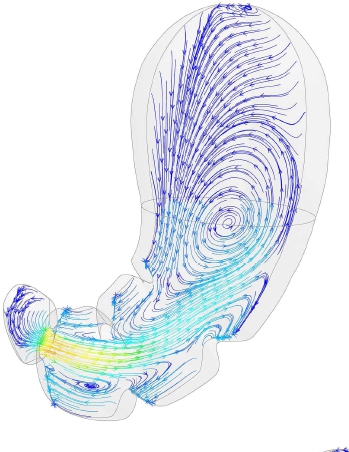}\label{Stomach_b}}

%\caption{3D advection in the stomach ... adapted from \cite{Ferrua2014}.}%

\caption{Gastric digestion as physiological instance of transport in time-periodic container flow: (a) configuration; (b) simulated instantaneous 3D streamline portrait (adapted from \cite{Ferrua2014});
}

\label{Stomach}
\end{figure}

A further physiological example of transport in 3D time-periodic container flows exists in the digestive process inside the stomach. The objective is efficient
extraction of nutrients by the mixing and ``kneading'' of the gastric contents via peristaltic movement of the stomach wall through
so-called ``antral contraction waves''. These waves periodically compress and expand the stomach and its contents and thus set up
a 3D time-periodic flow driven by movement at/of the boundary reminiscent of aneurysms and lung alveoli and, by the beforementioned analogy,
3D lid-driven cavity flows. The main differences with said examples are (i) that the stomach, though topologically again spheroidal, has
a more complex shape (\FIG{Stomach_a}) and (ii) involves elaborate forcing by the {\it entire} boundary. This results in the
formation of multiple families of vortical structures of varying extent and intensity as exemplified in \FIG{Stomach_b} by typical
instantaneous 3D streamlines \cite{Ferrua2014}.
The individual structures -- particularly in the central part of the stomach -- clearly resemble those in \FIG{SteadyCylinderFlow_a}.

The highly non-trivial interplay of the vortical structures in the course of a period are likely
to produce (at least local) chaotic advection and first exploratory studies on simplified systems indeed suggest this \cite{Arrieta2015}.
However, the contribution of advective transport to actual gastric mixing remains unclear yet is considered important for better
understanding of this process \cite{Bornhorst2017}. Hence, this is a further problem that may benefit considerably from insights in
3D Lagrangian transport and analogies with the above systems.

%\paragraph{Mixing in a champagne glass}
\paragraph{Champagne tasting}

A perhaps surprising (though appealing) instance of 3D transport in container flows emerges in champagne
tasting. Transport and mixing due to the internal flow driven by bubble formation upon pouring into the
glass is namely believed key to the exhalation of flavours and aromas and thus the taste
of the champagne \cite{Beaumont2013}.

\begin{figure}
\centering
\includegraphics[width=\columnwidth,height=0.75\columnwidth]{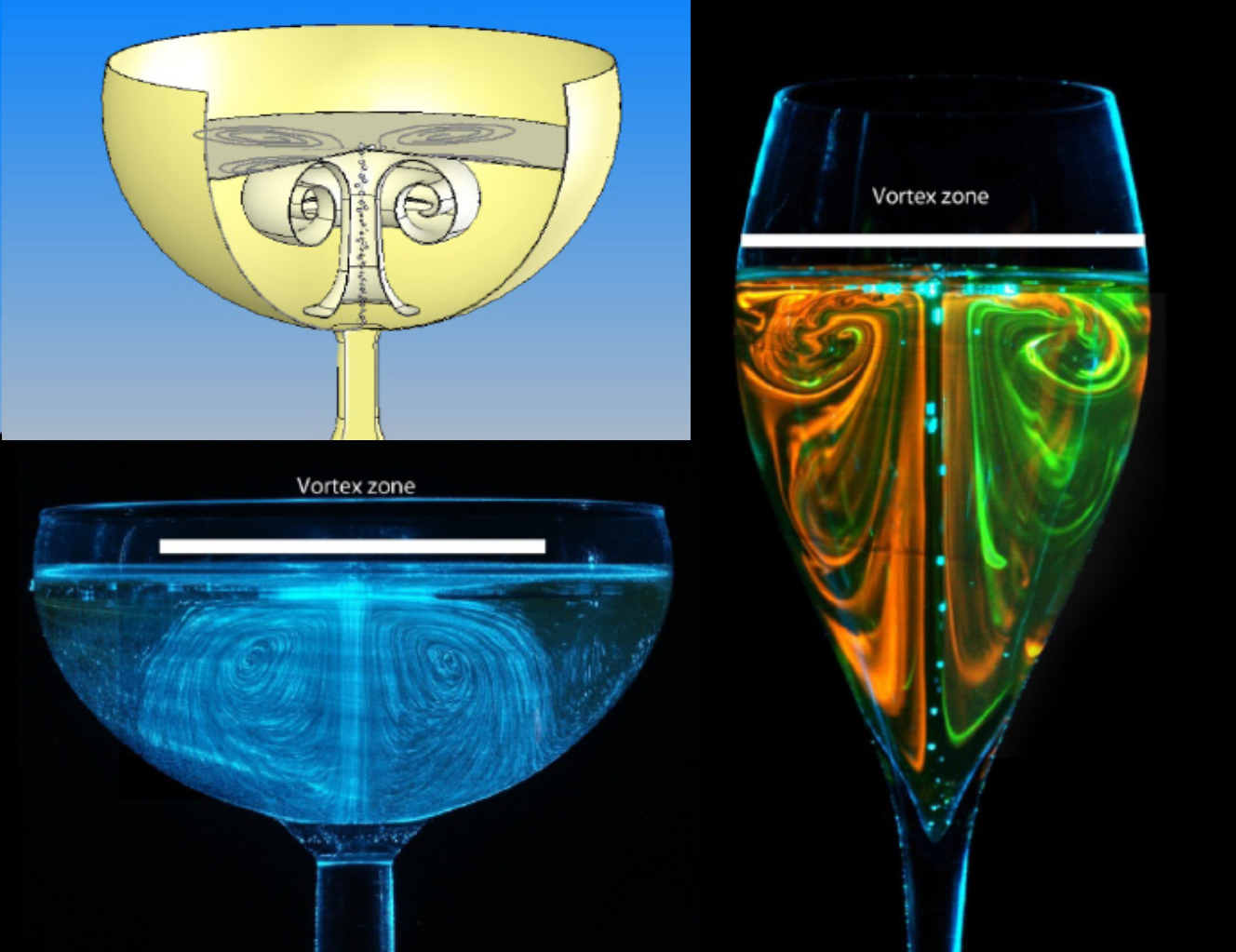}

\caption{Flow and mixing in champagne glass as culinary instance of (transport) in 3D container flows: (top, left)
cross-sectional circulation driven by central bubble column; (bottom, left) experimental visualisation in couple glass by fluorescent dye; (bottom, right) experimental visualisation in flute by polymer particles (adapted from \cite{Beaumont2013}).}
\label{Champagne}
\end{figure}

Champagne glasses typically have engravings at the bottom as nucleation site for a central column of rising CO$_2$ bubbles as driver of a cross-sectional circulation in the $rz$-plane following \FIG{Champagne} (top, left). This is visualised in \FIG{Champagne} by vertical cross-sections of fluorescent dye in a coupe glass
(bottom, left) and polymer particles in a flute (bottom, right). The vortical structures thus created are qualitatively similar yet their extent depends on the
type of glass \cite{Beaumont2013}.

The 3D flow constitutes a vortex ring with a corresponding 3D flow topology reminiscent of that below the disk impeller in \FIG{StirredTankFountain} yet with one fundamental difference.
The impeller actively imposes uni-directional azimuthal flow and, by virtue of the resulting analogy with periodic duct flows, gives rise to KAM tori and associated chaotic
regions (\SEC{ContainersIndustrial}).
The champagne flow, on the other hand, stems from passive internal forcing by the (approximately) axi-symmetric bubble column. Its small asymmetric fluctuations
induce weak azimuthal flow without one preferential direction and, in consequence, continuous reversal in the course of time.
%
%This precludes a single transformation following \eqref{Steady3DvsUnsteady2D} with the azimuthal coordinate $\theta$ acting as canonical time $\tau$ and thus implies essentially 3D unsteady
%dynamics.
%
%Dominant cross-sectional circulation in combination with weak fluctuations of the azimuthal flow around a zero mean substantiates the above findings. This namely are
%key ingredients for resonance-induced defects in tori and ensuing global dispersion as in \FIG{UnsteadyChaos2_a} (right). These considerations suggest that
%transport and mixing in the champagne glass may be highly non-trivial and of unexpected richness.
%In-depth Lagrangian analysis -- with a particular focus on resonance phenomena -- enables rigorous exploration of this.
%
Such dominant cross-sectional circulation in combination with weak azimuthal fluctuations are in fact key ingredients for resonance-induced defects in tori and, in consequence, RID as in \FIG{UnsteadyChaos2_a} (right). RID namely ensues from weak perturbation of an
idealised flow topology consisting of closed streamlines (\SEC{Conditions3Ddynamics}). These considerations suggest that transport and mixing in the champagne glass may be highly non-trivial and
of unexpected richness. In-depth Lagrangian analysis -- with a particular focus on resonance phenomena -- enables rigorous exploration of this.

\begin{figure}
\centering
%\includegraphics[width=\columnwidth]{./Cilia_on_Coral2.jpg}
%\includegraphics[width=0.5\columnwidth]{./Cilia_on_Coral3.jpg}

%\subfigure[]{\includegraphics[width=0.55\columnwidth,height=0.4\columnwidth]{Figures/Cilia_on_Coral4.jpg}\label{CiliaDrivenFlow_a}}
%\subfigure[]{\includegraphics[width=0.44\columnwidth,height=0.4\columnwidth]{Figures/Cilia_on_Coral3.jpg}\label{CiliaDrivenFlow_b}}

\includegraphics[width=0.53\columnwidth,height=0.4\columnwidth]{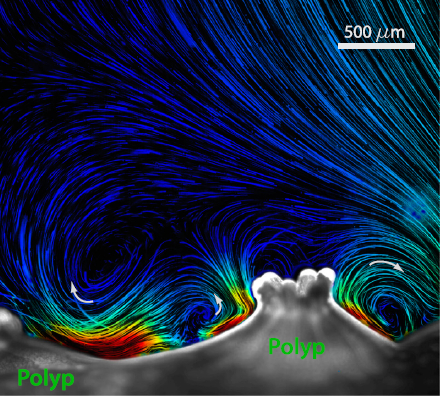}
\hspace*{3pt}
\includegraphics[width=0.43\columnwidth,height=0.4\columnwidth]{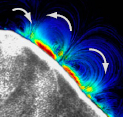}

\caption{Cilia-driven vortical flow at coral polyps as biological instance of (transport) in 3D lid-driven container
flows demonstrated by typical flow patterns above coral explants visualised experimentally by 2DPTV. Colour represents flow speed (blue: stagnant;
red: $\mathcal{O}(1\; {\rm mm/s})$); white arrows indicate circulation (adapted from \cite{Shapiro2014}).}

\label{CiliaDrivenFlow}
\end{figure}

\paragraph{Cilia-driven flows at coral reefs} Transport in the cilia-driven flow at polyps in coral reefs shown \FIG{fig:intro_examplesb}, while, as for the ACEO-driven flow (\SEC{ContainersIndustrial}), not immediately
apparent, intimately relates to flows in containers.
% in that this concerns a boundary-driven flow in Nature.
%
% as counterpart to the lid-driven technical flows.\\\\
%
The cilia, though strictly in an open domain, namely create a boundary-driven flow in a confined region adjacent to the coral surface consisting of
vortical structures comparable to that of the boundary-driven container flows shown in \FIG{SteadyCylinderFlow} and \FIG{ACEO1b}. This is
demonstrated in \FIG{CiliaDrivenFlow} for typical flow patterns above coral explants obtained by 2DPTV using powdered coral food as tracer
particles \cite{Shapiro2014}. The 2DPTV measurements reveal arrays of vortices with alternating circulation (indicated by white arrows) involving flow speeds
ranging from stagnant fluid (blue) to a maximum velocity of $\mathcal{O}(1\; {\rm mm/s})$ (red). Note in particular the striking resemblance of the pair
of counter-rotating vortices in \FIG{CiliaDrivenFlow} (right) with the ACEO-driven flow in \FIG{ACEO1b}.
These analogies imply similar Lagrangian transport as in said container flows.

\subsection{Flow in drops}
\label{Drops}

%\subsubsection{Introduction}
\subsubsection{General configuration}
\label{DropsConfig}

Neglecting two-phase effects as drop coalescence and breakup, the flow inside a drop is in a dynamical sense a subclass of flow in containers considered above. They namely both concern confined flows in (topologically) spheroidal domains and the corresponding Lagrangian transport thus is subject to the same
kinematic and topological constraints. The primary difference with container flows is that drops are bounded by a fluid-fluid interface instead of an
(elastic) solid wall and thus offer functionalities (particularly in micro-fluidic systems) such as immersion in a bulk flow and external forcing by
techniques unavailable to the former.
%
%(i) generically has a flexible shape and (ii) admits a non-zero parallel velocity \red{(so does lid-driven cavity)}.
%
Hence (potential) technological applications involving drops are entirely different from the above container flow and
this motivates its distinction as a separate flow class.

\subsubsection{Industrial and technological relevance}
\label{DropsIndustrial}

Lagrangian transport and mixing in drops is relevant to a range of micro-fluidic applications and emerging
technologies. Two basic configurations can be distinguished:
\begin{itemize}
\item[$\bullet$] {\it Microfluidic networks} Here droplets propagate in an ambient flow through (interconnected) micro-channels
and chambers and serve both as processing and transportation vessels.
This principle finds application in micro-fluidic devices for batch processing of samples in e.g. pharmaceutical manufacturing or
bio-engineering using droplets as chemical reactors or cell incubators so as to boost reaction rates, shield
against contamination and/or provide protected environments \cite{Bringer2004,Stone2005,Kintses2010,Levin2012,Lagus2013,Vitor2015}.

%diagnostic or bio-medical analyses involving step-wise treatment and examination of samples in consecutive modules %\cite{Bringer2004,Stone2005,Kintses2010,Levin2012,Lagus2013,Vitor2015}.

\item[$\bullet$] {\it Digital microfluidic platforms} Here discrete droplets are trapped (and displaced) on solid substrates.
Applications of this approach include processing and analysis of microscopic fluid quantities as e.g. chemical synthesis of drugs or
immunoassays for clinical diagnostics of minute blood samples \cite{Fair2007,Pollack2011,Choi2012}.

%via the surface tension of the fluid and ... internal processing ... internal flow driven/manipulated by .. surface by electro-wetting ... %actuation of internal flow by acoustic streaming ... e.g. examination of blood droplets ...  clinical diagnostic systems ... %\cite{Fair2007,Pollack2011,Choi2012} \blue{Application: small liquid quantities ... clinical diagnostic systems as immunoassays for examination %of minute blood samples ... chemical synthesis in micro-scale fluid quantities }

\end{itemize}
A great diversity of techniques is available for actuation and manipulation of the internal flow within microscopic drops.
%
%However, the corresponding Lagrangian transport and mixing is subject to essentially the same generic kinematic and topological constraints as the previous container flows and droplet-based %micro-fluidic systems, irrespective of the specific configuration, thus share fundamental properties with the latter as well as among each other. This is demonstrated below by way of examples.
%
The intimate connection with container flows nonetheless means that droplet-based micro-fluidic systems, irrespective of the specific configuration, share
fundamental properties with the former as well as among each other. This is elaborated below for several flow-forcing techniques.

%and can be exploited for (further) deepening insights and development of systems.

%\red{Can be exploited for (further) development of systems and methods.}

%\paragraph{Flow/mixing via ambient flow}
\paragraph{Internal flow induced by ambient flow}

\begin{figure}
\centering
%\subfigure[Inducing chaos in drops by propagation in curved micro-channels]{\includegraphics[width=\columnwidth,height=0.25\columnwidth]{Figures/ChaosMicroDroplets.jpg}\label{ChaosDroplets_a}}
%\includegraphics[width=\columnwidth,height=0.35\columnwidth]{Figures/ChaosMicroDroplets.jpg}
\subfigure[Transport in straight channels]{\includegraphics[width=\columnwidth,height=0.2\columnwidth]{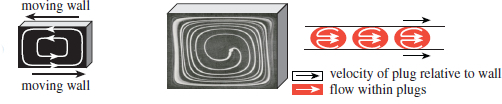}\label{ChaosDroplets_a}}
\subfigure[Transport in curved channels]{\includegraphics[width=\columnwidth,height=0.2\columnwidth]{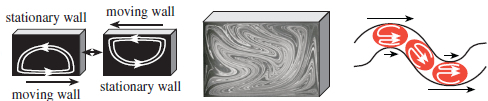}\label{ChaosDroplets_b}}

%\caption{Transport in droplets immersed in micro-channel flow versus channel geometry:
%\caption{Transport in droplets in micro-channel flow versus lid-driven cavity flow:
\caption{Transport in propagating droplets (``plugs'') in micro-channel flow versus channel geometry:
%
%(a) equivalence straight channels with steady cavity flow implies non-chaotic transport;
%(b) equivalence curved channels with time-periodic cavity flow paves way to chaotic transport
%
(a) straight channels yield non-chaotic transport due to equivalence with steady cavity flow;
(b) curved channels admit chaotic transport due to equivalence with time-periodic cavity flow
%
%(a) non-chaotic transport in straight channels due to equivalence with steady cavity flow; (b) chaotic transport in curved channels due to equivalence
%with time-periodic cavity flow
(adapted from \cite{Ottino_Kinematics_1989}).}
\label{ChaosDroplets}

%Straight channels, by virtue of constant droplet shape and propagation speed, compare to steady wall
%forcing and thus yield a steady global circulation with typically non-chaotic dynamics (Fig. 34(a)). Curved channels
%with a spatially periodic geometry, on the other hand, give rise to oscillating droplet shape and speed and, analogous to
%time-periodic wall forcing, result in a time-periodic internal flow that admits chaotic dynamics (Fig. 34(b))

\end{figure}

Droplets immersed in a micro-channel flow driven by e.g. a pressure gradient are subject to shear stresses at the fluid-fluid
interface (and generically travel slightly faster than the bulk) \cite{Stone2004}. This yields a non-zero surfacial velocity
in said interface and thus induces an internal flow within the drop akin to lid-driven cavities (\FIG{ChaosDroplets}).
Straight channels, by virtue of constant droplet shape and propagation speed, compare to steady wall forcing and thus
yield a steady global circulation with typically non-chaotic dynamics (\FIG{ChaosDroplets_a}).
Curved channels with a spatially periodic geometry, on the other hand, give rise to oscillating droplet shape and speed and, analogous
to time-periodic wall forcing, result in a time-periodic internal flow that admits chaotic dynamics (\FIG{ChaosDroplets_b}) \cite{Bringer2004}.

\begin{figure}

\subfigure[Serpentine micro-channel]{\includegraphics[width=\columnwidth,height=0.25\columnwidth]{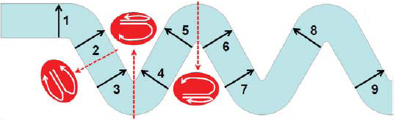}\label{ChaosDroplets1_a}}
\subfigure[Experimental mixing pattern]{\includegraphics[width=\columnwidth,height=0.35\columnwidth]{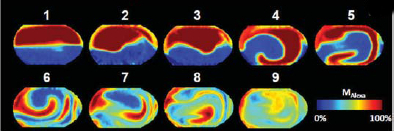}\label{ChaosDroplets1_b}}
\subfigure[Computational mixing pattern]{\includegraphics[width=\columnwidth,height=0.35\columnwidth]{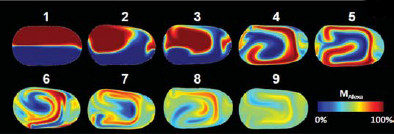}\label{ChaosDroplets1_c}}

\caption{Mixing inside propagation droplet immersed in serpentine micro-channel: (a) configuration; (b) experimental visualisation of mixing
pattern by fluorescent imaging microscopy; (c) computational visualisation of mixing pattern (adapted from \cite{Jiang2012}).}

\label{ChaosDroplets1}
\end{figure}

Practical implementation of the above working principle is found in a droplet travelling through a serpentine micro-channel
according to \FIG{ChaosDroplets1_a} \cite{Bringer2004,Jiang2012}. \FIG{ChaosDroplets1_b} visualises the internal mixing pattern by the
evolution of a binary concentration distribution in the droplet midplane using fluorescent lifetime imaging microscopy and reveals the characteristic stretching
and folding of chaotic advection. The corresponding simulated evolution is shown in \FIG{ChaosDroplets1_c} and closely agrees with the experiments.

\begin{figure}

%\subfigure[Theoretical topology buoyant (left) and non-buoyant (right) droplet.]{
\subfigure[Ideal topology buoyant (left) and non-buoyant (right) droplet.]{
\includegraphics[width=0.42\columnwidth,height=0.4\columnwidth]{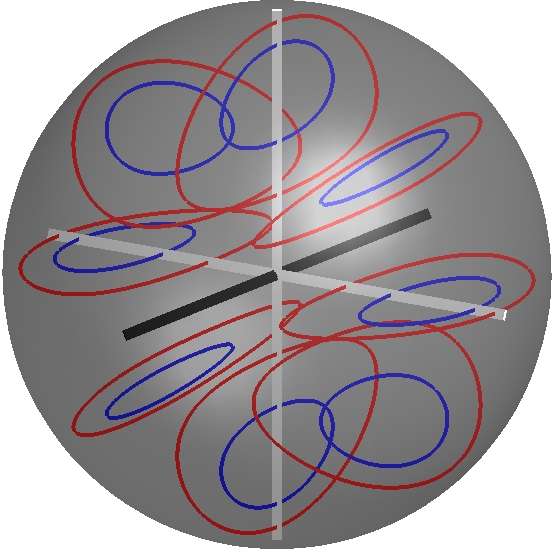}
\hspace*{6pt}
\includegraphics[width=0.42\columnwidth,height=0.4\columnwidth]{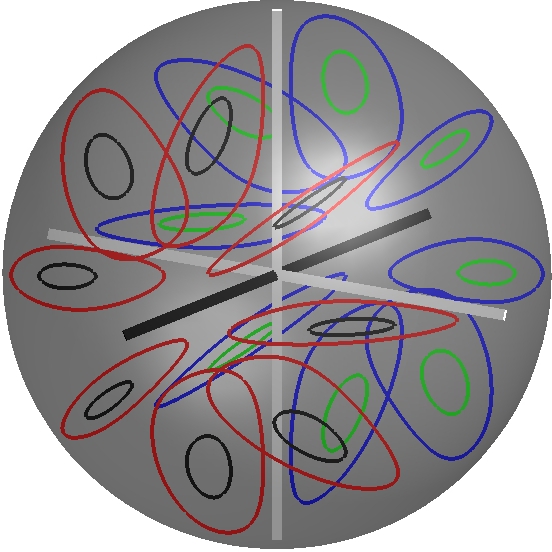}\label{ChaosDroplets2X_a}
}

\subfigure[Real topology buoyant (left) and non-buoyant (right) droplet.]{
\includegraphics[width=0.42\columnwidth,height=0.4\columnwidth]{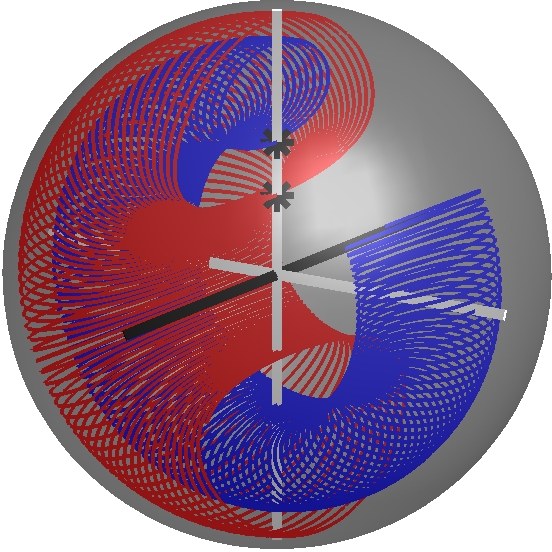}
\hspace*{6pt}
\includegraphics[width=0.42\columnwidth,height=0.4\columnwidth]{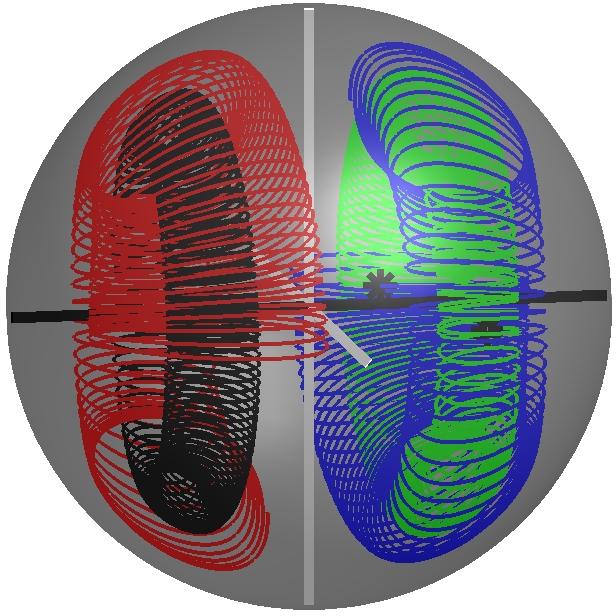}\label{ChaosDroplets2X_b}
}

\subfigure[Mixing pattern in serpentine-channel flow]{\includegraphics[width=\columnwidth,height=0.28\columnwidth]{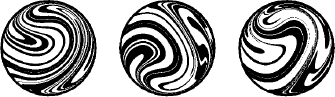}\label{ChaosDroplets2X_c}}

\caption{Buoyant gravity-driven and non-buoyant stress-driven drops as basic flow configurations: (a) 3D topologies for ideal
conditions (black indicates $z$-axis and direction of travel; gray indicates $(x,y)$-axes); (b) 3D topologies for realistic conditions; (c) simulated
mixing pattern in serpentine channel due to switching between buoyant and non-buoyant droplet flow. Panel (c) adapted from \cite{Stone2005}.
}
\label{ChaosDroplets2X}
\end{figure}

In-depth experimental analysis of 3D flow and transport within immersed (propagating) microscopic drops is highly challenging and largely
beyond current experimental methods. Hence its exploration to date primarily involves theoretical and computational
studies.
Two standard cases underlying such studies are: (i) buoyant drops driven by gravity; (ii) non-buoyant drops driven by viscous stresses at the
fluid-fluid interface. The non-dimensional 3D Stokes flow inside droplets of unit radius is in the co-moving reference frame, with $z$ the direction of
travel, described analytically by
\begin{eqnarray}
u_x = zx,\quad u_y = zy,\quad u_z = 1 - 2r_\ast^2 + z^2,
%\label{HillVortex}
\label{BuoyantDrop}
\end{eqnarray}
and
\begin{eqnarray}
u_x &=& x(r_\ast^2 + 2z^2 - 1),\nonumber\\
u_y &=& y(r_\ast^2 + 2z^2 - 1),\nonumber\\
u_z &=& 2z(1 - 2r_\ast^2 + z^2),
%\label{HillVortex}
\label{NonBuoyantDrop}
\end{eqnarray}
for former and latter case, respectively, with $r_\ast^2 = x^2 + y^2 + z^2$ \cite{Stone_steady_1999}.\footnote{Flow \eqref{BuoyantDrop} corresponds with Hill's spherical vortex and
is an analytical solution to both the Euler equations governing inviscid flow and the Stokes equations governing the flow inside a buoyant spherical droplet rising in a viscous
fluid \cite{Bajer_class_1990}. The latter, in turn, corresponds with the Hadamard-Rybczynski problem in the non-inertial limit $Re=0$ \cite{Stone2005}.}
\FIG{ChaosDroplets2X_a} gives the flow topologies for \eqref{BuoyantDrop} (left) and \eqref{NonBuoyantDrop} (right) comprising, by virtue of axi-symmetric flow ($u_\theta = 0$), of
two and four ``islands'' of closed streamlines in planes of constant $\theta$, respectively.
(The black line indicates the $z$-axis and direction of travel.)
However, these flow topologies -- analogous to the cylinder
flow in \FIG{SteadyCylinderFlow} -- are singular limits in that {\it any} perturbation induces transition into (multiple families of) tori (\SEC{DuctFlowIndustrial}).
Disturbance of \eqref{BuoyantDrop} and \eqref{NonBuoyantDrop} by e.g. a weak swirl $\vec{u}' = \vec{\omega}\times\vec{x}$ (mimicking arbitrary imperfections), with
$\vec{\omega}=(\omega_x,0,\omega_z)$ following \cite{Stone_steady_1999} and $\omega_x=\omega_z = 10^{-2}$, leads to the formation of one and two families of tori, respectively,
centred on the axis of motion (\FIG{ChaosDroplets2X_b}).
Thus the perturbation exposes the ``true'' non-chaotic baseline of the flow topology in droplets, which, by consisting of tori, is (dynamically) indeed equivalent to the above
container flows. This overall implies similar Lagrangian dynamics and ``routes to chaos''; the tori in \FIG{ChaosDroplets2X_b} generically give way to a 3D steady topology of KAM tori
and chaos as in e.g. \FIG{VortexBreakdownBubble} and \FIG{SteadyCylinderFlow} upon intensifying the perturbation of \eqref{BuoyantDrop} and
\eqref{NonBuoyantDrop} \cite{stone_nadim_strogatz_1991,Stone_steady_1999}. The specific behaviour may depend significantly on the relative strength of poloidal versus
toroidal flow, though.

The study in \cite{Stone2005} adopts base flows \eqref{BuoyantDrop} and \eqref{NonBuoyantDrop} to model 3D flow and Lagrangian transport within the travelling drop
in the serpentine channel (\FIG{ChaosDroplets1}). Flow \eqref{BuoyantDrop} represents the straight channel (\FIG{ChaosDroplets_a}) and, taking inevitable imperfections
into account, yields a non-chaotic steady topology comprising of a single family of tori following \FIG{ChaosDroplets2X_b} (left). The time-periodic flow
in the curved channel (\FIG{ChaosDroplets_b}) admits (in the presumed Stokes limit) construction from linear combinations of \eqref{BuoyantDrop} and \eqref{NonBuoyantDrop} and, by
effectively switching continuously
between the instantaneous base-flow topologies in \FIG{ChaosDroplets2X_b}, induces breakdown of tori and thus yields chaotic advection. \FIG{ChaosDroplets2X_c} demonstrates this by the characteristic stretching-and-folding pattern emerging (for 3 typical situations) in the cross-section $x=0$ of the 3D distribution of a tracer (white) initially occupying the lower droplet hemisphere \cite{Stone2005}.

%\paragraph{Internal flow induced by other forcings}
\paragraph{Internal flow induced by electro-hydrodynamic forcing}

A configuration intimately related to the above serpentine channel exists in a translating droplet in a straight channel subject to an axial electric field $\vec{E} = E\vec{e}_z$. The latter, given
the dielectric properties of droplet and bulk fluids are sufficiently apart, induces forces in the fluid-fluid interface that set up a so-called ``Taylor circulation'' \cite{Chang1982}. Its
topology corresponds (for a stationary droplet) with \FIG{ChaosDroplets2X_b} (right), where the black axis coincides with the
direction of travel, meaning that electro-hydrodynamic forcing has a comparable effect as shear due to curvature in the serpentine
channel. Droplets travelling in the $z$-direction retain this topology yet develop a $z$-wise asymmetry: the leading/trailing family of tori expands/diminishes and, inherently, the separatrix -- spanned
by the grey axes in \FIG{ChaosDroplets2X_b} (right) -- shifts towards the trailing side of the droplet \cite{Chang1982}.

\begin{figure}
\centering

\subfigure[Time-periodic forcing]{
\includegraphics[width=0.32\columnwidth,height=0.55\columnwidth]{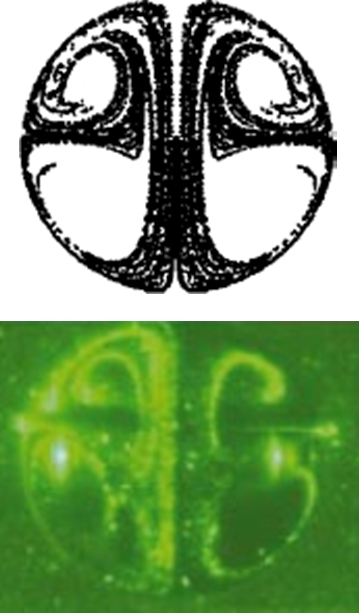}
\label{DropletEM_a}
}
\hspace*{12pt}
\subfigure[Steady misaligned forcing]{
\includegraphics[width=0.5\columnwidth,height=0.5\columnwidth]{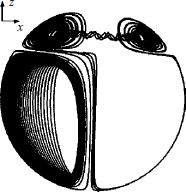}
\label{DropletEM_b}}

%\caption{
%Internal flow in droplets in microfluidic networks induced by electro-hydrodynamic forcing ... Panel a adapted from \cite{Ward_drops_2003}; panel %b adapted from \cite{ward_homsy_2006}.}

\caption{Lagrangian transport in $z$-wise falling droplet subject to electro-hydrodynamic forcing:
(a) stroboscopic map in $xz$-plane for time-periodic electric field along $z$-axis as simulated (top) and visualised by tracers (bottom) (adapted from \cite{Ward_drops_2003}); (b) 3D streamline for steady electric field misaligned with $z$-axis (adapted from \cite{ward_homsy_2006}).}

\label{DropletEM}
\end{figure}

Chaotic advection by destruction of the tori arrangement following \FIG{ChaosDroplets2X_b} (right) can be achieved basically in two ways: (i) unsteady field strength $E(t)$ so as to change the position of said separatrix in time; (ii) misalignment between electric field $\vec{E}$ and translation direction so as to break the symmetries of the topology.
The former is investigated for a falling droplet in a tank subject to a time-periodic electric field $\vec{E}(t) = E\sin(\omega t)\vec{e}_z$ in \cite{Ward_drops_2003}. \FIG{DropletEM_a} visualises
typical Lagrangian transport conditions by 2D temporal stroboscopic maps in the $xz$-plane of tracer particles according to simulations (top) and experiments
(bottom). This reveals erratic wandering of tracers around surviving KAM tori (white patches in top panel) and thus signifies chaotic advection.
%Moreover, the chaotic region proliferates mainly in the trailing (top) side of the droplet for reasons elaborated below.
%
The underlying mechanism intimately relates to RID (\SEC{ContainersIndustrial}) and is elaborated below in the context of thermo-capillary forcing,

Chaos by misalignment is investigated in the same configuration by imposing an electric field $\vec{E} = E\left(\cos\alpha\,\vec{e}_x + \sin\alpha\,\vec{e}_z\right)$ under an angle $\alpha$ with
the translation direction \cite{ward_homsy_2006}. This results in similar erratic wandering around KAM tori, as demonstrated by a typical 3D streamline in \FIG{DropletEM_b}, again implying chaotic
dynamics. Primary difference with the above is the fixed position of the separatrix, i.e. the imaginary plane in \FIG{DropletEM_b} between the vortical structures at the trailing (top) and
leading (bottom) droplet side. The slight tilt emanates from the misaligned electric field ($\alpha\neq 0$).

Reorientation of the electric field enables similar accomplishment of chaos for
a {\it stationary} droplet. This is demonstrated in \cite{Xu2007} by a time-periodic reorientation
of the above field via alternating angles $\alpha = 0$ and $\alpha = \alpha_w$.

\paragraph{Internal flow induced by thermo-capillary forcing}

Thermo-capillary effects in droplets immersed in a bulk fluid and exposed to an axial temperature gradient
\begin{eqnarray}
\vec{\nabla}T = (\kappa_0 + \kappa_1 z)\vec{e}_z,
\label{TempGrad}
\end{eqnarray}
with coefficients $\kappa_0$ and $\kappa_1$ parameterising its uniform and $z$-wise linear contributions, respectively, may induce forces in the fluid-fluid interface akin
to the above electro-hydrodynamic phenomena and thus drive an internal circulation \cite{Grigoriev_microdroplets_2006}. A uniform axial temperature gradient $(\kappa_0 > 0,\kappa_1 = 0)$ causes droplet motion in $+z$-direction and yields an internal circulation and corresponding topology following \eqref{BuoyantDrop} and \FIG{ChaosDroplets2X_b} (left), respectively. Chaotic advection again requires destruction of tori and can here be achieved by imposition of a non-uniform temperature gradient $(\kappa_0 > 0,\kappa_1 \neq 0)$ and a perpendicular vortical motion due to shearing by the ambient flow. The non-uniform gradient yields an entirely intact topology following
\FIG{ChaosDroplets2X_b} (right) yet with the same $z$-wise asymmetry (and shift of the separatrix) as the above electro-hydrodynamic case; superposition
of the shear initiates tori breakdown and results in chaotic streamlines wandering around surviving KAM tori in the same was as
shown in \FIG{DropletEM_b} \cite{Grigoriev_microdroplets_2006,Vainchtein_droplets_2007}.\footnote{These flow conditions can be generated experimentally
by e.g. differential heating via a traversing laser beam \cite{Grigoriev_microdroplets_2006}.}

The onset of 3D chaotic advection in the present steady flow stems from defects that develop in both tori families near the separatrix for weak perturbation of the non-chaotic baseline \cite{Vainchtein_droplets_2007}. This results in drifting and gradual chaotic dispersion reminiscent of RID in unsteady container flows (\FIG{UnsteadyChaos2_a}) for all streamlines that encounter these defects. Essential difference with RID -- and its  steady counterpart \cite{Vainchtein_resonant_2006} -- is that defects here emanate from significant slow-down of Lagrangian motion near the separatrix instead of ``true'' resonance following \eqref{Resonance2}. (Slow motion is a generalisation of resonance in that this affects the dynamics in a similar way;
see \SEC{ContainersIndustrial} and \SEC{Conditions3Ddynamics}.) The impact is therefore restricted to the ``outer'' tori
close to the separatrix; the ''inner'' tori survive unscathed. Hence the dispersion thus induced is restricted to a chaotic region surrounding the latter tori. Stronger perturbation overrides these phenomena and results in the ``conventional'' breakdown of tori and (further) proliferation of the chaotic region (\SEC{Towards3D}).

Dispersion due to local defects in LCSs of 3D steady flows has first been examined in \cite{Bajer_class_1990} and denoted ``trans-adiabatic drift''. Here significant slow-down of Lagrangian motion near certain isolated stagnation points caused this phenomenon. Analogous flow structures
strongly suggest that basically the same mechanisms are at play in the above electro-hydrodynamically forced flow \cite{ward_homsy_2006,Wu2015}. Hence the chaotic dynamics in \FIG{DropletEM} are representative of the weakly-perturbed state for both electro-hydrodynamic and thermo-capillary forcing. Moreover, comparable baseline topologies implicate this mechanism also in the separatrix crossing in convection cells (\SEC{WebsIndustrial}).

\paragraph{Internal flow induced by electro-wetting}

In contrast to forcing in microfluidic networks,
electro-wetting is a promising forcing technique for droplets in digital microfluidic platforms and relies
on the dependence of the contact angle between droplet surface and substrate on the electric potential between former and latter.
This enables actuation of an internal time-periodic flow through controlled oscillation of the droplet shape by variation of
the contact angle via an AC voltage applied between an electrode at the droplet pole and the substrate \cite{Mugele2006,Mugele2011}.

\begin{figure}
\centering

\subfigure[Cross-sectional stroboscopic maps]{\includegraphics[width=0.8\columnwidth,height=0.35\columnwidth]{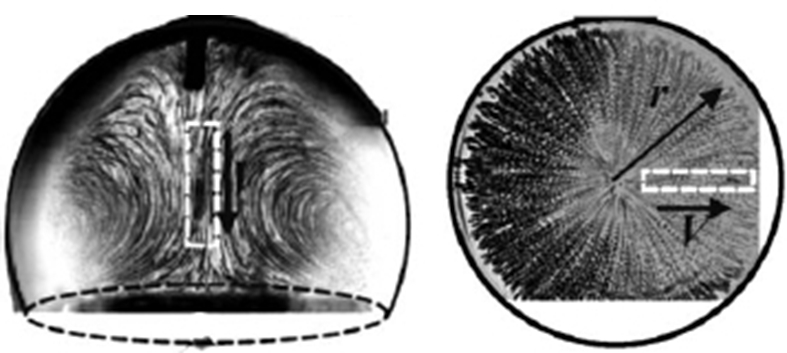}\label{ElectroWetting_a}}
\subfigure[Evolution of mixing pattern (from left to right) ]{\includegraphics[width=\columnwidth,height=0.3\columnwidth]{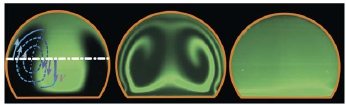}\label{ElectroWetting_b}}

%\caption{Chaotic advection inside a drop induced by electro-wetting. Reproduced/adapted from \cite{Mugele2011}.}

\caption{Lagrangian transport in time-periodic droplet flow on substrate induced by electro-wetting: (a) stroboscopic
map in vertical (left) and horizontal (right) cross-section visualised experimentally by tracers (arrows indicate
flow direction); (b) evolving mixing pattern visualised by fluorescent dye. Adapted from \cite{Mugele2011}.}

\label{ElectroWetting}
\end{figure}

\FIG{ElectroWetting_a} visualises the 3D Lagrangian transport in a droplet flow driven by electro-wetting by temporal
stroboscopic maps of tracer particles in a vertical cross-section through the axis (left) and a horizontal cross-section
near the substrate (right) \cite{Mugele2011}. This reveals a global toroidal circulation about the vertical droplet axis, directed downwards in the
centre and radially outward at the bottom, and implies a flow topology that is a perturbed state of a non-chaotic baseline
following \FIG{ChaosDroplets2X_b} (left). (The horizontal black axis in the latter coincides with the vertical droplet axis.)
%
%However, given weak toroidal circulation ... non-trivial ... may ... essentially 3D phenomena ... defects ... \\\\
%
The corresponding transport is visualised in \FIG{ElectroWetting_b} by the evolution (from left to right) of fluorescent
dye (green) and the symmetric counter-rotating ``swirls'' in the middle snapshot strongly suggest a topology that, similar
to \FIG{StirredTankFountain}, contains sizeable KAM tori. Hence, contrary to the findings in \cite{Mugele2011}, chaotic
advection seems to occur only locally, namely in the surroundings of said tori. Moreover, dominant poloidal
flow in conjunction with weak (fluctuating) toroidal flow closely resembles the situation in the champagne glass in \FIG{Champagne} and thus
suggests dynamics that, by considerations given before, may likely involve resonance phenomena.
%
%In-depth Lagrangian analysis (by e.g. Poincar\'{e} sectioning) may conclusively establish the nature of the transport.
%
%\begin{itemize}
%\item[$\bullet$]
%Following the motion of dye patches, we show that the number of oscillation cycles required
%to achieve mixing scales logarithmically with the Peclet number as expected for chaotic mixing \cite{Mugele2006,Mugele2011}.
%\end{itemize}

\begin{figure}
\centering

%\subfigure[]{\includegraphics[width=0.8\columnwidth,height=1.1\columnwidth]{Figures/StreamingMicrodroplet2.jpg}}
%\hspace{12pt}
%\subfigure[]{\includegraphics[width=0.43\columnwidth,height=0.39\columnwidth]{Figures/LidDrivenCavity2.jpg}}

\subfigure[Simulated streamlines]{\includegraphics[width=0.4\columnwidth,height=1.1\columnwidth]{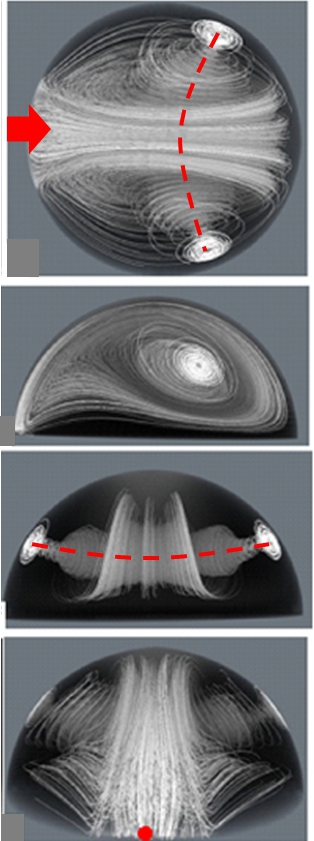}\label{StreamingMicrodroplet_a}}
\hspace*{3pt}
\subfigure[Experimental streamlines]{\includegraphics[width=0.4\columnwidth,height=1.1\columnwidth]{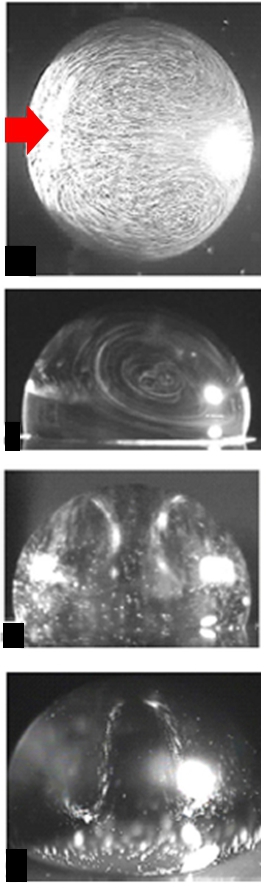}\label{StreamingMicrodroplet_b}}

%\caption{Internal flow induced by acoustic streaming: experiments (left) versus simulations (right).
%Shown are (from top) the top/side/downstream/upstream views. Adapted from \cite{Alghane2011}.}

\caption{Lagrangian transport in steady droplet flow on substrate induced by acoustic streaming: (a) simulated
3D streamline portrait (arrow indicates impingement of SAW; dashed curve indicates axis of
primary circulation); (b) experimental visualisation of 3D streamline portrait by tracer particles. Adapted from \cite{Alghane2011}.}

\label{StreamingMicrodroplet}

\end{figure}

\paragraph{Internal flow induced by acoustic streaming} A relatively novel forcing technique for droplet flow in digital platforms employs 3D streaming induced by
Rayleigh surface acoustic waves (SAWs) impinging on the surface \cite{Alghane2011}. \FIG{StreamingMicrodroplet} shows a typical internal streamline pattern driven
by a SAW (red arrow), obtained by simulations (panel a) and experimental visualisation (panel b), and reveals two adjacent toroidal
circulations about a common centre of rotation (red dashed curve).
%
%\FIG{StreamingMicrodroplet_a} shows a typical (simulated) internal streamline
%pattern driven by a SAW (red arrow), with its experimental visualisation by tracer particles in \FIG{StreamingMicrodroplet_b}, and reveals two adjacent toroidal
%circulations about a common centre of rotation (red dashed curve).
%
This implies (on grounds of symmetry) a baseline topology similar to \FIG{ChaosDroplets2X_b} (right) consisting of two families of tori, where the horizontal black axis coincides with the red dashed curve in \FIG{StreamingMicrodroplet_a},
%
%and thus 3D Lagrangian dynamics comparable to the above flows driven by electro-hydrodynamic and thermo-capillary forcing (\FIG{DropletEM}).
%
suggesting behaviour comparable to the above flows driven by electro-hydrodynamic and thermo-capillary forcing (\FIG{DropletEM}).
However, in contrast with the latter, the acoustic streaming exhibits a significant toroidal circulation (i.e. about the red
dashed curve in \FIG{StreamingMicrodroplet}), implying a 3D flow topology comparable to the steady inertial cylinder flow in \FIG{SteadyCylinderFlow}: two symmetric families of KAM tori surrounded by chaos. Thus here chaos likely results from the conventional breakdown of tori associated with a predominantly toroidal motion instead of the resonance-induced scenario for the (mainly poloidal)
flow driven by electro-hydrodynamic or thermo-capillary forcing.
%
%Such fundamental similarities offer ways to deepen insights into the transport properties in droplets and further develop micro-fluidic technologies
%relying on them. Thorough Lagrangian transport analyses are, as pointed out before, crucial to this end.
%

%\paragraph{Drop stability}
\paragraph{Integrity of immersed drops}

\begin{figure}
\centering

\includegraphics[width=0.8\columnwidth,height=\columnwidth]{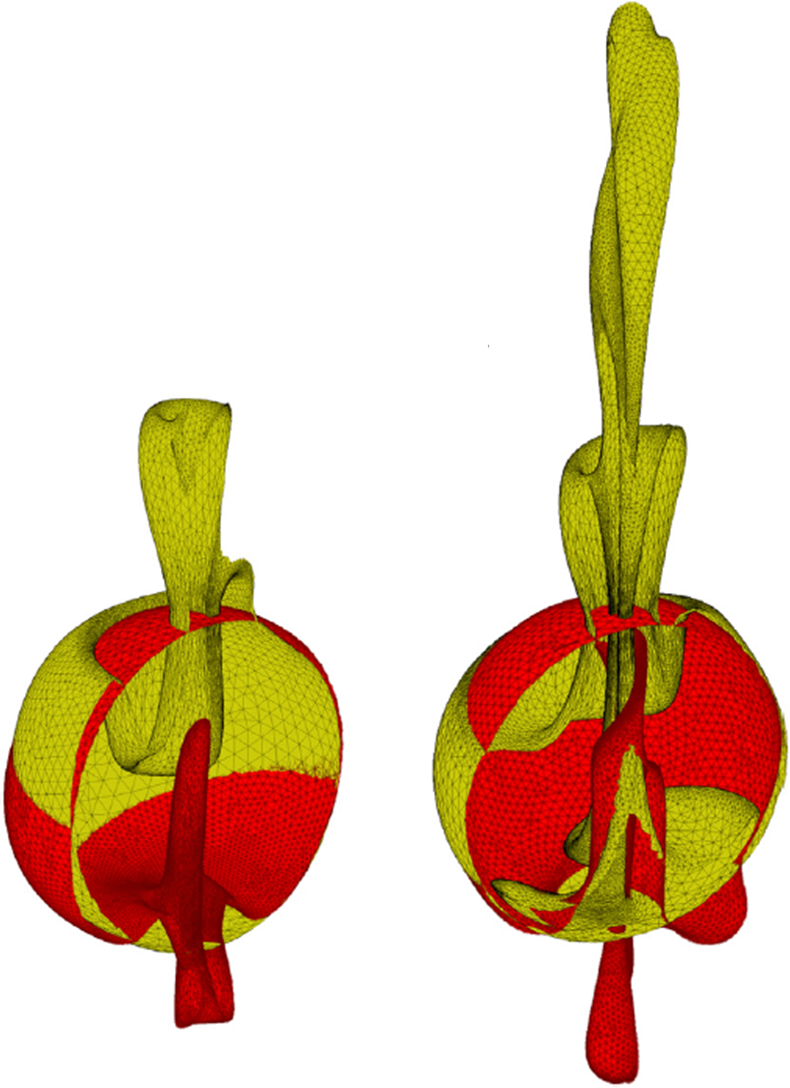}

%\caption{\blue{Droplet instability ... progression in time from left to right ... Chaotic advection induced by droplet instability visualised by %the heteroclinic interaction between a stable (red) and an unstable (olive green) manifold in the 3D unsteady flow of Hill's spherical vortex. %Reproduced from \cite{Branicki2009}.}}

\caption{Instability of immersed buoyant droplet due to perturbation of ambient flow by unsteady strain
demonstrated by simulated proliferation (from left to right) and interaction of stable (red) and unstable (green) manifold
of two DHTs (not shown) emerging from disintegration of (initially) spherical interface. Reproduced from \cite{Branicki2009}.}

\label{HillsVortex}
\end{figure}

Droplet-based microfluidic networks assume intact droplets and absence of transport between interior and ambient flow. However, droplets may exhibit leakage or even break-up due to e.g.
excessive shear stresses and thus compromise the functionality of micro-fluidic devices \cite{Debon2015,Zhu2016}. The fluid-fluid interface that bounds immersed droplets is a material surface
and thus constitutes an LCS in its own right. Hence the Lagrangian dynamics of the interface is crucial to the integrity and stability of droplets. Consider to this end the response of a buoyant
droplet following \FIG{ChaosDroplets2X_b} to perturbation of the ambient flow by an unsteady strain investigated computationally in \cite{Branicki2009}.
%
%\FIG{HillsVortex} shows
%the proliferation and interaction of two LCSs emerging from the disintegration of the (initially) spherical interface \cite{Branicki2009}. Said LCSs are the stable (red) and unstable (olive green)
%manifold of so-called ``distinguished hyperbolic trajectories'' (DHTs), that is, material curves on the droplet interface that are the counterparts in aperiodic
%flows to hyperbolic periodic lines in time-periodic flows.\\\\
%
\FIG{HillsVortex} shows the proliferation and interaction of the stable (red) and unstable (green) manifold of two so-called ``distinguished hyperbolic trajectories'' (DHTs) emerging
from the disintegration of the (initially) spherical interface. DHTs and their manifolds are in essence the counterparts in aperiodic flows to hyperbolic periodic points and manifolds
in time-periodic flows and thus dictate (chaotic) transport in basically the same way as illustrated in \FIG{OrderWithinChaos} (\SEC{ConditionsChaos}) \cite{Ide2002}. Here these LCSs determine the intricate
3D detrainment and entrainment of droplet and ambient fluid, respectively, that ensues from the droplet instability.

\subsubsection{Relevance beyond industry}
\label{DropsNonIndustrial}

Discussed below are instances of 3D Lagrangian (chaotic) transport in non-industrial flows that closely relate to flows in drops. The aim, as for duct (\SEC{DuctFlowNonIndustrial}) and container (\SEC{ContainersNonIndustrial}) flows, is again to expose similarities and analogies between (seemingly) different problems.

%\paragraph{Cytoplasmic streaming}
\paragraph{Cytoplasmic flow and transport}

%An intriguing example of ... found in Cytoplasmic flow. \red{Fluid body confined by membrane; carrier in ambient flow; non-trivial flow forcing via membrane.}
%\blue{Similarities with above drops in ambient flow:
%\begin{itemize}
%\item[$\bullet$] Semi-permeable/flexible boundary $\rightarrow$ cell membrane.
%\item[$\bullet$] Functionality/role of boundary: generator/control of internal flow via flexibility and non-zero parallel flow
%\item[$\bullet$] Encapsulating fluid body so as to enable transport in ambient flow.
%\end{itemize}
%}
%\red{Some overlapping elements with container flows particularly for non-technological systems: stomach also generates internal flow by flexible boundary (yet without non-zero
%parallel flow); alveoli wall is also permeable membrane (yet without fully enclosing fluid body).}

%Cells are the basic building blocks of living organisms and (regarding transport) the biological analogy to the above drops by acting as processing and transportation vessels.\\\\

%Migrating cells in living organisms are (regarding transport) the biological analogy to the above drops in ambient flows.

\begin{figure}
\centering

\subfigure[Schematic of yolk mixing]{\includegraphics[width=0.58\columnwidth,height=0.3\columnwidth]{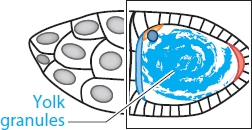}\label{CytoplasmicFlow_a}}
\hspace*{2pt}
\subfigure[Cross-sectional velocity]{\includegraphics[width=0.38\columnwidth,height=0.3\columnwidth]{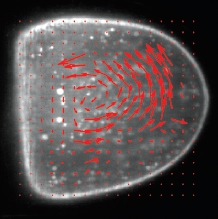}\label{CytoplasmicFlow_b}}
\subfigure[Yolk mixing visualised by tagged yolk granules]{
\includegraphics[width=0.32\columnwidth,height=0.3\columnwidth]{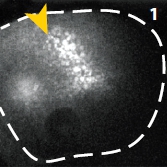}
\hspace*{2pt}
\includegraphics[width=0.32\columnwidth,height=0.3\columnwidth]{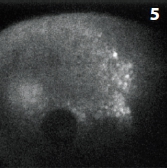}
\hspace*{2pt}
\includegraphics[width=0.32\columnwidth,height=0.3\columnwidth]{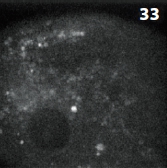}
\label{CytoplasmicFlow_c}}

%\caption{Cytoplasmic flow and mixing in a \red{drosophila oocyte} due to cytoplasmic streaming. Reproduced/adapted from \cite{Quinlan2016}.}

\caption{Cytoplasmic flow and mixing in drosophila oocyte driven by cytoplasmic streaming as physiological analogy
to droplet flow: (a) yolk mixing by cytoplasmic streaming; (b) cross-sectional velocity (red vectors) visualised experimentally by 2DPIV; (c) yolk mixing visualised
experimentally by motion and dispersion of yolk granules in occyte (dashed outline) tagged by tracer injected at yellow arrow (numbers
indicate time instance in minutes). Adapted from \cite{Quinlan2016}.}

\label{CytoplasmicFlow}
\end{figure}

Cytoplasmic flow and transport inside cells of living organisms is (regarding domain topology) the physiological analogy to the above drops. An important forcing mechanism is so-called ``cytoplasmic streaming'', that is, internal circulation within cells by flow actuation at the cell boundary via ``molecular motors'' in a manner akin to shear-driven flow in droplets. \FIG{CytoplasmicFlow_a} illustrates this for the cytoplasmic circulation inside a drosophila oocyte for the purpose of mixing of yolk granules as nourishment during cell development \cite{Quinlan2016}. The corresponding cross-sectional circulatory velocity
field in the (approximate) midplane obtained by 2DPIV is given in \FIG{CytoplasmicFlow_b}. Mixing is visualised in \FIG{CytoplasmicFlow_c} by the motion and dispersion of yolk granules in the occyte (indicated by dashed outline) tagged via a tracer injected at the yellow arrow (numbers indicate time instance in minutes). The tagged yolk initially migrates as a compact patch, subsequently moves away from the midplane and reappears in a well-mixed state. These observations in conjunction with the domain shape strongly suggest a baseline topology following \FIG{ChaosDroplets2X_b} (right), where the grey axes span the separating midplane.
The appreciable circulation in \FIG{CytoplasmicFlow_b}, in turn, suggests a flow resembling the droplet flow driven by acoustic
streaming (\FIG{StreamingMicrodroplet}). This, by analogy, implies a 3D flow topology comparable to the steady inertial cylinder flow
in \FIG{SteadyCylinderFlow} and, inherently, chaos due to conventional tori breakdown.

A forcing mechanism employed specifically by motile cells is flow induced via deformation during migration in an ambient flow \cite{Koslover2017}. This suggests cytoplasmic flow and transport comparable to droplets propagating a serpentine channel (\FIG{ChaosDroplets1}).

\paragraph{Delivery and mixing of drugs}

\begin{figure}
\centering

\subfigure[Cross-section of eye]{
%trim=0cm 0cm 0cm 0,clip,
\includegraphics[trim=0cm -2cm 0cm 0,clip,width=0.38\columnwidth,height=0.35\columnwidth]{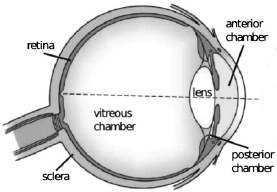}
\label{DrugDeliveryEye_a}}
%\hspace*{2pt}
\subfigure[Lagrangian trajectories]{\includegraphics[width=0.58\columnwidth,height=0.5\columnwidth]{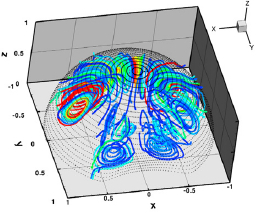}\label{DrugDeliveryEye_b}}

%\caption{Drug delivery in the vitreous chamber of the eye. ...  Reproduced/adapted from \cite{Bonfiglio2013}.}

\caption{Drug delivery in vitreous chamber of eye as physiological analogy to droplet flow: (a) cross-section of eye; (b) simulated Lagrangian
trajectories in upper hemisphere (distinguished by colour) using experimental 3D time-periodic velocity field constructed from
cross-sectional {\it in vitro} 2DPIV. Adapted from \cite{Bonfiglio2013}.}

\label{DrugDeliveryEye}
\end{figure}

A further physiological counterpart of Lagrangian transport in drops is found in the delivery and mixing of drugs in the vitreous chamber of the
eye (\FIG{DrugDeliveryEye_a}). Intravitreal drug delivery namely is a common treatment for retinal diseases and advective transport by
the internal flow induced through so-called ``saccades'' (rapid eye rotations during redirection of the visual axis) are considered key to
this process. The study by \cite{Bonfiglio2013} investigates the advection characteristics via reconstruction of the 3D time-periodic flow field in an
{\it in vitro} physical model of the vitreous chamber from cross-sectional 2DPIV measurements.
%
% and its employment for Lagrangian transport analyses by numerical particle tracking using the experimental velocity field.\\\\
%
Lagrangian transport analyses by numerical particle tracking using the experimental velocity field reveal two toroidal circulations about the vertical
axis and approximately symmetric about the midplane between upper and lower hemisphere.
%
%This implies a baseline topology in \FIG{ChaosDroplets2X_b} (right), where the grey axes span the $xy$-plane in \FIG{DrugDeliveryEye_b}.
%
\FIG{DrugDeliveryEye_b} gives typical trajectories in the upper hemisphere $z\geq 0$ (distinguished by colour) and demonstrates that said circulations consist of a predominantly poloidal circulation and a weak toroidal drift.
This implies a baseline topology following \FIG{ChaosDroplets2X_b} (right), where the grey axes span the $xy$-plane in \FIG{DrugDeliveryEye_b}.
%
%Moreover, the prevailing poloidal motion suggests non-trivial Lagrangian dynamics due to resonance-induced tori breakdown and separatrix crossing similar to the electro-hydrodynamic (and thermo-capillary) droplet
%flow in \FIG{DropletEM}.
%
The prevailing poloidal motion suggests non-trivial Lagrangian dynamics due to separatrix crossing similar to the electro-hydrodynamic (and thermo-capillary) droplet flow in \FIG{DropletEM} and, given time-periodicity, likely also RID as in \FIG{UnsteadyChaos2_a} (right).

%The prevailing poloidal motion suggests non-trivial Lagrangian dynamics due to resonance-induced tori breakdown and separatrix crossing similar to the electro-hydrodynamic (and thermo-capillary) droplet
%flow in \FIG{DropletEM}.

% where the black axis corresponds with the vertical $z$-axis in \FIG{DrugDeliveryEye_b}.

\subsection{Flow in webs}
\label{Webs}

%\subsubsection{Introduction}
\subsubsection{General configuration}
\label{WebsConfig}

\begin{figure}
\centering

\subfigure[Convection cells]{\includegraphics[width=0.43\columnwidth,height=0.42\columnwidth]{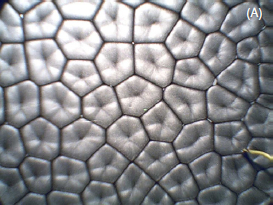}\label{CellFormationa}}
\hspace*{2pt}
\subfigure[Porous matrix]{\includegraphics[width=0.54\columnwidth,height=0.48\columnwidth]{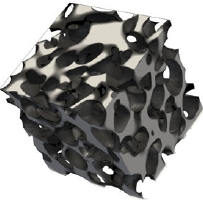}\label{CellFormationb}}

\caption{Partitioning of flow domains into interconnected elementary cells: (a) ``soft'' partitioning in e.g. convection cells (reproduced from \cite{Maroto2007});
(b) ``hard'' partitioning by e.g. interconnected cavities in porous matrix (adapted from \cite{Dentz2018}).}

\label{CellFormation}
\end{figure}

A final class of systems relevant in the present scope concerns flows in a domain with an internal partitioning into interconnected elementary cells that admit exchange of fluid and scalars. This may involve a ``soft'' partitioning by a mesh of stream surfaces or material interfaces as e.g. in convection cells of buoyancy-driven flows (\FIG{CellFormationa}) or a ``hard'' partitioning by a solid matrix enclosing interconnected cavities as e.g. in porous media (\FIG{CellFormationb}).
%
%The Lagrangian transport by flows in such ``webs'' of interconnected cells exhibits, on grounds of the structure of the domain, characteristics of both periodic duct flows (\SEC{Ducts}) and %confined flows (\SEC{Containers} and \SEC{Drops}).
%
The Lagrangian transport by flows in such ``webs'' of interconnected cells exhibits, on grounds of the domain structure, characteristics of both periodic
duct flows (transport {\it along} cells) and container flows (transport {\it within} cells). Hence flows in webs define in essence a hybrid class incorporating features of the
systems considered in \SEC{Ducts}--\SEC{Drops}.

\subsubsection{Industrial and technological relevance}
\label{WebsIndustrial}

The examples of flows in webs shown in \FIG{CellFormation} in fact constitute two important configurations in the context of technological applications:
\begin{itemize}

\item[$\bullet$] {\it Convection cells} Lagrangian transport by laminar flow in convection cells is relevant to materials processing and fabrication involving coatings and films \cite{Hare1994,Bassou2009,Saranjam2016,Bunz2006,Escale2012,Munoz2014,Zhang2015,Deng2015} as well as for emerging technologies as liquid metal batteries \cite{Kelley2014,Weber2015,Kelley2018}.

\item[$\bullet$] {\it Porous media} Lagrangian transport in porous media is key to both natural and engineered environments. The former primarily concerns subsurface flows involved in e.g. hydrology, enhanced oil and heat recovery and aquifer heat-storage systems \cite{Brown2012,Chen2015,Seredkin2016,Guo2016,Cunningham2002,Goltz2012}. Transport in
    engineered porous media is at the heart of e.g. thermo-chemical heat-storage systems, micro-fluidic devices consisting of micro-vascular networks and compact heat exchangers based on metal foams \cite{Pardo2014,Therriault2003,Boomsma2003,Hugo2011,Zhao2012}.

\end{itemize}
The general flow and transport characteristics of both configurations are discussed and illustrated below using similarities and analogies with the above flow
classes. Their manifestation and role in practical applications are subsequently exemplified by way of representative cases.

\paragraph{Convection cells: generic characteristics} Fluid confined by a hot bottom wall and cold top wall tends to rise and descend near former and latter, respectively, due to buoyancy and (upon overcoming viscous friction) thus sets up a pattern of circulations. This phenomenon is known as
Rayleigh-B\'{e}nard convection and the vortical structures thus created, commonly denoted convection (or B\'{e}nard) cells, form regular patterns in the present regime of laminar flows \cite{Manneville2006}. Thin fluid films on a hot horizontal wall and with a free
top surface behave essentially the same; here circulations emerge from an interplay of buoyancy and surface tension and result
in B\'{e}nard-Marangoni convection. Both Rayleigh-B\'{e}nard and B\'{e}nard-Marangoni convection yield (under 3D laminar conditions) hexagonal
patterns of convection cells as exposed in \FIG{CellFormationa} by flow visualisation in the free surface of
a silicon oil layer using aluminium tracer particles \cite{Maroto2007}.

\begin{figure}
\centering

\subfigure[3D streamline pattern in two neighbouring hexagonal cells]{\includegraphics[width=0.9\columnwidth,height=0.5\columnwidth]{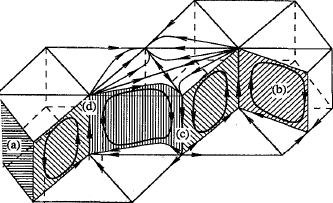}\label{ConvectionCells1b}}

\subfigure[Global diffusive transport]{
\includegraphics[width=0.42\columnwidth,height=0.4\columnwidth]{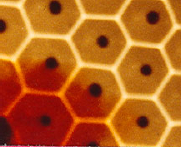}
%\label{ConvectionCells2a}}
%
\hspace*{6pt}
%
%\subfigure[due to diffusion]{
%\includegraphics[width=0.23\columnwidth,height=0.23\columnwidth]{./ConvectionCells4a.jpg}
%\includegraphics[width=0.23\columnwidth,height=0.23\columnwidth]{./ConvectionCells4b.jpg}
%\label{ConvectionCells2b}}
\label{ConvectionCells2a}}
\subfigure[Global advective transport]{
\includegraphics[width=0.42\columnwidth,height=0.41\columnwidth]{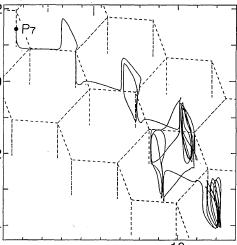}
\label{ConvectionCells2b}}

%\subfigure[]{
\includegraphics[trim=0cm 0cm 0cm 0,clip,width=0.48\columnwidth,height=0.21\columnwidth]{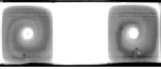}
\hspace*{3pt}
\includegraphics[trim=0cm 0cm 0cm 0,clip,width=0.48\columnwidth,height=0.21\columnwidth]{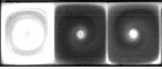}
%}
%\includegraphics[trim=0cm 0cm 0cm 0,clip,width=0.48\columnwidth,height=0.21\columnwidth]{Figures/SolomonFlow2b_left.jpg}
%\hspace*{3pt}
%\includegraphics[trim=0cm 0cm 0cm 0,clip,width=0.48\columnwidth,height=0.21\columnwidth]{Figures/SolomonFlow2b_right.jpg}
%
\subfigure[Restricted (left) versus global (right) advective transport]{
\includegraphics[trim=0cm 0cm 0cm 0,clip,width=0.48\columnwidth,height=0.21\columnwidth]{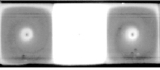}
\hspace*{3pt}
\includegraphics[trim=0cm 0cm 0cm 0,clip,width=0.48\columnwidth,height=0.21\columnwidth]{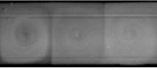}
\label{ConvectionCells2c}}

\caption{Global 3D transport in network of convection cells:
(a) 3D streamline pattern in neighbouring cells consisting (cell-wise) of closed streamlines within
vertical cross-sections (surface ``b'') centered around the hexagon axis (line ``d'') (adapted from \cite{Dykhne1994});
(b) global diffusive transport visualised experimentally by dye spreading (red) across cell interfaces (bright pattern) (adapted from \cite{Medale_hexagon_2007});
(c) global advective transport visualised by simulated 3D trajectory (starting at $P_7$) in perturbed flow (adapted from \cite{Sano1993});
(d) restricted (left) versus global (right) advective transport in unperturbed and perturbed array of vortices, respectively, visualised experimentally by initial (top) and final (bottom)
tracer distribution (adapted from \cite{Solomon_resonant_2003}).}
\label{ConvectionCells2}

\end{figure}

\FIG{ConvectionCells1b} schematically gives the 3D streamline pattern in two neighbouring hexagons consisting (within each cell) of
closed streamlines within vertical cross-sections (surface ``b'' in right cell) centered around the hexagon axis (line ``d'' in left
cell) \cite{Dykhne1994}. Thus the flow topology is equivalent to that in the droplet in \FIG{ChaosDroplets2X_a} (left), where
the {\it horizontal} black axis corresponds with the {\it vertical} hexagon axis.
%
%, yet, for reasons given before, is in fact a singular limit.\\\\
%
However, such topologies, for reasons given before, are singular limits; the real cell-wise topology emerges for arbitrarily
small perturbations and comprises of a family of tori centered on the (hexagon) axis similar to \FIG{ChaosDroplets2X_b} (left).

Flow topologies consisting of complete tori admit inter-cellular transport across separatrices (e.g. plane ``a'' in \FIG{ConvectionCells1b})
only by diffusion. \FIG{ConvectionCells2a} experimentally visualises such diffusion-enabled global transport by the spreading of dye (red) in silicon oil across the ``honeycomb'' demarcated by the cell interfaces (bright pattern) \cite{Medale_hexagon_2007}.
%
%The kinematic analogy with electro-hydrodynamic and thermo-capillary droplet flows (i.e. similar baseline topology and dominant poloidal %circulation) strongly suggests that, upon weak perturbation, the tori in convection cells generically develop defects by the same resonance %phenomena (i.e. due to slow motion as well as ``true'' resonance by synchronised poloidal/rotational rotations following \eqref{Resonance} and %\eqref{Resonance2}).
%
The kinematic analogy with electro-hydrodynamic and thermo-capillary droplet flows (i.e. similar baseline topology and dominant poloidal circulation) strongly suggests that, upon weak perturbation, the tori in convection cells generically develop defects by the same resonance phenomena.\footnote{That is, due to ``true'' resonance by synchronised poloidal/rotational rotations following \eqref{Resonance} and \eqref{Resonance2} and/or its generalisation, viz. significant local slow-down of Lagrangian motion (\SEC{ContainersIndustrial} and \SEC{Conditions3Ddynamics}).}
This facilitates crossing of separatrices following \FIG{DropletEM} and thus enables {\it advective} inter-cellular transport as demonstrated in \FIG{ConvectionCells2b} by a typical simulated 3D trajectory of a tracer released at position $P_7$ in a weakly-perturbed flow \cite{Sano1993}.

Experimental evidence of such resonance-induced inter-cellular transport is given in \FIG{ConvectionCells2c} by the dispersion of fluorescent tracer particles in an array of vortices set up in sulfuric acid by MHD forcing. (Each vortex corresponds with a hexagonal convection cell in \FIG{ConvectionCells1b} in top view.) Shown is the initial (top) and final (bottom) distribution of tracers released in one vortex (bright) in the unperturbed flow driven by steady MHD forcing (left) versus the perturbed flow by additional time-periodic forcing (right) \cite{Solomon_resonant_2003}. This reveals a dramatic difference. Tracers in the unperturbed flow remain trapped in the original vortex; tracers in the perturbed flow, on the other hand, exhibit global dispersion due to the resonance-induced separatrix crossing. The essential role of time-periodicity in this process implies that here (in particular) resonance according to \eqref{Resonance} -- and thus RID following \FIG{UnsteadyChaos2_a} (right) -- is the key player.

\paragraph{Convection cells: materials and fabrication technology}

\begin{figure}
\centering

\subfigure[drying polymer solution]{\includegraphics[width=0.45\columnwidth,height=0.42\columnwidth]{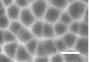}\label{SurfaceDefects_c}}
\hspace*{6pt}
\subfigure[drying paint films]{\includegraphics[width=0.45\columnwidth,height=0.42\columnwidth]{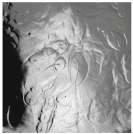}\label{SurfaceDefects_b}}

\caption{Surface defects in coatings and films due to B\'{e}nard-Marangoni convection cells: (a) corrugated pattern in polymer film on glass substrate visualised by shadowgraphy (bar represents 2.5 mm) (reproduced from \cite{Bassou2009}); (b) ``orange peel'' texture
in paint on stainless steel ( $23\times 23$ mm$^2$ area.) (reproduced from \cite{Saranjam2016});.}

\label{SurfaceDefects}
\end{figure}

Important for the fabrication and application of coatings as paints or epoxy resins is homogeneous distribution of solvents and agents. Inhomogeneity of certain additives may result in the formation of convection cells due to B\'{e}nard-Marangoni convection (\FIG{CellFormationa}) and, by the ensuing restriction of the
advective transport of (other) additives to toroidal LCSs following \FIG{ConvectionCells1b}, thus compromise the coating quality \cite{Hare1994,Bassou2009,Saranjam2016}. Typical surface defects arising from this are corrugated patterns, as visualised for a drying polymer film on a glass substrate in \FIG{SurfaceDefects_c} by shadowgraphy \cite{Bassou2009} and ``orange peel'' textures, as shown in \FIG{SurfaceDefects_b} for a drying paint film on stainless steel \cite{Saranjam2016}.
%
%Important for the fabrication and application of pigmented coatings as paints or epoxy resins is the homogeneous distribution of solid pigments within the liquid binder solution. Inhomogeneous %pigment distribution may namely result in the formation of convection cells due to B\'{e}nard-Marangoni convection (\FIG{CellFormationa}) and, by the ensuing restriction of advective transport of %e.g. solvents and agents to toroidal LCSs following \FIG{ConvectionCells1b}, thus compromise the coating quality \cite{Hare1994,Bassou2009,Saranjam2016}. Typical surface defects arising from this %are ``orange peel'' textures, as shown in \FIG{SurfaceDefects_b} for a drying paint film on stainless steel \cite{Saranjam2016}, and corrugated patterns, as visualised for a drying polymer film on a %glass substrate in \FIG{SurfaceDefects_b} by shadowgraphy \cite{Bassou2009}.
%
Fundamental insights into 3D Lagrangian transport in convection cells may deepen the understanding of these adverse effects and offer ways to mitigate or eliminate them. The global advective transport across convection cells in \FIG{ConvectionCells2c} (right) is systematically induced by weak time-periodic perturbation of the flow using a frequency resonant with the typical circulation frequencies. This means that, given the particular nature of the perturbation is immaterial for small magnitudes, basically {\it any} perturbation during (industrial) fabrication and application of coatings -- e.g. weak vibrations attuned to the natural frequencies -- may promote global transport as in \FIG{ConvectionCells2c} (right) and thus counteract the convection cells.

\begin{figure}
\centering

\subfigure[Micro-structured polymer film]{\includegraphics[width=0.47\columnwidth,height=0.4\columnwidth]{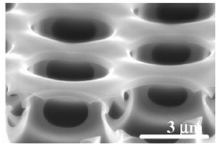}\label{UtiliseCells_a}}
\hspace*{6pt}
\subfigure[Micro-structured solar cell]{\includegraphics[width=0.47\columnwidth,height=0.4\columnwidth]{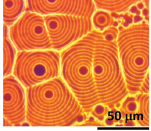}\label{UtiliseCells_b}}

\caption{Utilisation of B\'{e}nard-Marangoni convection cells for fabrication: (a) honeycomb porous micro-structure in polymer film visualised by scanning electron imaging (adapted from \cite{Munoz2014}); (b) micro-structured photonic crystals for solar cells in perovskite film visualised by microscopy (adapted from \cite{Deng2015}).}

\label{UtiliseCells}
\end{figure}

Cell formation due to B\'{e}nard-Marangoni convection may, rather than being an undesired phenomenon, also find utilisation in materials technology for the fabrication of
textured coatings and films \cite{Bunz2006,Escale2012,Munoz2014,Zhang2015}. Certain micro-structured polymer films e.g. rely on the formation of so-called ``breath figures'' due
to the condensation of moisture (contained in air) on cold solid or liquid surfaces. Running moist air over a polymer solution leads to the formation of water droplets
on the liquid surface that (while sinking) congregate into hexagonal patterns due to their Lagrangian transport by the convection cells. The subsequent evaporation
of solvent and water during solidification of the polymer solution leaves an imprint of the entrained water droplets as pores in the shape of the convection cells in the
solid polymer matrix. \FIG{UtiliseCells_a} demonstrates this principle for the honeycomb porous micro-structure thus created in a thin polymer film on a
glass substrate visualised by scanning electron imaging \cite{Munoz2014}. Lagrangian transport in the convection cells evidently is crucial to this fabrication technique
and utilisation of fundamental insights -- to e.g. actively manipulate droplet migration and congregation -- may enable its further development.

%\red{Engineered porous media: fabrication of porous polymer films/coatings.}

An approach intimately related to the above has been proposed for the micro-structuring of photonic crystals, which is key to the optical properties
and, inherently, the performance of solar cells \cite{Deng2015}. Here convection cells due to B\'{e}nard--Marangoni convection in a liquid perovskite film on a heated
substrate determine the advective solvent transport during solidification and thus ``assemble'' micro-structures in the crystals as visualised in \FIG{UtiliseCells_b} by
microscopy. The convection cells and, in consequence, the micro-structure admit manipulation by the substrate temperature and solution concentration. Further development
of this process may also greatly benefit from exploitation of fundamental insights into 3D Lagrangian transport in convection cells.

\paragraph{Porous media: generic characteristics}

\begin{figure}
\centering

\subfigure[Global and pore-scale transport]{\includegraphics[width=0.9\columnwidth,height=0.7\columnwidth]{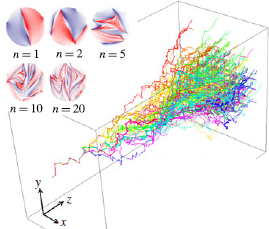}\label{PorousMedia1}}

\subfigure[Manifolds of skin-friction field]{\includegraphics[width=0.6\columnwidth,height=0.5\columnwidth]{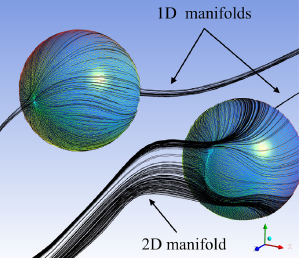}\label{PorousMedia1b}}

\caption{Transport in random porous media: (a) {\it global} transverse spreading
of $z$-wise throughflow visualised by simulated streamlines (distinguished by colour) starting
in single pore at inlet and corresponding {\it pore-scale} chaotic advection (inset) visualised by simulated mixing pattern; (b) manifolds emanating from stagnation points in skin-friction field on pore walls. Adapted from \cite{Lester2016b}
}
%\label{PorousMedia1}
\end{figure}

Porous media typically consist of a web of pores with random orientation and connectivity and throughflows, in
consequence, spread transverse to the main flow direction. \FIG{PorousMedia1} demonstrates this by the
$xy$-wise spreading of 3D streamlines in a steady $z$-wise throughflow in such a medium described by tracer particles (distinguished
by colour) released on a line in a single pore at the inlet $z=0$ \cite{Lester2013,Lester2016}. The global transverse
particle dispersion is dictated by the 3D structure of the pore network and, given its random
composition, converges on a Gaussian distribution with progressing stream-wise position $z$.

The local flow through consecutive pores undergoes repeated splitting and recombination similar as in partitioned duct flows (\SEC{Ducts}). A fundamental difference between these
systems and porous media is that the random (instead of systematic) reorientation in the latter generically results
in chaotic advection at the pore level \cite{Lester2013}. This is demonstrated in \FIG{PorousMedia1} (inset)
by the typical evolution of the transverse mixing pattern within pores (of circular cross-section) with stream-wise
position $z_n$ (starting from inlet $n=1$), revealing the characteristic stretching and folding that signifies chaotic
advection. Note the particular resemblance of the pore-scale mixing pattern with the patterns in the Quatro and serpentine mixers
in \FIG{Quatro2a} and \FIG{ChaoticSerpentineMicroMixer_c}, respectively, which also come about by splitting and recombination.

The LCSs associated with pore-scale chaotic advection in 3D steady flows are the manifolds emerging from stagnation
points $\vec{\tau} = \vec{0}$ of the skin-friction field
%
%$\vec{\tau} = (\vec{\nabla}\vec{u})\cdot\vec{n}_\Gamma$
%
\begin{eqnarray}
%\vec{\tau} = \left(\vec{\nabla}\vec{u}\right)\cdot\vec{n}_\Gamma,
%\vec{\tau} = (\vec{\nabla}\vec{u})\cdot\vec{n}_\Gamma,
\vec{\tau} = \vec{n}\cdot\left.\vec{\nabla}\vec{u}\right|_\Gamma = \left.\frac{\partial\vec{u}}{\partial n}\right|_\Gamma,
\label{SkinFrictionField}
\end{eqnarray}
on the no-slip pore wall $\Gamma$ with outward normal $\vec{n}$ \cite{Lester2013}. \FIG{PorousMedia1b} gives typical 1D and 2D manifolds emanating
from such points in a porous matrix consisting of disordered spheres (flow from right to left) \cite{Lester2016b}. These entities are the direct
counterpart to manifolds of internal stagnation or periodic points shown in e.g. \FIG{UnsteadyChaos1_b} and the 2D manifolds generically act as
separatrices between e.g. circulation zones or the diverging streams of impinging jets \cite{Surana2006}. Here the 2D manifolds of adjacent
pores, ensuing from impingement and flow separation, interact transversally in a manner akin to the stable (blue) and unstable (red) manifolds in \FIG{OrderWithinChaosa} and thus accomplish pore-scale
chaotic advection.
%
%The pore-scale chaotic advection results from non-trivial interaction of the 2D manifolds of adjacent pores in a manner akin to the stable (blue) and unstable (red) manifolds
%in \FIG{OrderWithinChaosa}.

Global {\it transverse} dispersion as in \FIG{PorousMedia1} originates from the random distribution of trajectories over pores comparable to the
downstream distribution over multiple blood vessels in the aneurysm flow in \FIG{CerebralAneurysm1b} and depends
primarily on the network geometry. Pore-scale chaos enhances transverse stretching of fluid elements
and thus promotes their split-up during change-over between consecutive pores, but its effect on global transverse dispersion, though
not yet conclusively established, seems marginal \cite{Lester2016b}.

Global {\it stream-wise} dispersion, on the other hand, is significantly affected by pore-scale chaos. It causes streamlines to ``sample'' the entire pore cross-section
and the full pore-scale velocity distribution, meaning that particles effectively migrate downstream by the cross-wise
{\it averaged} velocity, as opposed to the {\it local} velocity in the Poiseuille-like throughflows in 2D steady systems. This chaos-induced averaging of the particle velocity tends to homogenise the pore-wise residence time and thus retards stream-wise particle dispersion -- diminishing stream-wise ``diffusion'' of the global particle concentration -- relative to 2D porous media devoid of pore-scale chaos \cite{Lester2016,Lester2016b}.\footnote{
The same chaos-induced averaging occurs in periodic duct flows and, besides homogenisation of particle
velocities, furthermore promotes stream-wise throughflow of {\it all} particles, or equivalently, suppresses
particle entrapment in e.g. localised recirculation zones \cite{Speetjens2014}.
}

The above findings are relevant to randomly-structured porous media yet equally non-trivial Lagrangian transport is anticipated for 3D steady
throughflows in case of structure imparted by a natural heterogeneity (as e.g. anisotropy in fractured media) or engineered by design. This reinforces the link with periodic duct flows (which rely on {\it structural} reorientation) and thus likely yields pore-scale flow topologies
where, similar to \FIG{RAM2a}, chaotic regions coexist with LCSs as separatrices or stream tubes. Such intricate topologies are expected to cause fundamental departures of the global transport in 3D porous media from their 2D counterparts in a comparable way as before. However, essentially
3D pore-scale Lagrangian transport and its upscaling to global transport (as well as the interplay with diffusion) remain largely open problems \cite{Lester2016b,Dentz2018}.
This concerns in particular the proper upscaling of microscopic pore-scale behaviour to the mesoscopic level of ``representative elementary volumes'' (REVs). REVs are the smallest elementary volumes in a macroscopic system admitting averaged representations of flow and transport over both the fluid domain and porous matrix \cite{Bear1972}. Such averaged representations make complex problems
computationally tractable and thus are of great importance for practical analyses. This is elaborated below.

%\paragraph{Natural porous media: enhanced subsurface transport}
\paragraph{Natural porous media: subsurface transport}
%\paragraph{Random porous media: subsurface transport}
%\paragraph{Natural porous media}

\begin{figure}
\centering

%\subfigure[]{\includegraphics[width=0.48\columnwidth,height=0.4\columnwidth]{Figures/ContaminantPlume1.jpg}\label{ContaminantPlume_a}}
%\hspace*{6pt}
%\subfigure[]{\includegraphics[width=0.48\columnwidth,height=0.4\columnwidth]{Figures/ContaminantPlume2.jpg}\label{ContaminantPlume_b}}

\includegraphics[width=\columnwidth,height=0.65\columnwidth]{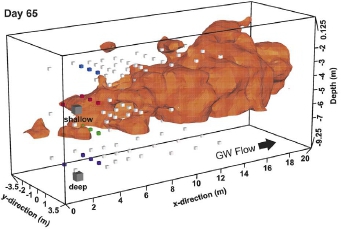}

\caption{Transport in natural porous media illustrated by subsurface spreading of methane plume through groundwater flow (``GW'') visualised
(via instantaneous concentration iso-surface) by ground-penetrating radar in field experiment following controlled injection at ``shallow'' and
``deep'' positions. Adapted from \cite{Steelman2017}.}

\label{ContaminantPlume}
\end{figure}

Natural porous layers in the subsurface typically posses a random structure and the above findings on both 3D pore-scale and 3D global transport demonstrated in \FIG{PorousMedia1} have
important ramifications for practical situations. Consider for example the contamination of soil and groundwater via the spreading of a methane plume due to leakage during enhanced oil recovery.
\FIG{ContaminantPlume} shows a typical instantaneous 3D methane gas plume (via a concentration iso-surface) visualised by ground-penetrating radar in a field experiment in an actual aquifier
following controlled injection at the indicated ``shallow'' and ``deep'' positions and spreading by the groundwater (``GW'') flow \cite{Steelman2017}. The analogy with the configuration in
\FIG{PorousMedia1} strongly
suggests similar 3D transport characteristics: pore-scale chaotic advection and, in consequence, suppressed downstream diffusion of the global plume front.
%
%\blue{However, conventional subsurface models do not account for these essentially 3D phenomena and thus likely provide incorrect predictions of %the plume spreading, which may have significant practical consequences.}

Crucial for large-scale practical problems such as the plume spreading in \FIG{ContaminantPlume} are tractable (computational) models for the global flow and transport. Full prediction and
simulation of pore-scale behaviour namely is (i) physically impossible due to a lack of detailed microscopic information and, even if available, (ii) prohibitively expensive. The concept of REVs for description of mesoscopic flow and transport introduced before is a common approach towards such macroscopic models \cite{Bear1972}. (The entire sample in \FIG{CellFormationa} constitutes an REV typical of sands or bead packs.)
The standard Darcy model is a well-known REV-based macroscopic model and assumes that the REV-averaged volumetric flux
density $\bar{\vec{q}}$ [m/s] (commonly denoted as the ``Darcy velocity'') is proportional to the driving pressure gradient following
\begin{eqnarray}
\bar{\vec{q}} = -\frac{\kappa}{\mu}\vec{\nabla} p,
\label{DarcyFlow}
\end{eqnarray}
with $\kappa$ and $\mu$ the permeability and fluid viscosity, respectively, at the mesoscopic REV scale \cite{Bear1972}. The average pore-scale velocity $\bar{\vec{v}}$ is given by the Dupuit-Forchheimer relation $\bar{\vec{v}} = \bar{\vec{q}}/\phi$, with $0<\phi<1$ the porosity, and e.g. corresponds with the mean stream-wise (``GW'') plume flow in \FIG{ContaminantPlume}.

The transverse and stream-wise spreading (of e.g. the plume in \FIG{ContaminantPlume}) by the dispersion of particles in the porous matrix is in conventional (Darcy-type) REV models incorporated via Fickian diffusion.
%
%superimposed upon the average
%
%characterised by diffusion coefficients ...
%
This expands the governing equations for the particle motion to an advection-diffusion equation of the form \eqref{ADE}, with $C$ representing the particle ``concentration'' \cite{Bear1972}. However, such upscaling from pore-scale to REV-scale transport does not account for phenomena as pore-scale chaos and its impact on global dispersion thus likely produces incorrect predictions of e.g. the plume spreading in \FIG{ContaminantPlume}, which may have significant practical consequences.
%
%\blue{Hence proper upscaling of non-trivial pore-scale transport is absolutely critical for the development of faithful REV models of complex porous media yet remains challenging %\cite{Lester2016b,Dentz2018}.}
%
%\blue{Hence upscaling primarily involves development of faithful REV models that adequately incorporate non-trivial 3D pore-scale transport \cite{Lester2016b,Dentz2018}.}
%
Hence development of faithful REV models that adequately incorporate non-trivial 3D pore-scale transport is absolutely critical for reliable analyses of practical problems
in complex porous media \cite{Lester2016b,Dentz2018}.

Subsurface REV models, besides mere analysis and prediction of transport in large-scale systems, furthermore enable utilisation of Lagrangian concepts for enhanced
subsurface transport.
%
% driven by injection and extraction wells.
%
This has great potential for a range of applications including enhanced geothermal systems
\cite{Brown2012,Chen2015}, {\it in situ} recovery of minerals or oil \cite{Seredkin2016,Guo2016} and groundwater remediation \cite{Cunningham2002,Goltz2012}.
%
% and basically always concerns one of two problems: (i) rapid extraction of a subsurface scalar (e.g. heat or contaminants) or (ii) {\it in situ} processing and remediation (e.g. of groundwater).
%
LCSs associated with the average pore-scale velocity $\bar{\vec{v}} = \bar{\vec{q}}/\phi$, with $\bar{\vec{q}}$ according to \eqref{DarcyFlow}, determine the global
subsurface transport characteristics and admit manipulation by the pumping schemes for the injection and extraction wells that must be used in such applications to drive the subsurface flow.
Thus purposeful conditions can (in principle) be systematically accomplished: global chaotic advection for efficient distribution of e.g. leaching
solutions for minerals extraction or creation of large-scale islands as {\it in situ} processing zones for groundwater remediation. First exploratory studies on 2D reservoirs serve as promising ``proof of principle'' of this Lagrangian approach and may pave the way to new technologies for subsurface flow engineering \cite{Bagtzoglou2007,Trefry2012,Piscopo2013,Neupauer2014,Varghese2017,Rodriguez2017,Speetjens2019}.

%\paragraph{Engineered porous media}
%\paragraph{Engineered porous media: enhanced (thermal) transport}
\paragraph{Engineered porous media: transport enhancement}
%\paragraph{Engineered porous media: microvascular networks}
%\paragraph{Micro-fluidic networks} Porous media:

\begin{figure}
\centering

\subfigure[Mixing of incoming fluid streams]{\includegraphics[width=0.8\columnwidth,height=0.4\columnwidth]{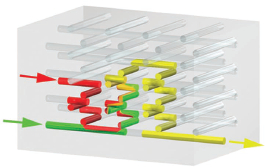}\label{MixingMicrovascularNetwork_a}}

\subfigure[Experimental visualisation (top view)]{\includegraphics[width=0.8\columnwidth,height=0.4\columnwidth]{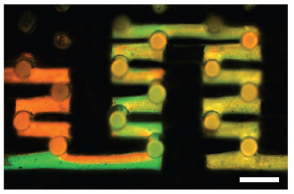}\label{MixingMicrovascularNetwork_b}}

\caption{Transport in engineered porous media illustrated by mixing in engineered 3D micro-vascular network: (a) mixing of incoming fluid streams (red/green) into
homogeneous outgoing fluid stream (yellow); (b) experimental visualisation by fluorescent dye (bar represents 500 $\mu$m). Adapted from \cite{Therriault2003}.}
\label{MixingMicrovascularNetwork}

\end{figure}

Engineering porous media with a given structure enables systematic manipulation of the (pore-scale) flow and transport and
%
%... functionalities.\\\\
%
%Engineering porous media with a given structure enables\blue{, besides or instead of external ... pumping schemes,} systematic manipulation and exploitation of the complex flow and
%transport inherent in such systems.
%
 e.g. finds application in engineered 3D micro-vascular networks for mixing purposes. These configurations in essence expand on inline partitioned-duct mixers (\SEC{DuctFlowIndustrial}) by progressive splitting and recombination of multiple fluid streams in a network of interconnected ducts to achieve
%both global transverse distribution and
``pore-scale'' chaos
%-- and, depending on the application, simultaneous global transverse distribution --
akin to that in random porous media (\FIG{PorousMedia1}).
\FIG{MixingMicrovascularNetwork_a} schematically demonstrates this
%
%for an internal architecture consisting of progressive branching of two ... reconnection in to one single outflow channel ...
%
by the interaction between two incoming fluid streams (red and green) and their mixing into a homogeneous outgoing
fluid stream (yellow) during progression through a micro-vascular network fabricated of cylindrical duct segments \cite{Therriault2003}. \FIG{MixingMicrovascularNetwork_b}
gives an experimental visualisation of the actual process using fluorescent dye.

The above example concerns a duct connectivity such that two incoming streams combine into one outgoing stream as per \FIG{MixingMicrovascularNetwork_a}; this renders each such connection within the porous matrix analogous to the T-shaped micro-mixer in \FIG{TshapedMicroMixer1} yet with a more complicated duct geometry. However, the same duct architecture admits more elaborate connectivity tailored to attain different transport characteristics as e.g. simultaneous ``pore-scale'' chaos and global transverse distribution as in \FIG{PorousMedia1} or a certain heterogeneity in global dispersion properties.
%Thus the considered microvascular network nicely demonstrates the potential of engineered porous media.

\begin{figure}
\centering

%\subfigure[ ... ]{\includegraphics[trim=0cm -0.5cm 0cm 0,clip,width=0.44\columnwidth,height=0.4\columnwidth]{Figures/MetalFoamHX1b.jpg}\label{MetalFoamHX_a}}
%\subfigure[ ... ]{\includegraphics[trim=0cm 0cm 0cm 0,clip,width=0.44\columnwidth,height=0.4\columnwidth]{Figures/MetalFoamHX1b.jpg}\label{MetalFoamHX_a}}
\subfigure[Fe-Cr-Al-Y etal foam]{\includegraphics[width=0.42\columnwidth,height=0.4\columnwidth]{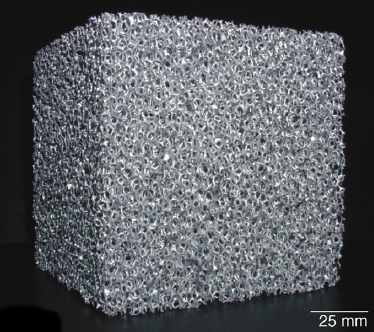}\label{MetalFoamHX_a}}
\subfigure[Cooling of hot fluid]{\includegraphics[width=0.55\columnwidth,height=0.4\columnwidth]{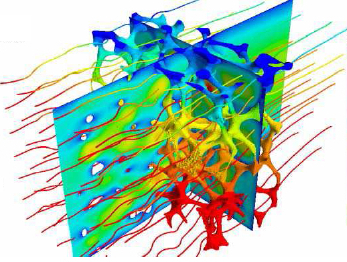}\label{MetalFoamHX_b}}

%\subfigure[Visualisation (top view)]{\includegraphics[width=0.8\columnwidth,height=0.4\columnwidth]{Figures/MixingMicrovascularNetwork2.jpg}\label{MixingMicrovascularNetwork_b}}

\caption{Transport in engineered porous media illustrated by compact heat exchanger based on metal foam: (a) typical metal foam (Fe-Cr-Al-Y alloy) for
heat-transfer purposes (reproduced from \cite{Zhao2012}); (b) simulated cooling of hot fluid flow entering (from the left) metal foam with cold
top/adiabatic bottom visualised by temperature of foam surface and streamlines (blue/red: min/max) and velocity magnitude (blue/red: zero/max) in planar cross-sections (adapted from \cite{Hugo2011}).
}

\label{MetalFoamHX}

\end{figure}

Engineered porous media may furthermore aim at enhancing scalar transfer {\it between} flow and porous matrix.
%
%Consider to this end hierarchically-structured porous materials for heat storage consisting of a metal foam for structural integrity ... with a
%phase-change material (PCM) for thermal functionality \cite{Li2012}. Inflow of hot air or water (heated in e.g. a
%solar-thermal collector) ``charges'' the PCM by a solid-to-liquid phase change; ...\\\\
%
% configuration: metal matrix embedded in PCM; heating causes melting of solid PCM; buoyancy-induced flow\\
%
An important application exists in compact heat exchangers for heating/cooling of fluid flows based on metal foams (\FIG{MetalFoamHX_a})
\cite{Boomsma2003,Hugo2011,Zhao2012}. The metal foam realises the favourable attributes of a heat exchanger in a small volume: high
interfacial area between solid and fluid; high thermal conductivity in the solid; complex flow geometry promoting mixing and, in consequence,
enhancing convective heat transfer. \FIG{MetalFoamHX_b} demonstrates a typical situation by the simulated cooling of a hot fluid
flow entering (from the left) a metal foam with an imposed cold top side and adiabatic bottom \cite{Hugo2011}. This reveals an essentially 3D flow
and temperature field and thus implies a non-trivial interplay of heat transfer by 3D pore-scale (chaotic) advection and diffusion in both
fluid and solid.
Foam-based heat exchangers generically involve dense structures comparable to \FIG{MetalFoamHX_a}, necessitating, similar to the above subsurface systems, REV representations for modelling and analysis of practical devices.
However, 3D pore-scale transport phenomena as demonstrated in \FIG{MetalFoamHX_b} are, for reasons given before, typically beyond conventional REV models, meaning that proper upscaling is again crucial for reliable description of the global transport characteristics.

\subsubsection{Relevance beyond industry}
\label{WebsNonIndustrial}

\paragraph{Plate tectonics}

\begin{figure}
\centering

%\subfigure[...]{\includegraphics[width=0.48\columnwidth,height=0.5\columnwidth]{Figures/MantleConvection4.jpg}\label{MantleConvection}}
%\subfigure[...]{\includegraphics[width=0.48\columnwidth,height=0.5\columnwidth]{Figures/MantleConvection3.jpg}\label{MantleConvection}}
%\hspace*{6pt}
%\subfigure[..]{\includegraphics[width=0.42\columnwidth,height=0.42\columnwidth]{Figures/TransportLivingTissue3.jpg}\label{TransportLivingTissue_b}}

%\subfigure[Plate tectonics]{\includegraphics[trim=0cm -3cm 0cm 0,clip,width=0.5\columnwidth,height=0.35\columnwidth]{Figures/MantleConvection5c.jpg}\label{MantleConvection_a}}
%\subfigure[...]{\includegraphics[width=0.49\columnwidth,height=0.49\columnwidth]{Figures/MantleConvection3b.jpg}\label{MantleConvection_b}}

\subfigure[Plate tectonics due to mantle convection]{\includegraphics[trim=0cm 0cm 0cm 0,clip,width=0.7\columnwidth,height=0.35\columnwidth]{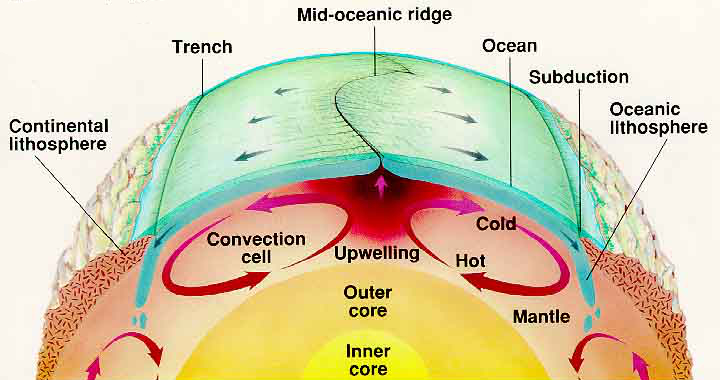}\label{MantleConvection_a}}
%\subfigure[...]{\includegraphics[width=0.6\columnwidth,height=0.6\columnwidth]{./MantleConvection3b.jpg}\label{MantleConvection_b}}
\subfigure[Up/downwellings demarcating convection cells]{\includegraphics[width=0.7\columnwidth,height=0.72\columnwidth]{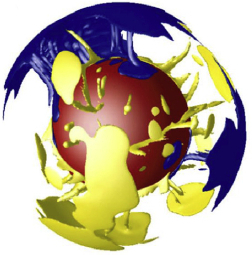}\label{MantleConvection_b}}

%\subfigure[Single-sided subduction]{\includegraphics[width=0.9\columnwidth,height=0.3\columnwidth]{Figures/MantleConvection7.jpg}\label{MantleConvection_c}}
\subfigure[Single-sided subduction]{\includegraphics[width=0.9\columnwidth,height=0.33\columnwidth]{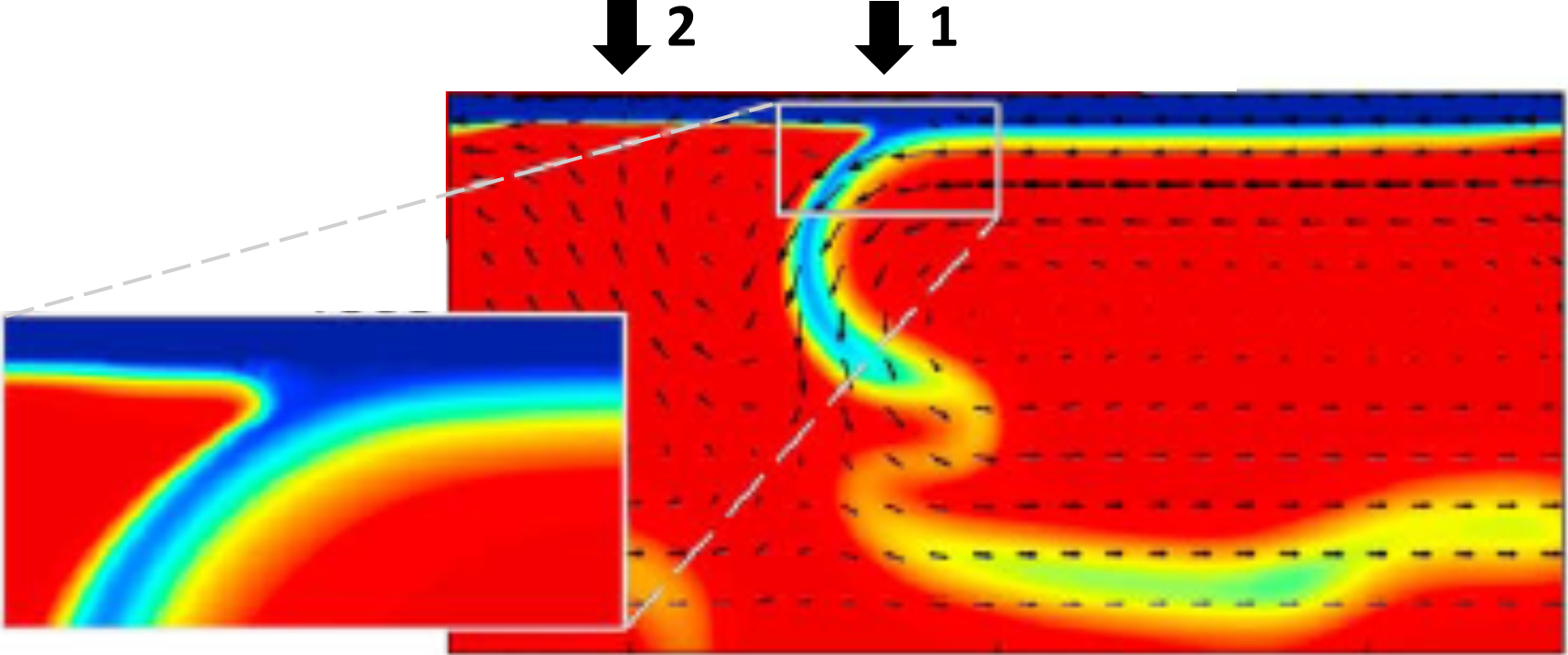}\label{MantleConvection_c}}

\caption{Plate tectonics induced by mantle convection as geophysical instance of convection cells: (a) mechanism (reproduced from \cite{Monroe1992}); (b) up- and downwellings demarcating convection cells visualised by ``low'' (blue) and ``high'' (yellow)
simulated temperatures (adapted from \cite{Zhong2007}); (c) single-sided subduction demonstrated by intrusion of low-temperature (blue) crust
material into high-temperature (red) mantle at position 1 and resulting circulation zone between positions 1 and 2 (adapted from \cite{Crameri2012}).}

\label{MantleConvection}
\end{figure}

Convection cells set up in the mantle of the Earth (consisting of viscous molten rock) by its hot inner core are a major driving mechanism behind plate
tectonics in the lithosphere (i.e. the outer crust) \cite{Monroe1992,COLTICE2017120}. The associated up- and downwellings by hot and cold thermal
plumes, respectively, namely cause the formation of ridges and trenches (via so-called ``subduction'') due to viscous friction exerted on the crust
in a way as illustrated in \FIG{MantleConvection_a} and thus contribute to shaping the topography of the Earth's surface.
This configuration is analogous to B\'{e}nard-Marangoni convection in a liquid film by concerning a fluid layer inside a 3D spherical shell
as in \FIG{UnsteadyChaos2_a} (left) that is heated from below by the hot inner sphere and subject to lithosphere-induced surface stresses at the outer sphere.
Hence the system tends to form a pattern of convection cells akin to \FIG{SurfaceDefects} with corresponding up- and downwellings due to thermal plumes centred on lines ``d'' and ``c'', respectively, in \FIG{ConvectionCells1b}.
However, different (material) compositions of oceanic versus continental lithospheres and complex mantle rheologies significantly complicate this process, resulting in highly irregular patterns of up- and
downwellings as illustrated in \FIG{MantleConvection_b} by level sets of ``low'' (blue) and ``high'' (yellow) temperatures, respectively \cite{Zhong2007}. This irregularity suggests comparably disordered surfacial patterns (i.e. in the lithosphere) tending to that of drying paint in \FIG{SurfaceDefects_b}.

The computational study by \cite{Crameri2012}, modeling mantle and lithosphere as a single continuum with an
appropriate rheology, reveals that crust deformation is essential to the single-sided subduction (i.e. one tectonic plate
sliding below another as in \FIG{MantleConvection_a}) that in reality occurs at downwellings.
%
%Entrainment and transport of crust material into the mantle is central to this process yet, remarkably, dedicated 3D Lagrangian studies are %non-existent.
%
\FIG{MantleConvection_c} demonstrates this process by a cross-section of the temperature field, representing (via its strong correlation with the viscosity) the material composition, where low (blue) and high (red) temperature corresponds with crust and mantle, respectively. This reveals the downward intrusion of a high-viscosity ``plume'' of crust material (blue) into the mantle (red) at the contact line (position 1) of two converging
plates driven by the convective circulation. The corresponding velocity (black arrows) exposes a localised circulation zone to the left of the intruding plume,
meaning that the latter primarily consists of material stemming from the right plate, signifying its single-sided subduction below the
left plate.

Entrainment and transport of crust material into the mantle is central to the subduction and is, by its essentially Lagrangian nature, determined by the 3D flow topology. This, by analogy with convection cells, implies a scenario as before: breakdown of separatrices as plane ``a'' in \FIG{ConvectionCells1b}
in response to some perturbation and onset of dynamics as e.g. global dispersion according to \FIG{ConvectionCells2b}. Shown behaviour strongly
suggests that here the non-trivial rheology renders these separatrices unstable and thus induces the development of an asymmetric flow field
and topology. The local circulation zone in \FIG{MantleConvection_c} implies, besides the contact line at position 1, a further stagnation point
near position 2 and, given the clock-wise flow, interaction of the unstable and stable manifolds associated with former and
latter, respectively. These considerations implicate LCSs emerging from the lithosphere at such stagnation points/lines in the subduction, which
in 3D likely yields (local) chaotic advection.
%
%Dedicated Lagrangian analyses enable further exploration of these phenomena.
%Dedicated 3D Lagrangian analyses are thus expected to deepen insights into this phenomenon.
%Dedicated 3D Lagrangian analyses may thus considerably deepen insights into this phenomenon.
Dedicated Lagrangian analyses enable further exploration of these phenomena.
%
%Configuration = in essence 3D spherical counterpart to B\'{e}nard-Marangoni convection: 3D spherical annulus as in \FIG{UnsteadyChaos2_a} (left); %inner sphere stationary; buoyancy by hot inner sphere; outer sphere = free-slip condition; partitioning into mantle (inner thick layer attached %to inner sphere), atmosphere (thin outer layer attached to outer sphere) and crust (intermediate layer) modelled by different density, viscosity %and initial temperature; accounts for the non-trivial interaction between Mantle and crust that underlies tectonics; creates non-trivial stress %conditions in thin outer layer not unlike those due to surface tension in B\'{e}nard-Marangoni convection.
%

\paragraph{Micro-circulation in living tissue}

\begin{figure}
\centering

\subfigure[Impact of vascular occlusion]{\includegraphics[width=0.43\columnwidth,height=0.6\columnwidth]{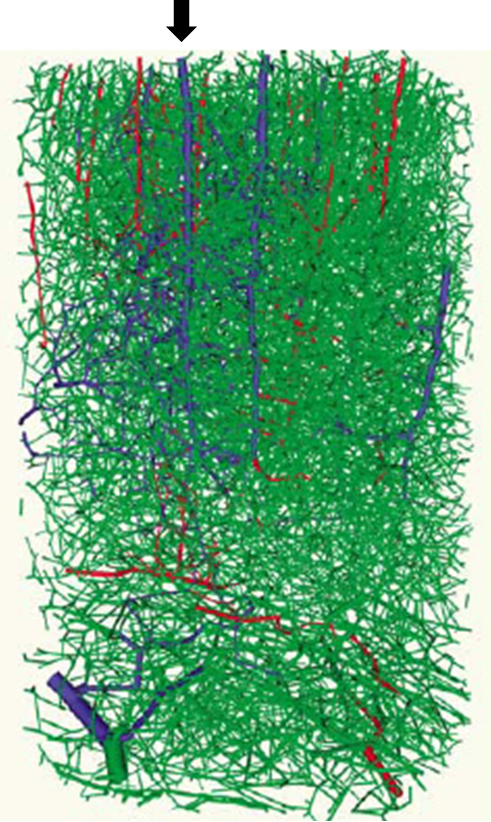}\label{TransportLivingTissue2a}}
\subfigure[Impact of tumor cells]{\includegraphics[width=0.56\columnwidth,height=0.6\columnwidth]{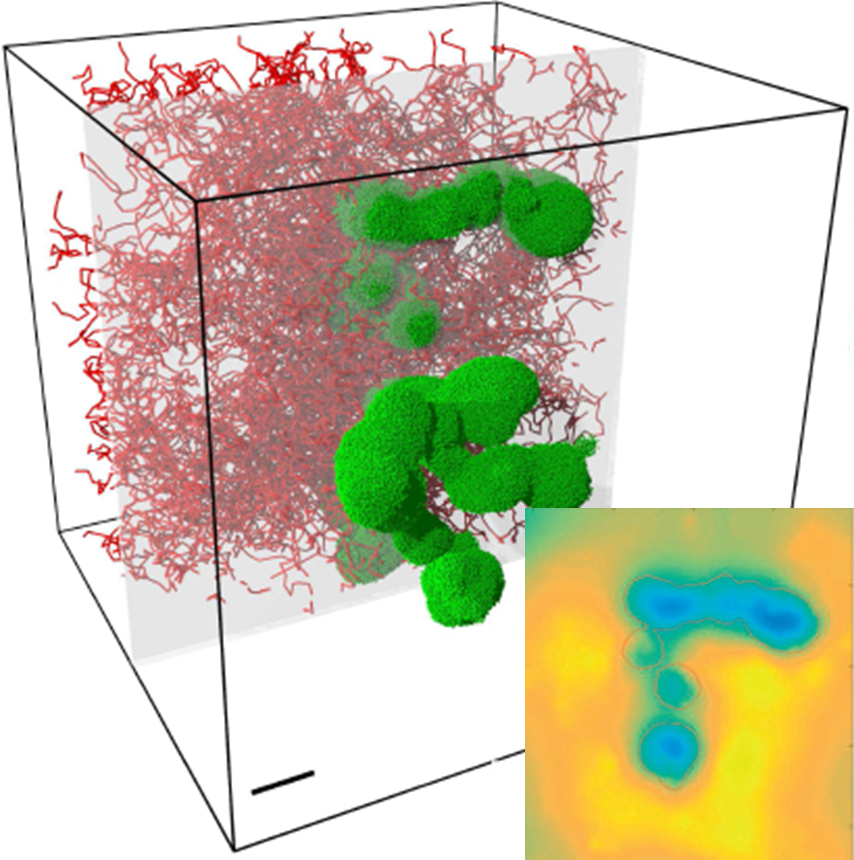}\label{TransportLivingTissue2b}}

\caption{Micro-circulation in living tissue as physiological instance of transport in porous networks: (a) impact of vascular occlusion (arrow) on global throughflow (from bottom) in
somatosensory cortex in rat brain visualised by simulated flow rate (red/blue: relatively larger/lower flow rate; green: unaffected region) (adapted from \cite{Reichold2009}); (b)
impact of tumor cells (green) on oxygen distribution via vascular network (red) visualised by simulated oxygen levels (yellow/blue: healthy/hypoxic)
in grey cross section (inset) (adapted from \cite{Ghaffarizadeh2016}).}

\label{TransportLivingTissue2}
\end{figure}

Exchange of oxygen, carbon dioxide and nutrients with living tissue occurs via so-called ``micro-circulation'' through a local network of microscopic blood vessels \cite{Pittman2013}. Each
organ has its own dedicated micro-circulation; e.g. lung alveoli (\SEC{ContainersNonIndustrial}) are covered by pulmonary capillaries for oxygen transfer into the blood stream.
\FIG{TransportLivingTissue2a} illustrates the complexity of micro-vascular networks by that of the somatosensory cortex in the brain of a rat visualised by 3D imaging using X-ray
tomographic microscopy \cite{Reichold2009}. The tissue and embedded network effectively constitutes a random porous medium, implying, by analogy with such media, generic transport
characteristics as demonstrated in \FIG{PorousMedia1}: chaotic advection locally within the vessels and, in consequence, relatively weak global stream-wise dispersion.

Important in a physiological context is the impact of local disruptions or malformations of the vascular network on the global transport by the micro-circulation. The computational
study by \cite{Reichold2009} investigates the effect of a vascular occlusion (i.e. blockage) of the blood vessel at the arrow in \FIG{TransportLivingTissue2a} by reduction of
its diameter by a factor 1,000 using a model based on graph representations of the micro-vasculature. Blue and red vessels in \FIG{TransportLivingTissue2a} have a significantly lower
and higher flow rate, respectively, compared to the unobstructed situation (green vessels are unaffected) and, by emerging throughout the entire network, expose a substantial impact
of the occlusion on the 3D global throughflow (from bottom to top). This is likely to cause departures from the transport properties of healthy networks. Typical vessel-wise residence
times, irrespective of the nature of the local transport, directly correlate with flow rates. Moreover, local transport depends, similar to periodic duct flows (\SEC{Ducts}),
strongly on the flow rate. Hence, though probably retaining its chaotic nature due to the random connectivity, quantitative changes may nonetheless alter the residence-time distributions
of affected blood vessels. The occlusion thus is expected to considerably affect the 3D global (stream-wise) dispersion properties.
%
%\blue{Treatment as anomalies in a random porous medium may offer a way for in-depth 3D transport analyses along the lines of %\cite{Lester2013,Lester2016}.}
%
Insights into both local and global impact of occlusions on 3D transport are crucial for deeper understanding of the physiology
of strokes, since reducted or disrupted cerebral oxygen delivery
%usually have major ramifications for the brain tissue and
is a major cause for this brain condition \cite{Deb2010}.
%
%The cerebrovascular system continuously delivers oxygen and energy substrates to the brain, which
%is one of the organs with the highest basal energy requirement in mammals. Discontinuities in the
%delivery lead to fatal consequences for the brain tissue.

Occlusions, by altering the flow rate, are malformations in the micro-vascular network that affect the (local) {\it permeability} (i.e. parameter $\kappa$ in \eqref{DarcyFlow}). Malformations in the tissue itself mainly alter another key mesoscopic transport property, namely the (local) {\it porosity} $\phi$. Notorious examples are cancerous tumors that, due to their rapid growth, typically have underdeveloped vascular networks and thus effectively reduce $\phi$. Hence tumors rely strongly on diffusion of oxygen from surrounding healthy tissue and, given this is a relatively slow process, tend to become hypoxic, which attributes to their malignant behaviour \cite{Spill2016}.
\FIG{TransportLivingTissue2b} demonstrates this for the simulated 3D oxygen transport in a realistic vascular network (red) containing several
tumors (green)
%\blue{using the diffusion-only computational model by} \cite{Ghaffarizadeh2016}.
using the computational model by \cite{Ghaffarizadeh2016,Macklin2016}.
The lower-right inset shows the local oxygen distribution in the grey cross-section and reveals a one-to-one correspondence between the oxygen-poor (blue) and
oxygen-rich (yellow) regions and the tumor and healthy tissue, respectively.
%
%\blue{Said model concerns only diffusion as transport mode and includes oxygen
%supply by the vascular network via line sources, however. Further model development aims at incorporating advection (and Lagrangian transport) \cite{Ghaffarizadeh2016}.
%}

Micro-vascular networks, unlike subsurface porous regions (\SEC{WebsIndustrial}), admit full characterisation to ever greater detail via medical imaging techniques, which paves
the way to
%flow and transport studies by
dedicated (REV-like) computational models incorporating the specific structure as demonstrated in \FIG{TransportLivingTissue2}.
%\blue{using the diffusion-only computational model by} \cite{Ghaffarizadeh2016}.)
This
%, given the continuous advancement in computer technology,
has great potential for computer-aided diagnosis and treatment based on numerical simulation of flow and transport in the patient-specific micro-vasculature and surrounding tissue \cite{Reorowicz2014,Rejniak2016,Jonasova2018}. Such computational medical tools may thus enable identification of regions affected by occlusions, prediction of the progression of hypoxic tumors or development of strategies for optimal drug delivery for chemotherapy.
The results shown in \FIG{TransportLivingTissue2} are obtained with the model by \cite{Ghaffarizadeh2016} that in fact incorporates the specific vascular structure. However, it
relies on a simplified diffusion-only transport model that represents the oxygen supply via the blood stream as line sources of oxygen at the vessel locations; advective and Lagrangian
transport associated with this stream is omitted. Hence, besides network structure and tissue composition, incorporation of the relevant (Lagrangian) transport phenomena in both modelling
and data interpretation is equally important.

%\blue{Incorporation of insights into 3D Lagrangian transport into models as well as data interpretation and treatment procedures is imperative for physical relevance and meaningfulness.}
%
%Powerful computational tools in concert with insights into local and global 3D (Lagrangian) transport in porous media e.g. enable identification of regions affected by occlusions, prediction of the %progression of hypoxic tumors or develop strategies for optimal drug delivery for chemotherapy.
%
%\red{... employ strategies similar to reservoir engineering to design injection protocols for drugs.}
%
%\blue{\bf Treatment as anomalies in the \red{permeability} or \red{porosity} in a random porous medium may offer a way for in-depth transport analyses along the lines of %\cite{Lester2013,Lester2016}.}\\\\

%\input{Theory_arXiv.tex}

%\section{Theoretical framework}
%\section{Theoretical framework for practical transport analyses}
%\section{Theoretical framework for 3D Lagrangian transport analyses}
%\section{Theoretical framework for 3D Lagrangian analyses}
%\section{Framework for 3D Lagrangian transport analyses}
%\section{Unified Lagrangian framework for 3D transport analyses}
%\section{Unified framework for 3D transport analyses}
%\section{Unified Lagrangian framework for transport studies}
\section{Unified Lagrangian framework for 3D transport}
\label{Theory}

The discussion in this section seeks to reconcile the above flow categorisation with theoretical/fundamental concepts and methods so as to (i) outline a unified framework
for systematic 3D Lagrangian transport analyses of practical flows and (ii) provide a ``gateway'' to
further literature on this subject matter. To this end the essentials of Lagrangian transport are treated in a tutorial-like approach via
{\it capita selecta} and exemplified with the systems highlighted in \SEC{Categorisation}.

\subsection{Critical points as organising entities}
\label{CriticalPoints}

Key organising entities of the Lagrangian flow topology are so-called ``critical points'' associated with the Lagrangian motion described
by kinematic equation \eqref{KinEq}. Critical points ``shape'' the flow topology by geometrically and topologically governing the proliferation
of fluid trajectories. These entities are common (or even inherent) in confined geometries both in the flow interior and on no-slip (internal)
walls and thus are of great relevance to practical flows.
The emergence and nature of critical points is elaborated below for 2D and 3D flows under steady and time-periodic conditions.

\subsubsection{Critical points in 2D flows}
%\paragraph{Stagnation points in steady flows}
%\subsubsection{Stagnation points in 2D steady flows}
\label{CriticalPoints2D}

%Consider to this end the linearisation $G(t)$ of flow $\vec{\Phi}_t$ near $\vec{x}_0$, i.e.

\begin{figure}
\centering

%\subfigure[elliptic]{\includegraphics[width=0.36\columnwidth,height=0.36\columnwidth]{Figures/EllipticPoint.jpg}\label{CriticalPoints_a}}
%\hspace*{12pt}
%\subfigure[hyperbolic]{\includegraphics[width=0.36\columnwidth,height=0.36\columnwidth]{Figures/HyperbolicPoint.jpg}\label{CriticalPoints_b}}

\subfigure[Elliptic (left) and hyperbolic (right) point in 2D flows]{
\includegraphics[width=0.4\columnwidth,height=0.4\columnwidth]{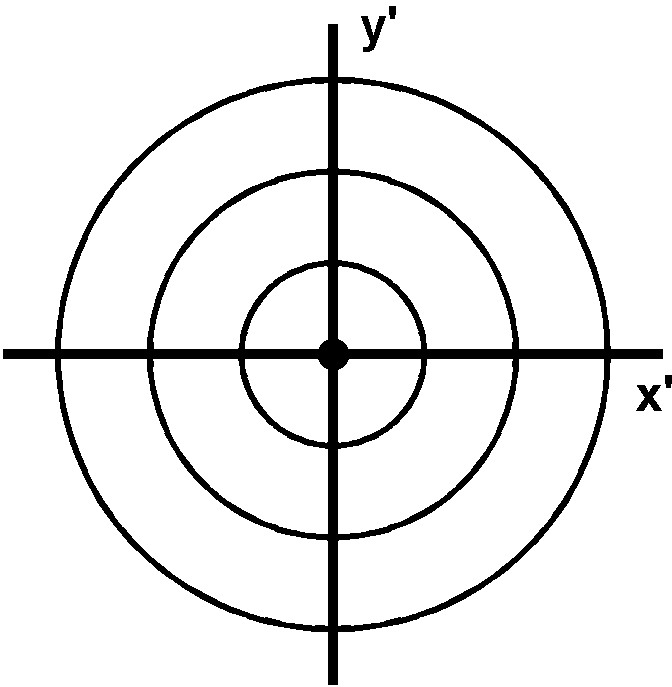}
\hspace*{12pt}
\includegraphics[width=0.4\columnwidth,height=0.4\columnwidth]{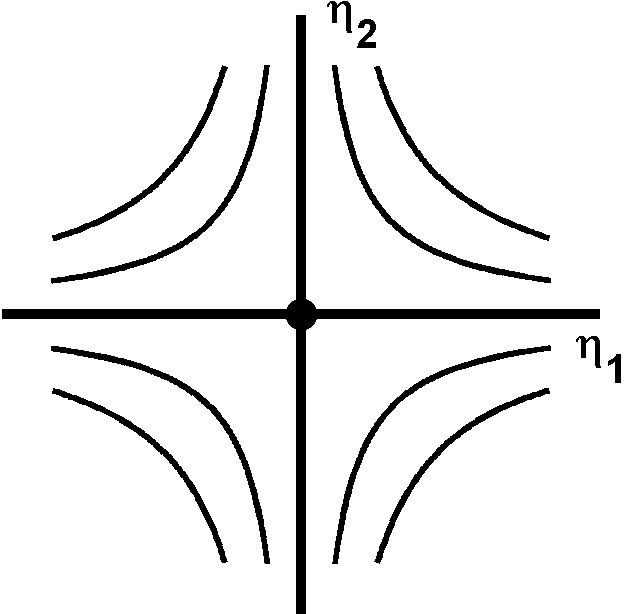}
\label{CriticalPoints_a}
}

\subfigure[Hyperbolic focus (left) and hyperbolic node (right) in 3D flows]{
\includegraphics[width=0.48\columnwidth,height=0.42\columnwidth]{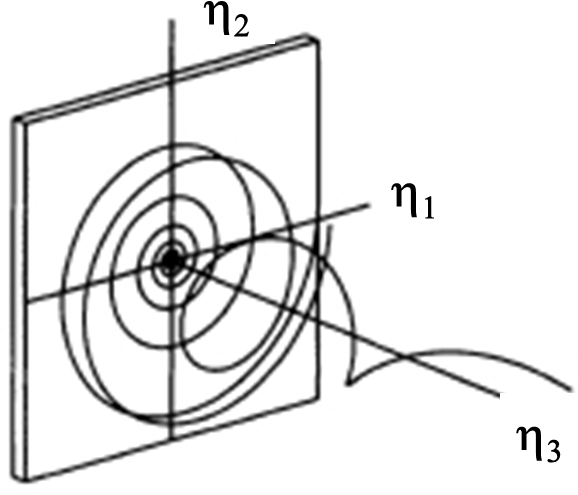}
\hspace*{3pt}
\includegraphics[width=0.48\columnwidth,height=0.42\columnwidth]{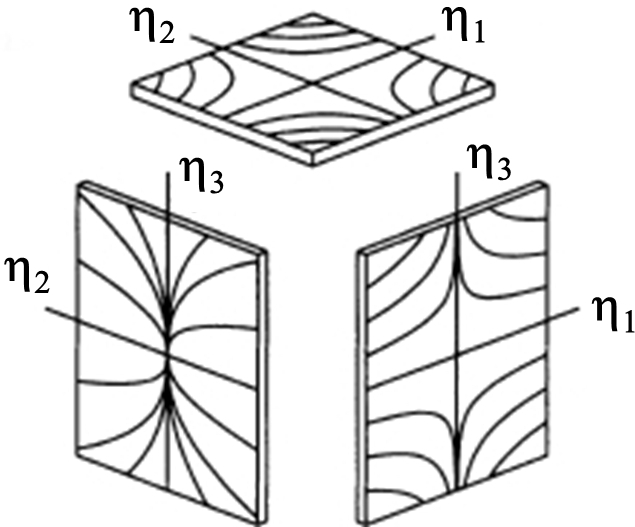}

\label{CriticalPoints_b}}

\caption{Flow topology near critical points $\vec{x}_0$ in the local reference frame $\vec{x}'= \vec{x} - \vec{x}_0$ with hyperbolic
points represented in the associated canonical reference frame $(\eta_1,\eta_2)$ and $(\eta_1,\eta_2,\eta_3)$ in 2D and 3D flows, respectively. Panel (b) adapted from \cite{Malyuga_cylinder_2002}.}
\label{CriticalPointsX}
\end{figure}

Critical points in {\it steady} flows $\vec{u} = \vec{u}(\vec{x})$ correspond with stagnation points, that is, positions $\vec{x}_0$ at which the fluid motion vanishes:
\begin{eqnarray}
\xvec{u}\left(\vec{x}_0\right) = 0\quad\Rightarrow\quad\vec{x}(t) = \vec{\Phi}_t(\vec{x}_0) = \vec{x}_0.
\label{StagPoint}
\end{eqnarray}
The flow topology near stagnation points $\vec{x}_0$ is determined by the linearisation $G(t)$ of flow $\vec{\Phi}_t$, i.e.
\begin{eqnarray}
%\frac{d\vec{x}'}{dt} = A\vec{x}'
%,\quad
%\quad\Rightarrow\quad
\vec{x}'(t) = G(t)\vec{x}'_0,\quad G(t) = \mbox{e}^{A t},\quad A = \left.\frac{\partial \vec{u}}{\partial \vec{x}}\right|_{\vec{x}_0},
\label{FlowLin}
\end{eqnarray}
with $\vec{x}'= \vec{x} - \vec{x}_0$ the local position vector and $A$ the strain-rate tensor. The linearised flow in spectral representation reads
\begin{eqnarray}
G(t)
%\mbox{e}^{A t}
= \sum_{i=1}^2 \mbox{e}^{\lambda_i t}\vec{v}_i,\quad
\lambda_{1,2} = \frac{J_1\pm\sqrt{J_1^2 - 4J_2}}{2},
\label{FlowLin2}
\end{eqnarray}
with $(\lambda_i,\vec{v}_i)$ the eigenvalue-eigenvector pairs of $A$ and
\begin{eqnarray}
J_1 =
%\mbox{tr}(A) =
\left.\vec{\nabla}\cdot\vec{u}\right|_{\vec{x}_0},\quad
J_2 =
%\mbox{det}(A) =
\left.\frac{\partial u_x}{\partial x}\frac{\partial u_y}{\partial y}\right|_{\vec{x}_0} - \left.\frac{\partial u_x}{\partial y}\frac{\partial u_y}{\partial x}\right|_{\vec{x}_0},
\label{FlowLin3}
\end{eqnarray}
the invariants of its characteristic equation $\lambda^2 - J_1\lambda + J_2 = 0$. Solenoidality through $J_1 = \mbox{tr}(A) = \sum_i\lambda_i =0$ yields $\lambda_1 = -\lambda_2 = \sqrt{-J_2}$ and thus
implies two kinds of stagnation points depending on the sign of $J_2 = \mbox{det}(A)$.
%
%
%\begin{itemize}
%
%\item[]
%
Case $J_2>0$ gives a complex-conjugate pair of eigenvalues $(\mbox{i}\lambda,-\mbox{i}\lambda)$, with $\mbox{i} = \sqrt{-1}$ and
$\lambda = \sqrt{J_2}$, and the local Lagrangian motion \eqref{FlowLin} in polar coordinates becomes
\begin{eqnarray}
r'(t) = r'_0,\quad \theta'(t) = \theta'_0 + \lambda t,
\label{FlowLin4a}
\end{eqnarray}
describing concentric circular streamlines around $\vec{x}_0$ following \FIG{CriticalPoints_a} (left).
Case $J_2<0$ gives a pair of real eigenvalues $(\lambda,-\lambda)$, with $\lambda = \sqrt{|J_2|} > 0$, resulting in
\begin{eqnarray}
\vec{x}'(t) = \sum_{i=1}^2\eta_i(t)\vec{v}_i,\quad \eta_i(t) = \eta_i(0)\mbox{e}^{\lambda_i t},
%\vec{x}'(t) = \eta_1\mbox{e}^{\lambda t}\vec{v}_1 + \eta_2\mbox{e}^{-\lambda t}\vec{v}_2,
%
%\eta_1(t) = \mbox{e}^{\lambda t}\eta_1(t),\quad \eta_2(t) = \mbox{e}^{-\lambda t}\eta_1(t),
%
\label{FlowLin4b}
\end{eqnarray}
as local Lagrangian motion, with $\vec{\eta} = (\eta_1,\eta_2)$ the canonical coordinates associated with the principal strain axes $(\vec{v}_1,\vec{v}_2)$, describing hyperbolic streamlines following \FIG{CriticalPoints_a} (right). Stagnation points ``shape'' the flow topology by imparting the topology of the linearised flow on their vicinity (i.e. the surrounding region devoid of other stagnation points). The step-wise steady flows in \FIG{LidDrivenCavity} in fact contain only a single stagnation point of the former kind ($J_2>0$) and the global streamline portrait, in consequence, is topologically equivalent to \FIG{CriticalPoints_a} (left).

%Former and latter kind of stagnation point are hereafter denoted {\it elliptic} and {\it hyperbolic} point, respectively, on the basis of the local streamline pattern.\footnote{\red{Note on terminology: elliptic and hyperbolic stagnation points are in literature commonly denoted ``center'' and ``saddle'', respectively. Here ... elliptic/hyperbolic is adopted to %underscore the essential similarity with periodic points and also for brevity ...}}

%\subsubsection{Periodic points in 2D time-periodic flows}

Critical points in {\it time-periodic} flows correspond with periodic points \eqref{PeriodicPoint} and are direct counterparts of stagnation points \eqref{StagPoint}. Local linearisation of mapping \eqref{StroboMap2} namely gives
\begin{eqnarray}
\vec{x}'_p = F^p\vec{x}'_0,\quad F = \left.\frac{\partial \vec{\Phi}_T}{\partial \vec{x}}\right|_{\vec{x}_0},
\label{MappingLin}
\end{eqnarray}
yielding an identical (spectral) structure as \eqref{FlowLin2} and \eqref{FlowLin3} upon substitution $\vec{u}\rightarrow\vec{\Phi}_T$ and
$\mbox{e}^{\lambda_i t}\rightarrow \lambda_i$, with $(\lambda_i,\vec{v}_i)$ here the eigenvalue-eigenvector pairs of deformation
tensor $F$.
Solenoidality through $J_2 = \mbox{det}(F) = \prod_i\lambda_i = 1$ yields
%
%$\lambda_1 = 1/\lambda_2 = J_1/2 + \sqrt{(J_1/2)^2-1}$
\begin{eqnarray}
\lambda_1 = 1/\lambda_2 = J_1/2 + \sqrt{(J_1/2)^2-1},
\label{MappingLin2a}
\end{eqnarray}
and
again implies two kinds of points yet here depending on $J_1 = \mbox{tr}(F)$ due to the different emergence of the
eigenvalues in the spectral structure. Case $|J_1|<2$ gives a complex-conjugate pair of eigenvalues $(\lambda,\lambda^\ast)$, with
$\lambda = J_1/2 + \mbox{i}\sqrt{1 - (J_1/2)^2}$ and $\lambda^\ast = 1/\lambda$; case $|J_1|>2$ gives a pair of real
eigenvalues $(\lambda,1/\lambda)$, with $\lambda = \lambda_1 > 1$. This results for former and latter case in
\begin{eqnarray}
%r'_p = r'_0,\quad \theta'_p = \theta'_0 + p\phi,
(r'_p,\theta'_p) = (r'_0,\theta'_0 + p\phi),\quad
\vec{x}'_p = \sum_{i=1}^2\eta_i\lambda_i^p\vec{v}_i,
\label{MappingLin2}
\end{eqnarray}
respectively, with $\phi = \mbox{arg}(\lambda) = \arctan(\sqrt{(2/J_1)^2-1})$, revealing local Lagrangian motion identical to that near the stagnation
points following \eqref{FlowLin4a} and \eqref{FlowLin4b}. (Note $|\lambda|=1$ for complex $\lambda$.) Hence the periodic points for $|J_1|<2$ and $|J_1|>2$ are indeed {\it direct}
counterparts to the above stagnation points for $J_2>0$ and $J_2<0$, respectively \cite{Speetjens2001}.

Stagnation/periodic points for complex-conjugate and real eigenvalue pairs are denoted {\it elliptic} and {\it hyperbolic} points, respectively, based on the structure of the local flow topology (\FIG{CriticalPointsX}).\footnote{Elliptic and hyperbolic stagnation points are in literature commonly denoted ``center'' and ``saddle'', respectively \cite{Guckenheimer}. The current nomenclature is adopted to underscore the essential similarity with periodic points.} Elliptic and hyperbolic points create LCSs that are topologically equivalent to \FIG{CriticalPoints_a} (left) and
\FIG{CriticalPoints_a} (right), respectively. Thus elliptic points constitute centres of the islands in e.g. \FIG{LidDrivenCavity_a} and \FIG{OrderWithinChaosa}; principal axes $\vec{v}_1$ and $\vec{v}_2$ of hyperbolic points are the origin of unstable and stable manifolds, respectively, as shown e.g. in \FIG{OrderWithinChaosa}.

The above reveals that elliptic and hyperbolic points are the only critical points possible in 2D solenoidal flows. This has the fundamental implication that the corresponding flow topologies are {\it always} composed of arrangements of LCSs corresponding with these points.
%
%Other entities as e.g. streamlines spiralling into a stagnation (implying shrinking of fluid parcels into a single point) are physically impossible.
%
Hence flow topologies with islands embedded in chaotic regions demarcated by manifolds following \FIG{OrderWithinChaosa} are characteristic for generic 2D time-periodic flows. However, important to note is that hyperbolic points and manifolds, though typically present, do not imply chaos {\it perse}; decisive for chaotic advection is the particular behaviour of the manifolds (\SEC{ConditionsChaos}).

%their presence is a necessary yet {\it not} sufficient for chaotic advection (\SEC{ConditionsChaos}).}

\subsubsection{Critical points in 3D flows}
\label{CriticalPoints3D}

%3D flows: similar analysis yields also here two fundamental kinds of stagnation points for solenoidal flows: focus and node as generalisations of elliptic and
%hyperbolic points.

Stagnation points and associated flow topologies are in 3D steady flows also determined by the spectral properties of strain-rate tensor $A$ following \eqref{FlowLin}, which
are governed by the characteristic equation $\lambda^3 - J_1\lambda^2 + J_2\lambda - J_3 = 0$, with invariants $J_1 = \mbox{tr}(A)$, $J_2 = (\mbox{tr}^2(A) - \mbox{tr}(A^2))/2$ and $J_3 = \mbox{det}(A)$.
Solenoidality again yields $J_1 = 0$ and real $A$ implies (at least) one real eigenvalue (say $\lambda_3$). Hence the dynamics in the $\eta_3$-direction (spanned by eigenvector $\vec{v}_3$) is always described
by
%
%$\eta_3(t) = \eta_3(0)\mbox{e}^{\lambda_3 t}$
%
\begin{eqnarray}
\eta_3(t) = \eta_3(0)\mbox{e}^{\lambda_3 t},
\label{Stagpoint3D0}
\end{eqnarray}
meaning that stagnation points differ solely by the behaviour in the $(\eta_1,\eta_1)$-plane. Two kinds can again be distinguished.

Case $J_3 > \lambda_3^3/4$ yields (besides real $\lambda_3$) a complex-conjugate pair of eigenvalues $(\lambda_R+\mbox{i}\lambda_I,\lambda_R-\mbox{i}\lambda_I)$, with
\begin{eqnarray}
\lambda_R = -\lambda_3/2,\quad \lambda_I = \sqrt{J_3/\lambda_3-(\lambda_3/2)^2},
\label{Stagpoint3D1}
\end{eqnarray}
resulting in local Lagrangian motion \eqref{FlowLin} in the $(\eta_1,\eta_2)$-plane (in polar coordinates) according to
\begin{eqnarray}
r'(t) = r'_0\mbox{e}^{\lambda_R t},\quad \theta'(t) = \theta'_0 + \lambda_I t,
\label{Stagpoint3D2}
\end{eqnarray}
describing spiralling streamlines about the $\eta_3$-axis following \FIG{CriticalPoints_b} (left). This kind of stagnation point is denoted
{\it hyperbolic focus} and is a 3D generalisation of the elliptic points of 2D flows. Principal difference with the latter is the (generically)
non-constant radius -- and ensuing spiralling motion -- due to $\lambda_R = -\lambda_3/2 \neq 0$. Two situations may occur: (i) distancing
from the $(\eta_1,\eta_1)$-plane while contracting around the $\eta_3$-axis for $\lambda_3>0$; (ii) approaching the $(\eta_1,\eta_2)$-plane
while diverging from the $\eta_3$-axis for $\lambda_3<0$. Said plane and axis define the 2D unstable (stable) and 1D stable (unstable) manifolds, respectively, in the former (latter) case.
The infinitesimal plane spanned by principal axes $\vec{v}_{1,2}$ is the origin of the 2D manifold in physical space; principal axis $\vec{v}_{3}$ is the origin of the corresponding 1D manifold.

Case $J_3 < \lambda_3^3/4$ gives (besides real $\lambda_3$) real eigenvalues
\begin{eqnarray}
\lambda_{1,2} = -\lambda_3/2 \pm \sqrt{(\lambda_3/2)^2 - J_3/\lambda_3},
\label{Stagpoint3D3}
\end{eqnarray}
and (upon properly redefining the canonical frame) can always be chosen such that $\lambda_1$ and $\lambda_{2,3}$ are of opposite sign. Thus Lagrangian
motion in both the $(\eta_1,\eta_2)$ and $(\eta_1,\eta_3)$-planes is essentially similar to the hyperbolic points of 2D flows according to \eqref{FlowLin4b} and yields 3D dynamics
following \FIG{CriticalPoints_b} (right). Hence the present kind of stagnation point is a generalisation of said points and denoted {\it hyperbolic node} hereafter.\footnote{The nomenclature of \cite{Malyuga_cylinder_2002} is adopted as uniform terminology for stagnation and periodic points. The classification for (instantaneous) 3D streamline patterns by \cite{Chong_general_1990} refers to the hyperbolic focus as ``focus'' and the hyperbolic node as ``node-saddle-saddle'' due to the saddle-like and node-like dynamics in the $(\eta_1,\eta_2)$,$(\eta_1,\eta_3)$-planes and $(\eta_2,\eta_3)$-plane, respectively.}
Here the $(\eta_2,\eta_3)$-plane and $\eta_1$-axis define the 2D unstable (stable) and 1D stable (unstable) manifolds, respectively, for $\lambda_1<0$ ($\lambda_1>0$).
The steady lid-driven cylinder flow for $Re>0$ in \FIG{SteadyCylinderFlow_0b} and \FIG{SteadyCylinderFlow_a} e.g. accommodates a hyperbolic focus (not shown) at the
intersection of the toroidal axis of the tori and the symmetry plane $y=0$. Said axis and plane in fact define the corresponding 1D and 2D manifolds, respectively.

The direct analogy between stagnation and periodic points is retained for 3D flows \cite{Malyuga_cylinder_2002}. The local behaviour near periodic points remains dependent
on the deformation tensor $F$ following \eqref{MappingLin} that, upon the substitution adopted before, again assumes the same spectral structure as $A$. Solenoidal flow
gives $J_3 = \mbox{det}(F) = 1$ and real $F$ implies (at least) one real eigenvalue (say $\lambda_3$) and, in consequence, $\eta_3$-wise dynamics similar
to \eqref{Stagpoint3D0}: $\eta_{3,p} = \eta_{3,0}\lambda_3^p$. Hence the kind of periodic points in 3D flows is also determined by the behaviour in the $(\eta_1,\eta_1)$-plane.
Case $J < 2/\sqrt{\lambda_3}$ and $J > 2/\sqrt{\lambda_3}$, where $J = \lambda_1 + \lambda_2 = J_1-\lambda_3$, define the counterparts to \eqref{Stagpoint3D1}
and \eqref{Stagpoint3D3}, respectively, with here
\begin{eqnarray}
\lambda_R = J/2,\quad \lambda_I = \sqrt{1/\lambda_3-(J/2)^2}.
\label{PeriodicPoint3D1}
\end{eqnarray}
as real and imaginary parts of the complex eigenvalue and
\begin{eqnarray}
\lambda_{1,2} = J/2 \pm \sqrt{(J/2)^2-1/\lambda_3},
\label{PeriodicPoint3D3}
\end{eqnarray}
as real eigenvalues. The local Lagrangian motion in the $(\eta_1,\eta_2)$-plane is a generalisation of \eqref{MappingLin2} and becomes
\begin{eqnarray}
(r'_p,\theta'_p) = (|\lambda|^pr'_0,\theta'_0 + p\phi),\quad \vec{x}'_p = \sum_{i=1}^2\eta_i\lambda_i^p\vec{v}_i,
\label{PeriodicPoint3D2}
\end{eqnarray}
with $\phi = \arctan(\sqrt{(2/J)^2/\lambda_3-1})$ and $|\lambda| = 1/\sqrt{\lambda_3}$. (Principal difference with 2D flows is that generically $|\lambda|\neq 1$ and $\lambda_1\lambda_2 = 1/\lambda_3\neq 1$.) Thus cases $J < 2/\sqrt{\lambda_3}$ and $J > 2/\sqrt{\lambda_3}$
result in dynamics according to \FIG{CriticalPoints_b} (left) and \FIG{CriticalPoints_b} (right), respectively, and indeed are entirely equivalent to the
above stagnation points. Hence former and latter are also denoted {\it hyperbolic focus} and {\it hyperbolic node}.
The isolated periodic point in the time-periodic cylinder flow in \FIG{UnsteadyChaos1_b} (not shown) e.g. is a hyperbolic focus. The highly-convoluted 2D and 1D manifolds emanate from the $(\eta_1,\eta_2)$-plane and $\eta_3$-axis, respectively, of the local canonical frame in \FIG{CriticalPoints_b} (left) and demonstrate the global complexity that these entities may assume.
However, key for the nature of the Lagrangian dynamics, that is, chaotic versus non-chaotic, is -- in steady and time-periodic 3D flows -- again the particular behaviour of these manifolds (\SEC{ConditionsChaos}).

Critical points of the skin friction field $\vec{\tau}$ of 3D steady flows following \eqref{SkinFrictionField} have a flow topology
that corresponds with the hyperbolic focus and emerge in a ``saddle'' and ``node'' configuration if the canonical $(\eta_1,\eta_2)$-plane and $(\eta_2,\eta_3)$-plane, respectively,
in \FIG{CriticalPoints_b} (right) coincides with the wall \cite{Lester2013,Lester2016b}. (The fact that generically $\vec{\nabla}\cdot\vec{\tau}\neq 0$ is immaterial here.)
The associated 1D and 2D manifold give rise to a stream{\it line} and stream {\it surface} extending from the wall into the domain in the node and saddle configurations, respectively, in a way as
demonstrated in \FIG{PorousMedia1b}.

%\subsubsection{Stagnation and periodic lines in 3D flows}
\subsubsection{Critical lines in 3D flows}
\label{CriticalLines}

Stagnation points may merge into continuous curves and thus form so-called ``stagnation lines'' that, by definition, are characterised by absence of motion -- and, inherently, deformation -- in
the tangent direction $\vec{e}_t$. This implies $G\vec{e}_t = \vec{e}_t$ for the linearised flow \eqref{FlowLin} and thus renders the tangent an eigenvector of the strain-rate
tensor $A$, say $\vec{v}_3 = \vec{e}_t$, with corresponding eigenvalue $\lambda_3 = 0$.
Tensor $A$ becomes singular for $\lambda_3=0$ (i.e. $J_3=0$) and invalidates expressions \eqref{Stagpoint3D1} and \eqref{Stagpoint3D3} for the remaining eigenvalues $\lambda_{1,2}$ due
to an undefined fraction $J_3/\lambda_3$. Alternative relations are found via the simplified characteristic equation in the singular state, i.e. $\lambda(\lambda^2 - J_1\lambda + J_2)=0$, that
through $J_1=0$ yields $\lambda_{1,2} = \sqrt{-J_2}$ and thus gives the same form as for stagnation points in 2D steady flows. Hence the dynamics in the canonical ($\eta_1,\eta_2$)-plane transverse
to each point on the stagnation line is either elliptic ($J_2>0$) or hyperbolic ($J_2<0$) according to \FIG{CriticalPoints_a} (left) and \FIG{CriticalPoints_a} (right), respectively.
This has the fundamental implication that stagnation lines in 3D flows signify (locally) essentially 2D dynamics.
The common centre (dots) of the global family of closed streamlines in the Stokes limit ($Re=0$) of the steady cylinder flow in \FIG{SteadyCylinderFlow_0a} e.g. is a stagnation line consisting entirely of elliptic points.

Periodic points may in a similar way merge into so-called ``periodic lines'' and are the counterpart to stagnation lines in 3D time-periodic flows. Here absence
of motion in tangent direction $\vec{e}_t$ implies $F\vec{e}_t = \vec{e}_t$ for the linearised stroboscopic map \eqref{MappingLin} and thus gives an eigenvalue-eigenvector pair $(\lambda_3,\vec{v}_3) = (1,\vec{e}_t)$.
Relations \eqref{PeriodicPoint3D1} and \eqref{PeriodicPoint3D3} for $\lambda_3=1$ collapse on relation \eqref{MappingLin2a} for $|J|<2$ and $|J|>2$, respectively, and consequentially
identify with the spectral signatures for elliptic and hyperbolic points in 2D time-periodic flows. The dynamics transverse to periodic lines is, entirely analogous with stagnation lines,
therefore also essentially 2D.

The periodic line in the time-periodic cylinder flow in \FIG{UnsteadyChaos2_b} e.g. is partitioned into segments of elliptic (normal) and hyperbolic (heavy) points, yielding local transverse dynamics following \FIG{CriticalPoints_a} (left) and \FIG{CriticalPoints_a} (right), respectively. The quasi-2D chaos in the time-periodic cylinder flow
driven by alternating $x$-wise translation of top and bottom walls in \FIG{UnsteadyChaos1_a} (centre) e.g. emanates from multiple periodic lines arranged transverse to the
moving walls and consisting entirely of elliptic or hyperbolic points (not shown) \cite{Malyuga_cylinder_2002}. The ``holes'' in the chaotic layer outline tubes formed by merger of islands of the constituent periodic points of the elliptic lines.
Both the stable and unstable manifolds of the hyperbolic lines (in time-periodic as well as steady flows) are 2D surfaces that, similarly, result from merger of the manifold pairs of its individual points.

\subsubsection{Closed trajectories in 3D flows}
%\subsubsection{Closed streamlines in 3D steady flows}
\label{ClosedTrajectories}

An important generalisation of critical points and lines exists in closed streamlines in 3D steady flows and closed trajectories in stroboscopic maps of 3D time-periodic flows.

\paragraph{Closed streamlines in 3D steady flows}
%Fluid parcels travelling along {\it closed streamlines}
%
%Fluid parcels travelling along these entities
%
are described by fluid parcels that return to their initial position after a finite cycle time $\tau$, i.e. $\vec{\Phi}_\tau(\vec{x}_0) = \vec{x}_0$, meaning that the
parcel position $\vec{x}(t)$ is a periodic solution of kinematic equation \eqref{KinEq}: $\vec{x}(t) = \vec{x}(t+\tau)$. The local dynamics near a fluid parcel with momentary position
$\vec{x}_0(t)$ is governed by a linearised flow of the form \eqref{FlowLin} in the co-moving reference frame $\vec{x}'(t) = \vec{x}(t) - \vec{x}_0(t)$. The periodic nature of the strain-rate tensor, i.e. $A(t+\tau) = A(t)$, admits a linear coordinate transformation
$\vec{y}(t) = P(t)\vec{x}'(t)$, with periodic transformation matrix $P(t) = P(t+\tau)$, such that the linearised kinematic equation \eqref{KinEq} near $\vec{x}_0(t)$ becomes
time-independent, i.e.
\begin{eqnarray}
\frac{d\vec{x}'}{dt} = A(t)\vec{x}'(t)
\quad\Rightarrow\quad
\frac{d\vec{y}}{dt} = B\vec{y}(t),
\label{ClosedStreamline1}
\end{eqnarray}
with $B$ the constant strain-rate tensor in the $\vec{y}$-frame \cite{Verhulst2006}. The equivalence with the flow \eqref{FlowLin} near stagnation points readily yields $\vec{y}(t) = \mbox{e}^{B t}\vec{y}_0$ and, in turn, leads to
%
%This results in a local flow in terms of $\vec{y}$ that is similar to \eqref{FlowLin}, i.e. $\vec{y}(t) = \mbox{e}^{B t}\vec{y}_0$,
%in turn yielding
%
\begin{eqnarray}
%\vec{y}(t) = \mbox{e}^{B t}\vec{y}_0
%\quad\Rightarrow\quad
\vec{x}'(t) = G(t)\vec{x}'_0,\quad G(t) = P(t)\mbox{e}^{B t}P^{-1}(t),
\label{ClosedStreamline2}
\end{eqnarray}
as linearised flow in the co-moving physical frame. Note that flow $G$ generically is non-periodic: $G(t) \neq G(t+\tau)$.

Relation \eqref{ClosedStreamline2} exposes the flow near a closed streamline as a local similarity transform of the steady flow in the $\vec{y}$-frame, meaning that
the dynamics is determined by the spectral properties of $B$, which has important ramifications.
Material line elements of the streamline return to their initial position after time $\tau$, implying $G(\tau)\vec{e}_t(\tau) = \vec{e}_t(\tau)= \vec{e}_t(0)$, with $\vec{e}_t(t) = \vec{e}_t(t+\tau)$ again
the tangent, resulting in an eigenvalue-eigenvector pair $(\lambda_3,\vec{v}_3) = \left(0,P^{-1}(\tau)\vec{e}_t(\tau)\right)$ for $B$. This situation is identical to that for stagnation lines and thus gives
%
%$\lambda_{1,2} = \sqrt{-J_2(B)}$
%
\begin{eqnarray}
\lambda_{1,2} = \sqrt{-J_2(B)},
\label{ClosedStreamline3}
\end{eqnarray}
for the eigenvalues in the direction transverse to the closed streamline, rendering also here the behaviour essentially 2D. Moreover, given $B$ characterises the entire closed
streamline, the dynamics is qualitatively the same everywhere (i.e. the behaviour remains either fully elliptic or hyperbolic during the excursion of a fluid parcel).

Closed streamlines are key LCSs in many of the practical steady flows considered in \SEC{Categorisation}. The tori in the impeller-driven flow in
\FIG{StirredTankFountain_b} are centred on elliptic closed streamlines and the surrounding chaotic sea emanates from accompanying hyperbolic closed streamlines. The
inherent 2D nature of the transverse dynamics established above explains the qualitative resemblance of the cross-sectional topology in \FIG{StirredTankFountain_c} and
\FIG{StirredTankFountain_d} with 2D time-periodic systems (\FIG{OrderWithinChaosa}).
The tori and chaotic regions in e.g. the rotating-lid and translating-lid cylinder flows in \FIG{VortexBreakdownBubble} and \FIG{SteadyCylinderFlow}, respectively, the ACEO-driven
flow in \FIG{ACEO1} as well as the baseline topologies of micro-droplets in \FIG{ChaosDroplets2X_b} correspond in a similar way with elliptic/hyperbolic streamlines.

Streamlines reconnecting via the periodic inlet-outlet of steady duct flows (denoted ``periodic streamlines'' hereafter)
must always exist in these systems and thus are crucial to the Lagrangian transport properties of e.g. industrial inline mixers \cite{Speetjens2014}. These LCSs namely invariably result in a flow topology as demonstrated for the RAM in \FIG{RAM2a}: stream-wise tori centered on elliptic streamlines embedded in chaotic regions ``driven'' by hyperbolic streamlines.
Former and latter emerge as elliptic and hyperbolic periodic points, respectively, in the cross-sectional topology in \FIG{RAM2b}.
%
%These LCSs also here explain the essentially 2D (chaotic) dynamics.
%
The magnetic flux surfaces in MHD representations of tokamak fusion reactors in \FIG{Tokamak} are, by analogy with periodic duct flows (\SEC{DuctFlowNonIndustrial}), centered on elliptic magnetic field lines. The ``o-points'' and ``x-points'' in \FIG{Tokamakb} are poloidal intersections of elliptic and hyperbolic field lines.
The invariable emergence of closed (magnetic) streamlines in these systems also here explains their essentially 2D (chaotic) dynamics.

%\subsubsection{Closed trajectories in 3D time-periodic flows}

\paragraph{Closed trajectories in 3D time-periodic flows} correspond with stroboscopic maps \eqref{StroboMap} of individual fluid parcels that densely fill closed
loops over infinitely many periods. To this end the associated mapping \eqref{StroboMap2} must be {\it aperiodic}, that is,
the individual parcel positions in \eqref{StroboMap} remain distinct at all times.
%
%{\it none} of the individual parcel positions in \eqref{StroboMap} ever coincide.
%
Such closed trajectories e.g. sit at the centre of the tori in the rheology-induced time-periodic cylinder flow in \FIG{Unsteady1a} and the
time-periodic annular flow in \FIG{UnsteadyChaos2_a}.

Closed trajectories thus described are, irrespective of the progression of fluid parcels along discrete positions, smooth and continuous closed curves
reminiscent of a closed streamline.
Continuity then suggests an equally smooth and continuous topology in their direct proximity
that, in consequence, may be considered equivalent to the streamline portrait of a fictitious steady solenoidal ``flow'' $\vec{\nu} = \vec{\nu}(\vec{x})$.
%
%Continuity then suggests that the structure of the stroboscopic map in the direct proximity of these entities
%is topologically equivalent to the streamline portrait of a fictitious steady ``flow'' $\vec{\nu} = \vec{\nu}(\vec{x})$ meeting $\vec{\nabla}\cdot\vec{\nu}=0$.
%
Fluid parcels are in this representation mapped along streamlines of $\vec{\nu}$ governed by kinematic equation $d\vec{x}/ds = \vec{\nu}$, with
$s$ a fictitious time, and the closed trajectory -- analogous to ``true'' closed streamlines -- is its periodic solution $\vec{x}(s) = \vec{x}(s+\tau)$.

Fictitious time $s$ acts as a spatial coordinate along the streamlines of $\vec{\nu}$ and thus admits expression of the positions $\vec{x}_p$ of a
fluid parcel on the closed trajectory in terms of a corresponding $s_p$. This, via \eqref{ClosedStreamline2}, readily yields
\begin{eqnarray}
%\vec{x}'(t) = G(t)\vec{x}'_0,\quad G(t) = P(t)\mbox{e}^{B t}P^{-1}(t),
\vec{x}'_p = F_p\vec{x}'_0,\quad F_p = P(s_p)\mbox{e}^{B s_p}P^{-1}(s_p),
\label{ClosedOrbit1}
\end{eqnarray}
as local mapping in the co-moving reference frame and, by analogy, implies the same behaviour as near closed
streamlines: essentially 2D dynamics governed by eigenvalue pair \eqref{ClosedStreamline3} of either fully elliptic or fully hyperbolic nature.

The extent of the tori in the beforementioned rheology-induced time-periodic cylinder flow in \FIG{Unsteady1a} demonstrates that the essentially
2D behaviour associated with closed trajectories may impact a sizeable region. However, the 2D dynamics imparted by these entities can also
be very localised, as exemplified by the defective tori -- signifying 3D dynamics -- in the time-periodic annular flow in \FIG{UnsteadyChaos2_a} due
to RID. Similar restriction of the 2D dynamics of closed trajectories by resonance-induced tori breakdown is suspected in e.g. time-periodic droplet flows driven
by electro-wetting in \FIG{ElectroWetting} and the time-periodic flow in the vitreous chamber of the eye (\FIG{DrugDeliveryEye}). The basic mechanisms
of RID are elaborated in \SEC{Conditions3Ddynamics}.

\subsubsection{Symmetries as organising mechanisms}
\label{Symmetries}

Symmetries play a critical role in the structure of the flow topology and, inherently, the Lagrangian dynamics of 3D (un)steady flows \cite{Aref2017}. Insight into this role is important especially for industrial flows, since manufactured devices and engineered flows often have symmetries due to the geometry and/or the repetitive nature of the process such as the systematic flow reorientation in periodic duct flows (\SEC{DuctFlowIndustrial}).

\begin{figure}
\centering
\subfigure[Time-reversal symmetry]{\includegraphics[width=0.4\columnwidth,height=0.45\columnwidth]{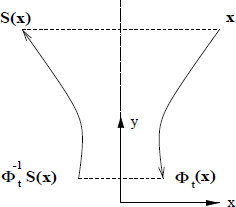}\label{Symmetry_a}}
\hspace*{6pt}
%\subfigure[...]{\includegraphics[width=0.52\columnwidth,height=0.45\columnwidth]{Figures/Symmetry2.jpg}\label{Symmetry_b}}
%\subfigure[...]{\includegraphics[width=0.52\columnwidth,height=0.45\columnwidth]{Figures/Symmetry2b.jpg}\label{Symmetry_b}}
\subfigure[Continuous symmetry]{\includegraphics[width=0.52\columnwidth,height=0.45\columnwidth]{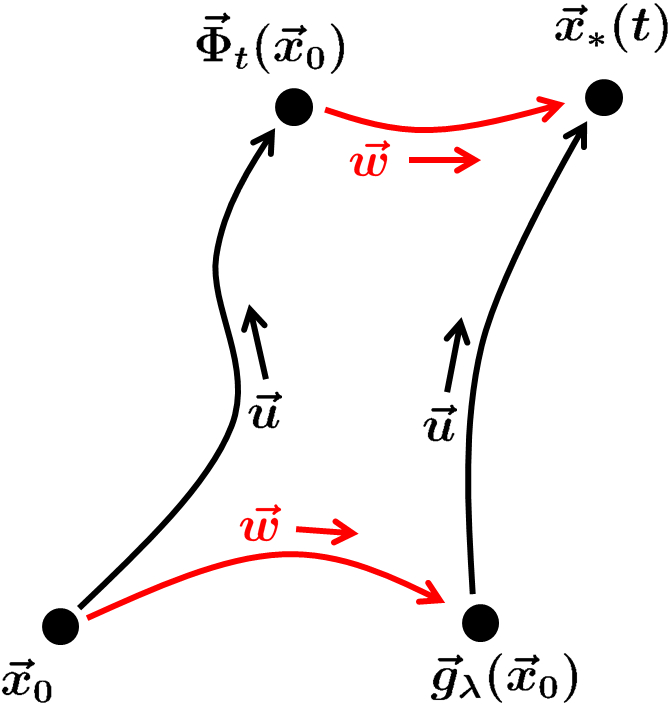}\label{Symmetry_b}}

\caption{Symmetries in the Lagrangian dynamics: (a) discrete time-reversal reflectional symmetry $S:(x,y,z)\rightarrow(-x,y,z)$ about plane $x=0$ (reproduced from \cite{Speetjens2001}); (b) continuous symmetry $\vec{w}$ commuting with flow $\vec{u}$ according to $\vec{x}_\ast(t) = \vec{g}_\lambda(\vec{\Phi}_t(\vec{x}_0)) = \vec{\Phi}_t\left(\vec{g}_\lambda(\vec{x}_0)\right)$.
}
\label{Symmetry}
\end{figure}

Discrete geometrical
and dynamical symmetries in particular are key organising mechanisms for critical points/lines and closed trajectories as well as their associated LCSs. An important
class exists in reflectional symmetries $S$ (i.e. mirroring positions or LCSs about some symmetry plane $\mathcal{P}$) and two important instances can be distinguished
within the present scope: {\it ordinary} and {\it time-reversal} reflectional symmetries. Former and latter are for both flow
$\vec{\Phi}_t$ and mapping $\vec{\Phi}_T$ following \eqref{KinEq} and \eqref{StroboMap2} given by
\begin{eqnarray}
\vec{\Phi} = S\vec{\Phi} S,\quad
\vec{\Phi} = S\vec{\Phi}^{-1}S,
\label{Symmetries2}
\end{eqnarray}
respectively, and relate the (reversed) dynamics on either side of the symmetry plane $\mathcal{P}$. \FIG{Symmetry_a} illustrates this for the time-reversal reflectional symmetry about plane $x=0$ (i.e. $S_x:(x,y)\rightarrow(-x,y)$):
the trajectory in the forward flow starting from $\vec{x}$ is the mirror image of that in the reversed flow starting from $S(\vec{x})$ about $x=0$.

Both ordinary and time-reversal reflectional symmetries following \eqref{Symmetries2} imply emergence of LCSs $\mathcal{L}$ as either entities symmetric about $S$, i.e. $S(\mathcal{L}) = \mathcal{L}$ or as symmetric pairs $\{\mathcal{L},S(\mathcal{L})\}$. Time-reversal reflectional symmetries furthermore imply stagnation lines (steady flow) and periodic lines (time-periodic flow) in symmetry plane $\mathcal{P}$ \cite{Malyuga_cylinder_2002,Speetjens_stokes_2004}. The steady cylinder flow for $Re=0$ in \FIG{SteadyCylinderFlow_0a} e.g. has an ordinary reflectional symmetry $S_y: (x,y,z)\rightarrow (x,-y,z)$ and time-reversal reflectional symmetry $S_x: (x,y,z)\rightarrow (-x,y,z)$ about plane $y=0$ and plane $x=0$, respectively, resulting in the (elliptic) stagnation line in the former and symmetric streamline pairs about the latter plane. Symmetry $S_y$ is preserved under inertial conditions ($Re>0$) and thus dictates a symmetric arrangement of tori and chaotic trajectories in \FIG{SteadyCylinderFlow_0b} and \FIG{SteadyCylinderFlow_a} about plane $y=0$ (mirror images not shown). The periodic line in plane $y=-x$ of the time-periodic cylinder flow in \FIG{UnsteadyChaos2_b} (right) in a similar way emanates from a time-reversal reflectional symmetry $S_{xy}: (x,y,z)\rightarrow (-y,-x,z)$ \cite{Speetjens_stokes_2004}.
Symmetries may also manifest themselves locally. The time-reversal reflectional symmetry $S_\alpha: (r,\theta,z) \rightarrow (r,\alpha - \theta,z)$ of the base flow $\vec{v}$ of the 2.5D RAM flow about the centerline $\alpha/2$ $(=\pi/8)$ of the first aperture results
in a reflectional symmetry $S_\gamma: (r,\theta) \rightarrow (r,\alpha - \Theta - \theta)$ about axis $\theta = (\alpha - \Theta)/2$ $(= -3\pi/40)$ in the cross-section (\FIG{RAM2b}) of the 3D flow topology in \FIG{RAM2b} \cite{Speetjens_symmetry_2006}.

Time-reversal reflectional symmetries furthermore imply that streamlines (within a confined flow domain) crossing the
symmetry plane $\mathcal{P}$ must cross twice and form closed streamlines \cite{Shankar1997,Shankar_cylinder_1998}. Thus symmetry $S_x$ of the non-inertial steady cylinder flow gives rise to a continuous family of closed streamlines centered on
said stagnation line (of which one specimen is shown in \FIG{SteadyCylinderFlow_0a}). Conversely, breaking of such a symmetry (if existent) is necessary for non-trivial (chaotic) dynamics; the emergence of tori embedded in chaos in \FIG{SteadyCylinderFlow_0b} and \FIG{SteadyCylinderFlow_a} is e.g. a direct consequence of the breakdown of $S_x$ by fluid inertia.

The above global organisation of the flow topology generalises to stroboscopic maps with a time-reversal reflectional symmetry.
Islands emerge either as symmetric pairs or as symmetric entities about $\mathcal{P}$ as demonstrated in \FIG{OrderWithinChaosa}
for the time-periodic 2D lid-driven flow, which has a time-reversal reflectional symmetry about the horizontal centerline
(not shown). The chaotic sea is subject to the same organisation in that the stable and unstable manifolds of hyperbolic
points/lines within $\mathcal{P}$ form symmetric pairs about this plane. Thus the stable (blue) and unstable (red)
manifolds in \FIG{OrderWithinChaosa} are mirror images about said centerline. The 2D manifolds associated with the
hyperbolic segments of the periodic line in \FIG{UnsteadyChaos2_b} (right) in a similar way are mirror
images about the symmetry plane of $S_{xy}$.

LCSs as islands/tori and manifolds are robust and therefore satisfy the global symmetry. Individual trajectories
within chaotic seas, on the other hand, are generically {\it asymmetric} due to their extreme sensitivity to
initial conditions. The slightest asymmetry in the initial positions of the forward/reversed
flows in \FIG{Symmetry_a}, i.e. $\vec{x}$ versus $S(\vec{x})+ \vec{\epsilon}$, with $|\vec{\epsilon}|\lll1$, eventually always leads to symmetry breaking.

Symmetries other than reflections may play important roles as well. An ordinary {\it non-reflectional} symmetry e.g.
relates the tube segments on the elliptic segments of the (perturbed) periodic line in \FIG{UnsteadyChaos2_b} (right)
and in fact dictates their merger with an inner/outer shell to form one LCS via RIM
\cite{Speetjens_inertia_2006,Speetjens_merger_2006}. Moreover, {\it continuous} symmetries according to \FIG{Symmetry_b}
may emerge with far-reaching consequences for the dynamics. This is elaborated in \SEC{ContinuousSymmetry}.

\subsection{Hamiltonian structure of 2D flows}
\label{Hamiltonian2D}

\SEC{Categorisation} revealed that Lagrangian transport in many 3D flows is (locally) similar to that of 2D unsteady flows and the
above advanced critical lines (\SEC{CriticalLines}) and closed trajectories (\SEC{ClosedTrajectories}) as LCSs underlying such
(effectively) 2D behaviour. Isolated critical points (\SEC{CriticalPoints3D}), on the other hand, are key indicators of
essentially 3D dynamics. Hence the emergence (or absence) of such LCSs offers important insights into the nature of the
Lagrangian motion.
However, a more rigorous distinction of true 3D from (essentially) 2D dynamics than just based on LCSs is relevant for analysis and design
of practical flows. Batch mixers using time-periodic stirring must e.g. operate at conditions yielding global 3D chaos such
as in \FIG{UnsteadyChaos1_a} (right); any (local) 2D behaviour implies sub-optimal performance of such devices.
The ability to unambiguously distinguish between 2D and 3D dynamics enables identification and accomplishment of favourable operating conditions.

The 2D--3D distinction may rely upon the fundamental property that kinematic equation \eqref{KinEq} for solenoidal 2D unsteady flows has a Hamiltonian structure, viz.
%
%
%Solenoidality for 2D flows namely imposes a particular structure upon kinematic equation \eqref{KinEq}, viz.
%
\begin{eqnarray}
\frac{dx}{dt} = u_x = \frac{\partial H}{\partial y},\quad\quad\frac{dy}{dt} = u_y = -\frac{\partial H}{\partial x},
\label{Hamiltonian1}
\end{eqnarray}
%
%and thus renders the 2D equations of motion a {\it Hamiltonian system},
with function $H = H(\vec{x},t)$ its so-called ``Hamiltonian'' \cite{Aref1983,Ottino_Kinematics_1989}.
The latter for steady conditions, i.e. $H = H(\vec{x})$, satisfies
\begin{eqnarray}
\frac{dH}{dt} = \vec{u}\cdot\vec{\nabla}H = 0,
%\frac{\partial H}{\partial y}\frac{\partial H}{\partial x} - \frac{\partial H}{\partial x}\frac{\partial H}{\partial y} = 0.
\label{Hamiltonian1b}
\end{eqnarray}
and thus restricts the motion of fluid parcels to its level sets. Property \eqref{Hamiltonian1b} renders $H$ a so-called ``constant of motion'' (COM) and here
defines the conventional stream function of 2D steady flows. Restriction of fluid motion to level sets of $H$ precludes chaotic advection and thus puts forth
unsteadiness, i.e. $H = H(\vec{x},t)$, as a {\it necessary} (though not {\it sufficient}) condition for chaos in 2D flows \cite{Aref1983,Ottino_Kinematics_1989}. This
namely invalidates property \eqref{Hamiltonian1b} by virtue of $dH/dt = \partial H/\partial t\neq 0$.\footnote{Steady/unsteady Hamiltonian systems are in dynamical-systems
terminology commonly denoted autonomous/non-autonomous systems \cite{Ott_chaos_2002}.} Flows devoid of chaotic behaviour due to the
restriction of Lagrangian dynamics to level sets of a COM $H$ are termed {\it integrable} \cite{Ott_chaos_2002}; $H$ namely implicitly gives the solution
$\vec{x}(t) = \vec{\Phi}_t(\vec{x}_0)$ to kinematic equation \eqref{KinEq} via $H\left(\vec{x}(t)\right) = H(\vec{x}_0)$.

The Hamiltonian structure furthermore dictates that chaos, if occurring, emanates from the breakdown of islands following specific
scenarios described by the well-known KAM and Poincar\'{e}-Birkhoff theorems \cite{Ottino_Kinematics_1989,Ott_chaos_2002}. This
results in the characteristic composition of flow topologies of systems exhibiting Hamiltonian chaos, i.e. large central
islands encircled by island chains and embedded in a chaotic sea, as happening in e.g. the cross-sectional dynamics
of the 3D steady flows shown in \FIG{StirredTankFountain_d}, \FIG{VortexBreakdownBubble} or \FIG{SteadyCylinderFlow_b}. Counterparts to these theorems exist for such 3D steady flows exhibiting essentially 2D behaviour \cite{Aref2017}.

%\subsection{Periodic duct flows}
%\subsection{Link 3D steady and 2D unsteady flows}
%\subsection{Embedded Hamiltonian structure in 3D steady flows}
\subsection{Hamiltonian dynamics in 3D steady flows}
\label{Hamiltonian3Dsteady}

\subsubsection{General}
\label{HamiltonianGeneral}

The fact that the Lagrangian dynamics in 2D solenoidal flows is always governed by Hamiltonian mechanics offers a way
to establish whether a 3D steady flow exhibits (essentially) 2D dynamics. This notion namely implies that any such
flow must admit a (formal) transformation $\mathcal{F}$ following \eqref{Steady3DvsUnsteady2D} into a
2D unsteady system with a Hamiltonian structure according to \eqref{Hamiltonian1}.
This transformation rests on three cornerstones. First, a coordinate transformation
\begin{eqnarray}
\mathcal{G}: \vec{x} \rightarrow \vec{\xi},
\label{CoordinateTransform}
\end{eqnarray}
from the Cartesian frame $\vec{x} = (x,y,z)$ to an appropriate curvilinear reference
frame $\vec{\xi} = (\xi_1,\xi_2,\xi_3)$ with scaling factors $h_i = |\partial\vec{x}/\partial \xi_i|$ and Jacobian
$J = |\partial\vec{x}/\partial\vec{\xi}| = h_1 h_2 h_3$ following \cite{Arfken1995}. Second, the vector potential $\vec{\mathcal{A}}$ of $\vec{u}$
for vanishing $\xi_2$-component, i.e. $\vec{\mathcal{A}} = (\mathcal{A}_1,0,\mathcal{A}_3)$, defined implicity as
\begin{eqnarray}
%u_1 &=& \frac{1}{h_2h_3}\frac{\partial\widetilde{\mathcal{A}}_3}{\partial \xi_2},\nonumber\\
%u_2 &=& \frac{1}{h_1h_3}\left[\frac{\partial\widetilde{\mathcal{A}}_1}{\partial \xi_3} - \frac{\partial\widetilde{\mathcal{A}}_3}{\partial \xi_1}\right],\nonumber\\
%u_3 &=& -\frac{1}{h_1h_2}\frac{\partial\widetilde{\mathcal{A}}_1}{\partial \xi_2},
%
h_2 h_3 u_1 &=& \widetilde{u}_1 = \frac{\partial\widetilde{\mathcal{A}}_3}{\partial \xi_2},\nonumber\\
h_1h_3 u_2 &=& \widetilde{u}_2 = \frac{\partial\widetilde{\mathcal{A}}_1}{\partial \xi_3} - \frac{\partial\widetilde{\mathcal{A}}_3}{\partial \xi_1},\nonumber\\
h_1h_2u_3 &=& \widetilde{u}_3 = -\frac{\partial\widetilde{\mathcal{A}}_1}{\partial \xi_2},
\label{VectorPotential}
\end{eqnarray}
with $\widetilde{\mathcal{A}}_i = h_i \mathcal{A}_i$ and $\widetilde{u}_i = J u_i/h_i$ the vector potential and flow, respectively, in the {\it Cartesian} frame spanned by $\vec{\xi}$.
Third, expression of kinematic equation \eqref{KinEq} in terms of $\vec{\xi}$ and $\vec{\mathcal{A}}$, i.e.
\begin{eqnarray}
\frac{d\xi_1}{dt} &=& \frac{u_1}{h_1} = \frac{\widetilde{u}_1}{J} = \frac{1}{J}\frac{\partial\widetilde{\mathcal{A}}_3}{\partial \xi_2},\nonumber\\
\frac{d\xi_2}{dt} &=& \frac{u_2}{h_2} = \frac{\widetilde{u}_2}{J}  = \frac{1}{J}\left[\frac{\partial\widetilde{\mathcal{A}}_1}{\partial \xi_3} - \frac{\partial\widetilde{\mathcal{A}}_3}{\partial \xi_1}\right],\nonumber\\
\frac{d\xi_3}{dt} &=& \frac{u_3}{h_3} = \frac{\widetilde{u}_3}{J}  = -\frac{1}{J}\frac{\partial\widetilde{\mathcal{A}}_1}{\partial \xi_2},
\label{KinEqCurvilinear}
\end{eqnarray}
which reveals that the Lagrangian motion corresponds with rescaled flow $\widetilde{\vec{u}}/J$ in said Cartesian $\vec{\xi}$-frame: $d\vec{\xi}/dt = \widetilde{\vec{u}}/J$.

%\paragraph{Quasi-2D flows ... $\xi_3$ independent ...}
%\paragraph{Quasi-2D flows}

The simplest class of 3D steady flows with an embedded Hamiltonian structure involves cases that admit a transformation \eqref{CoordinateTransform} such that the $\xi_3$-wise flow in $\vec{\xi}$-space is independent of $\xi_3$ and uni-directional: $\widetilde{u}_3 = \widetilde{u}_3(\xi_1,\xi_2) > 0$. This via \eqref{VectorPotential} implies $\mathcal{A}_1 = \mathcal{A}_1(\xi_1,\xi_2)$
and, in turn, a Hamiltonian structure for the $\xi_{1,2}$-wise component of \eqref{KinEqCurvilinear} according to
\begin{eqnarray}
%\frac{d\xi_1}{dt} = \frac{\widetilde{u}_1}{J} = \frac{1}{J}\frac{\partial\widetilde{\mathcal{A}}_3}{\partial \xi_2},\quad
%\frac{d\xi_2}{dt} = \frac{\widetilde{u}_2}{J} = -\frac{1}{J}\frac{\partial\widetilde{\mathcal{A}}_3}{\partial \xi_1},
%
\frac{d\xi_1}{dt} = \frac{\widetilde{u}_1}{J} = \frac{1}{J}\frac{\partial H}{\partial \xi_2},\quad
\frac{d\xi_2}{dt} = \frac{\widetilde{u}_2}{J} = -\frac{1}{J}\frac{\partial H}{\partial \xi_1},
\label{KinEqCurvilinear2}
\end{eqnarray}
with $H = \mathcal{A}_3(\xi_1,\xi_2,\xi_3)$ the corresponding Hamiltonian. Property $d\xi_3/dt = \widetilde{u}_3/J > 0$ (due to $\widetilde{u}_3>0$ and non-singular transformation \eqref{CoordinateTransform}) admits inversion of $\xi_3(t)$ and thus expression of the Hamiltonian as $H = H(\xi_1,\xi_2,t) = \mathcal{A}_3\left(\xi_1,\xi_2,\xi_3(t)\right)$. Hence
transformation \eqref{Steady3DvsUnsteady2D} here becomes
%
%with $H = \mathcal{A}_3(\xi_1,\xi_2,\xi_3)$ the corresponding Hamiltonian and $\xi_3$ a time-like variable. Property $d\xi_3/dt = \widetilde{u}_3/J > 0$ namely admits
%inversion of $\xi_3(t)$ and thus expression of \eqref{KinEqCurvilinear2} in terms of $(\xi_1,\xi_2,t)$. Hence transformation \eqref{Steady3DvsUnsteady2D} here becomes
%
\begin{eqnarray}
%\mathcal{F}: \zeta_1 = \xi_1,\quad\zeta_2 = \xi_2,\quad\tau = f^{-1}(\xi_3),
\mathcal{F}: \zeta_1 = \xi_1,\quad\zeta_2 = \xi_2,\quad\tau = t.
\label{Steady3DvsUnsteady2D_2a}
\end{eqnarray}
%
%with $\xi_3 = f(t)$. A steady Hamiltonian $H = H(\xi_1,\xi_2)$ yields
Moreover, a steady Hamiltonian $H = H(\xi_1,\xi_2)$ yields
\begin{eqnarray}
\frac{dH}{dt} = \vec{u}\cdot\vec{\nabla}H = \sum_{i=1}^3\frac{u_i}{h_i}\frac{\partial H}{\partial \xi_i} = \sum_{i=1}^3\frac{\widetilde{u}_i}{J}\frac{\partial H}{\partial \xi_i} = 0.
\label{Hamiltonian1c}
\end{eqnarray}
demonstrating consistency of \eqref{KinEqCurvilinear2} with \eqref{Hamiltonian1} and \eqref{Hamiltonian1b}.

3D steady flows satisfying only the relaxed condition of a uni-directional $\xi_3$-component of the physical flow, i.e. $u_3>0$ yet
permitting $\partial\mathcal{A}_1/\partial\xi_3\neq 0$ in \eqref{VectorPotential} and \eqref{KinEqCurvilinear}, define a second class with an
embedded Hamiltonian structure. Transformation \eqref{Steady3DvsUnsteady2D} following \cite{Bajer_3D_1994}, given by
%
%permitting $\partial\mathcal{A}_1/\partial\xi_3\neq 0$ in \eqref{VectorPotential} and \eqref{KinEqCurvilinear} and satisfying the relaxed condition of only a
%uni-directional $\xi_3$-component of the physical flow, i.e. $u_3>0$, define a second class with an
%embedded Hamiltonian structure. Transformation \eqref{Steady3DvsUnsteady2D} following \cite{Bajer_3D_1994}, given by
%
\begin{eqnarray}
\mathcal{F}: \zeta_1 = \xi_1,\quad\zeta_2 = -\widetilde{\mathcal{A}}_1(\vec{\xi}),\quad\tau = \xi_3,
\label{Steady3DvsUnsteady2D_2}
\end{eqnarray}
namely yields a Hamiltonian system according to \eqref{Hamiltonian1} in canonical space $\vec{\zeta} = (\zeta_1,\zeta_2)$ and time $\tau$, i.e.
\begin{eqnarray}
\frac{d\zeta_1}{d\tau} = \frac{\partial H}{\partial \zeta_2},\quad\quad\frac{d\zeta_2}{d\tau} = -\frac{\partial H}{\partial \zeta_1},
\label{Hamiltonian2}
\end{eqnarray}
with Hamiltonian $H = H(\zeta_1,\zeta_2,\tau) = \widetilde{\mathcal{A}}_3\left(\zeta_1,f(\zeta_1,\zeta_2,\tau),\tau\right)$ and $\xi_2 = f(\zeta_1,\zeta_2,\tau)$ implicitly defined
via \eqref{Steady3DvsUnsteady2D_2}. Transformation \eqref{Steady3DvsUnsteady2D_2} effectively translates the $\zeta_3$-wise flow into a time evolution, rendering $\zeta_3$ a fictitious time.

\subsubsection{Periodic duct flows}
\label{HamiltonianFormDuctFlows}

An important subclass of reoriented duct flows \eqref{SpatiallyPeriodic3D2} belonging to the above {\it first class} of flows with an embedded
Hamiltonian form are the so-called 2.5D duct flows, that is, simplified configurations where the base flow depends only on the
transverse coordinates: $\vec{v} = \vec{v}(r,\theta)$ and $\widetilde{u}_3(\xi_1,\xi_2) = v_z(r,\theta)>0$. Transformation \eqref{Steady3DvsUnsteady2D_2a} applied to the Cartesian frame
$\vec{\xi} = \vec{x}$ thus readily yields Hamiltonian form \eqref{KinEqCurvilinear2}, with $H = H(r,\theta) = A_z(r,\theta)$, and
here in fact identifies with \eqref{Hamiltonian1}.
This renders the duct flow $\vec{u}$ a cell-wise reorientation of a steady Hamiltonian system, meaning that chaotic advection
can occur only due to jumping of tracers between cell-wise reorientations of the KAM tori of the base flow at the cell interfaces $z=kL$.
The embedded Hamiltonian structure causes {\it all} 2.5D duct flows to
behave according to one universal Hamiltonian scenario \cite{Speetjens_symmetry_2006}.
The 2.5D approximation is a standard modelling approach for periodic duct flows and inline mixers (\SEC{DuctFlowIndustrial}) and is therefore adopted in many studies
on such devices \cite{Ottino_Kinematics_1989,Ottino2004}. The flow topology in the RAM in \FIG{RAM2} is e.g. simulated with the 2.5D model
from \cite{Speetjens_symmetry_2006} and computational analyses on the PPM commonly use the 2.5D models by \cite{Ottino_Kinematics_1989} and \cite{Meleshko1999}.

\begin{figure}
\centering
\subfigure[Canonical vs. physical space]{\includegraphics[width=0.48\columnwidth,height=0.47\columnwidth]{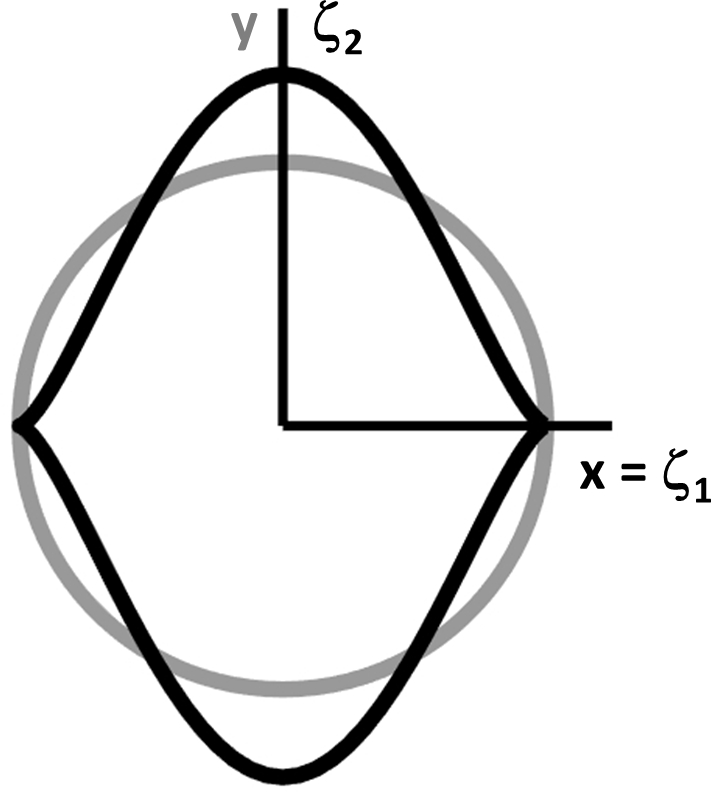}\label{RAM2canonical_a}}
\hspace*{2pt}
\subfigure[Canonical flow topology]{\includegraphics[width=0.49\columnwidth,height=0.47\columnwidth]{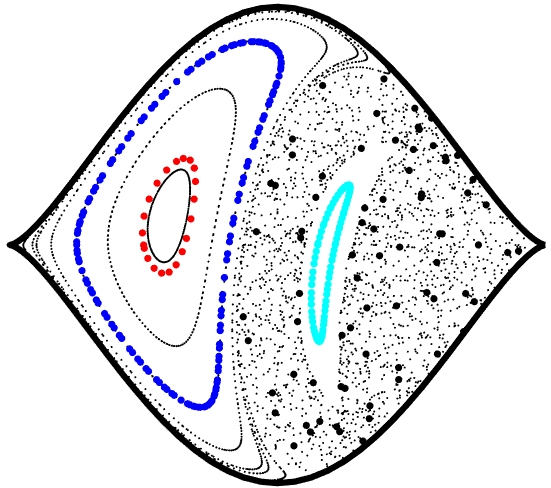}\label{RAM2canonical_b}}

\caption{Hamiltonian representation of the cross-sectional flow topology of the RAM in \FIG{RAM2b} following
transformation \eqref{Steady3DvsUnsteady2D_3}.
}
\label{RAM2canonical}
\end{figure}

Smooth transition between cells yields -- even in the Stokes limit -- an essentially 3D duct flow $\vec{u}(x,y,z)$ that precludes Hamiltonian
structure \eqref{KinEqCurvilinear2}. However, such flows readily admit the {\it second class} of embedded Hamiltonian forms in case of uni-directional axial
flow: $u_z(\vec{x}) > 0$ for all $\vec{x}$. This namely enables application of transformation \eqref{Steady3DvsUnsteady2D_2} directly to the Cartesian
coordinates $\vec{x}$, resulting in
\begin{eqnarray}
(\zeta_1,\zeta_2) = \left(x,-A_x(\vec{x})\right),\quad \tau = z,\quad H = A_z,
\label{Steady3DvsUnsteady2D_3}
\end{eqnarray}
for the associated Hamiltonian form \eqref{Hamiltonian2}.
Consider for example the 2.5D RAM flow in \FIG{RAM2}, which has an axial Poiseuille flow $v_z = u_z = 2U\left(1 - r^2\right)$, with $U$ the mean
throughflow. Performing transformation \eqref{Steady3DvsUnsteady2D_3} for the rescaled flow $\vec{u}' = \vec{u}/U$, which can be done without loss of
generality, via $u'_z = 2\left(1 - r^2\right)$ yields $A_x = 2y(y^2/3 + x^2 - 1)$ and a canonical versus physical cross-section
following \FIG{RAM2canonical_a}. The canonical representation of the cross-sectional topology in \FIG{RAM2b} is shown in \FIG{RAM2canonical_b}
and retains its structure, signifying consistency. The 2.5D RAM flow also admits form \eqref{Hamiltonian1} for reasons given above and can thus be reconciled with
two equivalent Hamiltonian systems that employ either real time $t$
or axial coordinate $z$ as canonical time.\footnote{Using the rescaled flow for transformation \eqref{Steady3DvsUnsteady2D_3} ensures a straightforward
link between the Hamiltonian forms. A uniform axial flow $u_z = U$ in that case namely results (via $u_z' = 1$) in $A_x = -y$ and thus identical canonical and physical spaces: $(\zeta_1,\zeta_2) = (x,y)$.
Moreover, Hamiltonian systems \eqref{Hamiltonian1} and \eqref{Hamiltonian2} then identify upon substituting $H = UA_z$ and $t = \tau/U$ in the former.}
%
%Consider as an introductory example a 3D duct flow with a uniform axial flow $u_z = U$. Performing the transformation for the rescaled flow $\vec{u}' = \vec{u}/U$, which
%can be done without loss of generality, via $u'_z = 1$ gives $\mathcal{A}_x = -y$ and thus a canonical space that coincides with the physical cross-section: $(\zeta_1,\zeta_2) = (x,y)$.
%Moreover, Hamiltonian system \eqref{Hamiltonian2} corresponds with the cross-sectional flow and identifies with \eqref{Hamiltonian1} upon substituting $H = U\mathcal{A}_z$ and $t = \tau/U$ in the latter.\\\\

Transformation \eqref{Steady3DvsUnsteady2D_3} applies to any duct flow with $u_z>0$ everywhere and translates into a canonical space/topology that is an $y$-wise deformed version of the physical cross-section/topology akin to \FIG{RAM2canonical}. The RAM flow in \FIG{RAM3a} and micro-mixer flow in \FIG{MicroMixerSuh2007} following \cite{Singh2008} and \cite{Suh2007}, respectively, e.g. qualify for this approach.

%The computational analyses of the RAM in \FIG{RAM3a}
%and the micro-mixer in \FIG{MicroMixerSuh2007} by the 3D CFD models of \cite{Singh2008} and \cite{Suh2007}, respectively, e.g. qualify for this approach.

3D steady duct flows with local axial backflow ($u_z < 0$) admit the second class of Hamiltonian forms \eqref{Hamiltonian2} upon
transformation \eqref{Steady3DvsUnsteady2D_2} in the net throughflow region following \SEC{DuctFlowIndustrial}, that is, the subregion connecting the periodic inlet-outlet allowing for a non-singular transformation \eqref{CoordinateTransform} such that $u_3>0$  \cite{Speetjens2014}. The remaining flow regions relate to essentially 3D LCSs as e.g. stagnation/periodic points due to internal recirculation zones or
separatices/manifolds emanating from critical points of the skin-friction field \eqref{SkinFrictionField} on internal walls (as demonstrated in \FIG{PorousMedia1b}
for the flow inside a porous matrix). This happens in the RAM under essentially 3D conditions with significant fluid inertia \cite{Speetjens2014}
and is suspected in 3D partitioned-duct flows as e.g. the Quatro mixer in \FIG{Quatro1} or the CSM in \FIG{ChaoticSerpentineMicroMixer}.

\subsubsection{Steady circulatory flows}
\label{HamiltonianFormCirculatoryFlows}

Steady confined circulatory flows as e.g. the impeller-driven flow in \FIG{StirredTankFountain} and the swirling flow in \FIG{VortexBreakdownBubble}
for appropriate conditions admit the above embedded Hamiltonian forms in the reordered cylindrical frame $\vec{\xi} = (z,r,\theta)$ ($h_1=h_2=1$, $h_3=J = r$) using
$\theta$ as time-like variable. Axi-symmetric angular velocity $\omega = u_\theta/r = \omega(r,z) > 0$ yields
\begin{eqnarray}
\frac{d r}{d\tau} = \frac{1}{r}\frac{\partial H}{\partial z},\quad\quad\frac{dz}{d\tau} = -\frac{1}{r}\frac{\partial H}{\partial r},
\label{Hamiltonian4}
\end{eqnarray}
as specific Hamiltonian form \eqref{KinEqCurvilinear2}, with $H = r\mathcal{A}_\theta$ the Stokes stream function.
Non-axi-symmetric angular velocity $\omega > 0$ gives Hamiltonian form \eqref{Hamiltonian2} for
\begin{eqnarray}
(\zeta_1,\zeta_2) = \left(z,-\mathcal{A}_z(\vec{x})\right),\quad \tau = \theta,\quad H = r A_\theta,
\label{Steady3DvsUnsteady2D_4}
\end{eqnarray}
and constitutes the swirling-flow counterpart to transformation \eqref{Steady3DvsUnsteady2D_3} for duct flows.
Transformation \eqref{Steady3DvsUnsteady2D_4}, rescaling the flow with the mean angular velocity $\Omega$, yields $\zeta_2 = -A_z = r^2/2$ for
a uniform angular velocity ($\omega' = 1$) and, upon substitution into \eqref{Hamiltonian2}, coincides with \eqref{Hamiltonian4}. Hence uniform $\xi_3$-wise flow
also here causes the first and second class of embedded Hamiltonian forms to identify.

Swirling flows with local angular backflow ($\omega < 0$) admit Hamiltonian form \eqref{Hamiltonian2} upon transformation \eqref{Steady3DvsUnsteady2D_2} in a ``net throughflow region'' as in the above duct flows. Moreover, any circulatory flow
can for appropriate conditions be reconciled with Hamiltonian form \eqref{Hamiltonian2} in a similar way.
%
% using the primary circulation as a canonical time.\\\\
%
The primary circulations about the toroidal axis in the tokamak in \FIG{Tokamak}, the stagnation line (dots) in the steady cylinder flow in \FIG{SteadyCylinderFlow_0a} as well as the black axis in the droplet flow in \FIG{ChaosDroplets2X_b} correspond
%(given uni-directional motion)
with canonical time $\tau$, meaning that these systems admit a transformation equivalent to \eqref{Steady3DvsUnsteady2D_4} upon connecting said axes with the $z$-axis.

%Canonical space $(\zeta_1,\zeta_2)$ and time $\tau$ e.g. correspond (given uni-directional toroidal flux) for the magnetic
%field in the tokamak in \FIG{Tokamak} with the (deformed) poloidal cross-section and toroidal coordinate $\theta$, respectively.
%The primary circulations about the stagnation line (dots) in the steady cylinder flow in \FIG{SteadyCylinderFlow_0a} and
%about the black axis in the droplet flow in \FIG{ChaosDroplets2X_b} for uni-directional are equivalent
%to the swirling flow about the $z$-axis in above cylindrical frame.\\\\

%\subsection{Link unsteady container flows with Hamiltonian systems}
%\subsection{Link 3D \underline{un}steady and 2D unsteady flows}
%\subsection{Embedded Hamiltonian structure in 3D unsteady flows}
\subsection{Hamiltonian dynamics in 3D unsteady flows}
\label{Hamiltonian3Dunsteady}

\subsubsection{Flows with continuous symmetry}
\label{ContinuousSymmetry}

%Flows with $u_3 = u_3(\xi_1,\xi_2,t)$ admit, expanding on the above, Hamiltonian form \eqref{KinEqCurvilinear2} with Hamiltonian $\mathcal{A}_3 = \mathcal{A}_3(\xi_1,\xi_2,\xi_3,t)$ and thus define %non-autonomous Hamiltonian systems. This situation e.g. occurs for unsteady counterparts of the above 2.5D duct flows. However, this situation is unlikely to occur in reality. Realistic, on the %other hand, are situations where the full flow becomes independent on a coordinate direction. \red{This implies a continuous symmetry and has fundamental consequences \cite{Mezic_symmetry_1994}:}
%

A generalisation of the above first class of steady flows with embedded Hamiltonian forms (\SEC{HamiltonianGeneral}) concerns unsteady
flows where $\widetilde{u}$ following \eqref{VectorPotential} becomes independent of one coordinate direction, say $\xi_3$, for an
appropriate transformation \eqref{CoordinateTransform}: $\widetilde{\vec{u}} = \widetilde{\vec{u}}(\xi_1,\xi_2,t)$. This happens if a
steady solenoidal ``flow'' $\vec{w}(\vec{x})$ exists, with $\vec{x}(\lambda) = \vec{g}_\lambda(\vec{x}_0)$ the flow along its ``streamlines'' parameterised by ``time'' $\lambda$, such that each fluid trajectory $\vec{x}(t) = \vec{\Phi}_t(\vec{x}_0)$ has a companion
\begin{eqnarray}
\vec{x}_\ast(t) = \vec{g}_\lambda(\vec{\Phi}_t(\vec{x}_0)) = \vec{\Phi}_t\left(\vec{g}_\lambda(\vec{x}_0)\right),
\label{Commute1}
\end{eqnarray}
that is also a fluid trajectory \cite{Mezic_symmetry_1994}. Thus $\vec{x}_\ast(t)$ can be
reached from $\vec{x}_0$ via two paths and the companion of a fluid trajectory follows from its advection by ``flow'' $\vec{w}$ according to \FIG{Symmetry_b}. This holds for any $\lambda\in\mathbb{R}$ and therefore gives rise to a continuous family of companion trajectories. Hence $\vec{w}$ defines a {\it continuous} symmetry
%-- as opposed to {\it discrete} symmetries \eqref{Symmetries} --
and its ``streamlines'' coincide with fixed positions $(\xi_1,\xi_2)$ in the curvilinear frame, or equivalently, the intersections of the coordinate surfaces of $\xi_{1,2}$ (implying $\vec{w} = w\vec{e}_3$ or, conversely, $\vec{e}_3 = \vec{w}/|\vec{w}|$).

Property \eqref{Commute1} is equivalent to
\begin{eqnarray}
%\vec{w}\cdot\vec{\nabla}\vec{u} = \vec{u}\cdot\vec{\nabla}\vec{w},
\vec{w}\cdot\vec{\nabla}\vec{u} - \vec{u}\cdot\vec{\nabla}\vec{w} = 0,
\label{Commute2}
\end{eqnarray}
and renders $\vec{u}$ and $\vec{w}$ commutative, meaning that a given position
can be reached via flow by, first, $\vec{u}$ and, second, $\vec{w}$ or in reversed
order as per \FIG{Symmetry_b}. Condition \eqref{Commute2} in conjunction with
solenoidality of $\vec{u}$ and $\vec{w}$ through vector identity
\begin{eqnarray}
\vec{\nabla}\times(\vec{u}\times\vec{w}) = \vec{w}\cdot\vec{\nabla}\vec{u} - \vec{u}\cdot\vec{\nabla}\vec{w} + \vec{u}(\vec{\nabla}\cdot\vec{w}) - \vec{w}(\vec{\nabla}\cdot\vec{u}),
\label{VecID}
\end{eqnarray}
yields
\begin{eqnarray}
\vec{\nabla}\times(\vec{u}\times\vec{w}) = 0\quad\Rightarrow\quad \vec{u}\times\vec{w} = -\vec{\nabla}H,
\label{COM1}
\end{eqnarray}
implying a COM $H$ due to $\vec{u}\cdot\vec{\nabla}H = -\vec{u}\cdot\vec{u}\times\vec{w} = 0$ that defines a
Hamiltonian for the $\xi_{1,2}$-wise dynamics \cite{Mezic_symmetry_1994,Mezic2009}.
Property $\widetilde{\vec{u}} = \widetilde{\vec{u}}(\xi_1,\xi_2,t)$ namely yields via vector potential \eqref{VectorPotential} a
Hamiltonian form \eqref{KinEqCurvilinear2}, with here $H = \mathcal{A}_3(\xi_1,\xi_2,t)$, and a corresponding structure for
$\vec{u}$ satisfying \eqref{COM1} for $\vec{w} = h_3\vec{e}_3$.\footnote{The coordinate transformation associated with the continuous symmetry in fact results in $d\vec{\xi}/dt = \widehat{\vec{u}}(\xi_1,\xi_2,t)$ for kinematic
equation \eqref{KinEqCurvilinear} and $J=J(\xi_1,\xi_2)$ \cite{Mezic_symmetry_1994}. Hence
relation $\widehat{\vec{u}} = \widetilde{\vec{u}}/J$ implies $\widetilde{\vec{u}} = \widetilde{\vec{u}}(\xi_1,\xi_2,t)$.}
The independence of the Hamiltonian $H$ on spatial coordinate $\xi_3$ has the important consequence that a continuous symmetry in a steady flow implies $H = H(\xi_1,\xi_2)$ and thus inherently non-chaotic dynamics.

Consider for example the RAM flow in \FIG{RAM2} without reorientation, i.e. $\Theta_R=0$ in \eqref{SpatiallyPeriodic3D2}, causing $\vec{u}$ to identify with base flow $\vec{v}$ and, given $\vec{v}=\vec{v}(x,y)$ here, resulting in $\vec{u} = \vec{u}(x,y)$. This puts forth uniform axial ``flow'' $\vec{w} = \vec{e}_z$ as symmetry and a Hamiltonian form of the cross-sectional flow following \eqref{Hamiltonian1}.
The same symmetry and Hamiltonian form emerge for diminishing rotation rate of the outer cylinder; tracers then only ``feel'' the averaged flow $\vec{u} = N^{-1}\sum_{k=0}^{N-1} \vec{v}(r,\theta - k\Theta) = \vec{u}(r,\theta)$ \cite{Speetjens_symmetry_2006}.
The impeller-driven steady flow in \FIG{StirredTankFountain} becomes axi-symmetric, i.e. $\vec{u} = \vec{u}(r,z)$, for tilt angle $\alpha=0$ and thus in a similar way leads to a symmetry $\vec{w} = r\vec{e}_\theta$ and corresponding Hamiltonian form \eqref{Hamiltonian4}. Steady conditions preclude chaos in these flows for reasons mentioned before.

The time-periodic flow in \FIG{UnsteadyChaos1_a} (left) due to reorientations of the steady lid-driven Stokes flow
in \FIG{SteadyCylinderFlow_0a} about the $z$-axis accommodates a continuous symmetry as above yet in a highly non-trivial manner.
Time-reversal and ordinary reflectional symmetries according to \eqref{Symmetries2} about planes $x=0$ and $y=0$, respectively,
impose the structure
\begin{eqnarray}
u_{r,z}(\vec{x}) = \breve{u}_{r,z}(r,z)\cos\theta,\quad
u_\theta(\vec{x}) = \breve{u}_\theta(r,z)\sin\theta,\quad
\label{SteadyCylinderFlow2}
\end{eqnarray}
upon the steady base flow \cite{Malyuga_cylinder_2002,Speetjens_stokes_2004}. This, via separation of variables of $dF_{1,2}/dt$ following \eqref{Hamiltonian1b}, yields two COMs, i.e.
\begin{eqnarray}
F_1(\vec{x}) = f_1(r,z),\quad F_2(\vec{x}) = f_2(r,z)\sin\theta,
\label{COMcylflow}
\end{eqnarray}
that can be shown to relate to \eqref{SteadyCylinderFlow2} through
\begin{eqnarray}
\frac{\breve{u}_{r}}{f_2} = \frac{1}{r}\frac{\partial f_1}{\partial z},\quad
\frac{\breve{u}_{z}}{f_2} = -\frac{1}{r}\frac{\partial f_1}{\partial r},
%\breve{u}_{r} = \frac{f_2}{r}\frac{\partial f_1}{\partial z},\quad
%\breve{u}_{z} = -\frac{f_2}{r}\frac{\partial f_1}{\partial r},
%\frac{u_r}{f_2\cos\theta} = \frac{1}{r}\frac{\partial f_1}{\partial z},\quad
%\frac{u_z}{f_2\cos\theta} = -\frac{1}{r}\frac{\partial f_1}{\partial r},
\label{SteadyCylinderFlow3}
\end{eqnarray}
exposing $f_1$ as the Stokes stream function of flow $(\breve{u}_{r}/f_2,\breve{u}_{z}/f_2)$
in the $rz$-plane.
%
%This suggests an embedded Hamiltonian structure, which can be found as follows.\\\\
%
This suggests an embedded Hamiltonian structure and associated continuous symmetry.
Its isolation embarks on expressing kinematic equation $d\vec{x}/dt = \vec{u}$ in terms of the curvilinear
frame $\vec{\xi} = (z,r,\alpha)$, with $\alpha = \ln(\sin\theta)$ substituting
%
%with $\alpha = \alpha(\theta)$ following $d\alpha = d\theta/\tan\theta$ substituting
%
angular coordinate $\theta$, yielding
\begin{eqnarray}
%\frac{dz}{dt} = u_z,\quad
%\frac{dr}{dt} = u_r,\quad
\frac{d(r,z)}{dt} = \breve{u}_{r,z}\cos\theta(\alpha),\quad
\frac{d\alpha}{dt} = \frac{u_\theta}{h_\alpha} = \frac{\breve{u}_\theta\cos\theta(\alpha)}{r},
\label{KinEq3}
\end{eqnarray}
where $h_\alpha = J = r\tan\theta$, $\vec{e}_\alpha = \vec{e}_\theta$ and $u_\alpha = \vec{u}\cdot\vec{e}_\alpha = u_\theta$.\footnote{Considering range $0<\theta<\pi/2$ is sufficient
due to symmetries; this gives $\sin\theta > 0$ and $\tan\theta > 0$ and thus a non-singular transformation.}
%
%Thus said kinematic equation assumes the form $d\vec{\xi}/dt = \vec{g}(r,z)\cos\theta(\alpha)$.
%
%Introducing the rescaled flow $\vec{u}^\star = \vec{u}/f_2\cos\theta$ via \eqref{SteadyCylinderFlow3}
%
Introducing the rescaled flow $\vec{u}^\star = \vec{u}/F_2$ via \eqref{SteadyCylinderFlow3} gives
\begin{eqnarray}
%\frac{dr}{dt} =
%\frac{1}{r}\frac{\partial f_1}{\partial z},\quad
%\frac{dz}{dt} =
%-\frac{1}{r}\frac{\partial f_1}{\partial r},\quad
%\frac{d\alpha}{dt} = \frac{\breve{u}_\theta}{r f_2},
%
%\frac{dr}{dt} =
%\frac{\partial f_1}{\partial z},\quad
%\frac{dz}{dt} =
%-\frac{\partial f_1}{\partial r},\quad
%\frac{d\alpha}{dt} = \frac{\breve{u}_\theta}{f_2},
%
\frac{dz}{dt} = u_1^\star = \frac{\widetilde{u}_1^\star}{J},\quad
\frac{dr}{dt} = u_2^\star = \frac{\widetilde{u}_2^\star}{J},\quad
\frac{d\alpha}{dt} = \frac{u_3^\star}{h_\alpha} = \frac{\widetilde{u}_3^\star}{J},
\label{KinEq4}
\end{eqnarray}
as counterpart to \eqref{KinEq3}, with
\begin{eqnarray}
\widetilde{\vec{u}}^\star = \left(-\frac{\partial f_1}{\partial r},\frac{\partial f_1}{\partial z},\frac{\breve{u}_\theta}{f_2}\right) = \widetilde{\vec{u}}^\star(\xi_1,\xi_2),
\end{eqnarray}
the corresponding flow \eqref{VectorPotential} in the Cartesian $\vec{\xi}$-frame. Its independence of $\xi_3$ signifies indeed a continuous symmetry and, in consequence, a Hamiltonian
structure \eqref{KinEqCurvilinear2} exists. The underlying relation \eqref{COM1} follows in the Cartesian frame spanned by $\vec{\xi}$ from vector identity \eqref{VecID} as before: flow $\widetilde{\vec{u}}^\star$ commutes with field $\widetilde{\vec{w}}_1^\star = \vec{e}_\alpha$
according to \eqref{Commute2} and, given both fields are solenoidal, thus satisfy \eqref{COM1} for $H = -f_1 = -F_1$. However, counterparts
$\vec{u}^\star$ and $\vec{w}_1^\star = h_\alpha\vec{e}_\alpha$, or equivalently, $\vec{u}$ and $\vec{w}_1 = (h_\alpha/F_2)\vec{e}_\alpha$
in {\it physical} space are non-commutative and, instead, via property
\begin{eqnarray}
\vec{w}\cdot\vec{\nabla}\vec{u} - \vec{u}\cdot\vec{\nabla}\vec{w} = -\vec{u}(\vec{\nabla}\cdot\vec{w}),
\label{Commute3}
\end{eqnarray}
yield
%(in conventional cylindrical coordinates) yield
%
%\eqref{COM1}
%
\begin{eqnarray}
\vec{u}^\star\times\vec{w}_1^\star = \vec{u}\times\vec{w}_1 = \vec{\nabla} F_1,
%= -\vec{\nabla} H,
%\quad H = -F_1,
\label{COM2}
\end{eqnarray}
from \eqref{VecID}, since $\vec{w}_1^\star$ and $\vec{w}_1$ are non-solenoidal.
Both $\vec{u}$ and $\vec{u}^\star$ satisfying \eqref{COM2} reflects the fact that they relate by
a scalar multiplication factor $F_2$ and thus have coinciding streamlines and, in turn, the same embedded
Hamiltonian structure.\footnote{The embedded Hamiltonian structure associated with transformation
\eqref{Steady3DvsUnsteady2D_2} is equally invariant to rescaling of the flow field \cite{Bajer_3D_1994}.}
%
%\red{
%Hence the symmetry manifests itself in a non-trivial manner, namely in the canonical flow $\widetilde{\vec{u}}^\star$ associated with the %rescaled physical flow $\vec{u}^\star$.
%}

COM $F_2$ in \eqref{COMcylflow} emanates from interaction between $\vec{u}$ and a second (non-solenoidal) field
$\vec{w}_2$ following \eqref{Commute3}, i.e.
\begin{eqnarray}
\vec{u}\times\vec{w}_2 = \vec{\nabla} F_2,\quad\vec{w}_2 = \breve{w}_r\vec{e}_r + \breve{w}_z\vec{e}_z,
\label{COM3}
\end{eqnarray}
with
\begin{eqnarray}
\breve{w}_{r} = \frac{1}{\breve{\omega}r}\frac{\partial f_2}{\partial z},\quad
\breve{w}_{z} = -\frac{1}{\breve{\omega}r}\frac{\partial f_2}{\partial r},\quad,
\label{COM3b}
\end{eqnarray}
advancing $f_2$ as the Stokes stream function of $\breve{\omega}\,\vec{w}_2$. The analogy with \eqref{SteadyCylinderFlow3} implies
an embedded Hamiltonian structure \eqref{KinEqCurvilinear2}, with $H = -f_2$, in the ``flow topology'' of ``flow'' $\vec{w}_2$. The latter emerges from essentially the same continuous symmetry as for axi-symmetric steady
circulatory flows. Rescaled field $\vec{w}_2^\star = (\breve{\omega}\breve{w}_{z},\breve{\omega}\breve{w}_{r},0) = \vec{w}_2^\star(r,z)$
in the reordered cylindrical frame $\vec{\xi} = (z,r,\theta)$ namely commutes with $\vec{g} = r\vec{e}_\theta$ following \eqref{Commute2} and via \eqref{VecID} results in
\begin{eqnarray}
%\vec{w}_2^\star\times\vec{g}^\star = \vec{w}_2\times\vec{g} = \vec{\nabla} f_2,
\vec{w}_2^\star\times\vec{g} = \vec{\nabla} f_2,
%= -\vec{\nabla} H,
%\quad H = -F_1,
\label{COM4}
\end{eqnarray}
assuming form \eqref{COM1} for $H = -f_2$. This, given streamlines of both $\vec{u}$ and $\vec{w}_2$ (and thus $\vec{w}_2^\star$) are restricted to level
sets of $F_2$ by virtue of \eqref{COM3}, indirectly constitutes an embedded Hamiltonian structure for the fluid motion.

The above reveals that COMs $F_{1,2}$ following \eqref{COMcylflow} can be reconciled with continuous symmetries according
to \cite{Mezic_symmetry_1994}, albeit in a non-trivial way. Solenoidal symmetries $\vec{w}$ {\it directly} commuting with $\vec{u}$
following \eqref{Commute2} are non-existent in the lid-driven cylinder flow; COMs $F_{1,2}$ instead originate from interaction of $\vec{u}$ with
non-solenoidal fields $\vec{w}_{1,2}$ following \eqref{Commute3} and thus via \eqref{VecID} lead to \eqref{COM2} and \eqref{COM3}.
Hence solenoidal symmetries $\vec{w}$ commuting with $\vec{u}$ as per \eqref{Commute2} in fact constitute a subclass of systems that yield COMs via relation \eqref{COM1}.

%Introducing the rescaled flow $\vec{u}^\star = h_\alpha\vec{u}/F_2$ via \eqref{SteadyCylinderFlow3} subsequently results in
%
%\begin{eqnarray}
%
%\frac{dz}{dt} = \widetilde{u}_1^\star =
%-\frac{\partial f_1}{\partial r},\quad
%\frac{dr}{dt} = \widetilde{u}_2^\star = \frac{\partial f_1}{\partial z},\quad
%\frac{d\alpha}{dt} = \frac{\breve{u}_\theta}{f_2},
%
%\label{KinEq4}
%\end{eqnarray}
%
%as counterpart to \eqref{KinEq3}, revealing that the associated kinematic equation $d\vec{x}/dt = \vec{u}^\star$ has the same embedded Hamiltonian structure \eqref{Hamiltonian4}
%as steady circulatory flows for axi-symmetric $\omega$, with $H=f_1$ the Stokes stream function. This has the fundamental implication that, given $\vec{u}^\star$ and $\vec{u}$ are
%parallel, the corresponding streamlines coincide and the flow topology of $\vec{u}$ thus effectively also has the Hamiltonian structure \eqref{KinEq4}.\footnote{\red{Similar as ... %\cite{Bajer_3D_1994} ... scalar multiplication factor ...}}
%

Relation \eqref{SteadyCylinderFlow3} identifies the level sets of $F_1$ with surfaces of revolution of the
level sets of Stokes stream function $f_1$, or equivalently, the streamlines of flow $(\breve{u}_r,\breve{u}_z)$ in the $rz$-plane.
Conversely, this causes the projection of 3D streamlines in the $rz$-plane to coincide with the level sets of $f_1$ \cite{Malyuga_cylinder_2002,Speetjens_stokes_2004}. Axi-symmetry preserves COM $F_1$ in time-periodic flows involving reorientations of the bottom wall and the spheroids in \FIG{UnsteadyChaos1_a} (left) and \FIG{UnsteadyChaos2_b} (left) thus correspond with its
level sets. COM $F_2$, on the other hand, exists only in the steady base flow and the intersections of its level sets with those
of $F_1$ define the closed streamlines in \FIG{SteadyCylinderFlow_0a}.

The restriction of the dynamics to invariant surfaces (i.e. the spheroidal level sets of $H_1 = -f_1$) implies (in terms of the intra-surface coordinates) an alternative Hamiltonian structure, with $H_2^\ast = F_2$, to that associated with $H_2 = -f_2$ in the base flow. This is elaborated in \SEC{InvariantSurfaces}. Hamiltonian $H_2^\ast$ {\it directly} (instead of {\it indirectly} through $\vec{w}_2$ following \eqref{COM3}) governs the intra-surface dynamics and, via the reorientation of the base flow, becomes {\it non-autonomous} in the time-periodic flow, thus leading to intra-surface Hamiltonian chaos as demonstrated in \FIG{UnsteadyChaos1_a} (left).

The two COMS $F_{1,2}$, or equivalently, the autonomous Hamiltonians $H_1$ and $H_2^\ast$ in the steady base flow (by confining
the dynamics to individual streamlines) implicitly define the full solution to kinematic equation \eqref{KinEq} via $F_{1,2}\left(\vec{x}(t)\right) = F_{1,2}(\vec{x}_0)$. This renders the
system {\it fully integrable} and, in consequence, entirely non-chaotic. The remaining single COM $F_1$ (or Hamiltonian $H_1$) in the corresponding time-periodic flow constitutes an only partial solution to \eqref{KinEq}, implying {\it partially integrability}, and permits 2D chaos within its level sets.

\subsubsection{Flows with invariant surfaces}
\label{InvariantSurfaces}

The flow topology of 3D unsteady flows with a single COM according to \eqref{Hamiltonian1b} is foliated into a continuous family of
invariant surfaces defined by its level sets (as e.g. the spheroids associated with COM $F_1$ following \eqref{COMcylflow}). The
restriction of the Lagrangian dynamics to invariant surfaces implies an embedded Hamiltonian structure that may coexist with a
possible Hamiltonian structure underlying these LCSs (as e.g. occurring in the cylinder flow).

Consider to this end kinematic equation \eqref{KinEqCurvilinear} in a curvilinear frame $\vec{\xi} = (\xi_1,\xi_2,\xi_3)$ such
that coordinates $(\xi_1,\xi_2)$ and $\xi_3$ are tangent and normal, respectively, to said surfaces (i.e. the latter coincide
with coordinate surfaces of $\xi_3$). Restriction to invariant surfaces implies $u_3 = 0$ and thus via $\vec{\nabla}\cdot\vec{u} = \sum_i\partial\widetilde{u}_i/\partial\xi_i = 0$ a solenoidal intra-surface 2D flow $(u_1,u_2)$ with, inherently, a Hamiltonian structure:
\begin{eqnarray}
%\vec{\nabla}\cdot\vec{u} =
\frac{\partial\widetilde{u}_1}{\partial \xi_1} + \frac{\partial\widetilde{u}_2}{\partial \xi_2} = 0
\quad\Rightarrow\quad
%\widetilde{u}_1 = \frac{\partial H}{\partial\xi_2},\quad
%\widetilde{u}_2 = -\frac{\partial H}{\partial\xi_1},
(\widetilde{u}_1,\widetilde{u}_2) = \left(\frac{\partial H}{\partial\xi_2},-\frac{\partial H}{\partial\xi_1}\right).
\label{InvariantSurfaces2}
\end{eqnarray}
This identifies with vector potential \eqref{VectorPotential} for $\widetilde{\mathcal{A}}_1 = 0$ and $H = \mathcal{A}_3$, leading
to Hamiltonian form \eqref{KinEqCurvilinear2} for the intra-surface dynamics described by $(\xi_1,\xi_2)$. Hamiltonian $H=H(\vec{\xi},t)$ can be interpreted in two ways. First, as the union of intra-surface Hamiltonians $H=H(\xi_1,\xi_2,t;\xi_3)$ parameterised by normal coordinate $\xi_3$. Second, as the global Hamiltonian $H=H(\xi_1,\xi_2,\xi_3;t)$ for the
system ``frozen'' at a given time instance $t$. Here $\xi_3$ acts as a time-like variable reminiscent of the 3D steady duct and circulatory flows in \SEC{Hamiltonian3Dsteady},
where $\xi_3$-wise progression through the 3D flow topology is similar to uni-directional ``flow'' $u_3>0$.

\begin{figure}
\centering

\subfigure[Intra-surface dynamics in ``smaller'' spheroids.]{
\includegraphics[trim=1cm 0cm 1cm 2cm,clip,width=0.48\columnwidth,height=0.5\columnwidth]{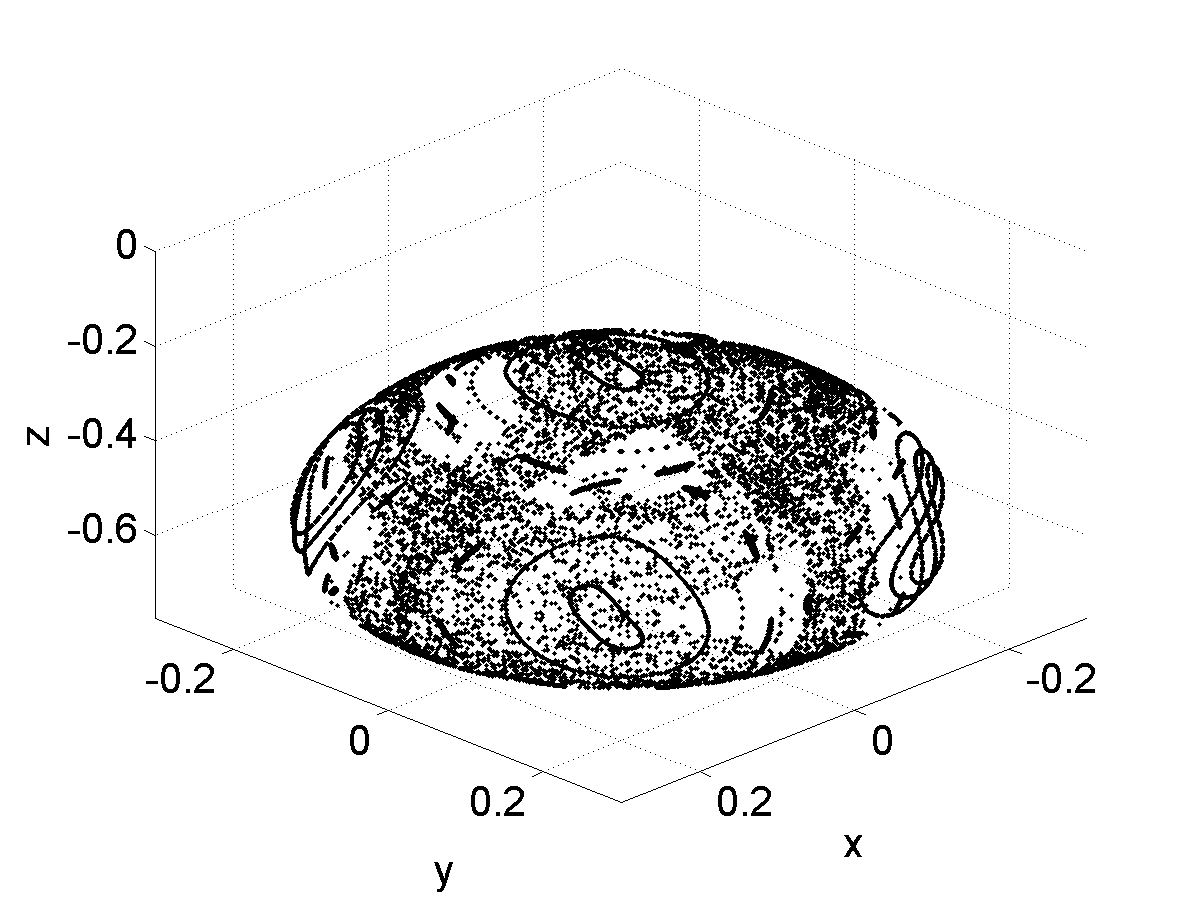}
\includegraphics[trim=0cm 0cm 1cm 2cm,clip,width=0.48\columnwidth,height=0.45\columnwidth]{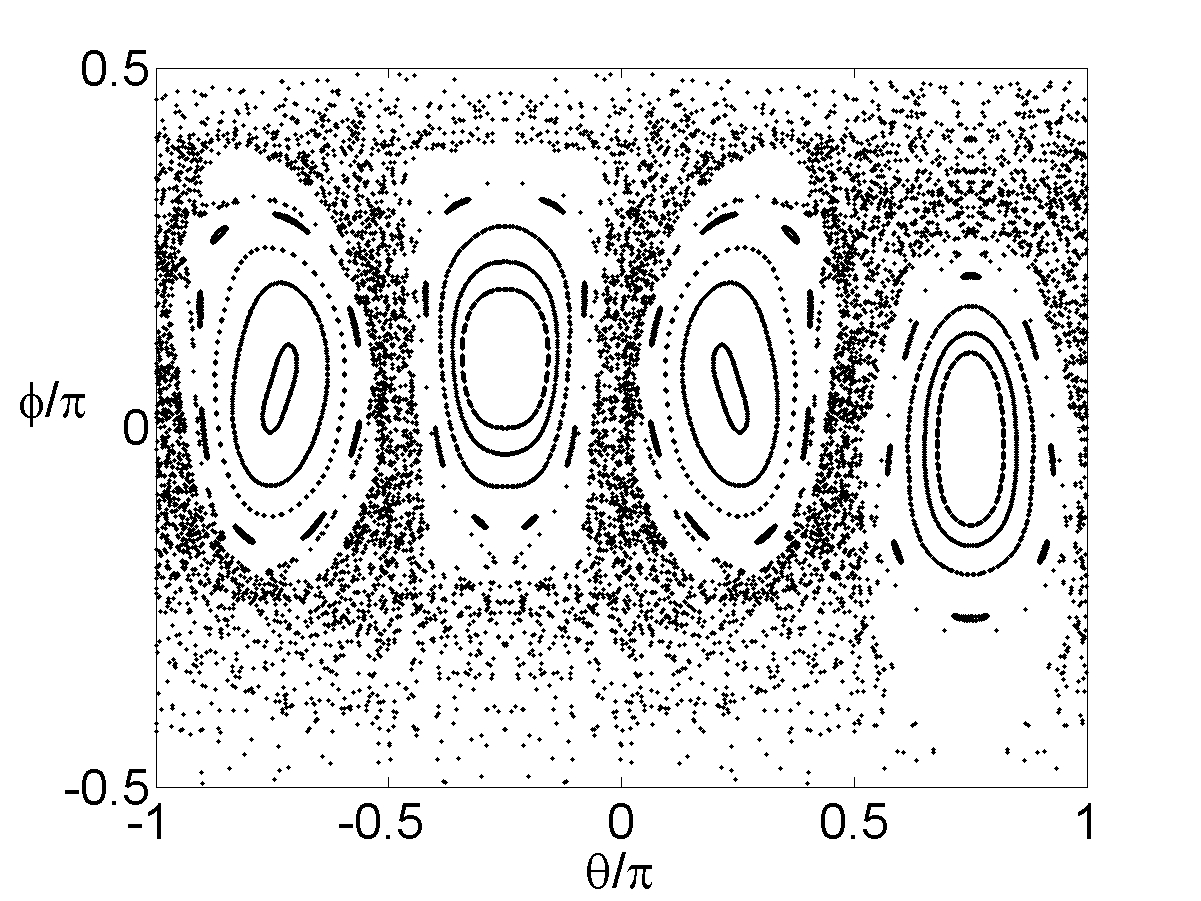}
\label{InvariantSurface_a}
}

\subfigure[Intra-surface dynamics in ``larger'' spheroids.]{
\includegraphics[trim=1cm 0cm 1cm 2cm,clip,width=0.48\columnwidth,height=0.48\columnwidth]{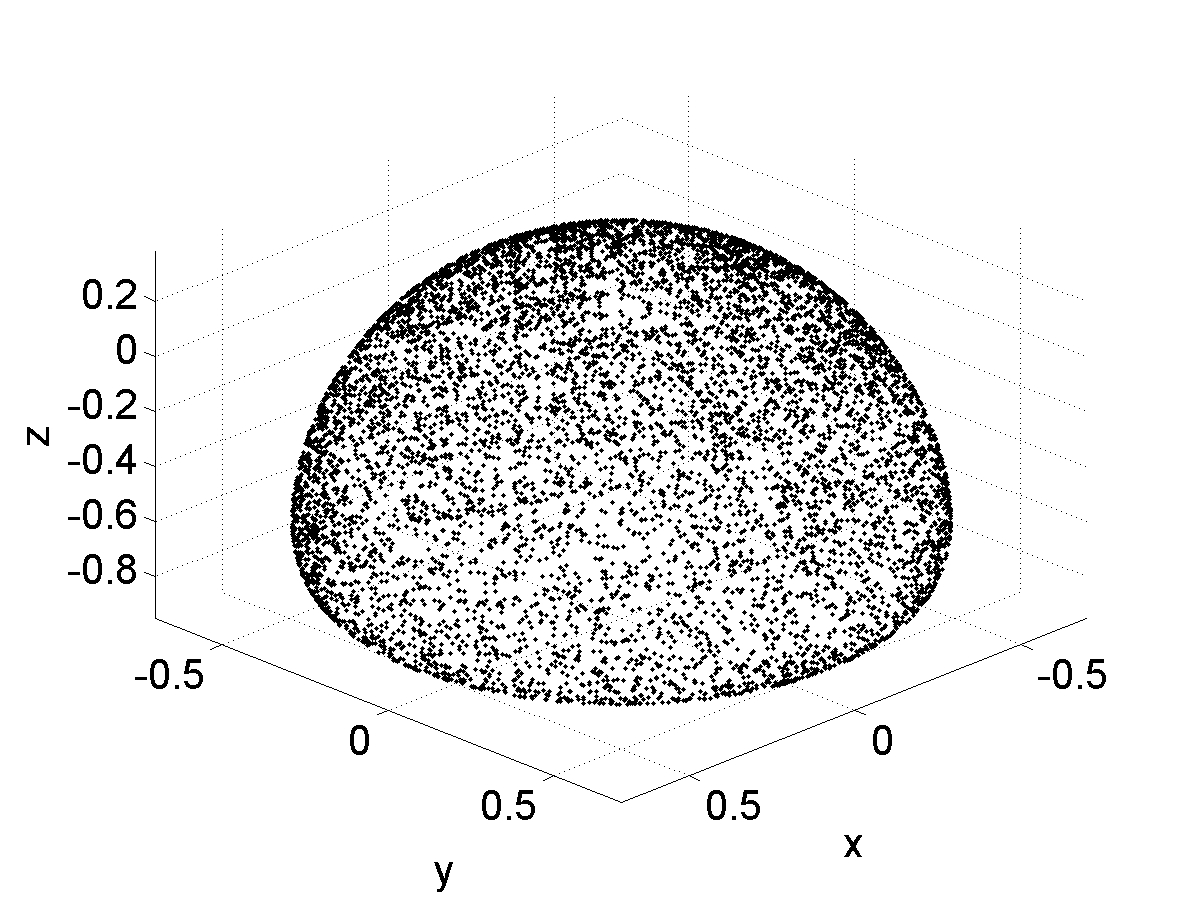}
\includegraphics[trim=0cm 0cm 1cm 2cm,clip,width=0.48\columnwidth,height=0.45\columnwidth]{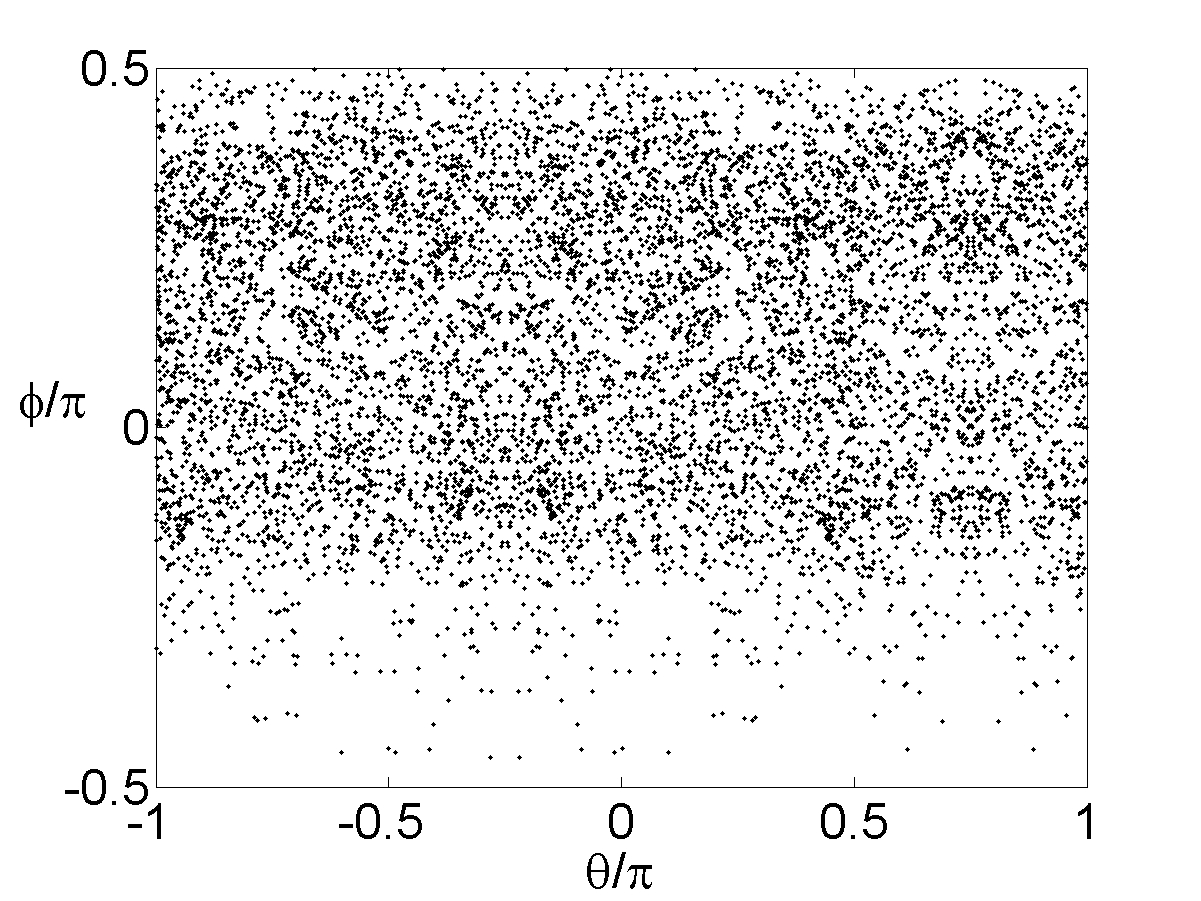}
\label{InvariantSurface_b}
}

\caption{Hamiltonian dynamics within invariant spheroids of non-inertial ($Re=0$) time-periodic cylinder flow following \FIG{UnsteadyChaos1_a} (left).}
\label{InvariantSurface}
\end{figure}

The beforementioned spheroids define the invariant surfaces in the flow topology of the time-periodic cylinder flow in \FIG{UnsteadyChaos2_b} (left). Tracers are within these LCSs restricted to streamlines defined by intersections of level sets of COMs $F_1$ and $F_2$. This
implies $H = F_2(\xi_1,\xi_2;\xi_3)$ as Hamiltonian in \eqref{InvariantSurfaces2} for the corresponding base flow and a non-autonomous counterpart $H = H(\xi_1,\xi_2,t;\xi_3)$ constructed from its reorientation for the time-periodic flow. The typical Hamiltonian dynamics in ``smaller'' spheroids is demonstrated in \FIG{InvariantSurface_a}
by the 3D (left) and corresponding intra-surface (right) stroboscopic map expressed in terms of the angular coordinates $(\xi_1,\xi_2)=(\theta,\phi)$ of
the spherical reference frame attached to the common centre of the spheroids. (The cylinder mantle and top/bottom walls correspond
with $r=1$ and $z=\pm 1$, respectively. The common spheroid centre sits at $z\approx -0.56$.) This reveals a characteristic Hamiltonian topology
consisting of (chains of) islands embedded in a chaotic sea. The intra-surface dynamics tends to become more chaotic upon progressing outwards
in $\xi_3$-direction as demonstrated in \FIG{InvariantSurface_b} for a ``larger'' spheroid. (Note that $\phi = \pm \pi/2$ corresponds with
the spheroid poles; hence the lower tracer density in these regions.)

\FIG{InvariantSurface} exposes a substantial variation in
intra-surface dynamics and this in fact reflects a fundamental difference with the abovementioned 3D steady flows. The latter are $\xi_3$-wise
periodic and thus invariably accommodate closed trajectories, which (besides possible localised phenomena) precludes qualitative topological
changes in $\xi_3$-direction (\SEC{ClosedTrajectories}). Hence e.g. the tubes in the RAM in \FIG{RAM2a}. The flow topology of 3D
unsteady flows with invariant surfaces as the time-periodic cylinder flow, on the other hand, is generically {\it aperiodic} in $\xi_3$ and
thus permits such changes. Hence the intra-surface dynamics in \FIG{InvariantSurface} and, inextricably linked with this, the partitioning of the
periodic line in \FIG{UnsteadyChaos2_b} into elliptic and hyperbolic segments.

The flow topology of the spherical tumbler for mixing of granular media in \SEC{ContainersIndustrial} for certain operating conditions also consists of invariant spheroids, meaning that the intra-surface dynamics shown
in \FIG{SphericalTumbler_b} (left) is of essentially the same Hamiltonian nature as in \FIG{InvariantSurface}.
Transport studies using kinematic models for stroboscopic maps associated with 3D time-periodic flows reveal similar
intra-surface Hamiltonian dynamics and chaos for a variety of surface topologies \cite{Gomez_invariant_2002}. This
substantiates its universality.

%\red{Link $\vec{w}_{1,2}$ with flexion field \cite{Aref2017}?}

%\subsection{Volume-preserving maps}
\subsection{Link with 3D volume-preserving flows and maps}

Steady solenoidal fluid flows and the stroboscopic maps of time-periodic solenoidal flows belong to the broader class of volume-preserving
flows and maps,\footnote{Volume-preserving maps are also denoted ``Liouvillian maps'' \cite{Cartwright1994}.}
respectively. Such generic flows and maps are widely used for studies on 3D (chaotic) advection, since they typically
rely on simple kinematic models that are easy to treat computationally and often even admit analytical
investigations. The well-known ABC flow e.g. first demonstrated the emergence of chaos in 3D steady flows and its discrete
counterpart enabled first investigations on RID shown in \FIG{UnsteadyChaos2_a} \cite{Dombre1986,Feingold1988,Cartwright1994}. Moreover, both
the ABC flow and map offered insights into the mechanisms underlying the particle clustering in \FIG{StirredTankParticleClustering}
\cite{Cartwright2002,Haller_clustering_2010}. Similar kinematic models established the generic nature of Hamiltonian chaos
within invariant surfaces as shown for the time-periodic cylinder flow in \FIG{UnsteadyChaos1_a} (left) and the emergence of RIM upon
their perturbation as per \FIG{UnsteadyChaos2_b} \cite{Gomez_invariant_2002,Moharana2013}.

%Other phenomena thus investigated include essentially 3D break-up
%mechanisms of invariant surfaces \cite{Mezic_invariant_2001,Meiss2012}.

Important in the present context is that kinematic models for volume-preserving flows and maps generically do {\it not} satisfy momentum conservation
and thus, inherently, can capture only kinematic features of 3D Lagrangian transport. This includes the above embedded Hamiltonian
structures yet excludes the fluid-dynamical mechanisms that facilitate such behaviour. Necessary for said structures is namely only the vector potential
%
%This includes the above embedded Hamiltonian structures. Necessary for this is namely only the vector potential
%
of the flow following \eqref{VectorPotential}, which exists solely by the grace of solenoidality, in
combination with a kinematic condition as e.g. uni-directional flow or a continuous symmetry.
%Hence satisfaction of momentum conservation is non-essential for adequate description of a Hamiltonian structure and corresponding dynamics.
Actual occurrence of a
kinematic condition for a given Hamiltonian structure, on the other hand, depends essentially on dynamic conditions governed by momentum conservation. The continuous symmetry underlying
%COM $F_1$ %following \eqref{COMcylflow} and associated
the intra-surface Hamiltonian dynamics within the spheroids of the time-periodic cylinder flow in \FIG{InvariantSurface}
e.g. hinges on the particular flow structure \eqref{SteadyCylinderFlow2} that, in turn, results from geometric symmetries
imparted on the flow due to the linearity of the momentum equation in the Stokes limit. Hence comprehensive 3D transport studies on realistic fluid flows must at some point always employ descriptions that satisfy both mass and momentum conservation.
Refer to \cite{Aref2017} for further discussion of kinematic versus dynamic conditions.

%The ABC flow e.g. has certain limitations regarding ... \cite{Mezic2002b}.\\\\

Furthermore relevant is that volume-preserving maps in the present scope derive from continuous time-periodic fluid flows and thus incorporate their dynamics.
A stroboscopic map therefore has an embedded Hamiltonian structure only if this already exists in the underlying flow. Conversely, an essentially
non-Hamiltonian time-periodic flow cannot produce a stroboscopic map with Hamiltonian dynamics. The intra-surface Hamiltonian structure within the
spheroids of the beforementioned time-periodic cylinder flow e.g. is a direct consequence of the presence of these LCSs in the base flow and their
invariance to reorientations of the driving wall. However, time-periodic flows thus constructed from reorientations of an essentially Hamiltonian
base flow must not necessarily adopt the Hamiltonian nature. Said time-periodic cylinder flow for $Re=100$ is essentially non-Hamiltonian, as
reflected by the emergence of an isolated periodic point with global manifolds shown (\FIG{UnsteadyChaos1_b}), notwithstanding the Hamiltonian
behaviour of the base flow (\FIG{SteadyCylinderFlow_a}).

%\subsection{Breakdown of link: essentially 3D dynamics}
\subsection{Towards essentially 3D (chaotic) dynamics}
\label{Towards3D}

\subsubsection{Conditions for chaos}
\label{ConditionsChaos}

The inherent Hamiltonian structure \eqref{Hamiltonian1} of 2D flows makes a non-autonomous Hamiltonian $H=H(\vec{x},t)$ a
necessary kinematic condition for chaos (\SEC{Hamiltonian2D}). Essentially 3D steady flows (i.e. dependent
on all spatial coordinates) are in their simplest form dynamically equivalent to a 2D unsteady flow and thus a non-autonomous
Hamiltonian system (\SEC{Hamiltonian3Dsteady}). This has the fundamental implication that {\it any} 3D (un)steady flow
meets the necessary kinematic criteria for chaos.\footnote{
This more formally follows from the Poincar\'{e}--Bendixson theorem, which states that continuous dynamical systems -- here kinematic equation \eqref{KinEq} -- admit chaos only
for a corresponding phase space -- here the space-time domain -- that is at least 3D \cite{Ott_chaos_2002}. 3D flows evidently meet this condition.
}
Whether (and to what degree) chaos actually happens is dictated by momentum conservation.
%
%Steady Euler flows ... due to momentum conservation ... Bernoulli function $\zeta = p + \rho g z + \vec{u}\cdot\vec{u}/2$ as COM, with $p$
%the pressure and $g$ the gravitational acceleration acting in $-z$ direction, ... Beltrami flows ...
%
The structure of the momentum equation implies that realistic laminar flows (i.e. subject to both viscous and inertial forces) generically lack a universal dynamical restriction to chaos \cite{Aref2017}. Hence such flows in principle always exhibit chaos in (at least) some regions and typically have a flow topology consisting of LCSs surrounded by chaos as illustrated by the RAM in \FIG{RAM2}. Global restrictions to chaos occur only for case-specific dynamic conditions as e.g. the symmetry-induced confinement to spheroids
and, inherently, essentially 2D (chaotic) dynamics, in the non-inertial time-periodic cylinder flow following \FIG{UnsteadyChaos1_a} (left).\footnote{Steady inviscid flows (i.e. so-called ``Euler flows'') are, in stark contrast, generically {\it non-chaotic} due to the restriction of Lagrangian dynamics to level sets of the Bernoulli function $\zeta = p + \rho g z + \vec{u}\cdot\vec{u}/2$, with $p$ the pressure, $\rho$ the density and $g$ the gravitational
acceleration acting in $-z$ direction \cite{Aref2017}.
}

The necessary {\it topological condition for chaos} is the transversal intersection of stable and unstable manifolds associated with hyperbolic points/lines/trajectories as per \SEC{CriticalPoints} \cite{Ottino_Kinematics_1989,Ott_chaos_2002,Aref2017}. Such interactions are denoted {\it homoclinic} and {\it heteroclinic} if involving (un)stable manifolds of the same or two different hyperbolic LCSs, respectively. \FIG{OrderWithinChaosa} e.g. shows the homoclinic intersection of the stable (blue) and unstable (red) manifolds of the central hyperbolic point (cross) in the archetypal 2D time-period flow; these LCSs furthermore interact heteroclinically with manifolds (not shown) of the two adjacent hyperbolic points. The actual mechanism leading to exponential stretching of fluid parcels and, in consequence, chaotic advection is the accumulation of intersections upon approaching a periodic point along the stable/unstable manifold during forward/backward progression in time.

\paragraph{3D steady flows}

Transversal manifold interaction may in 3D steady flows occur via various scenarios. This generically excludes
stagnation lines, since these entities are atypical of realistic flows. The elliptic stagnation lines that
e.g. shape the idealised streamline topology of the cylinder flow in \FIG{SteadyCylinderFlow_0a} and droplet
flows in \FIG{ChaosDroplets2X_a} for basically any perturbation give way to a hyperbolic focus and elliptic
closed streamlines, respectively, in the corresponding non-ideal (i.e. realistic) topologies shown in
\FIG{SteadyCylinderFlow_a}, \FIG{SteadyCylinderFlow_0b} and \FIG{ChaosDroplets2X_b}. Hence only isolated stagnation
points and closed/periodic hyperbolic streamlines are of practical relevance. (This includes critical points of the
skin-friction field on grounds of their equivalence to hyperbolic nodes; \SEC{CriticalPoints}.) The 1D manifolds of
stagnation points can immediately be ruled out from transversal interaction scenarios; they can intersect 2D manifolds
namely only at stagnation points. However, this prevents progression of tracers along these intersections and thus
violates the nature of manifolds. This constraint e.g. prohibits intersection of the 1D and 2D manifolds of the
skin-friction field in \FIG{PorousMedia1b}.

\begin{figure}
\centering
\includegraphics[width=\columnwidth,height=0.9\columnwidth]{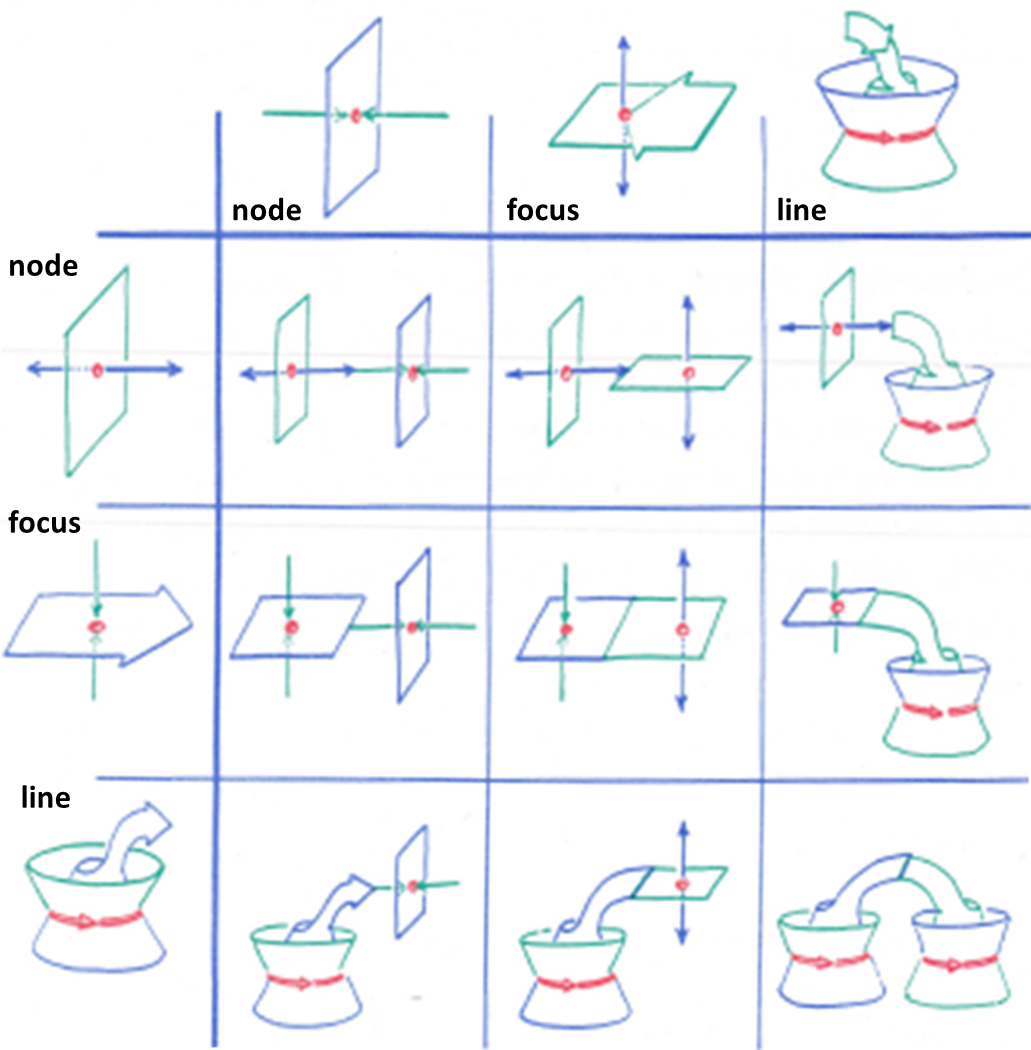}
\caption{Scenarios for heteroclinic interaction between stagnation points (red dots) and hyperbolic closed
streamlines (red circles) in 3D steady flows. Blue/green lines/surfaces indicate unstable/stable 1D/2D manifolds. Adapted from \cite{Abraham1982}.}
\label{CanonicalCases}
\end{figure}

Heteroclinic interaction between stagnation points and hyperbolic {\it closed} streamlines is possible via 9 scenarios following \cite{Abraham1982} (\FIG{CanonicalCases}).
%
%Fig. 4.5.8 of \cite{Abraham1982}. \red{... \FIG{CanonicalCases} ...}
%
Heteroclinic {\it transversal}  interaction is restricted to 2D manifolds and, given the latter are stream surfaces, thus defines streamlines (denoted ``heteroclinic trajectories'') with corresponding dynamics
as follows:\footnote{The flow topology of the convection cells in \FIG{ConvectionCells1b} e.g. results from heteroclinic {\it non-transversal} interactions between hyperbolic nodes in the stress-free fluid surface and free-slip bottom wall.}
\begin{itemize}
\item[$\bullet$] {\bf Point--point interaction} (Sec. 4.1 of \cite{Abraham1982}): a single heteroclinic
trajectory connects both points; exponential stretching is absent and thus dynamics are non-chaotic.

\item[$\bullet$] {\bf Point--line interaction} (Sec. 4.3 of \cite{Abraham1982}): two heteroclinic trajectories
emerge from the point and asymptotically spiral towards a limit cycle defined by the closed streamline; accumulation of
windings of the spiralling trajectories upon approaching the closed streamline and elongation of the 2D manifold of the
line along the 1D manifold of the point signifies exponential stretching of both 2D manifolds and thus chaos.\footnote{This
describes the topology of the interaction; the progression in time depends on the manifold stabilities. Interaction of
the 2D unstable and 2D stable manifold of a point and line, respectively, yields spiralling forward and elongation
backward in time. This reverses for opposite stabilities.
}

\item[$\bullet$]  {\bf Line--line interaction} (Sec. 4.4 of \cite{Abraham1982}): the intersecting (topologically
cylindrical) manifolds yield two heteroclinic trajectories that asymptotically spiral both forward and backward
in time to limit cycles defined by the closed streamlines; accumulation of windings upon approaching the latter
again produces chaos.
%
%The cylindrical shape of the manifolds in fact implies intermediate heteroclinic
%trajectories (i.e. ``Birkhoff's signature'') that greatly increase the topological complexity and the
%rate of stretching (Sec. 4.5 of \cite{Abraham1982}). The intersecting manifolds demarcate structures
%denoted ``lobes'' that grow increasingly
%
%due to said accumulation of windings
%
The cylindrical shape of the manifolds in fact implies intermediate heteroclinic
trajectories (i.e. ``Birkhoff's signature'' \cite{Abraham1982}) that, through the formation of so-called ``lobes'' \cite{Beigie1994},
greatly increase the rate of stretching.

\end{itemize}
The 2D manifolds of single closed streamlines admit {\it homoclinic} intersection similar to the heteroclinic
line--line interaction (Ch. 5 in \cite{Abraham1982}). These findings put forth point--line and, by their
topological complexity, particularly line--line interactions as ``drivers'' of chaos in 3D steady flows.
However, 2D manifolds may also impede chaos in their capacity as barriers to 1D/2D manifolds of the same stability.

Hyperbolic {\it periodic} streamlines are also closed in the sense of reconnecting via a periodic inlet-outlet
and thus admit the same point--line and line--line interactions as elaborated above. (Heteroclinic trajectories
here ``spiral'' towards the limit cycle(s) via repeated reconnection.) However, heteroclinic interaction between
{\it closed} and {\it periodic} streamlines is impossible (Conjecture 1 in \cite{Speetjens2014}).
Moreover, point--line interaction is possible only if the 2D stable (unstable) and 2D unstable (stable) manifolds of
the point and streamline are exclusively interacting with each other (which follows from Conjecture 3 in
\cite{Speetjens2014}). These topological constraints are e.g. crucial to the emergence of a net throughflow
region in periodic duct flows and, in consequence, the persistence of an embedded Hamiltonian structure in this
subregion \cite{Speetjens2014}.

\paragraph{3D unsteady flows}

Topological freedom -- and, inherently, the scenarios for transversal interaction -- increases substantially for 3D unsteady flows. The 1D manifolds of isolated periodic points in stroboscopic maps can, for instance, intersect with 2D manifolds, thus permitting the homoclinic transversal manifold interaction for the hyperbolic focus in the time-periodic cylinder flow in \FIG{UnsteadyChaos1_b}.

Unsteady conditions furthermore allow for non-trivial point--point interactions leading to chaos. Consider e.g. the ideal buoyant droplet in \FIG{ChaosDroplets2X_a} (left); its spherical surface and internal (black) axis are the merged 2D and 1D manifolds, respectively,
of hyperbolic nodes at the axis endpoints. Aperiodic unsteady perturbation triggers transversal interaction of the manifolds of
DHTs (aperiodic counterparts of hyperbolic nodes) following \FIG{HillsVortex} and thus chaotic advection due to lobe
formation akin to transversal line--line interactions.
Similar behaviour occurs upon subjecting a droplet immersed in a steady swirling flow (its surface and axis emanate in a similar way
as before from interacting hyperbolic foci) to time-periodic perturbations. This is demonstrated in Fig. 15 of \cite{Beigie1994} (originally from \cite{Holmes1984}).

The above examples concern chaos directly resulting from relaxation of topological restrictions of 3D steady flows. However, insights
into the full set of transversal manifold interactions possible under essentially 3D unsteady conditions remain incomplete.

%... is in fact the frontier of Lagrangian transport in fluid flows \cite{Aref2017}.\\\\

%\begin{itemize}
%\item[$\bullet$] \red{3D unsteady flows:} greater topological freedom; only incomplete insight into possible interactions.
%Truly 3D chaos: e.g. homoclinic transversal interaction of manifolds of isolated periodic point in time-periodic cylinder flow for $Re=100$ in \FIG{UnsteadyChaos1_b}. True 3D chaos in 3D unsteady systems is the frontier %\cite{Aref2017}.
%\end{itemize}

\subsubsection{Conditions for 3D dynamics}
\label{Conditions3Ddynamics}

Lagrangian transport in 2D solenoidal flows is governed by Hamiltonian mechanics due to the Hamiltonian structure \eqref{Hamiltonian1} of
kinematic equation \eqref{KinEq} (\SEC{Hamiltonian2D}). This notion enables rigorous distinction of essentially 2D dynamics from truly 3D
dynamics in 3D solenoidal flows. The former namely implies an embedded Hamiltonian structure according to \SEC{Hamiltonian3Dsteady} and
\SEC{Hamiltonian3Dunsteady} in kinematic equation \eqref{KinEq} relating to the physical flow via (formal) transformation \eqref{Steady3DvsUnsteady2D}; the latter implies (local) absence of such a structure.

The essentially 2D nature of the dynamics imparted by an embedded Hamiltonian structure suggests that the flow topology of such 3D flows
is basically a ``3D extrusion'' of the typical Hamiltonian topology of 2D flows illustrated in \FIG{OrderWithinChaosa} in the $\xi_3$-direction of
an appropriate curvilinear frame \eqref{CoordinateTransform} following \SEC{Hamiltonian3Dsteady} and \SEC{Hamiltonian3Dunsteady}. Thus
islands become tubes/tori centered on elliptic lines/closed trajectories defined in \SEC{CriticalLines} and \SEC{ClosedTrajectories}
surrounded by chaotic streamlines/trajectories as e.g. for the RAM in \FIG{RAM2}. Hence the emergence of such LCSs in the various
configurations in \SEC{Categorisation} are clear indicators of (local) intrinsically 2D Hamiltonian dynamics. The breakdown
of tubes/tori into chaos is described by 3D counterparts to the beforementioned KAM and Poincar\'e--Birkhoff theorems (\SEC{Hamiltonian2D}) \cite{cheng90,cheng90b}.

The above advances (local) departure from or {\it a priori} absence of an embedded Hamiltonian structure in \eqref{KinEq} as necessary kinematic condition for truly 3D (chaotic) dynamics in 3D (un)steady flows. Two important mechanisms in this respect are elaborated below: singularities and resonances.

\paragraph{Singularities}

%The above advances (local) breakdown or {\it a priori} absence of an embedded Hamiltonian structure in \eqref{KinEq} as necessary kinematic condition for truly 3D (chaotic) dynamics in 3D (un)steady %flows.

Actual local breakdown of an embedded Hamiltonian structure manifests itself mathematically in singularities in the corresponding transformation \eqref{Steady3DvsUnsteady2D}.
%
%Local breakdown manifests itself mathematically in singularities in the transformation \eqref{Steady3DvsUnsteady2D} from a 3D flow to an embedded Hamiltonian structure.
%
Consider as an instructive example the 2.5D duct flows introduced before, which (for $\vec{\xi} = \vec{x}$) adopt both Hamiltonian form \eqref{KinEqCurvilinear2} and \eqref{Hamiltonian2} via transformation \eqref{Steady3DvsUnsteady2D_2a} and \eqref{Steady3DvsUnsteady2D_2}, respectively, due to the axial component of the base flow meeting $v_z(x,y)>0$ (\SEC{HamiltonianFormDuctFlows}). The latter changing
sign at $v_z=0$ implies a non-monotonic relation $z(t)$, prohibiting its global inversion, and thus a singularity in
transformation \eqref{Steady3DvsUnsteady2D_2a}. Vanishing $u_3 = v_z$, similarly, yields a singular Jacobian $J_\ast = |\partial\vec{\zeta}/\partial\vec{\xi}| = -\partial\widetilde{A}_1/\partial\xi_2 = \widetilde{u}_3$, with $\vec{\zeta}=(\zeta_1,\zeta_2,\tau)$, for transformation \eqref{Steady3DvsUnsteady2D_2}. Hence both cases only admit a local transformation
into the respective Hamiltonian forms in subregions $v_z>0$ and $v_z<0$ that, given continuity, are separated by (a) smooth surface(s) $v_z=0$.
Superposition of e.g. the previous 2.5D RAM flow with a uniform axial backflow of relative strength $0\leq \alpha \leq 1$, i.e.
\begin{eqnarray}
v_z = u_z = 2U(1 - r^2) - 2 \alpha U = 2U(r_\ast^2 - r^2),
\end{eqnarray}
results in flow reversal from $v_z>0$ to $v_z<0$ at the cylindrical surface with radius $r_\ast = \sqrt{1 - \alpha}$.
Transformation \eqref{Steady3DvsUnsteady2D_3}, again using $\vec{u}'= \vec{u}/U$, gives $J_\ast = 2(r_\ast^2 - r^2)$ and
indeed becomes singular at $r = r_\ast$, implying a partitioning of the global dynamics into two coexisting Hamiltonian systems.
Property $v_z=0$ thus precluding a single global Hamiltonian structure, despite local dynamics retaining their Hamiltonian
nature, is already an essentially 3D situation. Tracers generically namely can cross surfaces $v_z=0$ and, within one
global flow, switch between local Hamiltonian systems.\footnote{This means that here $v_z\neq 0$, or generically, $u_3\neq 0$
is sufficient for a non-singular transformation \eqref{Steady3DvsUnsteady2D_2}; the condition $u_{1,2,3}\neq 0$ (i.e. absence
of stagnation points) according to \cite{Bajer_3D_1994} is unnecessarily stringent.}

Partitioning into multiple Hamiltonian systems and/or restriction of Hamiltonian dynamics to subregions may furthermore stem from
singularities in coordinate transform \eqref{CoordinateTransform}. Periodic duct flows, by expression in the curvilinear frame
aligned with the 3D streamlines according to \cite{Speetjens2014}, e.g. in principle always meet condition $u_3>0$ necessary
for a non-singular transformation \eqref{Steady3DvsUnsteady2D_2}. However, stagnation points and (further) topological complexity
of the streamlines often rule out a single global transformation \eqref{CoordinateTransform} and thus restrict embedded
Hamiltonian structures to subregions admitting non-singular $\mathcal{G}$. Topological considerations strongly suggest that
periodic duct flows accommodate at least one such subregion, viz. the beforementioned net throughflow region (possibly
partitioned into multiple throughflow regions). Hence absence of a single global Hamiltonian structure again reflects an
essentially 3D situation. Moreover, the net throughflow region (though itself accommodating Hamiltonian dynamics) may assume
a complex structure particularly in partitioned-duct flows. Stagnation points and corresponding manifold interactions in
the stagnation area of the T-shaped micro-mixer (\FIG{TshapedMicroMixer2}b) are e.g. likely to have a major impact on the
dynamics and structure of the (multiple) throughflow region(s).

Generic 3D (un)steady flows are likely to develop singularities in the transformation from physical to canonical space as
well as in that from the Cartesian to an appropriate curvilinear frame -- causing breakdown/restriction of (local)
Hamiltonian structures -- as exemplified above. Moreover, the number and/or region of influence of singularities generically
grow with increasing departure from an originally Hamiltonian structure and thus progressively diminish the latter in favour of
essentially 3D dynamics. The global topological complexity of the transversally-interacting 1D/2D manifolds of the isolated
periodic point in the time-periodic cylinder flow for $Re=100$ in \FIG{UnsteadyChaos1_b} e.g. strongly suggests full
breakdown of the Hamiltonian structure of the corresponding non-inertial limit ($Re=0$) in \FIG{InvariantSurface}.

\paragraph{Resonances} Resonances are a further mechanism that may lead to 3D dynamics by causing essentially 3D departures from 2D Hamiltonian
dynamics. This may promote 3D chaos as well as formation of intricate LCSs as demonstrated in \FIG{UnsteadyChaos2} by RID and RIM, respectively. The essence of these fundamentally distinct resonance phenomena is exemplified below by way of simplified maps.

\begin{figure}
\centering
\subfigure[Fully resonant torus]{\includegraphics[trim=3cm 2cm 3cm 1.8cm,clip,width=0.45\columnwidth,height=0.48\columnwidth]{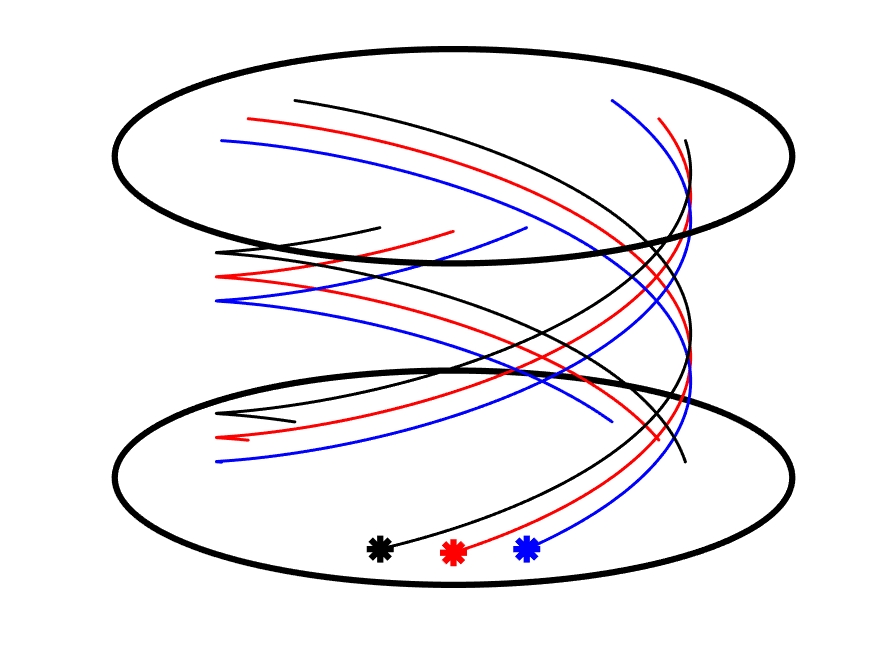}\label{ResonanceX_a}}
\hspace*{6pt}
\subfigure[Local resonance within torus]{\includegraphics[trim=3cm 2cm 3cm 1.8cm,clip,width=0.45\columnwidth,height=0.48\columnwidth]{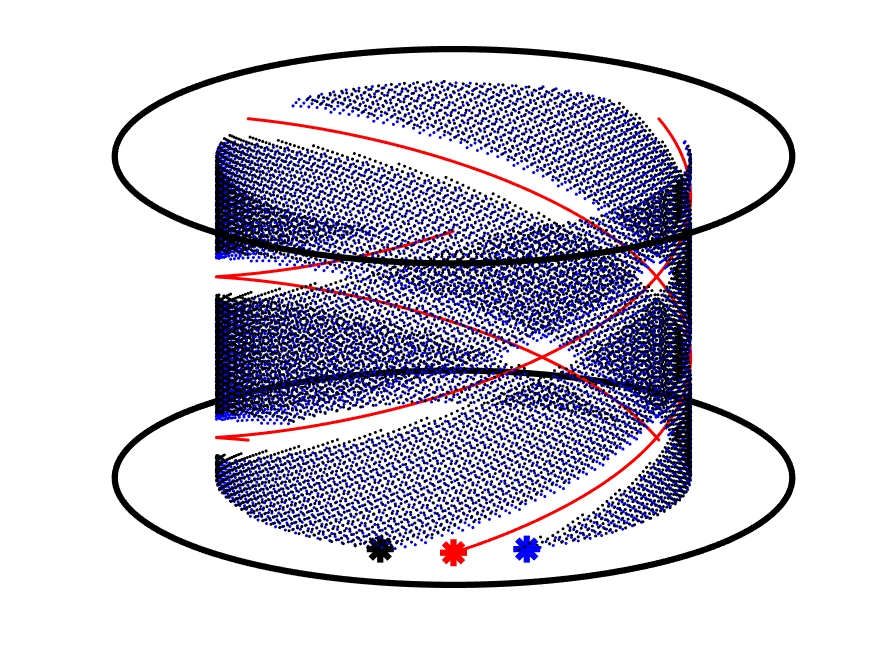}\label{ResonanceX_b}}

\caption{Resonance demonstrated by 3 trajectories in a periodic cylindrical duct starting at the inlet $z=0$ (bottom) from the marked
positions: (a) fully resonant torus for $\omega = m\gamma/n$, $\gamma = 0.3$ and $(m,n) = (2,3)$; (b) isolated resonant trajectory (red) separating torus segments delineated by drifting tracers (black/blue).}
\label{ResonanceX}
\end{figure}

{\it RID} may occur in (local) flow topologies consisting of tori as in \FIG{UnsteadyChaos2_a} (left) described by trajectories spiralling around the toroidal axis. The Lagrangian motion is qualitatively
%similar to the helical trajectories along concentric tubes in a cylindrical periodic duct of length $L=2\pi$ following
similar to the helical trajectories along concentric tubes in a cylindrical periodic duct following
\begin{eqnarray}
r_{p+1} = r_p,\quad\theta_{p+1} = \theta_p + \omega,\quad z_{p+1} = z_p + \gamma,
\label{MappingRID}
\end{eqnarray}
with $\omega = \omega(\vec{x}_0)$ and $\gamma = \gamma(\vec{x}_0)$ the period-wise azimuthal and axial displacement, respectively. The transverse ($xy$-wise) mapping is identical to the 2D elliptic mapping in \eqref{MappingLin2} and $(\theta,z)$ represents the local poloidal/toroidal coordinates $(\phi,\varphi)$ following \eqref{Resonance}. Reconnection -- and thus resonance -- occurs for $\omega$ and $\gamma$ forming rational fraction $\omega/\gamma = m/n$ as per \eqref{Resonance}. Generically $\omega = \omega(r)$ and $\gamma = \gamma(r)$, meaning that tori are either entirely non-resonant or resonant. Non-resonant tori survive weak perturbations according to the (3D counterpart of the) KAM theorem; resonant tori disintegrate following the corresponding Poincar\'{e}-Birkhoff theorems due to violation of the ``averaging principle'' first mentioned in \SEC{ContainersIndustrial} \cite{Ott_chaos_2002,cheng90b}. (This violation in fact
occurs both for ``true'' resonance and slow-down of motion along trajectories to the same order of magnitude as the perturbation.\footnote{Consider for illustration the 2D simplification of \eqref{MappingRID} subject to a weak perturbation $\epsilon \ll r_0$: $(r_{p+1},\theta_{p+1}) = (r_p + \epsilon\cos\theta_p,\theta_p + \omega)$. Non-resonant circulation (e.g. $\omega = \sqrt{2}$) yields angular positions $\theta_p$ that densely fill the range $0\leq \theta \leq 2\pi$ and corresponding radii $r_p$ that oscillate about $r_0$ with a bandwidth of $\mathcal{O}(\epsilon)$.
Thus the tracer positions $(r_p,\theta_p)$ gradually describe a closed orbit that is approximately circular and has {\it average} radius $\bar{r} = \lim_{P\rightarrow\infty}\sum_{p=0}^{P-1} r_p/P = \int_0^{2\pi} r(\theta)d\theta/2\pi = r_0$. This averaging process effectively neutralises the perturbations and the original circular orbit $r_p=r_0$ survives up to $\mathcal{O}(\epsilon)$. Resonant circulation $\omega = 2\pi$, on the other hand, yields $(r_p,\theta_p) = (r_0 + p\epsilon\cos\theta_0,\theta_0)$ and thus causes a progressive radial drift of $\mathcal{O}(p\epsilon)$ and, in consequence, breakdown of the original circular orbit.
This behaviour generalises to slow motion. Slow circulation $\omega = \epsilon \ll 2\pi$ via $\cos\theta_p = \cos(\theta_{p-1} + \epsilon)\approx \cos\theta_{p-1}$ namely gives $(r_p,\theta_p) \approx (r_0 + p\epsilon\cos\theta_0,\theta_0 + p\epsilon)$ and indeed results in a similar radial drift.
}) Such torus-wise survival {\it or} breakdown is the ``conventional'' Hamiltonian route to chaos.
\FIG{ResonanceX_a} gives 3 neighbouring trajectories (distinguished by colour) in a periodic duct of unit radius and length $2\pi$ starting at the inlet $z=0$ (bottom) from the marked positions on a typical resonant torus for $\gamma = 0.3$ and $(m,n) = (2,3)$; thus reconnection occurs after 3 cycles and single trajectories appear as 3 segments. This torus disintegrates for any perturbation in an actual flow. Irrational $\omega/\gamma$ (e.g. $\omega/\gamma=\sqrt{2}$) yields a non-resonant (and surviving) torus densely filled by the trajectories (not shown).

%\footnote{\red{Except for ``slow'' circulation $\omega = \epsilon \ll 2\pi$. Reconsidering the above example via $\cos\theta_p = \cos(\theta_{p-1} + \epsilon)\approx \cos\theta_{p-1}$ namely gives %$(r_p,\theta_p) \approx (r_0 + p\epsilon\cos\theta_0,\theta_0 + p\epsilon)$ and results in a similar radial drift. Hence slow motion indeed has the same effect as resonance.}}

However, 3D flows also admit resonances in subregions within tori, causing only local interruption of the averaging process and, in consequence, local defects in tori (rather than
full breakdown). This may occur upon weak perturbation of a flow topology consisting entirely of reconnected trajectories following \FIG{ResonanceX_a} and, inherently, resonant tori. Consider for illustration an azimuthal
displacement $\omega(r,\theta) = \omega_0(r) + \epsilon\sin(\theta)$ such that $\omega_0$ gives a rational fraction with $\gamma$ as before and $\epsilon=0$ thus represents the
unperturbed state. Perturbation $0<\epsilon\ll\omega_0$ breaks the rationality, except for isolated rational trajectories starting at $\theta_0 = 0$ and $\theta_0 = \pi$, and thus triggers dynamics
as shown in \FIG{ResonanceX_b} for $\omega_0/\gamma = m/n$, with $(m,n)$ and $U$ as before, and $\epsilon = 5\times 10^{-4}$: azimuthal drifting of tracers (black/blue) away from the rational
trajectory (red). The drifting tracers delineate the torus segments partitioned by the rational trajectory; local defects develop at the latter and enable random jumps between tori and thus
promote global dispersion of tracers (i.e. RID) as demonstrated in \FIG{UnsteadyChaos2_a} (right) and observed or suspected in a range of other (un)steady 3D flows discussed
in \SEC{Categorisation}.\footnote{Resonances are often considered singularities as well. However, this is in the sense of interrupting the averaging process that underlies the formation/persistence of LCSs; they generically do {\it not} constitute singularities in the transformations to embedded Hamiltonian structures.}

%This is in essence the mechanism behind RID as demonstrated in \FIG{UnsteadyChaos2_a} (right) and observed or suspected in a range of other (un)steady 3D flows discussed
%in \SEC{Categorisation}. The drifting tracers delineate the torus segments partitioned by the rational trajectory ...\\\\

%
%\footnote{\red{Such local defects are in principle also possible for system \eqref{Hamiltonian1} via non-monotonic $\phi(r)$ (i.e. so-called ``non-twist maps'' \cite{Morrison2000}). However, %resonances generically depend on all spatial coordinates,
%i.e. $\phi = \phi(\vec{x})$, and correspond with a very specific (and thus non-generic) time-dependence of Hamiltonian. Hence resonances essentially 3D.}}

RID occurs if motion along unperturbed trajectories becomes of the same order as the perturbation due to resonance or local slow-down. This gives
perturbed motion that is comparable in all coordinate directions and, in consequence, no longer restricted to the original tori. Behaviour akin to RID occurs if the (otherwise weak) perturbation has localised peaks of the same order as the typical motion along the unperturbed trajectories \cite{Smith_localized_2017}. (Such peaks
can e.g. develop in shear layers in granular media or non-Newtonian fluids.) The effect
is similar as before: perturbed motion comparable in all coordinate directions that again causes local breakdown of tori. Thus both resonance/slow-down
and perturbation peaks enable tracer dispersion across tori; differences primarily concern the intensity in that dispersion by former and
latter mechanism is relatively slow and fast, respectively.\footnote{Reconsider for illustration the 2D simplification of \eqref{MappingRID} subject
to a generally weak perturbation $\epsilon \ll r_0$ as before yet with a localised peak of $\mathcal{O}(r_0\omega)$ around radius $r=r_\ast$. Trajectories outside
the vicinity of the peak (given non-resonant circulation) remain within a narrow band $|r_{p+1}-r_p|\sim\mathcal{O}(\epsilon)$ and thus again survive
in an averaged way. Tracers on trajectories near this peak, on the other hand, undergo a radial ``push'' $|r_{p+1}-r_p|\sim\mathcal{O}(r_0\omega)$ comparable to
the azimuthal displacement of $\mathcal{O}(r_0\omega)$ and thus are deflected. This deflection is essentially similar as for resonance/slow-down yet of greater intensity, i.e. $|r_{p+1}-r_p|\sim\mathcal{O}(r_0\omega)$ versus $|r_{p+1}-r_p|\sim\mathcal{O}(\epsilon)$.}

{\it RIM} may occur upon perturbation of flow topologies accommodating invariant surfaces ``pierced'' by (a) periodic line(s) as e.g.
the spheroids in the Stokes limit $Re=0$ of the time-periodic cylinder flow in \FIG{UnsteadyChaos2_b} (left). Fluid inertia induces drift across the spheroids and yields Lagrangian
motion transverse and parallel to the periodic line qualitatively similar to that near periodic points in 2D systems as per \eqref{MappingLin} and an
axial uni-directional motion, respectively, i.e.
\begin{eqnarray}
%\vec{x}_{n+1} = F(\vec{x}_n)\vec{x}_n,\quad z_{n+1}> z_n,
\vec{x}_{p+1}^\perp = F\vec{x}_p^\perp,\quad z_{p+1}> z_p,
\label{MappingRIM}
\end{eqnarray}
with $\vec{x}^\perp = (x,y)$. Key to RIM is a $z$-dependent map $F = F(z)$ that smoothly changes its qualitative nature, determined by $J_1 = \mbox{tr}(F)$, at some axial position $z_0$ from $|J_1|<2$ to $|J_1|>2$, thus transiting from elliptic ($z<z_0$) to hyperbolic ($z>z_0$) dynamics, respectively. This happens in actual flows for (perturbed) periodic lines partitioned into elliptic ($|J_1|<2$) and hyperbolic ($|J_1|>2$) segments and thus merges tubes and shells -- created on former and latter segments, respectively, by the averaging process -- into intricate LCSs as demonstrated in \FIG{UnsteadyChaos2_b} (right).

The degenerate points $|J_1|=2$ separating elliptic and hyperbolic segments (denoted ``parabolic points'') act as
resonances by terminating and disconnecting the averaging processes that underlie the tube and shell
formation and thus pave the way to their merger (i.e. RIM). Perturbations of the cylinder flow generically yield a
uni-directional drift along the periodic line yet isolated periodic points may also emerge near the parabolic point
as observed for some instances of RIM in the sphere-driven time-periodic flow studied in \cite{Moharana2013}. However,
absence/presence of such periodic points is non-essential for RIM. The decisive kinematic condition is a qualitative
change in dynamics due to segmentation of periodic lines.

Resonances as exemplified above occur for small departures from an intact embedded Hamiltonian structure comprising
of complete LCSs and, unlike singularities, thus affect Lagrangian dynamics only under such conditions. Moreover,
the topology of the LCSs is crucial for RID and RIM; former and latter phenomenon, if occurring, are inextricably linked
to toroidal and spheroidal invariant surfaces, respectively \SEC{ContainersIndustrial}.
Chaotic advection
resulting from RID and RIM (via e.g. ``leaky shells; \SEC{ContainersIndustrial}) therefore is relatively
weak compared to a system far away from a Hamiltonian state as e.g. a flow topology dominated by manifolds
of isolated periodic points following \FIG{UnsteadyChaos1_b}. Hence RID and RIM are (in a practical sense)
primarily precursors to global chaos and of only limited use for (industrial) mixing purposes. The delicate
and non-trivial LCSs formed by RID and RIM may, on the other hand, find practical applications other than
mixing.
Large departures from embedded Hamiltonian structures are in any case imperative to accomplish truly
3D (chaotic) dynamics yet insights into the response of 3D flow topologies -- and their potential applications -- to
progressive perturbations are far from complete.

\section{Commercialisation of chaotic advection}
%\section{Commercialisation of chaotic advection}
%\section{Commercial applications of Lagrangian (chaotic) transport}
%\section{Commercialisation of Lagrangian (chaotic) transport}
\label{Commercialisation}

This section considers how the extensive scientific knowledge and insights of 3D Lagrangian transport developed over the last 30 years have migrated to industrial practice and technology.   The success of translation is measured by the uptake of fluid chaos in patentable ideas, new devices that move beyond university laboratories and ultimately to the commercialisation and use of these ideas and devices in application. However,  the  process of taking a novel idea to a successful outcome is an arduous journey that is familiar to all who have attempted it.  Although the
practical uses of chaos are myriad, there is a general reluctance in traditional industries to adopt new technologies that are seen as untried and untested.  Even in new industries and applications, the tendency to fall back on existing technology is strong.

Obtaining reliable information on commercialisation of fluid chaos has proven difficult due to the sensitivity around intellectual property and a lack of clarity around ownership of ideas.  Consequently, primarily the patent literature is examined.  Although an application for patent protection is often a precursor for  commercial applications,  patent  activity must be viewed cautiously from this viewpoint and can be seen only as an indicator.  Most patents are never successfully commercialised and, conversely, some commercially successful ideas
are never patented. This section finishes with a brief discussion of a number of technologies that have progressed to commercialisation.

\subsection{Patents}
\label{subsec:patents}

%\begin{figure}
%  \centering
%\includegraphics[width=\columnwidth]{Figures/patents.png}
%  \caption{Cumulative numbers of published patents about or
%    consciously using chaotic advection.  The gray curve shows the
%    numbers relating to microfluidic applications.}
%  \label{fig:patents}
%\end{figure}

%\blue{Patents may serve as an indicator for commercial potential. Below: Murray's modified text.}\\\\

 An internet search using {\it Google Patent}  (https://patents.google.com) with a search term of {\tt ``(chaos OR chaotic) AND (mixer OR mixing) AND fluid''} in October 2018 yielded a list of approximately 560 patents.  This list was consolidated by removing duplicate patents in different jurisdictions, patents for identical devices claimed for different applications, devices that had nothing to do with fluid mixing and devices that claimed to use chaotic mixing as a loose terminology for turbulent mixing.  Patents that mentioned in passing the use of unspecified ``chaos/chaotic mixers'' as one of many possibilities in a list of mixers were also discounted. Thus 293 patents remained and consisted of 173 patents for ``stand alone'' 3D chaos mixers and 120 patents in which chaotic mixing was explicitly mentioned as part of at least some embodiments of the patent.  This latter group included patents where specific types of chaotic mixers were mentioned as well as those in which an unspecified ``chaotic mixer'' was explicitly mentioned as an integral part in at least some embodiments.

\begin{figure}
  \centering
\includegraphics[width=0.95\columnwidth]{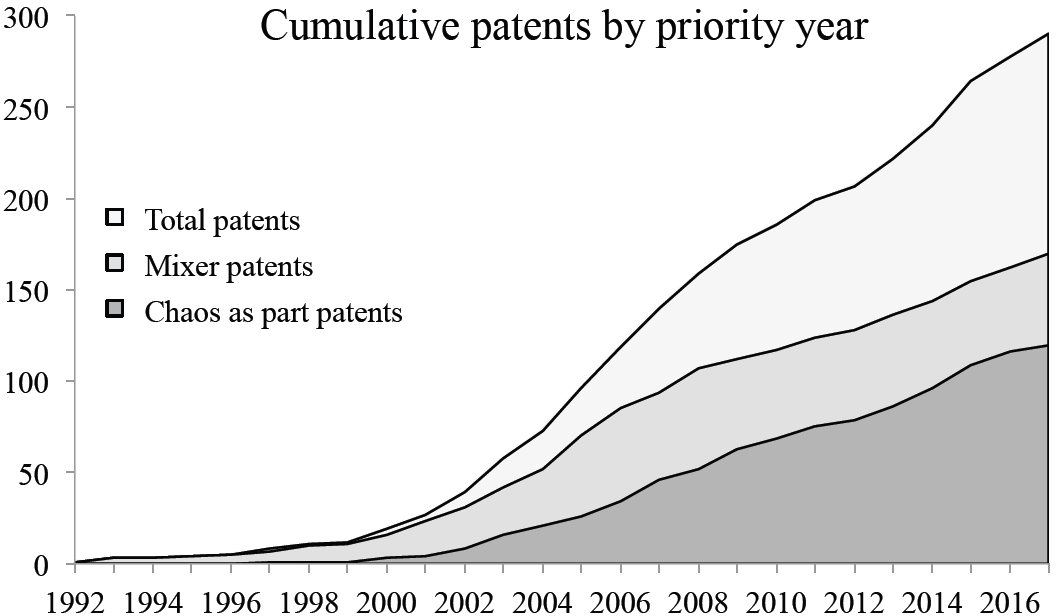}
  \caption{Cumulative numbers of published chaotic mixing patents by priority year and category .  Top curve is total patents, the middle curve is stand-alone mixers and the bottom curve is use of chaotic mixers as part of the invention in at least some embodiments}
  \label{fig:PatentType}
\end{figure}

The cumulative number of chaotic mixing patents as a function of year is shown in Fig.~\ref{fig:PatentType}, where mixer patents are clearly separated from those in which chaotic mixing is claimed only as part of the invention.    The year plotted in this figure is based on the priority date, not the filing, publication or grant date because the priority date determines which is the ``first application'' in practice.   The median time difference between priority and publication dates in this list was approximately 21 months, although time differences as high as 15 years were noted.  The early history consists of patents that are fluid mixers only.  In 1997 the first patent appears that specified the use of chaotic mixing as just part of the invention and by around 2005 the  annual number of new mixers versus the use of chaos ``in part'' is approximately equal.

Several of the early patents are worth a special mention.  Although the published patent record of chaotic advection begins in 1994 with the publication of four patents, the majority of these had earlier priority dates.

The patent with the earliest priority date (June 1992), and hence the honour of the first patented use of chaotic advection, belongs to Sen and Chang \cite{Sen_patent_1994} of the Gas Research Institute in Illinois. Entitled ``Process and Apparatus for enhancing In-Tube Heat Transfer by Chaotic Mixing'', their device exploits a twisted-pipe type of the duct flows
following \SEC{Ducts} used for heat exchange.  The patent was allowed to lapse in 2002 and no record has been found regarding its use in application.
With a priority date just 5 months later (Nov 1992), Castelain, Le Guer and Peerhossaini \cite{Castelain_patent_1994} also patented a twisted-pipe device for heat exchange. However, the record indicates the patent was disallowed and was withdrawn in 1998,  possibly because the claims in Sen and Chang~\cite{Sen_patent_1994} encompassed those in~\cite{Castelain_patent_1994}.

Mackley, Skelton and Smith \cite{Mackley_patent_1994} patented a mixing vessel with a priority date in March 1993.  It consisted of a duct with radial baffles that divides the tube into chambers and uses oscillatory flow (the first example of patenting time-dependent chaotic flow) to generate time-periodic separated flow with crossing streamlines in different phases of the period (similar to
the reorientation of a base flow in the time-periodic cylinder flow in \FIG{UnsteadyChaos1}) and mixing throughout the tube. The record suggests this patent did not make it into the PCT stage and was abandoned in 1998; however, the technology has progressed as we detail later.

Tjahjadi and Foster of the General Electric company patented modifications to extruder screws (also belonging to the duct flows following \SEC{Ducts}) that generated chaotic motion in polymer melts to improve mixing and heat/mass transfer characteristics of the extruder.  This patent had a priority date of September 1993, and appeared to have been maintained by General Electric until 2008 when it as assigned to Sabic Innovative Plastics B.V. of the Netherlands.

Also having a 1993 priority date, Logan \cite{Logan_patent_1994} filed on a device to be inserted into ducts. Logan's device has three long rods welded into an open triangular prism with sine wave baffles welded to the faces of the triangle,
which the patent claims produces stretching, folding and chaotic mixing for viscous liquids.  Most curiously, down the center of the triangle are acoustic horns claimed to provide additional mixing.  We have been unable to find evidence that Logan's device was ever tested and although payment was continued until 2006, the fate of the technology is unknown.

The only chaotic mixing patent with a 1994 priority date was that of Kwon and San~\cite{Kwon_patent_1994} of the Pohang Iron and Steel company for another extruder screw that generated chaotic motion in polymers melts.  Once again this patent was allowed to lapse in 2006, and no information has been found as to its commercialisation.

The early history of chaotic mixing patents is entirely filled by devices that are termed here ``macro-scale'' devices, operating at length scales from centimetres to metres.  This changed in July 1997 with the first micromixing patent  by Evans, Liepman and Pisano~\cite{Evans_patent_2000}.    Unlike most subsequent micromixer designs, this device is not a duct.  Instead it has a thin, planar chamber, into which fluid streams are pumped from alternating directions, making this system a container flow according to \SEC{Containers}.  The time-periodic pulsation and their reorientation accomplish the mixing within the chamber.  Fee payments on this patent were up to date in 2011.

\begin{figure}
  \centering
\includegraphics[width=0.95\columnwidth]{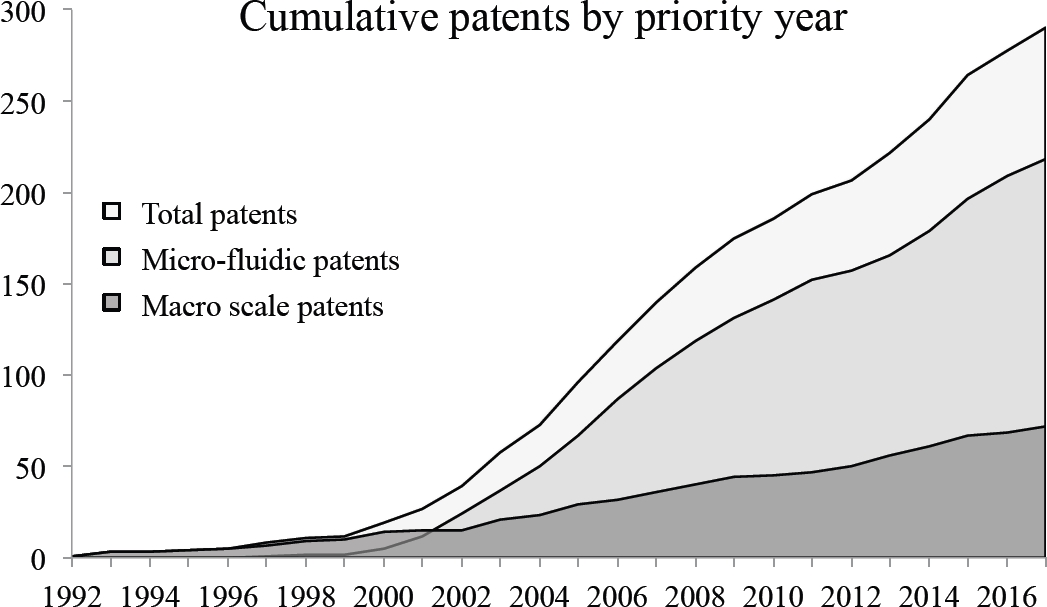}
  \caption{Cumulative numbers of published chaotic mixing patents priority year and scale.  Top curve is total patents,  the middle curve is micromixers and the bottom curve is for macro-scale mixing devices.}
  \label{fig:PatentScale}
\end{figure}

The cumulative number of chaotic mixing patents is shown in
Fig.~\ref{fig:PatentScale}, this time distinguishing between
macro-scale and micro-scale patents.  Again the year plotted in this
figure is based on priority date.  The most noticeable feature of this
plot is that since approximately 2002, the number of patents filed
annually at each scale is approximately constant, with micro-fluidic
patents being filed approximately three times more frequently as
macro-scale.  A significant part of the micro-fluidic abundance is due
to the ``chaos as part of the working principle'' patents seen in Fig.~\ref{fig:PatentType}.
With very few exceptions, these ``chaos in part'' patents involve the
use of micromixers.  Clearly the drive for miniaturisation and the
concomitant drastic reduction in Reynolds number mentioned in \SEC{Intro} has been one of the
main drivers of new applications using chaotic advection. Moreover,
the reduction of length scale has opened up the practical utilisation
of diverse surface forces drive fluid motion on the micro scale, forces
that are negligible at the macro scale.

Classification of patents according to the flow categories described in \SEC{Categorisation} is also instructive and is shown in Table~\ref{tab:Class}.  Here screw extruders (10\% of total patents) have been categorised as ``partitioned-duct flows'', the majority of wall-driven container flows also involved unsteady driving and were categorised as ``unsteady container flows'' and that different inventions for mixing inside droplets were lumped together due to the small numbers -- the majority (4\% of total patents) being background-flow driven.  At 40\% of total patents, open-duct flows have a clear majority, with approximately 50\% of open-duct flow patents involving a means to drive a spatially or temporally varying cross flow either via patterned walls (e.g. a herringbone pattern) or via direct forcing (electro-kinetic, magnetic, varying wettability, mechanical).  The other 50\% of open-duct flow patents involve the use of duct geometry including 2D and 3D twisted and serpentine channels and/or wall baffles and/or varying channel cross section to generate secondary flows and streamline chaos.  Worth noting is that very few fluid chaos patents involve flows that are spatially fully 3D, mirroring the nascent state of knowledge in such systems~\cite{Aref2017}.

\begin{table}[h]
\begin{center}
\caption{Patent classification by percentage in terms of the flow categorisation following \SEC{Categorisation}.}
\label{tab:Class}
\begin{tabular}{c | c }
 \hline
Category & Percentage  \\ [0.5ex]
 \hline
Open-duct flows (\SEC{DuctFlowIndustrial}) & 40\%	\\
Partitioned-duct flows (\SEC{DuctFlowIndustrial}) 		& 13\% 	\\
Unsteady container flows (\SEC{ContainersIndustrial})		& 8\%	\\
Branching and connecting (\SEC{DuctFlowIndustrial})		& 8\%	\\
Forced container flows (\SEC{ContainersIndustrial})		& 7\%	\\
Mixing inside droplets (\SEC{DropsIndustrial})		& 7\%	\\
Other (various)				& 17\%
 \end{tabular}
 \end{center}
 \end{table}

%\subsection{Case Studies}
%\subsection{Commercially-available devices}
\subsection{Commercialisation attempts}

The following presents a brief discussion of a few technologies that have
attempted, and in some case succeeded, in commercialisation.  Making
connections between patents, inventors and current commercial activity
has proven difficult, especially for patents that include chaotic
mixing as part of the process.  There are likely a number of other
commercially available devices that have not been uncovered for this
review\footnote{Please contact the authors with information about successful
commercial applications of chaos.}.

\subsubsection{Kenics mixer}
\label{kenics}

The Kenics mixer \cite{Kenicspatent1965}  is an an-line mixer with internal twisted, offset plates.  It has been a commercial success for Chemineer Inc. ({\it chemineer.com}) since 1965 and has spawned numerous variations on the original theme (PPM, SMX, Quatro and a number of micro-fluidic equivalents).  The inventors correctly identified the secondary flows induced by the twisted elements and the
split-and-recombine nature of the flow (\SEC{DuctFlowIndustrial}) as the source of effective mixing (``This results in an eddy current motion in each partial stream, which causes some mixing of components ... As the fluid meets the upstream edge of the second element it is forced to split again ...'').  Although the device has been demonstrated numerous time to mix via the process of Lagrangian chaos, e.g.~\cite{Hobbs1998}, the patent appeared well before the concept of Lagrangian fluid chaos had been described.  Regardless, it is one of the stand-out examples of fluid chaos in practical application.  The range of similar-style mixers to appear after the original patent lapsed points towards the difficulty in achieving successful commercialisation in the allowable 20-year window.  It also suggests that opening up a concept to other minds provides the freedom to develop alternative, and potentially better, embodiments of the idea that could contribute to longer term success.

\subsubsection{Oscillating Baffled Reactor}
\label{OBR}

The patent by Mackley, Skelton and Smith \cite{Mackley_patent_1994} for ``Processing of mixtures by means of pulsation'' resulted a processes later termed  Oscillatory Flow Mixing  ({\it malcolmmackley.com/innovation/oscillatory-flow-mixing}) which is embodied in the Oscillatory Baffled Reactor (OBR).  Although the original patent lapsed in 1998, at least two companies are commercialising the concept, Cambridge Reactor Design, (with the ``Rattlesnake'' continuous flow reactor ({\it cambridgereactordesign.com}) and NiTech Solutions Ltd. ({\it nitechsolutions.co.uk}) with chemical reactors and crystallisation vessels.  Each company has more recent separate patents covering different aspects and applications of OBRs.  NiTech Solutions has had commercial units operating in pharmaceutical manufacturing since
2007.\footnote{ Personal communication with Prof. Xiongwei Ni.})  The advantages of the technology are clear when it is considered that one 2m$^{3}$ NiTech reactor operating at 100$^+$\,l/min has replaced two 150\,m$^{3}$ batch stirred tank reactors in one application.  Other NiTech installations in a range of chemical manufacturing plants have been completed and around 50 lab-scale reactors have been installed in laboratories for pharma and chemical companies as well as in universities.  The consistent message to emerge from installations of this type of reactor technology is that it makes the correct product reproducibly, creates less waste, there are less downstream unit operations required for separation and the process is generally greener and safer as a result.

\subsubsection{Rotated Arc Mixer}
\label{RAM}

As introduced in \SEC{Ducts}, the Rotated Arc Mixer (RAM) is an
in-line mixer without internal baffles designed on the basis of
scientific insights into Lagrangian transport and the notion of
chaotic advection (refer to \blue{www.youtube.com/watch?v=j0owt8XD6xM}) The
design was determined using {\it a priori} numerical simulation and
the images shown in Fig.~\ref{PPM_RAMa} are taken from the first
experiment undertaken in the device in 1998.  The poor and good mixing
conditions and dye trace injection locations were identified {\em
  before} the experiment was run to highlight the robustness of the
design methodology.  The RAM can substantially outperform conventional
inline mixers in terms of energy consumption and mixing quality,
especially for highly viscous materials or materials in which fouling
and scaling could rapidly clog internal mixing elements.

The elegance
of the idea earned it recognition as one of the top designs in the
``AIChE 10x Design Challenge'' \cite{RAM2010} in 2010.  Two patents
have been granted, one for flow
mixing~\cite{RAM_FluidMixer_patent_2006} and one for heat
exchange~\cite{RAM_HeatExchanger_patent_2010}.  Confidential trials of
a full-scale RAM were undertaken in a plant owned and operated by a
multinational food manufacturer.  The outcome of the trial showed that
the continuous-flow RAM increased line productivity by 25\% (by fully
accomplishing the desired material transformation so that downstream
equipment could be operated at capacity) and reduced mixing energy by
95\%, replacing two conventional mixers.  Further, a sensory
evaluation panel concluded that the that the RAM produced product had
a ``step change'' improvement in product quality.  Despite the
apparent benefits of the RAM, implementation has not
progressed beyond trials.  This is a case-in-point of where a
technology has been proven superior in application yet the
conservatism inherent in large, traditional industries still results in a
reluctance to adopt this.

%mixed more rapidly, at lower shear and with
%less energy than their conventional batch process
% produced in the RAM had better "mouthfeel" than their standard product

Rights to develop the technology currently reside with Tasweld
Engineering Pty Ltd. ({\it tasweldengineering.com.au}).  Although
the patents were allowed to lapse in 2018, in the case of the RAM,
knowledge of the underlying Lagrangian transport is essential when
designing a unit for a given application.  It is this know-how -- much
of it encapsulated in this review -- that will result in success or
failure, not a knowledge of how to manufacture the physical device.
This statement will be true of most devices based on Lagrangian chaos.

\subsection{The future of commercial application}

The approximately linear increase in fluid chaos patents year-on-year suggests that the technology field is still in the ascendancy or growth phase of the innovation S-curve.  The difficulty in identifying successful commercialisations suggest it is likely in early growth.   Significant take-off will require existing mixing processes to be increasingly replaced by those based on 3D deterministic Lagrangian chaos or will require emerging technologies that are created with fluid chaos as an integral part of their design.

Although some industries have used chaotic mixing devices for decades (for example screw extruders  in polymer processing), most have not.  The experience of the authors, their colleagues and many companies they have talked to indicates there is a reluctance to utilise ideas and equipment that has not already been proven in other similar applications, even when pilot-scale testing has proven the utility and improved performance of fluid chaos in their applications.
The difficulty in bootstrapping the use of fluid chaos is perhaps particularly disappointing given that it is generally possible to accurately predict the underlying deterministic flow which in turn allows good designs to be provided {\it a priori} that do not require significant iteration.

The proliferation of mixers spawned by the concepts embodied in the Kenics mixer might also suggest that a rush of commercialisation may arise once original patents have lapsed.  This is potentially good news for processes where chaos has advantages, but less beneficial for the inventors and their financial backers.  However, it seems likely that fluid chaos will make its impact felt in emerging technologies where it can be built in from the ground up.   Indeed this appears to be the case in micro-fluidic applications, a large number of which require the use of 3D Lagrangian chaos in order to be viable. ``Watch this space'' seems to be an appropriate concluding instruction.

%\clearpage
%\input{Conclusions}

%\section{Conclusions and outlook}
\section{Concluding remarks}
\label{Conclusions}

The scope of this review is transport and mixing of scalar quantities such as additives, chemical species, heat and nutrients in realistic three-dimensional (3D) fluid flows under laminar flow conditions. This is motivated by its ubiquity in many systems and processes both in industry and Nature. Transport and mixing is considered in terms of the Lagrangian motion of fluid parcels (``advection'')
and thus admits description and investigation by the geometry, topology and coherence of fluid trajectories. This ``Lagrangian flow
topology'' and corresponding advective transport has strong similarities and analogies across a wide range of
practical flows in industry and beyond. This underlying universality of practical
flows enables categorisation of laminar transport problems into four canonical configurations: flows in (i) ducts, (ii) containers, (iii) drops and (iv) webs.

%The connections between practical flows established via the proposed categorisation
%
The fundamental connections between many different instances of practical flows established by
the proposed categorisation is the central outcome of the present review. This
%
%addresses its principal motivation, viz. bridging the gap between fundamentals and applications,
%
contributes to reaching its principal aim. viz. the stimulation of further development and
utilisation of know-how on 3D Lagrangian transport,
in three following ways.
{\it First}, by exposing the ubiquity and diversity of Lagrangian transport phenomena and creating awareness of their broad relevance.
{\it Second}, by enabling transfer of insights and knowledge between transport problems both within and across scientific disciplines. For example, the same mechanism of ``chaotic advection'' that yields efficient mixing in industrial inline mixers is also
implicated in vascular diseases such as thrombosis and atherosclerosis. {\it Third}, by reconciling practical flows with fundamental
and theoretical studies on Lagrangian transport and chaotic advection so as to bridge the still considerable gap between
practice and theory. Studies and designs namely often insufficiently use the available scientific knowledge and expertise on laminar transport phenomena. Hence much can be gained by further cross-disciplinary research and these efforts should
concentrate strongly on the fundamentals of Lagrangian transport in 3D realistic flows and its translation to design and understanding
of practical applications.

Many challenges remain within this context. Those that have been exposed and suggested by this review include:
%
%Major challenges within this context as exposed and suggested by this review include:
%
%Major challenges within this context as exposed and suggested by this review include:
%
%Theoretical and fundamental studies on Lagrangian transport should focus more strongly on 3D realistic flows and then particularly
%under unsteady conditions.
%
%Theoretical and fundamental studies on Lagrangian transport should focus more strongly on 3D realistic flows and then particularly
%under unsteady conditions.
%
\begin{itemize}

\item[$\bullet$] {\it Lagrangian transport in realistic geometries} involving no-slip walls and non-periodic coordinates. The complex
geometry of the flow domain is e.g. critical to transport in inline static (micro-)mixers or porous media.

\item[$\bullet$] {\it Formation and interaction of Lagrangian coherent structures (LCSs) in 3D unsteady flows.} Transversal manifold
interaction for critical points and lines
%
%associated with critical points and lines
in such flows remains elusive yet is key to truly 3D chaos.

\item[$\bullet$] {\it Lagrangian (chaotic) transport far away from unperturbed states.} Fundamental studies often consider the
response of non-chaotic systems to small perturbations and the resulting chaos due to e.g. RID and RIM is relatively weak and thus of limited use for practical mixing purposes. Hence {\it full} exploration of the routes that (via resonances and/or bifurcations) lead to strongly-3D (chaotic) conditions is of great practical relevance.

\item[$\bullet$] {\it Lagrangian transport in aperiodic and finite-time (transient) flows.} Many realistic
(particularly non-industrial) flows are aperiodic in space and/or time and thus beyond the existing Lagrangian machinery.
This motivates the (ongoing) development of dedicated Lagrangian methods for such systems, which basically are
generalisations of LCSs \cite{Peacock2013,Haller2015,Budisic2016,Ide2002,Beigie1994,Froyland2010,Mosovsky2013}.
Important for practical usefulness is (at least) establishing consistency between the various definitions and interpretations of LCSs or, ideally, development of one comprehensive methodology for aperiodic/finite-time flows.

\item[$\bullet$] {\it Lagrangian transport for non-mixing purposes.} The majority of Lagrangian transport
studies attempt to destroy LCSs that obstruct transport in order to accomplish chaotic advection. However, LCSs
may also be instrumentalised for non-mixing purposes such as e.g. ``unmixing'' of particle suspensions or the deliberate creation
of coexisting mixing and entrapment regions \cite{Wang_clustering_2014,Speetjens2019}.
%
% Moreover,
%LCS-based concepts as ``burning invariant manifolds'' to describe the propagation of reaction fronts may greatly deepen insights
%into complex transport phenomena in chemically-reacting flows \cite{Mitchell2012,Gowen2015}. \\\\
%
%
%RID and RIM potentially useful ... delicate interplay of ... partial confinement to (remnants of) LCSs ... and chaos ...
%
%\red{Capture into resonance?}
%
Moreover, LCS-based concepts such as ``burning invariant manifolds'' to describe the propagation of reaction fronts may deepen insights
into complex transport phenomena in chemically-reacting flows \cite{Mitchell2012,Gowen2015}.
%
%Further investigation of exploration of utilisation of LCSs for non-mixing purposes as e.g. separation, shielding or confinement may have %promising (non-industrial) practical applications.

\item[$\bullet$] {\it Lagrangian transport coupled to other physics.} Transport processes often involve significant contributions from diffusion and/or chemical reaction. Random-walk(-like) effects may come into play during advection of small particles or particle dispersion in random porous media. This enables crossing of topological barriers and thus tends to promote chaotic dynamics \cite{Metcalfe_beyond_2012,Lester2016,Dentz2018}. Such diffusive behaviour combines with advection
into a net advective-diffusive flux that sets up scalar transport via well-defined ``scalar transport paths''. This notion enables generalisation of the concept of Lagrangian flow topology and LCSs to advective-diffusive transport -- describing e.g. heat transfer by a ``thermal topology'' -- and in fact any transport problem involving a continuous
scalar flux \cite{Speetjens2012,Balasuriya2018}. This approach may pave the way to a unified Lagrangian formalism for scalar transport in fluid flows yet further development into a mature methodology is necessary for practical usefulness.

\item[$\bullet$] {\it Lagrangian transport in discontinuous media.} Flowing media generically deform continuously due to the fact that the entire body of fluid remains smoothly connected. However, the flowing medium may under certain conditions be ``cut up'' into disconnected regions that are redistributed within the flow domain
akin to the shuffling of a deck of cards. This happens, for instance, in case of extraction/injection of fluid
via valves and discontinuous (i.e. ``avalanche-like'') sliding within a granular medium. Here the front of the
injected fluid and the sliding surface define material discontinuities that may significantly impact the
Lagrangian flow topology and associated transport \cite{Smith_discontinuous_2017,Smith2017}.

%
%in that they, though particularly for chaotic advection becoming extremely convoluted, remain smooth
%and connected at all times. However, ...
%

\item[$\bullet$] {\it Experimental studies} Experimental studies must be an integral part of research efforts addressing these challenges so as to ensure physical meaningfulness and practical relevance of new insights.

\end{itemize}
A major challenge in its own right is instilling the ``Lagrangian mindset'' in scientists and practitioners in engineering and life sciences dealing with
laminar transport phenomena in one form or another. This is also clearly reflected in the reluctance in industry to implement a proven principle as chaotic advection
in actual mixing equipment, as suggested by the survey on relevant patents and commercial applications.
This challenge can be overcome by, first, creating awareness of the relevance of these phenomena and the existence of dedicated machinery to describe and analyse them and, second, developing the necessary know-how and skills
for its employment. The present review may be a first incentive to this end. Crucial in the long run is the structural integration of the ``Lagrangian mindset'' into education and training so as to expose students and practitioners alike to this approach and encourage its employment for analysis and design. Moreover, practical application necessitates its incorporation in handbooks and design strategies as well as in dedicated engineering and analysis tools such as CFD modules for Lagrangian transport studies.

\bigskip

Two of the authors (GM and MR) invented the RAM but
retain no economic rights in its commercialisation.

\bibliography{References_AMR}

\end{document}